\documentclass[preprint]{elsarticle}

\usepackage{amssymb}
\usepackage{indentfirst}
\usepackage{graphicx}
\usepackage{epstopdf}
\usepackage{epsfig}
\usepackage{overpic}
\usepackage{array}
\usepackage{color}
\usepackage{multirow}
\usepackage{hyperref}
\usepackage{float}
\usepackage{subfigure}
\usepackage{amssymb}
\usepackage{amsmath,amssymb, amsthm}
\usepackage[mathscr]{euscript}
\usepackage{latexsym,bm}
\usepackage{booktabs}
\usepackage{bbm}
\usepackage{diagbox}
 \usepackage[all]{xy}
\usepackage{diagbox}
\usepackage{algorithm}  
\usepackage{algorithmicx} 
\usepackage{algpseudocode} 
\usepackage{mathtools}

\newtheorem{remark}{Remark}
\newtheorem{example}{\textup{\textbf{Example}}}

\newcommand\D{\textup{d}}

\def\mi{\mathrm{i}}
\def\me{\mathrm{e}}

\def\br{\bm{x}}

\def\bk{\bm{k}}
\def\bp{\bm{p}}
\def\bP{\bm{P}}
\def\bri{\bm{s}}
\def\by{\bm{y}}

\def\bn{\bm{n}}
\def\bR{\bm{R}}
\def\bri{\bm{\xi}}
\def\bmm{\bm{m}}

\def\br_1i{\bm{\xi}}

\def\br{\bm{r}}
\def\bl{\bm{l}}
\def\bs{\bm{s}}
\def\pdo{{\rm \Psi} \textup{DO}}
\makeatletter
\@addtoreset{equation}{section}
\makeatother

\begin{document}

\newsavebox{\tablebox}

\begin{frontmatter}



\title{Beyond quantum mean-field approximation: Phase-space formulation of many-body time-dependent density functional theory and efficient spectral approximations}

\author[label1]{Jiong-Hang Liang}
\ead{liangjionghang@sjtu.edu.cn}
%\author[label1]{D. Wu}

\author[label]{Yunfeng Xiong\corref{cor}}
\ead{yfxiong@bnu.edu.cn}

\affiliation[label1]{organization={Key Laboratory for Laser Plasmas and School of Physics and Astronomy, and Collaborative Innovation Center of IFSA, Shanghai Jiao Tong University},
            postcode={200240},
            state={Shanghai},
            country={China}}
\affiliation[label]{organization={School of Mathematical Sciences, and Laboratory of Mathematics and Complex Systems, Ministry of Education, Beijing Normal University, Beijing},
            postcode={100875},
            state={Beijing},
            country={China}}
%Department and Organization
 %           addressline={}, 
   %         city={},
%            postcode={100871}, 
 %           state={Beijing},
 %           country={China}
\cortext[cor]{To whom correspondence should be addressed}

\begin{abstract}
As a universal quantum mechanical approach to the dynamical many-body problem, the time-dependent density functional theory (TDDFT) might be inadequate to describe crucial observables that rely on two-body evolution behavior, like the double-excitation probability and two-body dynamic correlation. One promising remedy is to utilize the time-dependent 2-reduced density matrix (2-RDM) that directly represents two-body observables in an $N$-particle system, and resort to the extended TDDFT for multibody densities to break the confines of spatial local on one-body density  [Phys. Rev. Lett. 26(6) (2024) 263001]. However, the usage of 2-RDM is prohibitive due to the augmented dimensionality, e.g., 4-D space for unidimensional 2-RDM.  This work  addresses the high-dimensional numerical challenges by using an equivalent Wigner phase-space formulation of 2-RDM and seeking efficient spectral approximations to nonlocal quantum potentials. For spatial periodic case, a pseudo-difference operator approach is derived for both the Hartree-exchange-correlation term and two-body collision operator, while the discretization via the Chebyshev spectral element method is provided for non-periodic case. A massively parallel numerical scheme, which integrates these spectral approximations with a distributed characteristics method, allows us to make the first attempt for real simulations of 2-RDM dynamics. Numerical experiments demonstrate  the  two-body correction to the quantum kinetic theory, and show the increase in the system's entropy induced by the two-body interaction. Thus it may pave the way for an accurate description of 2-RDM dynamics and advance to  a practical application of many-body TDDFT.
\end{abstract}


%Research highlights

%This in turn provides a reasonable reference solutions for our numerical experiments.


\begin{keyword}
Wigner equation \sep quantum BBGKY hierarchy \sep quantum many-body system \sep density functional theory \sep spectral methods



% MSC codes here, in the form: \MSC code \sep code
\MSC[2020]  
81S30 \sep %Phase-space methods including Wigner distributions, etc. applied to problems in quantum mechanics
81V70 \sep %Many-body theory; quantum Hall eﬀect
82M36 \sep %Computational density functional analysis in statistical mechanics
82M22 \sep %Spectral, collocation and related (meshless) methods applied to problems in statistical mechanics
65M22  %Numerical solution of discretized equations for initial value and initial-boundary value problems involving PDEs

% or \MSC[2008] code \sep code (2000 is the default)

\end{keyword}

\end{frontmatter}

\section{Introduction}

The description of matter that is driven out of equilibrium by arbitrary time-dependent perturbation has become an important research topic in nuclear \cite{Ullrich2014,WenBartonRiosStevensen2018,MoldabekovShaoBellenbaum2025_arXiv}, plasma \cite{Haas2019,ZamanianMarklundBrodin2013,HuLiangShengWu2022}, condensed-matter physics \cite{ShapirHamoPeckerMocaLegezaZarandIlani2019,VuSarma2020,KnightQuineyMartin2022} and semiconductor devices \cite{bk:NedjalkovQuerliozDollfusKosina2011,BenamBallicchiaWeinbubSelberherrNedjalkov2021,BallicchiaEtlNedjalkovFerryKosinaWeinbub2025_arXiv}.
As a universal quantum mechanical approach to the dynamical many-body problem, the time-dependent density functional theory (TDDFT) becomes a work horse for these situations for its capability to deal with the system that is initially in a stationary state and is then acted upon by a perturbation, or that is in a non-eigenstate and propagates freely in time. Despite its theoretical advantage, the accuracy of TDDFT  is generally limited to the inexact form of exchange–correlation (XC) functional  \cite{Ullrich2014} and the mean-field approximation \cite{KnightQuineyMartin2022}, making it inadequate to describe crucial properties and observables that rely on the two-body evolution behavior, including correlation entropy, double-excitation probability \cite{AppelGross2010}, dissipative collisions of two nuclei \cite{WenBartonRiosStevensen2018} and quantum entanglement triggered by electric interaction \cite{BallicchiaEtlNedjalkovFerryKosinaWeinbub2025_arXiv}. 
%As a consequence, it requires to break the confinement of spatial local on single-body density.

One promising candidate to treat these long-standing problems in TDDFT is to replace the single-body electron density by 2-reduced density matrix (2-RDM) \cite{AppelGross2010}, which is derived by integrating out the degrees of freedom of all but two of the fermions in $N$-particle density matrix \cite{KnightQuineyMartin2022}.  A key advantage in employing 2-RDMs lies in a fundamental result that the expectation value of a two-body operator can be expressed as a linear functional of the 2-RDM \cite{Husimi1940}, so that two-body correlation is explicitly incorporated. The evolution of 2-RDM can be obtained by reformulating the many-body von-Neumann equation in the BBGKY hierarchy, say,  the density matrix of order $N+1$ terms of matrices with order less than or equal to $N$. The system can be closed via neglecting three-body and higher-order correlations \cite{OlmstedCurtiss1975,SchroedterBonitz2024}. As the many-body density evolution within the binary interacting system can be constructed by a unique nonlocal potential for any order of the
BBGKY equation \cite{LiangHuWuShengZhang2024}, it constitutes a framework of many-body TDDFT, with the higher-order truncation within the hierarchy is expected to describe more accurately the time evolution of the strongly correlated systems \cite{WenBartonRiosStevensen2018,WangCassing1985}.

The computational aspects  of time-dependent 2-RDMs, however, still confront two fundamental challenges. First, the numerical cost for solving the many-body TDDFT is demanding: The degrees of freedom are quadrupled, e.g., 12-dimensional coordinates for two electrons \cite{WenBartonRiosStevensen2018}. It should be noted that the equations of motion for the 2-RDMs are theoretically equivalent to those for the effective two-particle orbitals \cite{LiangHuWuShengZhang2024}. The key distinction lies in the fact that the latter projects the problem into a 6-dimensional space spanned by two-particle orbitals, and its computational cost scales with the number of such orbitals. This leads to different scenarios for their application: the former serves as a complete theoretical starting point but is computationally complex, whereas the latter is more computationally feasible, particularly suited for describing low-temperature, few-body quantum systems. Second, the exact XC functionals that can accurately contain the dynamical exchange effects induced by the Pauli principle and  the electron-electron correlation effects are still unknown  \cite{WangCassing1985,ZamanianMarklundBrodin2013,Haas2019}. This issue is tied to the famous N-representability problem, delineating the thresholds at which the time-dependent 2-RDM can depict observables in an $N$-particle system \cite{BlanchardGraciaBondiaVarilly2012,LiangHuWuShengZhang2024}. To complete a thorough understanding of the dynamical correction to the Kohn-Sham theory, it requires to further explore the virial theorem for the electron system \cite{HolasMarch1995,LiangHuWuShengZhang2024} and incorporate momentum distributions and history-dependence in the XC functionals \cite{RajamHesslerGaunMaitra2009}.

This work makes the first and pivotal step towards the practical usage of 2-RDMs by addressing the high-dimensional numerical challenges. Instead of 
solving the von-Neumann equation, we resort to the equivalent phase-space formulation of density matrix, known as the Wigner function, to depict  the quasi-distribution in position and momentum simultaneously \cite{Wigner1932,OlmstedCurtiss1975,bk:NedjalkovQuerliozDollfusKosina2011}. The Wigner function is aesthetically more appealing due to its symmetry, being real as a consequence of the
hermiticity of the RDM \cite{SchmiderDahl1996}, and expresses quantum observables as statistical averages. Moreover, the dynamics of the Wigner function is analogous to the classical Boltzmann equation, whereas the collision operator is replaced by the nonlocal and highly-oscillating pseudo-differential operator ($\pdo$). Therefore, solving the Wigner dynamics shares the same kind of numerical difficulties as in the von-Neumann equation and urgently demands efficient numerical techniques capable of handling a large variety of wavelengths \cite{FilbetGolse2025,GaniuLoesingJaegerSchulz2025,FilbetGolse2025_arXiv}.  Fortunately, various of spectral methods have been proposed to attain highly accurate approximations to $\pdo$. Nowadays, the well-crafted Fourier spectral methods are capable to resolve the Wigner-Poisson system \cite{Ringhofer1990,SuhFeixBertrand1991,ArnoldLangeZweifel2000}, the unbounded potentials \cite{ChenShaoCai2019}, the spatial variation of effective mass \cite{SchulzSchulz2020} and $\pdo$ with singular symbols, including the Coulomb scattering \cite{XiongZhangShao2023} and Dirac-delta interaction \cite{ShaoSu2023}, while the discontinuous potential in the resonant tunneling diode can be accurately solved by the sinc-Galerkin spectral method \cite{jiangLuYaoZhang2023}. The Chebyshev spectral methods are proposed to efficiently resolve the two-body scattering \cite{ShaoLuCai2011,XiongChenShao2016}.  The Hermite spectral methods \cite{FilbetGolse2025,FilbetGolse2025_arXiv,CaiFanLiLuWang2012} and the spectral decomposition of force field  \cite{VandePutSoreeMagnus2017} are particularly suitable for  $\pdo$ in the semi-classical regime ($\hbar \ll 1$).

In view of more complicated $\pdo$ in many-body TDDFT, involving the combination of nonlinear Hartree potential, XC functionals, two-body scattering term and the external potential, this work provides a unified framework of spectral approximations, and the key lies in an appropriate mesh strategy. When the Wigner function is periodic or has a compact support in the spatial space, a pseudo-difference approach to  $\pdo$ is derived by directly expanding the nonlinear potential as the Fourier series and utilizing the completeness relation. When the spatial density is non-periodic,  it is readily to extend the the Chebyshev spectral approximations in \cite{ShaoLuCai2011,XiongChenShao2016} to $\pdo$ with nonlinear symbols as they can be exempt from the constraints on the spatial and momental mesh sizes. By further integrating with the distributed characteristics method \cite{MalevskyThomas1997,XiongZhangShao2023}, it builds up a massively parallel characteristic-spectral-mixed (CHASM) scheme to overcome  the numerical burdens of one-dimensional 2-RDMs (and also 3-RDMs). This allows to include double-electron repulsive interaction in quantum kinetic simulations to  validate the importance of two-body correction, as well as paves the way for evaluating the effects of dynamical XC functionals and two-body interaction in one-dimensional strongly correlated system \cite{BottiSchindlmayrDelSoleReining2007,RajamHesslerGaunMaitra2009,BlanchardGraciaBondiaVarilly2012,VuSarma2020,KnightQuineyMartin2022}. Moreover, the reliable deterministic scheme produces prefect benchmark solutions for the Wigner Monte Carlo methods \cite{KosinaNedjalkovSelberherr2003,MuscatoWagner2016,ShaoXiong2019}. The latter may potentially break the curse of dimensionality and resolve realistic models in 12-dimensional phase space \cite{XiongShao2024}.

%A system of confined charged electrons interacting via the long-range Coulomb force can form a Wigner
%crystal due to their mutual repulsion \cite{VuSarma2020}. 

%The quantum nature of the crystal emerges
%in the observed collective tunneling through a potential barrier \cite{ShapirHamoPeckerMocaLegezaZarandIlani2019}.

The rest of this paper is organized as follows. Section \ref{sec.basic} presents the framework of the many-body TDDFT from the density matrix representation to the Wigner approach. It follows by the construction of efficient spectral methods in Section \ref{sec.spectral_pdo}, including the pseudo-difference approach for the spatial periodic case and the Chebyshev spectral element methods for non-periodic case. Section \ref{sec.chasm} integrates the spectral approximations to $\pdo$ with a distributed characteristics method, yielding the framework of CHASM scheme. Typical numerical experiments are provided in Section \ref{sec.num}, with conclusion drawn in Section \ref{sec.conclusion}.

%he wave vector is briefly referred
%to as (normalized) momentum in this contribution \cite{SchulzSchulz2020},

%Wigner functions
%will be negative if the contributions of one-particle
%functions with odd parity around a given point in
%phase space outweigh the ones with even parity.
%For an atom or molecule with an inversion center,
%this means that, if the sum of the occupation num-
%bers of ”ungerade” natural orbitals is greater than
%the sum of the ones with “gerade” symmetry, a
%region of negativity will appear close to the inver-
%sion center in position, and close to the origin in
%momentum space. This is a direct consequence of
%the property of the Wigner function as the expecta-tion value of a parity operator in phase space
%\cite{SchmiderDahl1996}

\section{Many-body TDDFT under the density matrix representation}
\label{sec.basic}

We start from an $N$-body quantum system under the density matrix representation
\begin{equation}
\mi \hbar \partial_t \hat \rho_{1, \dots, N} = \left[\hat{H}_{1, \dots, N}, \hat \rho_{1, \dots, N}\right],
\end{equation}
where $\hbar$ is the reduced Planck constant,  $\hat{\rho}_{1, \dots, N}$ is $N$-body density matrix operator and $[A, B] = AB - BA$ is the commutator.  The $N$-body Hamiltonian operator reads that
\begin{equation}\label{N_Hamiltonian}
\hat{H}_{1,\dots N}(\hat{\bp}_1, \dots, \hat{\bp}_N, \hat{\br}_1, \dots, \hat{\br}_N) = \sum_{i=1}^N \frac{\hat{\bp}_i^2}{2m}  + \frac{1}{2} \sum_{i=1, i \ne j}^N \sum_{j =1}^N  {V}_{i j}(\hat{\br}_i, \hat{\br}_j) + {W}_{1, \dots, N}(\hat{\br}_1, \dots, \hat{\br}_N),
\end{equation}
${V}_{ij} = {V}(|\br_i - \br_j|)$ are the binary interactions and ${W}_{1, \dots, N}$ are the external potential. 
%For instance, for a many-body Hydrogen system under the Born-Oppenheimer approximation,
%\begin{equation}
%\hat{H}_{1, \dots, N} = - \frac{\hbar^2}{2m_e} \sum_{i}^N \nabla_i^2 + \sum_{i, l}^N \frac{e^2}{ |\hat{\br_i} - {\bR}_l^A|} + \frac{1}{2} \sum_{i, j \ne i}^N \frac{e^2}{ |\hat{\br}_i - \hat{\br}_j|}. 
%\end{equation}

Assume that particles are indistinguishable and define the $k$-RDM: $\hat \rho_{1, \dots, k} = \textup{Tr}_{k+1} \hat \rho_{1, \dots, N}$, where $\textup{Tr}_{k+1}$ denotes taking trace of the degrees of freedom of $k+1$ to $N$. Then we have the quantum BBGKY hierarchy  
\begin{equation}
\mi \hbar\partial_t \hat \rho_{1, \dots, k} = \left[\hat{H}_{1, \dots, k}, \hat{\rho}_{1, \dots, k}\right] + \frac{1}{k+1}\textup{Tr}_{k+1}\left[ \sum_{i=1}^{k+1} \hat{V}_{i, k+1}, \hat{\rho}_{1, \dots, k+1}\right] + \left[ \hat{W}_{1, \dots, k},  \hat{\rho}_{1, \dots, k} \right], 
\end{equation}
in which the $k$-th RDM $\hat{\rho}_{1, \dots, k}$ depends on the $(k+1)$-th  RDM $\hat{\rho}_{1, \dots, k+1}$. 

\subsection{Truncation of BBGKY hierarchy and the k-RDM dynamics }

In order to incorporate the many-body quantum effect into TDDFT \cite{BottiSchindlmayrDelSoleReining2007}, it is suggested to truncate the above BBGKY hierarchy by seeking effective $k$-body Hamiltonian depending on $k$-body density \cite{LiangHuWuShengZhang2024}:  
\begin{equation}\label{RDM_truncation}
\mi \hbar\partial_t \hat{\rho}_{1, \dots, k} = \left[ \hat{H}_{1, \dots, k}^{\textup{eff}}([n_{1, \dots, k}], t), \hat{\rho}_{1, \dots, k}(t) \right],
\end{equation}
where $n_{1,\dots, k}$ is the $k$-reduced electron density which can be obtained by
\begin{equation}
n_{1, \dots, k}(t) = \langle \bR_k | \hat{\rho}_{1, \dots, k} | \bR_k \rangle, \quad \int {n}_{1, \dots, k}(\bR_k, t) \D \bR_k  = 1
\end{equation}
and $\bR_k = (\br_1, \dots, \br_k)$ is short for the multidimensional spatial coordinates.

The general form of the effective $k$-body Hamiltonian is
\begin{equation}
 \hat{H}_{1, \dots, k}^{\textup{eff}}[n_{1, \dots, k}](t) = \hat{T}_{1, \dots, k}  + \hat{U}^{rc}_{1, \dots, k}[n_{1, \dots, k}](t) + \hat{W}_{1, \dots, k},
\end{equation}
where $ \hat{T}$, $ \hat{U}^{rc}$ and $ \hat{W}$ denotes the kinetic part, the reduced potential and the external potential, respectively. In particular, the effective reduced potential can be further represented as
\begin{equation*}
\begin{split}
&\hat{U}^{rc}_{1, \dots, k}[n_{1, \dots, k}](\bR_k, t) = \underbrace{\int \D \bR_k^{\prime} ~ \hat{K}_{1, \dots, k}^{rc}(\bR_k, \bR_k^{\prime}) n_{1 \dots, k}(\bR_k^{\prime}, t)}_{\textup{Hartree part and exchange-correlation part}} + \underbrace{\frac{1}{2} \sum_{i = 1, i \ne j}^k \sum_{j = 1}^k \hat{V}_{ij}(\br_i, \br_j)}_{\textup{two-body interaction}} \\
& = \frac{N-k}{N-1}\sum_{i=1}^k  \sum_{j=1}^k  \int \D \bR_k^{\prime} ~\hat{K}_{HXC}(\br_i, \br_j^{\prime}) {n}_{1, \dots, k}(\br_1^{\prime}, \dots, \br^{\prime}_{j},\dots, \br_k^{\prime}, t) + \frac{1}{2} \sum_{i = 1, i \ne j}^k \sum_{j = 1}^k\hat{V}_{ij}(\br_i, \br_j),
\end{split}
\end{equation*}
with $\bR_k^{\prime} = (\br_1^{\prime}, \dots, \br_k^{\prime})$. The first term corresponds to the mean-field component, and the second term is the self-interaction.  A general form of the Hartree-exchange-correlation kernel $ \hat{K}_{HXC}$ reads 
\begin{equation}
\hat{K}_{HXC}(\br_i, \br_j^{\prime}) =  \hat{V}_{H}(\br_j^{\prime})  + \hat{K}_{XC}(\br_i, \br_{j}^{\prime}),
\end{equation}
where $ \hat{V}_{H}(\br_j^{\prime}) $ is the Coulomb potential and $\hat{K}_{XC}(\br_i, \br_{j}^{\prime})$ is the exchange-correlation kernel in DFT. 
%In Section \ref{sec.mean_field}, we will give detailed examples of the Hartree potential $\hat{V}_{H}$.
%with the reconstructed potential $\hat{U}^{rc}_{1, \dots, k}([n_{1, \dots, k}], t)$
%\begin{equation}
%\hat{U}^{rc}_{1, \dots, k}([n_{1, \dots, k}], t) = \frac{1}{2} \sum_{i, j}^k \hat{V}_{i, j} + \hat{W}^{\prime}_{1, \dots, k}([n_{1, \dots, k}], t)
%\end{equation}

%As a particular case, for single-body  truncation
%\begin{equation}
%\mi \hbar \partial_t \hat{\rho}_{1} = \left[ \hat{H}_{1}^{\textup{eff}}[n_{1}](t), \hat{\rho}_{1}(t) \right],
%\end{equation}
%the reduced potential reads that
%\begin{equation}\label{reduced_potential_1}
%\hat{U}^{rc}_{1}[n_{1}](t) = \int_{\mathbb{R}^d} \D \br_2 \hat{V}_{12}(|\br_1 - \br_2|) n_{1}(\br_2, t),
%\end{equation}
%which gives rise to the Hartree approximation.

%Define the normalized $k$-reduced density with $n_0$ the equilibrium number density,
%\begin{equation}
%\tilde{n}_{1, \dots, k}= \frac{n_{1, \dots, k}}{n_0^k}, \quad \int \tilde{n}_{1, \dots, k}(\bR_k, t) \D \bR_k  = 1.
%\end{equation}

In principle, the reconstructed potentials $ \hat{U}^{rc}_{1, \dots, k}([n_{1, \dots, k}], t)$ can be determined in  a self-consistent way from the BBGKY hierarchy 
owing to the generalized van Leeuwen theorem developed in \cite{LiangHuWuShengZhang2024}, albeit with an intractable form \cite{WangCassing1985}. 

%Thus
%\begin{equation}
%\begin{split}
% \hat{K}_{12}^{rc}(\br_1, \br_2, \br_3, \br_4) = &\frac{1}{2} \left(\hat{V}_{13} + \hat{V}_{14} + \hat{V}_{23} + \hat{V}_{24}\right)\\
% & + \frac{\mu}{2k_B T}\left( \hat{G}_{13} + \hat{G}_{14} +  \hat{G}_{23} + \hat{G}_{24}\right)\left( \hat{V}_{13} + \hat{V}_{14} +  \hat{V}_{23} + \hat{V}_{24}\right) \\
% & + \textup{higher-order terms}.
%\end{split}
%\end{equation}

\subsection{The Wigner phase-space formulation of the k-RDM dynamics}

To solve the $k$-RDM, it is more convenient to use the corresponding Wigner function as all densities are presented as its statistical averages. By taking the Weyl-Wigner transform of Eq.~\eqref{RDM_truncation}, 
\begin{equation}
f_{1,\dots, N}(\bR, \bP, t) = \frac{1}{(2\pi \hbar)^{Nd}} \int_{\mathbb{R}^{Nd}} \D \bri~e^{\frac{\mi}{\hbar} \bri \cdot \bP} \Big \langle \bR - \frac{\bri}{2} \Big | \hat{\rho}_{1, \dots, N}(t) \Big |  \bR + \frac{\bri}{2} \Big  \rangle,
\end{equation}
 then the $(2Nd+1)$-dimensional Wigner dynamics for the $N$-body Hamiltonian \eqref{N_Hamiltonian} reads that
\begin{equation}\label{N_body_Wigner}
\frac{\partial }{\partial t}f_{1, \dots, N}+  \frac{\bP}{m} \cdot \nabla_{\bR} f_{1, \dots, N}=\frac{1}{2} \sum_{i=1, i \ne j}^N \sum_{j=1}^N\Theta_{V_{i, j}}[f_{1, \dots, N}] +  \Theta_{W_{1, \dots, N}}[f_{1, \dots, N}],
\end{equation}
where $\pdo$ for the interacting potentials can be represented by the linear combination of two-body scattering operators. 
\begin{equation}
\begin{split}
\Theta_{V_{i, j}}[f_{1, \dots, N}](\bR, \bP, t) = \frac{1}{\mi\hbar}  &\left(\frac{1}{2\pi \hbar}\right)^d \int_{\mathbb{R}^d} \D \bp^{\prime}  \me^{-\frac{\mi}{\hbar} \bp^{\prime} \cdot (\br_i - \br_j)} \mathcal{F}_{\br \to \frac{1}{\hbar}\bp^{\prime}}{V}_{i, j}(|\br|) \\
\times \Big \{ &f_{1, \dots, N}(\br_1, \dots \br_N, \bp_1, \dots, \bp_i - \frac{\bp^{\prime}}{2}, \dots, \bp_j +  \frac{\bp^{\prime}}{2}, \dots, \bp_N, t) \\
&  - f_{1, \dots, N}(\br_1, \dots \br_N, \bp_1, \dots, \bp_i + \frac{\bp^{\prime}}{2}, \dots, \bp_j - \frac{\bp^{\prime}}{2}, \dots, \bp_N, t) \Big \},
\end{split}
\end{equation}
and
$\pdo$ for the external field reads that
\begin{equation}
 \begin{split}
 \Theta_{W_{1, \dots, N}}[f_{1, \dots, N}](\bR, \bP, t)  = &\frac{1}{\mi\hbar} \left(\frac{1}{2\pi\hbar}\right)^{Nd} \int_{\mathbb{R}^{Nd}} \D \bP^{\prime} \int_{\mathbb{R}^{Nd}} \D \bri  ~\me^{-\frac{\mi}{\hbar} (\bP - \bP^{\prime}) \cdot \bri} \\
 &\times \left[ W_{1, \dots, N}(\bR - \frac{\bri}{2}, t) - W_{1, \dots, N}(\bR + \frac{\bri}{2}, t) \right] f_{1, \dots, N}(\bR, \bP^{\prime}, t).
\end{split}
 \end{equation}
In spite of the mathematical elegance, it is essentially difficult to solve the many-body Wigner equation \eqref{N_body_Wigner} due to the curse of dimensionality.  

A more feasible way is to adopt the truncation of quantum BBGKY hierarchy from Eq.~\eqref{RDM_truncation}. For indistinguishable particles (either boson or fermion), the Wigner function is invariant under any $N$-permutation $\sigma$, namely,
\begin{equation}\label{def.indistinguishable}
f(\br_1, \br_2, \dots, \br_N, \bp_1, \bp_2, \dots, \bp_N, t) = f(\br_{\sigma(1)}, \br_{\sigma(2)}, \dots, \br_{\sigma(N)}, \bp_{\sigma(1)}, \bp_{\sigma(2)}, \dots, \bp_{\sigma(N)}, t).
\end{equation}
Now it is convenient to define the $k$-reduced Wigner function \cite{OlmstedCurtiss1975}
\begin{equation}
f_{1, \dots, k}(\bR, \bP, t) = \frac{1}{(N-k)!}\int_{\mathbb{R}^{(N-k)d}} \D \br_{k+1} \dots \D \br_{N}  \int_{\mathbb{R}^{(N-k)d}} \D \bp_{k+1} \dots \D \bp_{N} ~f_{1,\dots, N}(\bR, \bP, t),
\end{equation}
then obtain the $k$-reduced Wigner dynamics
\begin{equation}\label{k_body}
\begin{split}
\frac{\partial }{\partial t}&f_{1, \dots, k}(\bR, \bP, t) +  \frac{\bP}{m} \cdot \nabla_{\bR} f_{1, \dots, k}(\bR, \bP, t)= \frac{\mi}{\hbar} \left(\frac{1}{2\pi\hbar}\right)^{kd} \int_{\mathbb{R}^{kd}} \D \bP^{\prime} \int_{\mathbb{R}^{kd}} \D \bri  ~  \me^{- \frac{\mi}{\hbar}~ \bri \cdot (\bP - \bP^{\prime})}   \\
& \times f_{1,\dots, k}(\bR, \bP^{\prime}, t) \underbrace{\Big \{\int_{\Omega}\D \bR^{\prime}~\left[K_{1,\dots, k}^{rc}(\bR - \frac{\bri}{2}, \bR^{\prime}) - K_{1,\dots, k}^{rc}(\bR + \frac{\bri}{2}, \bR^{\prime}) \right] n_{1,\dots, k}(\bR^{\prime}, t)}_{\textup{Hartree part and exchange-correlation part}} \\
 &  + \underbrace{\left[{W}_{1,\dots, k}(\bR - \frac{\bri}{2}) - {W}_{1,\dots, k}(\bR + \frac{\bri}{2})\right]  \Big \}}_{\textup{external potential}}  + \underbrace{\frac{1}{2}\sum_{i=1, i\ne j}^k \sum_{j=1}^k \Theta_{V_{i, j}}[f_{1, \dots, k}](\bR, \bP, t)}_{\textup{two-body interaction}}, 
\end{split}
\end{equation}
where $K_{1,\dots, k}^{rc}(\bR, \bR^{\prime})$ is the Weyl symbol of $\hat{K}_{1,\dots, k}^{rc}(\bR, \bR^{\prime})$. For brevity, we still use multidimensional notation $\bP = (\bp_1, \bp_2, \dots, \bp_k), \bR = (\br_1, \br_2, \dots, \br_k)$, $\Omega = \mathbb{R}^{kd}$, and
\begin{equation}
n_{1, \dots, k}(\bR, t) = \int_{\mathbb{R}^{kd}} f_{1, \dots, k}(\bR, \bP, t)  \D \bP.
\end{equation}
In this way, the $N$-particle distribution function is approximated in the precollision region by a functional of lower order distribution functions \cite{OlmstedCurtiss1975}. Like the density matrix representation, the $k$-reduced Wigner dynamics \eqref{k_body} should have a correspondence to the $N$-body Wigner dynamics \eqref{N_body_Wigner}, although it is difficult to obtain the exact formula of the reduced potential $K_{1,\dots, k}^{rc}$, i.e., the dynamical exchange-correlation term. This motivates the adoption of phenomenological effective potentials for the exchange and also for the correlation effects at the current stage \cite{Haas2019}.

The simplest choice is the adiabatic local density approximation (ALDA), which ignores all memory effects and depends only on the local single-body  particle density $n_i(\br_i)$, $i=1, \dots, N$. 
\begin{equation}
n_i(\br_i, t) = \int_{\mathbb{R}^{(N-1)d}}\left[\int_{\mathbb{R}^{Nd}}  f_{1, \dots, k}(\bR, \bP, t) \D \bP\right] \D \br_1 \dots \D \br_{i-1} \D \br_{i+1} \dots \D \br_k. 
\end{equation}
For instance, a popular parameterization of XC term is the Hedin–Lundqvist (H-L) potential
\begin{equation}\label{mean_field_approximation}
\begin{split}
 V_{XC}[n_i](\br_i, t) & =  \sum_{j=1}^N\int \D \bR_k^{\prime} ~{K}_{XC}(\br_i, \br_{j}^{\prime}) n_{1^{\prime}, \dots, k^{\prime}}(\br_1^{\prime}, \dots \br_j^{\prime}, \dots, \br_k^{\prime}, t)\\
 &\approx \underbrace{g_D \left(\frac{n_i(\br_i, t)}{n_0}\right)^{1/3}}_{\textup{exchange}} + \underbrace{g_D\frac{0.034}{a_B n_0^{1/3}} \ln (1 + 18.37 a_B n_i(\br_i, t)^{1/3})}_{\textup{correlation}}.
\end{split}
\end{equation}
Here $n_0$ is the equilibrium number density, $\omega_e = [n_0 e^2/(m\varepsilon_0)]^{1/2}$ is the electron plasma frequency, $\varepsilon_0$ is the vacuum permittivity, $v_F = (\hbar/m)(3\pi^2 n_0)^{1/3}$ is the Fermi velocity, $a_B = 4\pi \varepsilon_0 \hbar^2/(me^2)$ is the Bohr radius, and 
\begin{equation}
g_D = 0.985 \frac{(3\pi^2)^{2/3}}{4\pi} \frac{\hbar^2 \omega_e^2}{m v_F^2}.
\end{equation}

%So far we have excluded the reduced external potential $\hat{W}_{1, \dots, k}$. Despite the exact form cannot be trivially derived, an effective potential for the ion-electron interacting can be introduced by $n_{\textup{ion}}(\bR)$ as the ion background,  so that 
%\begin{equation}
%n_{1, \dots, k}(\bR, t) = \int_{\mathbb{R}^{kd}} f_{1, \dots, k}(\bR, \bP, t)  \D \bP - n_{\textup{ion}}(\bR), \quad \bR \in \Omega.
%\end{equation}

%Of the greatest importance is to incorporate two-body quantum effect to correct the Hartree approximation. It requires to study two-body truncation

%where the general form of $\hat{U}^{rc}_{12}$ is
%\begin{equation}
%\hat{U}^{rc}_{12}[n_{12}](t) =  \int_{\mathbb{R}^d} \D \br_3 \int_{\mathbb{R}^d} \D \br_4  \hat{K}_{12}^{rc}(\br_1, \br_2, \br_3, \br_4) n_{12}(\br_3, \br_4, t)
%\end{equation}
%
%
%In particular, 
%\begin{equation}
%\hat{V}_{HXC}(\br_i, \br_j^{\prime}) =  -\frac{Z}{|\br_i - \br_j^{\prime}|} + \frac{1}{|\br_i - \br_j|} + V_{XC}[n](\br) 
%\end{equation}

\subsection{The mean-field Wigner dynamics}
\label{sec.mean_field}

When $k=1$, Eq.~\eqref{k_body} reduces to the single-body Wigner dynamics (see Eq.~(11) in \cite{Haas2019})
\begin{equation}\label{sWigner}
\frac{\partial}{\partial t} f_1 + \frac{\bp_1}{m} \cdot \nabla _{\br_1} f_1 =  \frac{\mi}{\hbar} \left(\frac{1}{2\pi\hbar}\right)^{d}\int_{\mathbb{R}^d} \D \bp^{\prime} \int_{\mathbb{R}^d} \D \bri  \me^{-\frac{\mi}{\hbar} (\bp_1 - \bp^{\prime}) \cdot \bri} \left[ V(\br_1 - \frac{\bri}{2}, t) - V(\br_1 + \frac{\bri}{2}, t) \right] f_1(\br_1, \bp^{\prime}, t),
\end{equation}
with both Hartree part $V_H[n_1]$ and XC potential $V_{XC}[n_1] $:
\begin{equation}
V(\br_1, t) =  V_{HXC}[n_1](\br_1, t) =  V_H[n_1](\br_1, t)  +V_{XC}[n_1](\br_1, t) - V_{XC}[n_0],
\end{equation}
and $n_1(\br_1, t) = \int_{\mathbb{R}^d} f_1(\br_1, \bp_1, t)  \D \bp_1$, $V_{XC}[n_0]$ is the normalized constant.  The H-L potential \eqref{mean_field_approximation} is adopted for the XC term. 

%A formulation is given in terms of the energy 
%\begin{equation}
%E[n] = \int_{\mathbb{R}^d} n(\br) \epsilon_{\alpha}[n(\br)] \D \br.
%\end{equation}
%This leads to the famous Fermi-Thomas model
%\begin{equation}
%\begin{split}
%E_{TF}[n] = & C_1 \int_{\mathbb{R}^d} n(\br)^{5/3} \D \br + C_2 \int_{\mathbb{R}^d} n(\br)^{4/3} \D \br +\int_{\mathbb{R}^d} n(\br) V_{ext}(\br) \D \br + \frac{1}{2} \int_{\mathbb{R}^d} \int_{\mathbb{R}^d} \frac{n(\br) n(\br^{\prime})}{|\br - \br^{\prime}|} \D \br \D \br^{\prime}.
%\end{split}
%\end{equation}

%In particular, when the mean-field approximation \eqref{mean_field_approximation} is adopted, the H-L XC potential reads that
%\begin{equation}
%\begin{split}
%V_{XC}[n_1](\br_1, t) &= \int_{\mathbb{R}^{d}} \D \bR^{\prime} K_{XC}(\br_1, \bR^{\prime})  n_{1}(\bR^{\prime}, t)\\ 
%&\approx   \frac{g_D}{n_0^{1/3}}  (n_1(\br_1, t))^{1/3}   + \frac{0.034 g_D}{a_B n_0^{1/3}} \left[\ln (1 + 18.37 a_B (n_1(\br_1, t))^{1/3}) \right].
%\end{split}
%\end{equation}

Define the Fourier transform and its inverse
\begin{equation}
\mathcal{F}_{\bP \to \bri} f(\bR, \bP, t) = \int_{\mathbb{R}^{D}} \me^{-\mi \bP \cdot \bri} f(\bR, \bP, t) \D \bP, \quad \mathcal{F}^{-1}_{\bri \to \bP} \varphi(\bR, \bri, t) = \frac{1}{(2\pi)^{D}}\int_{\mathbb{R}^{D}} \me^{\mi \bP \cdot \bri} \varphi(\bR, \bri, t) \D \bri.
\end{equation}
For  the Hartree part, several examples are given as follows. 
\begin{example}[3-D Coulomb potential]
In 3-D case, we usually consider the Coulomb interaction, 
\begin{equation}
V_{i,j}(\br_i, \br_j)  = {V}_{ee}(|\br_i - \br_j|) =  \frac{e^2}{4\pi m_e \epsilon_0} \frac{1}{|\br_i - \br_j|}.
%\quad \hat{V}_{ext}(\br_j^{\prime}) = -\frac{Z}{|\bR_A - \br_j^{\prime}|}, \quad \br_i, \br_j \in \mathbb{R}^3,
\end{equation}
%with $\bR_A$ the atomic position and $Z$ is the atomic charge. 
Thus the Hartree potential is $\hat{V}_{H}(\br_1, t) =  \hat{V}_H[n_1](\br_1, t)$ depending on the single-body density $n_1$,
%\begin{equation}
% \hat{V}_H[n_j](\br_j) =  \int \D \bR_k^{\prime}~ {V}_{H}(\br_i, \br_{j}^{\prime}) n_{1^{\prime}, \dots, k^{\prime}}(\br_1^{\prime}, \dots \br_j^{\prime}, \dots, \br_k^{\prime}, t),
%\end{equation}
%where $\hat{V}_H[n_j]$ satisfies the Poisson equation
\begin{equation}
-\Delta_{\br_j} \hat{V}_H[n_1](\br_1, t) = \frac{e^2}{4\pi m_e \epsilon_0} n_1(\br_1, t).
\end{equation}
\end{example}

For 1-D problem, to get rid of the singularity, the screened or smoothed Coulomb potential is usually adopted \cite{VuSarma2020,KnightQuineyMartin2022}.
\begin{example}[1-D screened Coulomb potential]
The first is the screened Coulomb potential,
\begin{equation}
V_{i,j}(r_i, r_j)  = {V}_{ee}(|r_i - r_j|) =  \frac{e^{-\kappa|r_i - r_j|}}{2\kappa}, 
\end{equation}
which is derived by the screened Poisson equation with $\kappa$ the screening constant
\begin{equation}
(-\Delta_{r_1} + \kappa^2)V_H[n_1](r_1, t) = n_1(r_1, t), \quad r_1 \in \mathbb{R}.
\end{equation}
The Hartree part can be solved by the Fourier transform method
\begin{equation}
V_H[n_1](r_1, t)  = \int_{\mathbb{R}} \frac{e^{-\kappa|s|}}{2\kappa} n_1(r_1 - s, t) \D s = \int_{\mathbb{R}} \frac{1}{|k|^2 + \kappa^2} \widehat n_1(k, t) \me^{\mi k \cdot r_1} \D k. 
\end{equation}
%Since $ \frac{1}{|k|^2 + \kappa^2}$ decays very slowly, it requires to impose zero-padding on $n_1(r_1, t)$ to eliminate the artificial oscillation from the  periodic extension. 
\end{example}

\begin{example}[1-D smoothed Coulomb potential]
The second is the smoothed Coulomb kernel
\begin{equation}
V_{i,j}(r_i, r_j)  = {V}_{ee}(|r_i - r_j|) =  \frac{1}{\sqrt{|r_i - r_j|^2 + \epsilon}},
\end{equation}
with a small smoothing constant $\epsilon$. The Hartree part potential reads that
\begin{equation}\label{Hartree_smoothed_Coulomb}
V_H[n_1](r_1, t) = \int_{\mathbb{R}} \frac{1}{\sqrt{s^2 + \epsilon}} n_1(r_1 - s, t) \D s = \int_{\mathbb{R}} \left(\mathcal{F}_{r_1 \to k} \frac{1}{\sqrt{r_1^2 + \epsilon}}\right)  \widehat n_1(k, t) \me^{\mi k \cdot r_1} \D k.
\end{equation}
\end{example}

\subsection{The Wigner dynamics for 2-RDM}

\begin{figure}[!h]
\centering
        \includegraphics[width=3.6in,height=1.8in]{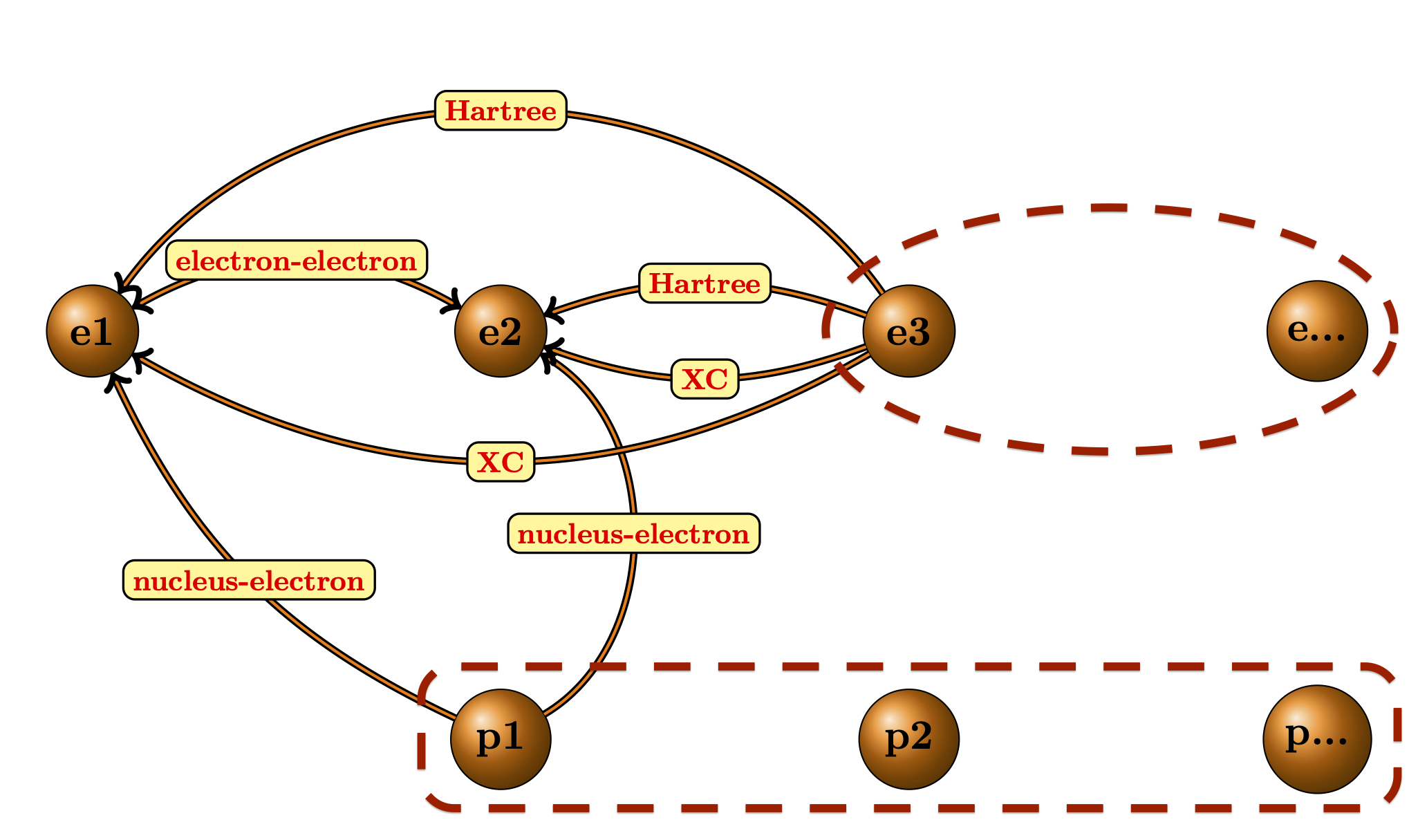}
     \caption{The picture of the two-body truncation of the quantum BBGKY hierarchy.\label{two_body_truncation}}  
\end{figure}

Of the most interest is to incorporate quantum two-body interactions to correct the Hartree approximation. This motivates to discuss the two-particle Wigner function to incorporate the two-body correlation is  in TDDFT. As presented in Figure \ref{two_body_truncation}, $\pdo$ is composed of three parts. 
\begin{equation}\label{two_body_truncated_Wigner}
\begin{split}
\Theta_V[f_{12}](\bR, \bP, t)  \approx ~ & \underbrace{\Theta_{V_{HXC}[n_1]}[f_{12}](\bR, \bP, t) +  \Theta_{V_{HXC}[n_2]}[f_{12}](\bR, \bP, t)}_{\textup{Hartree and exchange-correlation potential}} \\
&+ \underbrace{\Theta_{V_{12}}[f_{12}](\bR, \bP, t)}_{\textup{two-body interaction}} +\underbrace{\Theta_{W_{12}}[f_{12}](\bR, \bP, t)}_{\textup{external potential}}.
\end{split}
\end{equation}
In \ref{sec.appendix}, it is shown that Eq.~\eqref{two_body_truncated_Wigner} is compatible with the direct truncation of the triplet Wigner function \cite{OlmstedCurtiss1975}.

For the mean-field part, $\pdo$ can be cast into the convolution form,
\begin{equation}
\begin{split}
& \Theta_{V_{HXC}[n_1]}[f_{12}](\br_1, \br_2, \bp_1, \bp_2, t) = \int_{\mathbb{R}^d} V_W[n_{1}](\br_1, \bp_1 - \bp^{\prime}, t) f_{12}(\br_1, \br_2, \bp^{\prime}, \bp_2, t) \D \bp^{\prime}, \\
& \Theta_{V_{HXC}[n_2]}[f_{12}](\br_1, \br_2, \bp_1, \bp_2, t) = \int_{\mathbb{R}^d} V_W[n_{2}](\br_2, \bp_2 - \bp^{\prime}, t) f_{12}(\br_1, \br_2,  \bp_1, \bp^{\prime}, t) \D \bp^{\prime},
\end{split}
\end{equation}
where the Wigner kernel for the Hartree and exchange-correlation potential reads
\begin{equation}\label{decomposition}
\begin{split}
V_W[n_{j}](\br_j, \bp_j, t) = & \frac{\mi }{\hbar} \left(\frac{1}{\pi\hbar}\right)^{d} \left[\me^{-\frac{2\mi }{\hbar}\br_j \cdot \bp_j} \mathcal{F}_{\br_j \to -\frac{2}{\hbar} \bp_j}V_{HXC}[n_{j}](\br_j, t) - \me^{\frac{2\mi }{\hbar}\br_j \cdot \bp_j}  \mathcal{F}_{\br_j \to \frac{2}{\hbar} \bp_j} V_{HXC}[n_{j}](\br_j, t)\right].
\end{split}
\end{equation}

For the self-interaction $V_{12}(\br_1, \br_2) = V_{ee}(|\br_1 - \br_2|)$, the corresponding $\pdo$ can be cast into a scattering operator with oscillatory weights:
\begin{equation}\label{two_body_scattering}
\begin{split}
 \Theta_{V_{12}}[f_{12}](&\br_1, \br_2, \bp_1, \bp_2, t) = \frac{1}{\mi\hbar} \left(\frac{1}{\pi \hbar}\right)^d \int_{\mathbb{R}^d} \D \bp^{\prime}  \me^{-\frac{2\mi}{\hbar} \bp^{\prime} \cdot (\br_1 - \br_2)} \underbrace{\mathcal{F}_{\br \to \frac{2}{\hbar}\bp^{\prime}}{V_{ee}}(|\br|)}_{\textup{collision kernel}} \\
 &\times \underbrace{\left\{ f_{12}(\br_1, \br_2, \bp_1 - \bp^{\prime}, \bp_2 + \bp^{\prime}, t) -  f_{12}(\br_1, \br_2, \bp_1 + \bp^{\prime}, \bp_2-  \bp^{\prime}, t)\right\}}_{\textup{two-body scattering}},
\end{split}
\end{equation}
and for the external field $W_{12}(\br_1, \br_2) = V_{ext}(\br_1) + V_{ext}(\br_2)$, 
\begin{equation}\label{external}
\begin{split}
 \Theta_{W_{12}}[f_{12}]&(\br_1, \br_2, \bp_1, \bp_2, t) =\frac{1}{\mi \hbar} \left(\frac{1}{\pi \hbar}\right)^d \sum_{i=1}^2 \int_{\mathbb{R}^d} \D \bp^{\prime}  \me^{-\frac{2\mi}{\hbar} \bp^{\prime} \cdot \br_i } \underbrace{\mathcal{F}_{\br \to \frac{2}{\hbar}\bp^{\prime}}{V}_{ext}(\br)}_{\textup{collision kernel}} \\
 &\times \underbrace{\left\{ f_{12}(\br_1, \br_2, \bp_1 - \bp^{\prime} \delta_{1i}, \bp_2- \bp^{\prime} \delta_{2i}, t) -  f_{12}(\br_1, \br_2, \bp_1 + \bp^{\prime} \delta_{1i}, \bp_2 + \bp^{\prime} \delta_{2i}, t)\right\}}_{\textup{single-body  scattering}}
\end{split}
\end{equation}
%\begin{equation}
%\begin{split}
% \Theta_{V_{HXC}}&[f_{12}](\br_1, \br_2, \bp_1, \bp_2, t) = \frac{Z}{\mi \hbar} \left(\frac{1}{\pi \hbar}\right)^d \sum_{i=1}^2 \int_{\mathbb{R}^d} \D \bp^{\prime}  \me^{-\frac{2\mi}{\hbar} \bp^{\prime} \cdot \br_i } \underbrace{\mathcal{F}_{\br \to \frac{2}{\hbar}\bp^{\prime}}{V}_{HXC}(\br)}_{\textup{collision kernel}} \\
% &\times \underbrace{\left\{ f_{12}(\br_1, \br_2, \bp_1 - \bp^{\prime} \delta_{1i}, \bp_2- \bp^{\prime} \delta_{2i}, t) -  f_{12}(\br_1, \br_2, \bp_1 + \bp^{\prime} \delta_{1i}, \bp_2 + \bp^{\prime} \delta_{2i}, t)\right\}}_{\textup{single-body  scattering}}
%\end{split}
%\end{equation}
with $\delta_{ij}= 1$ for $i=j$ and $\delta_{ij}= 0$ for $i \ne j$.

\section{Spectral approximations to $\pdo$}
\label{sec.spectral_pdo}

As seen in Eqs.~\eqref{decomposition}, \eqref{two_body_scattering} and \eqref{external}, both the highly oscillating structure of $\pdo$ and the nonlinear symbol pose substantial computational difficulties. Therefore, one of the primary challenges is the development of efficient numerical techniques capable of describing a wide range of wavelengths in $\pdo$, as well as reducing the computational cost. This inspires us to seek efficient spectral approximations with an appropriate mesh strategy. In this section, two kinds of spectral discretization to $\pdo$s will be discussed for both mean-field parts and two-body scattering, depending on whether the Wigner function is periodic (or has a compact support) or non-periodic  in the spatial space. 

In fact, the spatial periodization inherently leads to a uniform discretization of the momentum variable, and $\pdo$ reduces to the pseudo-difference operator. The coupling of spatial and momental variables, however, results in some constraints on the spacings $\Delta x$, $\Delta p$ and the numbers of grid points $N_x$, $N_p$  \cite{SchulzSchulz2020}. To be more specific, for nonlinear $\pdo$, it requires $\Delta x \Delta p = \frac{\hbar \pi}{N_x}$, and  $N_x = 2N_p$ for the acceleration of calculation via FFTs. The correlation between time steps, mesh size, and the physical constant $\hbar$ gives rise to formidable challenges in the semi-classical regime ($\hbar \ll 1$). One can refer to \cite{FilbetGolse2025,FilbetGolse2025_arXiv} for an insightful discussion on this issue.

Such constraints can be exempted when the Wigner function is defined in the free space. The Chebyshev spectral method can provide an analytical representation of $\pdo$ \cite{ShaoLuCai2011,XiongChenShao2016}, in which the reduced Planck constant $\hbar$ is absorbed into the precomputed oscillatory integrals. Nevertheless, the Fourier spectral method seems to be more advantageous for the two-body scattering term, as the double integrals can be greatly simplified by exploiting the Fourier orthogonal relation.

\subsection{Spatial periodic case: Pseudo-difference operator}
\label{sec.periodic_spectral}

For the computational domain $\Omega = \mathcal{X}_1 \times \mathcal{X}_2 \times \mathcal{P}_1 \times \mathcal{P}_2 $, suppose the spatial periodic boundary condition is imposed over the domain $ \mathcal{X}_1 \times \mathcal{X}_2$, 
\begin{equation}
 \mathcal{X}_1 \times \mathcal{X}_2 = \left[-\frac{L_x}{2}, \frac{L_x}{2}\right]^{d} \times \left[-\frac{L_x}{2}, \frac{L_x}{2}\right]^{d},
\end{equation} 
and assume that the potential is also periodic, which holds for the Hartree and exchange-correlation term. Then $\pdo$ reduces to a semi-discretized version known as the pseudo-difference operator. In this situation, it is natural to adopt the Fourier spectral method in the momentum space.

We adopt  a uniform grid $[-N_x \Delta x/2, N_x \Delta x/2]$ with spacing $\Delta x = L_x /N_x$ in each direction. Moreover, the spacing of momentum space should depend on $L_x$, that is, $\Delta p = \frac{\hbar \pi}{L_x}$. When $N_p$ points are adopted in each direction, the truncated momentum space is 
\begin{equation}
\mathcal{P}_1 \times \mathcal{P}_2 = \left[-\frac{N_p \Delta p}{2}, \frac{N_p \Delta p}{2}\right]^d \times \left[-\frac{N_p \Delta p}{2}, \frac{N_p \Delta p}{2}\right]^d , 
\end{equation}
provided that the Wigner function vanishes outside $\mathcal{P}_1 \times \mathcal{P}_2$.

\subsubsection{Single-body component}

Under the spatial periodic boundary condition, the potential $V_{HXC}$ allows a Fourier series expansion
\begin{equation}
V_{HXC}[n_1](\br_1, t) = \sum_{\bmm \in \mathbb{Z}^d} \widehat V_{\bmm}[n_1](t) e_{\bm{m}}(\br_1), \quad e_{\bm{m}}(\br) =  \me^{-\frac{2\pi \mi}{L_x}   \bm{m} \cdot \br}.
\end{equation}
%where
%\begin{equation}
%\widehat V_{HXC}[n_1]\left(\bk, t\right) \approx (\Delta x)^d  \sum_{\bn} V_{HXC}[n_1](\bn \Delta x, t)\me^{- 2\mi \bk \cdot \bn \Delta x}.
%\end{equation}

Thus the Wigner kernel reads that
\begin{equation}\label{semi_discrete_Wigner_potential}
\begin{split}
&\frac{1}{\hbar^d}\int_{\mathbb{R}^d} \me^{-\frac{\mi}{\hbar}  \bp^{\prime} \cdot \bri}  \left[V_{HXC}[n_1](\br_1  - \frac{\bri}{2}, t) - V_{HXC}[n_1](\br_1  + \frac{\bri}{2}, t)\right] \D \bri \\
& = \sum_{\bmm \in \mathbb{Z}^d}  \widehat V_{\bmm}[n_1](t) e_{\bmm}(\br_1)\frac{1}{\hbar^d}\int_{\mathbb{R}^d} \me^{-\frac{\mi}{\hbar}  \bp^{\prime} \cdot \bri}  \left[e_{\bmm}\left(-{\bri}/{2}\right) - e_{\bmm}\left({\bri}/{2}\right)\right] \D \bri \\
& =  (2\pi)^d \sum_{\bmm \in \mathbb{Z}^d}  \widehat V_{\bmm}[n_1](t) e_{\bmm}(\br_1)  \left[ \delta(\bp^{\prime} - \frac{\bmm \hbar \pi}{L_x}) - \delta(\bp^{\prime} + \frac{\bmm \hbar \pi}{L_x}) \right],
\end{split}
\end{equation}
and it attains at the pseudo-difference operator 
\begin{equation}
\begin{split}
\Theta_{V_{HXC}[n_1]}[f_{12}]&(\br_1, \br_2, \bp_1, \bp_2, t) =  \frac{1}{\mi \hbar} \sum_{\bmm \in \mathbb{Z}^d}  \widehat V_{\bmm}[n_1](t) e_{\bmm}(\br_1) \\
& \times \left\{f_{12}(\br_1, \br_2, \bp_1 -  \frac{\bmm \hbar \pi}{L_x}, \bp_2, t) - f_{12}(\br_1, \br_2, \bp_1 +  \frac{\bmm \hbar \pi}{L_x}, \bp_2, t) \right\}.
\end{split}
\end{equation}

Since we only consider discrete momentum $\bp_1  = \bn \Delta p$, $\Delta p = \frac{\hbar \pi}{L_x}$, it yields
\begin{equation}\label{Hartree_FSM}
\begin{split}
&\Theta_{V_{HXC}[n_1]}[f_{12}](\br_1, \br_2, \bn \Delta p, \bp_2, t)\\
 &=  \frac{1}{\mi \hbar} \sum_{\bmm \in \mathbb{Z}^d}  \widehat V_{\bmm}[n_1](t) e_{\bmm}(\br_1) \left[f_{12}(\br_1, \br_2, (\bn -\bmm) \Delta p, \bp_2, t) - f_{12}(\br_1, \br_2, (\bn+\bmm) \Delta p, \bp_2, t) \right] \\
&\approx  \frac{1}{\mi \hbar} \sum_{\bmm (2N_p)}  \widehat V_{\bmm}[n_1](t) e_{\bmm}(\br_1) \left[f_{12}(\br_1, \br_2, (\bn -\bmm) \Delta p, \bp_2, t) - f_{12}(\br_1, \br_2, (\bn+\bmm) \Delta p, \bp_2, t) \right],\\
%& \approx  \frac{1}{\mi \hbar} \sum_{\bmm (N_p)}  \left[ \widehat V_{\bmm- \bn}[n_1](t)  e_{- \bn}(\br_1) - \widehat V_{\bmm + \bn}[n_1](t)  e_{\bn}(\br_1)\right]  f_{12}(\br_1, \br_2, \bmm \Delta p, \bp_2, t) e_{\bmm}(\br_1),\\
%& \approx \frac{1}{\mi \hbar} \sum_{\bmm (N_p)}  \left[ \widehat V_{\bn- \bmm}[n_1](t) - \widehat V_{\bn +  \bmm}[n_1](t) \right] e_{\bmm}(\br_1) f_{12}(\br_1, \br_2, \bmm \Delta p, \bp_2, t) .
\end{split}
\end{equation}
where $\sum_{\bmm (2N_p)}  = \sum_{m_1 = -N_p}^{N_p} \cdots \sum_{m_d = -N_p}^{N_p}$. Eq.~\eqref{Hartree_FSM} can be accelerated by FFTs, provided that $N_x = 2N_p$ and $f_{12} = 0$ outside $\mathcal{P}_1 \times \mathcal{P}_2$.

%In particular, when $N_x = 2N_p$, we can further expand $f_{12}$ as a Fourier series in $\br_1$-direction
%\begin{equation}
%f_{12}(\br_1, \br_2, \bn \Delta p, \bp_2, t) \approx \sum_{\bme (N_x)} a_{\bme}( \br_2, \bn \Delta p, \bp_2, t) e_{\bme}(\br_1).
%\end{equation}
%Substituting it into Eq.~\eqref{Hartree_FSM} yields that
%\begin{equation*}
%\begin{split}
%&\Theta_{V_{HXC}[n_1]}[f_{12}](\br_1, \br_2, \bn \Delta p, \bp_2, t)\\
%&\approx  \frac{1}{\mi \hbar} \sum_{\bme (N_x)} \sum_{\bmm (2N_p)}  \widehat V_{\bmm}[n_1](t) \left[ a_{\bme}(\br_2, (\bn +\bmm) \Delta p, \bp_2, t) - a_{\bme}(\br_2, (\bn-\bmm) \Delta p, \bp_2, t) \right] e_{\bmm}(\br_1)  e_{\bme}(\br_1),
%\end{split}
%\end{equation*}
%which can be accelerated by FFT.

\begin{example}[1-D Coulomb potential under the neutralized ion background]

For the quantum electron-ion plasma, assume that the system is neutralized by the external ion background $n_I(\br_1, \br_2)$, namely, 
\begin{equation}\label{neutralized_condition}
\iint_{\Omega} n_{12}(\br_1, \br_2, t) \D \br_1 \D \br_2 = \iint_{\Omega} n_I(\br_1, \br_2) \D \br_1 \D \br_2,
\end{equation}
with $n_{12}(\br_1, \br_2, t) = \iint_{\mathbb{R}^{2d}} f_{12}(\br_1, \br_2, \bp_1, \bp_2, t) \D \bp_1 \D \bp_2$.  The Hartree part $V_H[n_j]$ can be obtained by solving the 1-D Poisson equation directly,
\begin{equation} 
-\Delta_{\br_j} V_H[n_j](\br_j, t) = \frac{e}{\varepsilon_0} \int_{\mathbb{R}^{d}} \left[ n_{12}(\br_1, \br_2, t) - n_I(\br_1, \br_2)\right] \D \br_j.
\end{equation}

Now take $j=1$ as an example. Since $n_{12}$ allows a Fourier series expansion, it yields  
\begin{equation}
\int_{\mathbb{R}^d} \left[ n_{12}(\br_1, \br_2, t) - n_I(\br_1, \br_2)\right] \D \br_2 = \sum_{\bm{m} \in \mathbb{Z}^d\setminus \{\bm{0}\}} a_{\bm{m}}(t) e_{\bm{m}}(\br_1), \quad e_{\bm{m}}(\br) =  \me^{-\frac{2\pi \mi}{L_x}   \bm{m} \cdot \br }.
\end{equation}
In this situation, the zero mode vanishes due to the neutralized condition \eqref{neutralized_condition}, and consequently the singularity of the Coulomb potential is removed. It is convenient to obtain $ V_{H}[n_1](\br_1, t)$
\begin{equation}
 V_{H}[n_1](\br_1, t) =   \sum_{\bm{m} \setminus \{\bm{0}\} (N_x)}  \widehat V_{\bmm}[n_1](t)  e_{\bm{m}}(\br_1), \quad  \widehat V_{\bmm}[n_1](t) =  \frac{L_x^2}{4\pi^2 |\bm{m}|^2} a_{\bm{m}}(t).
\end{equation}
Thus we have a truncated Fourier series expansion
\begin{equation}
 V_{H}[n_1](\bn \Delta x, t) \approx   \sum_{\bm{m} \setminus \{\bm{0}\} (N_x)}  \frac{L_x^2 }{4\pi^2} \frac{a_{\bm{m}}(t)}{|\bm{m}|^2} \me^{-\frac{2\pi \mi}{L_x}   \bmm \cdot \bn \Delta x} =  \sum_{\bm{m} \setminus \{\bm{0}\} (N_x)}   \frac{L_x^2 }{4\pi^2} \frac{a_{\bm{m}}(t)}{|\bm{m}|^2} \me^{-\frac{2\pi \mi}{N_x}   \bmm \cdot \bn}.
\end{equation}
%which can be computed efficiently by the fast Fourier transform.

\end{example}

\subsubsection{Two-body interaction}

For two-body interaction like Eq.~\eqref{two_body_scattering}, assume the interacting potential allows a Fourier series
\begin{equation}
V_{ee}(\br) = \sum_{\bmm \in \mathbb{Z}^d} \widehat V_{\bmm} e_{\bm{m}}(\br), \quad e_{\bm{m}}(\br) = \me^{-\frac{2\pi \mi}{L_x}   \bm{m} \cdot \br}.
\end{equation}
Since
\begin{equation}
 \left(\frac{2}{ \hbar}\right)^d \mathcal{F}_{\br \to \frac{2\bp^{\prime}}{\hbar}}{V_{ee}}(\br) =  (2\pi)^d  \sum_{\bmm \in \mathbb{Z}^d} \widehat V_{\bmm}\delta\left(\bp^{\prime} + \frac{ \hbar \pi }{L_x} \bmm\right),
\end{equation}
substituting it into Eq.~\eqref{two_body_scattering}  yields that
\begin{equation}\label{scattering_representation}
\begin{split}
 \Theta_{V_{12}}&[f_{12}](\br_1, \br_2, \bp_1, \bp_2, t) = \frac{1}{\mi \hbar} \sum_{\bmm \in \mathbb{Z}^d}  \me^{\frac{2\pi \mi}{L_x} \bmm \cdot (\br_1 - \br_2)} \widehat V_{\bmm}\\
 &\times \left\{ f_{12}(\br_1, \br_2, \bp_1 - \frac{\hbar \pi \bmm}{L_x}, \bp_2 + \frac{\hbar \pi \bmm}{L_x}, t) -  f_{12}(\br_1, \br_2, \bp_1 + \frac{\hbar \pi \bmm}{L_x}, \bp_2 - \frac{\hbar \pi \bmm}{L_x}, t)\right\}.
\end{split}
\end{equation}

To evaluate the Fourier modes $\widehat V_{\bmm}$, it is observed that for $\Delta p = \frac{\hbar \pi}{L_x}$,
\begin{equation}
\begin{split}
\widehat V_{\bmm}  &= \frac{1}{L_x^d}\int_{[-\frac{L_x}{2}, \frac{L_x}{2}]^d} V_{ee}(\br)  \me^{\frac{2 \mi }{\hbar}   (\bm{m} \Delta p) \cdot \br} \D \br = \frac{1}{L_x^d} \int_{[-\frac{L_x}{2}, \frac{L_x}{2}]^d} V_{ee}(\br)  \me^{\frac{2\pi \mi}{L_x}   \bm{m} \cdot \br} \D \br.
\end{split}
\end{equation}
We can choose $\Delta p \Delta y = \frac{\hbar \pi}{2N_p}$, so that
\begin{equation}
\widehat V_{\bmm} \approx \left(\frac{\Delta y}{L_x}\right)^d \sum_{\bn (2 N_p)} V_{ee}(\bn \Delta y) \me^{\frac{2 \pi \mi}{2N_p}  \bmm \cdot \bn}.
\end{equation}

The infinite summation over $\bmm$ can be truncated to $ \sum_{\bmm (2N_p)}  = \sum_{m_1 = -N_p}^{N_p} \cdots \sum_{m_d = -N_p}^{N_p}$ when $f_{12}$ decays at $\mathcal{P}_1 \times \mathcal{P}_2$, with zero padding outside $\mathcal{P}_1 \times \mathcal{P}_2$. Then it arrives at the spectral approximation for $\bp_1 = \bn_1 \Delta p$, $\bp_2 = \bn_2 \Delta p$, $\Delta p = \frac{\hbar \pi}{L_x}$,
\begin{equation}\label{two_body_FSM}
\begin{split}
%&\Theta_{V_{12}}[f_{12}](\br_1, \br_2, \bn_1 \Delta p, \bn_2 \Delta p, t) \approx \frac{1}{\mi \hbar \pi^d} \sum_{\bmm (N_p)}  \me^{-\frac{2 \pi \mi}{L_x} \bmm \cdot (\br_1 - \br_2)} \widehat V_{\bmm}\\
% & \quad \times\left[ f_{12}(\br_1, \br_2, (\bn_1 + \bmm)\Delta p, (\bn_2 - \bmm)\Delta p, t) -  f_{12}(\br_1, \br_2, (\bn_1 - \bmm)\Delta p, (\bn_2 + \bmm)\Delta p, t)\right].
 &\Theta_{V_{12}}[f_{12}](\br_1, \br_2, \bn_1 \Delta p, \bn_2 \Delta p, t) \approx \frac{1}{\mi \hbar} \sum_{\bmm (2N_p)}  \me^{\frac{2\pi \mi}{L_x} \bmm \cdot (\br_1 - \br_2)}  \widehat V_{\bmm}\\
 &~~  \times\left\{ f_{12}(\br_1, \br_2, (\bn_1 - \bmm)\Delta p, (\bn_2 + \bmm)\Delta p, t) -  f_{12}(\br_1, \br_2, (\bn_1 + \bmm)\Delta p, (\bn_2 - \bmm)\Delta p, t)\right\}.
\end{split}
\end{equation}
 For $N_x^{2d} N_p^{2d}$ grid points, one shall evaluate Eq.~\eqref{two_body_FSM} for $\br_1 - \br_2 = \bm{\ell} \Delta x, \bm{\ell} \in [-2N_x, 2N_x]^d$, so that the total computational complexity of direct summation is $\mathcal{O}(N_x^{2d} N_p^{3d})$. 

In particular, when $N_x = 2N_p$, $\Delta x_1 = \Delta x_2 = \Delta x$,  then $ \me^{-\frac{2\pi \mi}{L_x} \bmm \cdot (\br_1 - \br_2)}  =  \me^{-\frac{2\pi \mi}{2 N_p} \bmm \cdot \bm{\ell}}$ for $\br_1 - \br_2 = \bm{\ell} \Delta x, \bm{\ell} \in [-2N_x, 2N_x]^d$. This implies that Eq.~\eqref{two_body_FSM} can be accelerated by FFT. For $N_x^{2d} N_p^{2d}$ grid points, the total computational complexity can be further reduced to $\mathcal{O}(N_x^{2d}  N_p^{2d} \log N_p)$.

\subsection{Spatial non-periodic case: Chebyshev spectral approximation}
\label{sec.non_periodic_spectral}

For the non-periodic case, suppose the number density $n_{12}(\br_1, \br_2, t)$ vanishes outside the spatial domain 
\begin{equation}
\mathcal{X}_1 \times \mathcal{X}_2 = \left[-\frac{L_x}{2}, \frac{L_x}{2}\right]^{d} \times \left[-\frac{L_x}{2}, \frac{L_x}{2}\right]^{d},
\end{equation} and a uniform grid $[-N_x \Delta x/2, N_x \Delta x/2]$ with spacing $\Delta x = L_x /N_x$ is adopted in direction. 

For the momental domain $\mathcal{P}_1 \times \mathcal{P}_2$, in the spirit of the spectral element method (SEM) \cite{ShaoLuCai2011,XiongChenShao2016}, we can first split $\mathcal{P}_1 \times \mathcal{P}_2$ into $M_1 \times  M_2$ cells
\begin{equation}
\mathcal{P}_1 \times \mathcal{P}_2 = \bigcup_{s_1 = 1}^{M_1}   \bigcup_{s_2 = 1}^{M_2} \mathcal{P}_{s_1} \times \mathcal{P}_{s_2}, \quad \mathcal{P}_{s} = \prod_{k=1}^d  \mathcal{P}_{s}^{(k)}  = \prod_{k=1}^d \left[d_{s}^{(k)}, d_{s+1}^{(k)}\right].
\end{equation}
Take the element $\mathcal{P}_{s_1} \times \mathcal{P}_{s_2}$ as an example. The Chebyshev spectral approximations of the Wigner function in $\bp_1$- or $\bp_2$-coordinate read
\begin{equation}\label{spectral_representation}
\begin{split}
f_{12}\left(\br_1 ,\br_2, \bp_{1}, \bp_{2}, t\right) &\approx \sum_{\bl_{1}} \sum_{\bl_{2}}
a_{s_1, s_2, \bl_1, \bl_2}
\left(\br_1,\br_2, t\right)C_{\bl_{1}}
\left(\bp_{1}\right) C_{\bl_{2}}
\left(\bp_{2}\right), ~~ \bp_{1} \in \mathcal{P}_{s_{1}}, \bp_{2} \in \mathcal{P}_{s_{2}},\\
f_{12}\left(\br_1 ,\br_2, \bp_{1}, \bp_{2}, t\right) &\approx \sum_{\bl_{1}}
 b_{s_1, \bl_{1}}
\left(\br_1,\br_2, \bp_2, t\right)C_{\bl_{1}}
\left(\bp_{1}\right), ~~ \bp_{1} \in \mathcal{P}_{s_{1}}, \\
f_{12}\left(\br_1 ,\br_2, \bp_{1}, \bp_{2}, t\right) &\approx \sum_{\bl_{2}}
 c_{s_2, \bl_{2}}
\left(\br_1,\br_2, \bp_1, t\right)C_{\bl_{2}}
\left(\bp_{2}\right), ~~ \bp_{2} \in \mathcal{P}_{s_{2}}, 
\end{split}
\end{equation}
where $C_{\bl_1}(\bp) = \prod_{k=1}^d C_{l_1^{(k)}}(p_k)$ with
\begin{equation}\label{eq:base}
C_{l_1^{(k)}}\left(p_k\right)  =
T_{l_1^{(k)}}\left(\eta_k\right), \,\,\,
p_k =\hat{d}_{s_1}^{(k)}+\frac{|\mathcal{P}_{s_1}^{(k)}|}{2} \eta_k,
\,\,\,
\hat{d}_{s_1}^{(k)}  = d_{s_1}^{(k)}+\frac{|\mathcal{P}_{s_1}^{(k)}|}{2},
\end{equation}
and $T_{l_1^{(k)}}(\eta_k)$ is the Chebyshev polynomial of the first kind.

\subsubsection{Single-body  component}
For convenience, we first take the coordinate transformation $\bp^{\prime}  = \hbar \bk^{\prime}$, so that
\begin{equation}
 \Theta_{V_{HXC}[n_1]}[f_{12}](\br_1, \br_2, \bp_1, \bp_2, t) = \int_{\mathbb{R}^d} V_W[n_{1}](\br_1, \frac{\bp_1}{\hbar} -  \bk^{\prime}, t) f_{12}(\br_1, \br_2, \hbar \bk^{\prime}, \bp_2, t) \D \bk^{\prime},
\end{equation}
where the Wigner kernel $V_W[n_1]$ reads that
\begin{equation}
V_W[n_1](\br_1, \bk^{\prime}, t) =  \frac{1}{\mi \hbar} \frac{1}{\pi^d} \left[\me^{-2\mi \br_1 \cdot \bk^{\prime}} \mathcal{F}_{\br_1 \to -2\bk^{\prime}}V_{HXC}[n_1](\br_1, t) -   \me^{2 \mi \br_1 \cdot \bk^{\prime}}  \mathcal{F}_{\br_1 \to 2 \bk^{\prime}} V_{HXC}[n_1](\br_1, t)\right].
\end{equation}

 We can approximate the Hartree-exchange-correlation term by $V_W[n_{1}](\br_1, \bk^{\prime}, t)$
\begin{equation}
V_W[n_1](\br_1, \bk^{\prime}, t) \approx \frac{1}{\mi \hbar \pi^d}  \left[\me^{-2 \mi \br_1 \cdot \bk^{\prime}} \widehat V_{HXC}[n_1](-\bk^{\prime}, t) -   \me^{2\mi \br_1 \cdot \bk^{\prime}}  \widehat V_{HXC}[n_1](\bk^{\prime}, t)\right],
\end{equation}
with the Fourier coefficients denoted by $\widehat V_{HXC}[n_1](\bk, t)$,
\begin{equation}
\widehat V_{HXC}[n_1](\bk, t) =  \int_{\mathbb{R}^d}  V_{HXC}[n_1](\br_1, t) e_{\bk}(\br_1) \D \br_1, \quad e_{\bk}(\br) = \me^{- 2 \mi \bk \cdot \br}.
\end{equation}

Using the Poisson summation formula \cite{BenedettoZimmermann1997}, it has
\begin{equation}
\begin{split}
\sum_{\bmm \in \mathbb{Z}^d}\widehat V_{HXC}[n_1]\left(\bk + \frac{2\pi \bmm}{2\Delta x}, t\right)  = (\Delta x)^d \sum_{\bn \in \mathbb{Z}^d}  V_{HXC}[n_1](\bn \Delta x, t)\me^{- 2\mi \bk \cdot \bn \Delta x}, 
\end{split}
\end{equation}
where $\Delta x$ is the uniform spacing. By ignoring the periodic images of $\widehat V_{HXC}[n_1]$, it yields that 
\begin{equation}
\widehat V_{HXC}[n_1]\left(\bk, t\right) \approx (\Delta x)^d  \sum_{\bn} V_{HXC}[n_1](\bn \Delta x, t)\me^{- 2\mi \bk \cdot \bn \Delta x},
\end{equation}
where $\sum_{\bn}$ is short for $\sum_{n_1 = -\frac{N_x}{2}}^{\frac{N_x}{2}} \cdots \sum_{n_d = -\frac{N_x}{2}}^{\frac{N_x}{2}}$. Therefore, the Wigner potential $V_W[n_1](\br_1, \bk^{\prime}, t)$ can be approximated by 
\begin{equation}\label{nonlinear_Wigner_potential}
V_W[n_1](\br_1, \bk^{\prime}, t) \approx \frac{(\Delta x)^d}{\mi \hbar \pi^d} \sum_{\bn}  V_{HXC}[n_1](\bn \Delta x, t) \left[ \me^{2\mi \bk^{\prime} \cdot \left(\bn \Delta x - \br_1\right)} - \me^{- 2\mi \bk^{\prime} \cdot \left(\bn \Delta x - \br_1\right)} \right].
\end{equation}

Combining Eq.~\eqref{nonlinear_Wigner_potential} with Eq.~\eqref{spectral_representation}, we attain the Chebyshev spectral approximation to the mean-field part of $\pdo$ 
\begin{equation}
\begin{split}
 \Theta_{V_{HXC}[n_1]}&[f_{12}](\br_1, \br_2, \bp_1, \bp_2, t) \approx  \frac{(\Delta x)^d}{\mi \hbar \pi^d} \sum_{s_1 = 1}^{M_1}  \sum_{\bl_{1}}  \sum_{\bn} b_{s_1, \bl_{1} }(\br_1, \br_2, \bp_2, t) V_{HXC}[n_1](\bn \Delta x, t) \\
& \hspace{-0.2cm}\times \left[ \int_{\mathcal{P}_{s_1}}  \me^{ \frac{2\mi }{\hbar} (\bp_1 - \bp^{\prime}) \cdot \left(\bn \Delta x - \br_1\right)} C_{\bl_1}\left(\bp^{\prime}\right) \D \bp^{\prime} - \int_{\mathcal{P}_{s_1}}  \me^{-\frac{2\mi }{\hbar} (\bp_1 - \bp^{\prime}) \cdot \left(\bn \Delta x - \br_1\right)} C_{\bl_1}\left(\bp^{\prime}\right) \D \bp^{\prime}\right].
\end{split}
\end{equation} 

For $\bp^{\prime} = (p^{\prime}_1, \dots, p^{\prime}_d)$, $\bn = (n_1, \dots, n_d)$ and $\br_1 = (r_{1, 1}, \dots, r_{1, d})$, since
\begin{equation}
\begin{split}
\int_{\mathcal{P}_{s_1}}  &\me^{\pm \frac{2\mi }{\hbar} \bp^{\prime} \cdot \left(\bn \Delta x - \br_1\right)}  C_{\bl_j}\left(\bp^{\prime}\right) \D \bp^{\prime} = \prod_{k=1}^d \int_{\mathcal{P}_{s_1}^{(k)}}\me^{\pm \frac{2\mi }{\hbar} p^{\prime}_k  \cdot \left(n_k \Delta x - r_{1, k}\right)} C_{l_1^{(k)}}\left(p_k^{\prime}\right) \D p_k^{\prime} \\
& = \prod_{k=1}^d \frac{|\mathcal{P}_{s_1}^{(k)}|}{2} \me^{\pm \frac{2\mi }{\hbar} \hat{d}_{s_1}^{(k)}  \cdot \left(n_k \Delta x - r_{1, k}\right)}  \int_{-1}^{1} \me^{\pm \frac{\mi }{\hbar} |\mathcal{P}_{s_1}^{(k)}|  \eta_k^{\prime} \cdot \left(n_k \Delta x - r_{1, k}\right)} T_{l_1^{(k)}}\left(\eta_k^{\prime}\right) \D \eta_k^{\prime} \\
& =  \prod_{k=1}^d \me^{\pm \frac{2\mi }{\hbar} \hat{{d}}_{s_1}^{(k)}  \cdot \left(n_k \Delta x - r_{1, k}\right)} \cdot  \prod_{k=1}^d \frac{|\mathcal{P}_{s_1}^{(k)}| }{2}  \int_{-1}^{1} \me^{\pm \frac{\mi }{\hbar} |\mathcal{P}_{s_1}^{(k)}|  \eta_k^{\prime} \cdot \left(n_k \Delta x - r_{1, k}\right)} T_{l_1^{(k)}}\left(\eta_k^{\prime}\right) \D \eta_k^{\prime}, 
\end{split}
\end{equation}
we arrive at the Chebyshev spectral approximation to $\pdo$
\begin{equation}\label{Chebyshev_PDO}
\begin{split}
 \Theta_{V_{HXC}[n_1]}[f_{12}](\br_1, &\br_2, \bp_1, \bp_2, t) \approx   \frac{(\Delta x)^d}{\mi \hbar \pi^d} \sum_{s_1 = 1}^{M_1}  \sum_{\bl_{1}}  \sum_{\bn} b_{s_1, \bl_{1} }(\br_1, \br_2, \bp_2, t) V_{HXC}[n_1](\bn \Delta x, t) \\
 & \times \Big \{  \me^{\frac{2\mi }{\hbar} (\bp_1 - \hat{\bm{d}}_{s_1}) \cdot \left(\bn \Delta x - \br_{1}\right)}\prod_{k=1}^d \frac{|\mathcal{P}_{s_1}^{(k)}| }{2} O_{l_1^{(k)}}\left(-\frac{|\mathcal{P}_{s_1}^{(k)}|}{\hbar}(n_k \Delta x -r_{1,k})\right)  \\
& \quad ~~ - \me^{- \frac{2\mi }{\hbar} (\bp_1 - \hat{\bm{d}}_{s_1})  \cdot \left(\bn \Delta x - \br_{1}\right)} \prod_{k=1}^d \frac{|\mathcal{P}_{s_1}^{(k)}| }{2} O_{l_1^{(k)}}\left(\frac{|\mathcal{P}_{s_1}^{(k)}|}{\hbar}(n_k \Delta x -r_{1,k}) \right)  \Big\},
\end{split}
\end{equation}
where the oscillatory integral is defined by
\begin{equation}\label{basic_osc_int}
O_{l}(z) = \int_{-1}^{1} \me^{i z \eta} T_l(\eta) \D \eta,
\end{equation}
which can be represented as a linear combination of spherical Bessel functions of the first kind (see Appendix in \cite{ShaoLuCai2011}).

Now define the tensor 
\begin{equation}
\mathcal{T}^{\pm}_{\bs_1, \bl_1}(\br_1, \bp_1) = \sum_{\bn}  V_{HXC}[n_1](\bn \Delta x, t) \me^{\pm \frac{2\mi }{\hbar} (\bp_1 - \hat{\bm{d}}_{s_1}) \cdot \left(\bn \Delta x - \br_{1}\right)}\prod_{k=1}^d |\mathcal{P}_{s_1}^{(k)}| O_{l_1^{(k)}}\left(\mp \frac{|\mathcal{P}_{s_1}^{(k)}|}{\hbar} (n_k \Delta x -r_{1,k})\right),
\end{equation}
then Eq.~\eqref{Chebyshev_PDO} can be simplified as
\begin{equation}
\begin{split}
& \Theta_{V_{HXC}[n_1]}[f_{12}](\br_1, \br_2, \bp_1, \bp_2, t) \approx \frac{(\Delta x)^d }{\mi \hbar \pi^d}  \sum_{s_1 = 1}^{M_1}  \sum_{\bl_{1}} b_{s_1, \bl_{1} }(\br_1, \br_2, \bp_2, t)  \left[\mathcal{T}^{+}_{\bs_1, \bl_1}(\br_1, \bp_1) - \mathcal{T}^{-}_{\bs_1, \bl_1}(\br_1, \bp_1)\right], \\
&  \Theta_{V_{HXC}[n_2]}[f_{12}](\br_1, \br_2, \bp_1, \bp_2, t) \approx \frac{(\Delta x)^d }{\mi \hbar \pi^d} \sum_{s_2 = 1}^{M_2}  \sum_{\bl_{2}} c_{s_2, \bl_{2} }(\br_1, \br_2, \bp_1, t)  \left[\mathcal{T}^{+}_{\bs_2, \bl_2}(\br_2, \bp_2) - \mathcal{T}^{-}_{\bs_2, \bl_2}(\br_2, \bp_2)\right].
 \end{split}
 \end{equation}
One can see that the constant $\hbar$ be absorbed into $O_{l}(z)$, which can be precomputed and stored.

\begin{remark}
The Hartree potential can be approximated by the Poisson summation formula. For long-range potentials, however, one should  impose zero-padding on $n(r_1, t)$ to avoid the overlap of periodic images. For Eq.~\eqref{Hartree_smoothed_Coulomb}, we can choose a uniform $r$-spacing $\Delta x$ and enlarge the $x$-domain to $[-\widetilde {N} \Delta x/2, \widetilde N \Delta x/2]$ and set $n(r_1, t) = 0$ outside $\mathcal{X}_1$. Then for $r_1 = m\Delta x$, it reads that
\begin{equation}\label{Hartree_approximation}
\begin{split}
V_H[n_1](m \Delta x, t) &\approx \Delta k \sum_{\ell = -\widetilde N/2}^{\widetilde N/2-1} \left(\mathcal{F}_{r_1 \to \ell \Delta k} \frac{1}{\sqrt{r_1^2 + \epsilon}}\right)  \widehat n_1(\ell \Delta k, t) \me^{\mi \ell \cdot m \Delta k \Delta x} \\
& \approx \Delta x \Delta k \sum_{\ell = - \widetilde N/2}^{\widetilde N/2-1}  \left(\sum_{\nu = -\widetilde{N}/2}^{\widetilde{N}/2-1} \frac{1}{\sqrt{(\nu \Delta x)^2  + \epsilon}} \me^{-\mi \ell \cdot \nu \Delta x \Delta k } \right) \widehat n_1(\ell \Delta k, t) \me^{\mi \ell \cdot m \Delta k \Delta x},
\end{split}
\end{equation}
where $ \Delta k = {2\pi}/{(\widetilde{N} \Delta x)}$. As suggested in \cite{LiuTangZhangZhang2024}, a three-fold zero padding, say, $\tilde N = 3N_x$ is enough to eliminate the artificial overlap, which is confirmed in our numerical examples (see Section \ref{sec_convergence}).
\end{remark}

\subsubsection{Two-body interaction}

For the electron-electron interaction, it is not easy to derive the Chebyshev spectral approximation to the scattering operator \eqref{scattering_representation} due to the lack of shifted invariance property of basis. Instead, we adopt the approach in \cite{XiongChenShao2016}. 

Start from the Poisson summation formula 
\begin{equation}
\begin{split}
&\sum_{\bmm_1 \in \mathbb{Z}^d}\sum_{\bmm_2 \in \mathbb{Z}^d} \mathcal{F}_{(\by_1, \by_2) \to (\bk_1, \bk_2)}D_V \left(\br_1, \br_2, \bk_1 + \frac{2\pi \bmm_1}{2\Delta x_1}, \bk_1 + \frac{2\pi \bmm_2}{2\Delta x_2}\right)  \\
&= (\Delta y_1)^d (\Delta y_2)^d \sum_{\bn_1 \in \mathbb{Z}^d}  \sum_{\bn_2 \in \mathbb{Z}^d}D_V(\br_1, \br_2, \bn_1 \Delta y_1, \bn_2 \Delta y_2)\me^{- 2\mi \bk_1 \cdot \bn_1 \Delta y_1 - 2\mi \bk_2 \cdot \bn_2 \Delta y_2} , 
\end{split}
\end{equation}
where
\begin{equation}
D_V(\br_1, \br_2, \by_1, \by_2) = V_{12}(\br_1 - \frac{\by_1}{2}, \br_2 - \frac{\by_2}{2}) - V_{12}(\br_1 + \frac{\by_1}{2}, \br_2 + \frac{\by_2}{2}).
\end{equation}
and $\Delta y_1 = \Delta y_2 = \hbar \pi/L_k$. By ignoring the periodic images and let $\bp_1 = \hbar \bk_1$, $\bp_2 = \hbar \bk_2$, $\pdo$ can be approximated by
\begin{equation}
\begin{split}
\Theta_{V_{12}}[&f_{12}](\br_1, \br_2, \bp_1, \bp_2, t) \approx \frac{(\Delta y_1 \Delta y_2)^d}{\mi \hbar \pi^d} \sum_{\bn_1 (N_x)}  \sum_{\bn_2 (N_x)}
\me^{- \frac{2\mi}{\hbar} \bp_1 \cdot \bn_1 \Delta y_1 - \frac{2\mi}{\hbar}\bp_2 \cdot \bn_2 \Delta y_2} \\
&\times D_V(\br_1, \br_2, \bn_1 \Delta y_1, \bn_2 \Delta y_2)  \sum_{s_{1}} \sum_{s_{2}} \sum_{\bl_{1}} \sum_{\bl_{2}} a_{s_1, s_2, \bl_1, \bl_2}\left(\br_1,\br_2, t\right)  O_{s_1, s_2, \bl_1, \bl_2}(\bn_1 \Delta y_1, \bn_2 \Delta y_2),
\end{split}
\end{equation}
where the two-body tensor is defined as (with $O_l(z)$ given by Eq.~\eqref{basic_osc_int})
\begin{equation*}
\begin{split}
O_{s_1, s_2, \bl_1, \bl_2}(\by_1 , \by_2 )  &= \int_{P_{s_1}} \D \bp_1 \int_{P_{s_2}} \D \bp_2~\me^{\frac{2\mi}{\hbar} \bp_1 \cdot \by_1  + \frac{2\mi}{\hbar}\bp_2 \cdot \by_2 } C_{\bl_{1}}\left(\bp_{1}\right)  C_{\bl_{2}}\left(\bp_{2}\right) \\
&= \me^{\frac{2\mi }{\hbar} \hat{\bm{d}}_{s_1} \cdot \by_1 + \frac{2 \mi }{\hbar} \hat{\bm{d}}_{s_2} \cdot \by_2} \prod_{k=1}^d \frac{|\mathcal{P}_{s_1}^{(k)}| |\mathcal{P}_{s_2}^{(k)}|}{4}   O_{l_1^{(k)}}\left(\frac{|\mathcal{P}_{s_1}^{(k)}|y_1^{(k)}}{\hbar} \right) O_{l_2^{(k)}}\left(\frac{|\mathcal{P}_{s_2}^{(k)}| y_2^{(k)}}{\hbar} \right). 
\end{split}
\end{equation*}
Again, the oscillation parts are absorbed into the precomputed oscillatory integrals. The price to pay is the computational complexity for the tensor product of oscillatory integrals, which is significantly higher than the Fourier spectral method.

\section{CHASM: A massively parallel framework}
\label{sec.chasm}

To further address the challenges induced by the storage of high-dimensional Wigner function and the stiffness inherited in the spatial advection, we employ a distributed semi-Lagrangian scheme. Combining it with efficient spectral approximations to $\pdo$ constitutes our characteristic-spectral-mixed  (CHASM) scheme \cite{XiongZhangShao2023}. In the following, we will briefly review the key ingredients in the distributed implementation of CHASM.

\subsection{Local spline interpolation}

For time integration, we use the Lawson predictor-corrector scheme
\begin{equation*}
\begin{split}
\textup{Predictor}:\widetilde{f}^{n+1}_{12}(\bR, \bP) &= f^{n}_{12}(\mathcal{A}_\tau(\bR, \bP)) + \tau \Theta_{V}[f^{n}_{12}](\mathcal{A}_\tau(\bR, \bP)), \\
\textup{Corrector}: f^{n+1}_{12}(\bR, \bP) &=  f^{n}_{12}(\mathcal{A}_\tau(\bR, \bP)) + \frac{\tau}{2} \Theta_V[\widetilde{f}^{n+1}_{12}](\bR, \bP) + \frac{\tau}{2} \Theta_{V}[f^{n}_{12}](\mathcal{A}_\tau(\bR, \bP)),
\end{split}
\end{equation*} 
where $\mathcal{A}_\tau(\bR, \bP) = (\bR - ({\tau}/{m})\bP, \bP)$, $\bR \in \mathbb{R}^D$,  $\bP \in \mathbb{R}^D$ with $D = 2d$. In order to evaluate the solutions $ f^{n}_{12}(\bR - (\tau/m)\bP, \bP))$ and  $ \Theta_V[f^{n}_{12}](\bR - (\tau/m)\bP, \bP)$ on the shifted grid, it resorts to the cubic spline interpolation, which strikes a well balance between accuracy and cost.  

The solution can be expanded as the tensor product of unidimensional splines $B_{\nu_k}$, $k = 1, \dots, D$, $\nu_k = -1, \dots, N_x^{(k)}+1$,
\begin{equation}
f^n_{12}(\bR, \bP) \approx \sum_{\nu_1 = -1}^{N_{x}^{(1)}+1} \sum_{\nu_2 = -1}^{N_{x}^{(2)}+1}  \cdots \sum_{\nu_{D} = -1}^{N_{x}^{(D)}+1} \eta^n_{\nu_1, \dots, \nu_D}(p_1, \dots, p_D) \prod_{k=1}^D B_{\nu_k}(r_k), 
\end{equation}
and $B_\nu$ is the cubic B-spline has compact support over four grid points, 
\begin{equation}
B_{\nu}(r) = 
\left\{
\begin{split}
&\frac{(r - r_{\nu-2})^3}{6h^3}, \quad  r \in [r_{\nu-2}, r_{\nu-1}],\\
&-\frac{(r - r_{\nu-1})^3}{2h^3} + \frac{(r - r_{\nu-1})^2}{2h^2} + \frac{(r - r_{\nu-1})}{2h} + \frac{1}{6}, \quad r \in [r_{\nu-1}, r_{\nu}],\\
&-\frac{(r_{\nu+1} - r)^3}{2h^3} +\frac{(r_{\nu+1} - r)^2}{2h^2} + \frac{(r_{\nu+1} - r)}{2h} + \frac{1}{6}, \quad r \in [r_{\nu}, r_{\nu+1}],\\
&\frac{(r_{\nu+2} - r)^3}{6h^3}, \quad r \in [r_{\nu+1}, r_{\nu+2}],\\
&0, \quad \textup{otherwise},
\end{split}
\right.
\end{equation}
implying that $B_{\nu - 1}, B_{\nu}, B_{\nu+1}, B_{\nu+2}$ overlap a grid interval $(r_{\nu}, r_{\nu+1})$. 

Suppose the $r$-domain in each direction is $[r_{\min}, r_{\max}]$ containing $N_x +1$ grid points with uniform spacing $h = \frac{1}{N_x}(r_{\max} - r_{\min})$, and we use $f(r_\nu)$ to denote the solution at $r_{\nu}$ for short. It requires to solve $N_x+3$ coefficients $\bm{\eta} = (\eta_{-1}, \dots, \eta_{N_x+1})$ by the following relations
\begin{equation}\label{three_term_relation}
f(r_{\nu}) = \frac{1}{6} \eta_{{\nu}-1} + \frac{2}{3} \eta_{{\nu}} + \frac{1}{6} \eta_{{\nu}+1}, \quad 0 \le {\nu} \le N_x,
\end{equation}
subject to either the periodic boundary condition $f(r_0) = f(r_{N_x})$ or the natural boundary condition $f^{\prime\prime}(r_0) = f^{\prime\prime}(r_{N_x}) = 0$ for the non-periodic case.

\begin{description}

\item[(1)] For the periodic case, $\eta_{k} = \eta_{N_x+ k}$, and it requires to solve $N_x+3$ coefficients by
\begin{equation}\label{spline_cofficient_eq}
\begin{split}
A_{per}(\eta_{-1}, \eta_{0}, \eta_{1}, \dots, \eta_{N_x}, \eta_{N_x+1})^T &= (0, f(r_0), f(r_{1}), \dots, f(r_{N_x}), 0)^T,
\end{split}
\end{equation}
where $(N_x+3)\times (N_x+3)$ coefficient matrix $A_{per}$ reads 
\begin{equation}\label{coefficient_matrix_periodic}
A_{per} = \frac{1}{6}
\begin{pmatrix}
1 & 0 & 0 & 0 & 0 & \cdots & -1 &0 &0 \\
1 & 4 & 1 & 0 & 0 & \cdots & 0 & 0 & 0\\
0 & 1     & 4 & 1     & 0 &           & 0 & 0 & 0\\
%0 &0     & 1 & 4     & 1 &  & 0 & 0 & 0\\
\vdots &\vdots & \vdots & \vdots & \vdots & \ddots & \vdots & \vdots & \vdots\\ 
%0 &           & 1 & 4 & 1 & 0 \\
0 &0 &      0     & 0 & 1 & 4 & 1 & 0 & 0\\
0 &0         &  0       & 0 & 0 & 1 & 4 & 1 & 0\\
0 & 0 &  0 & 0 & 0 & \cdots & 1 & 4 &1 \\
0 & 0 & -1 & 0 & 0 & \cdots & 0 & 0 &1 
\end{pmatrix}.
\end{equation}

\item[(2)] For the non-periodic case, the natural boundary condition $f^{\prime\prime}(r_0) = f^{\prime\prime}(r_{N_x})$  is equivalent to 
\begin{equation}\label{natural_boundary}
\frac{1}{h^2}  \eta_{-1} - \frac{2}{h^2}  \eta_{0} +  \frac{1}{h^2}  \eta_{1} = 0, \quad  \frac{1}{h^2}  \eta_{N_x-1} - \frac{2}{h^2}  \eta_{N_x} +  \frac{1}{h^2}  \eta_{N_x+1} = 0.
\end{equation}
All $N_x+3$ coefficients can be obtained by solving the algebraic equation 
\begin{equation}\label{spline_cofficient_eq}
\begin{split}
A (\eta_{-1}, \eta_{0}, \eta_{1}, \dots, \eta_{N_x}, \eta_{N_x+1})^T &= (f(r_{-1}), f(r_0), f(r_{1}), \dots, f(r_{N_x}), f(r_{N_x+1}))^T,
\end{split}
\end{equation}
where $(N_x+3)\times (N_x+3)$ coefficient matrix $A_{nat}$ reads 
\begin{equation}\label{coefficient_matrix_natural}
A_{nat} = \frac{1}{6}
\begin{pmatrix}
{6}/{h^2} & -12/h^2 & {6}/{h^2} & 0 & \cdots & 0 \\
1     & 4 & 1     & 0 &           & 0 \\
0     & 1 & 4     & 1 &  & 0 \\
\vdots & \vdots & \vdots & \vdots & \ddots & \vdots \\ 
\vdots &           & 0 & 1 & 4 & 1 \\
0         &  0       & 0 & {6}/{h^2} & -12/h^2 & {6}/{h^2} \\
\end{pmatrix}.
\end{equation}
Two ghost points $f(r_{-1})$ and $f(r_{N_x+1})$ must be added. Here we simply set $f(r_{-1}) = f(r_{N_x+1}) = 0$.

\end{description}

\subsection{Distributed parallelization}

The spline expansion is essentially global as it requires solving global algebraic equations with tridiagonal coefficient matrice. Nonetheless, we will show that the cubic spline can be reconstructed by imposing effective inner boundary conditions on the junctions of local patches \cite{MalevskyThomas1997,CrouseillesLatuSonnendrucker2009,XiongZhangShao2023}.

\begin{figure}[!h]
\centering
\subfigure[Approximation of $A^{-1}$ by $(A_M^{(l)})^{-1}$. (left: natural boundary, right: periodic boundary)\label{domain_decomposition}]{
{\includegraphics[width=0.49\textwidth,height=0.24\textwidth]{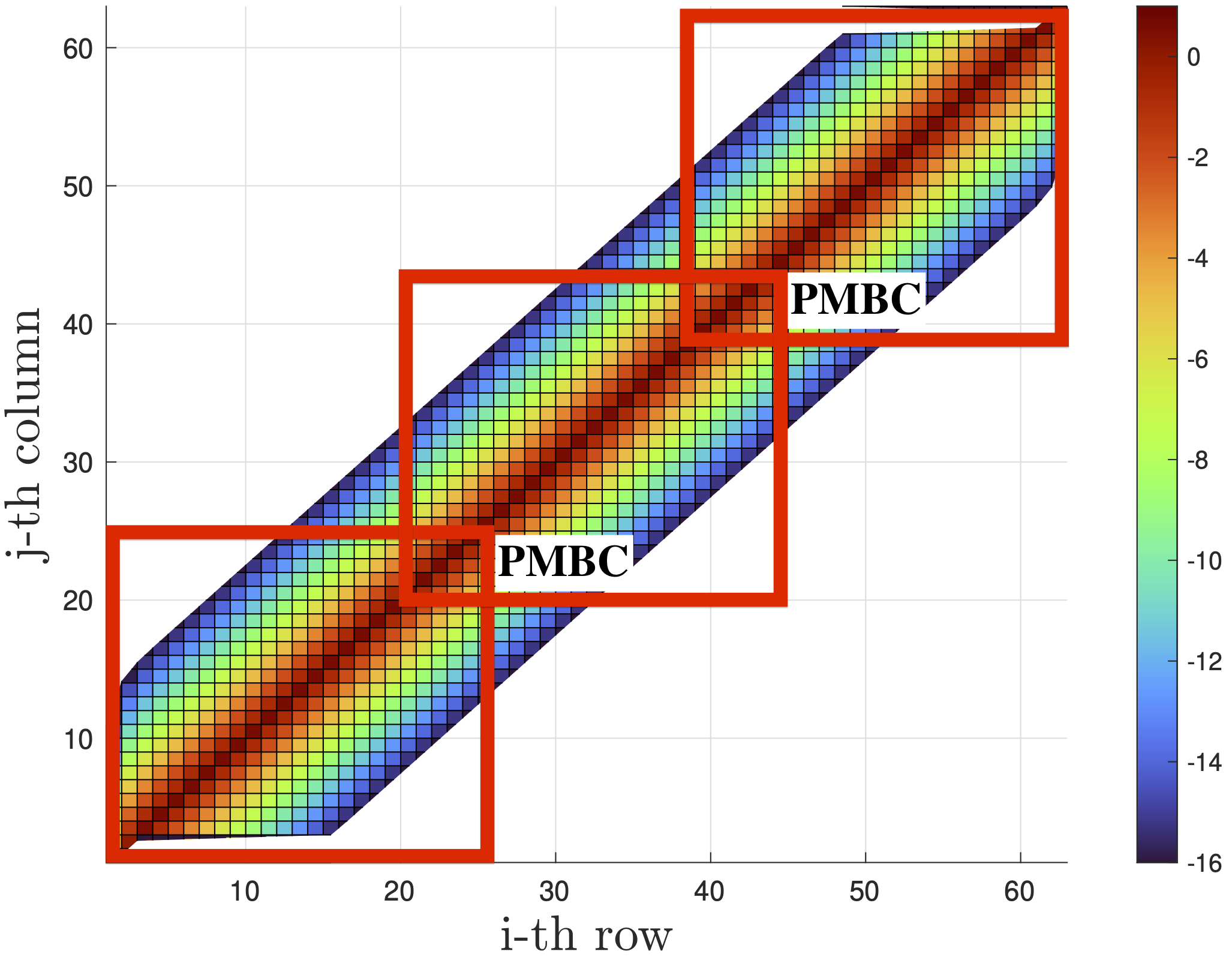}}
{\includegraphics[width=0.49\textwidth,height=0.24\textwidth]{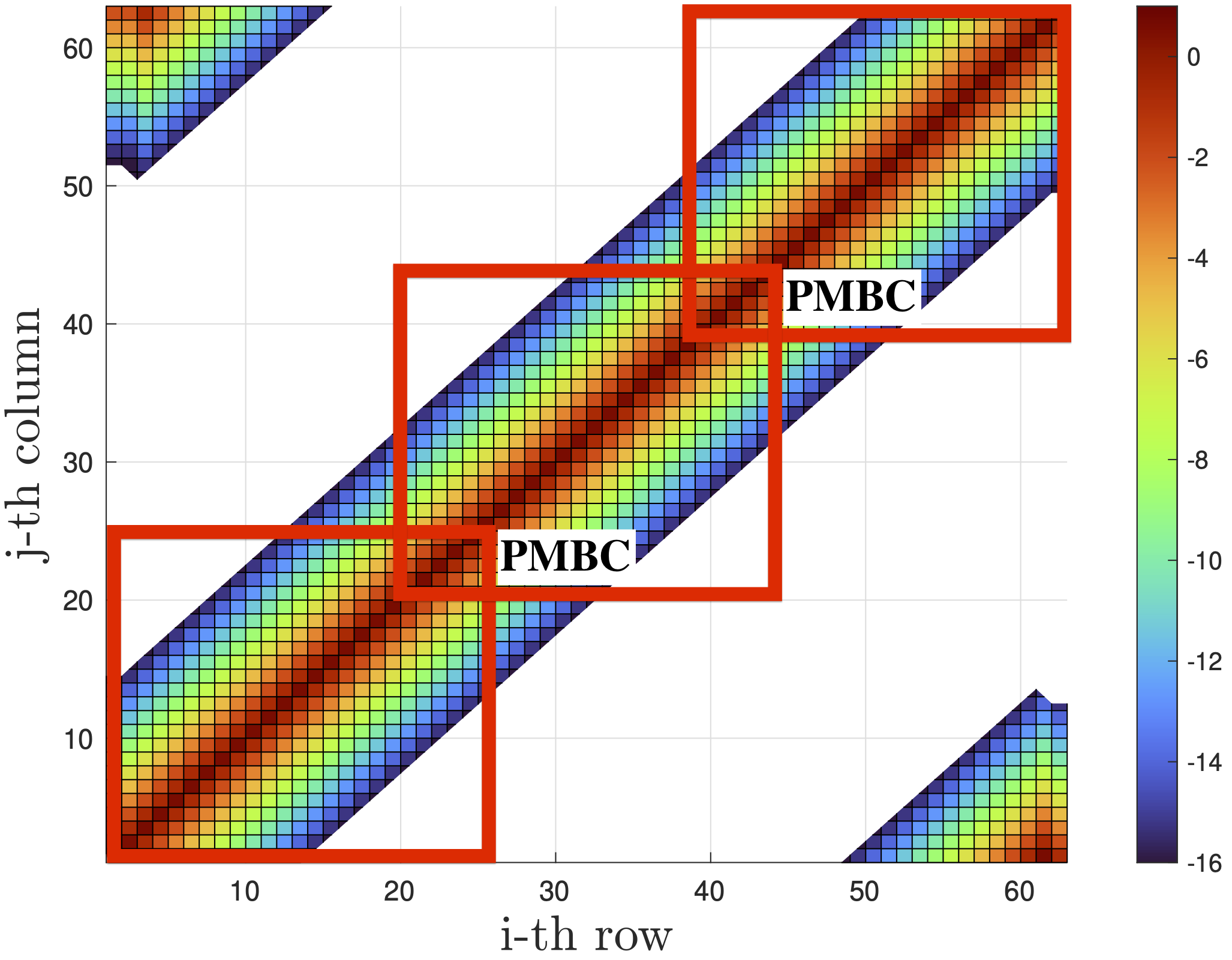}}}
\\
\centering
\subfigure[The element $|b_{ij}|$ of $A^{-1}$ in log10 scale. (left: natural boundary, right: periodic boundary)\label{element_decay}]{
\includegraphics[width=0.49\textwidth,height=0.24\textwidth]{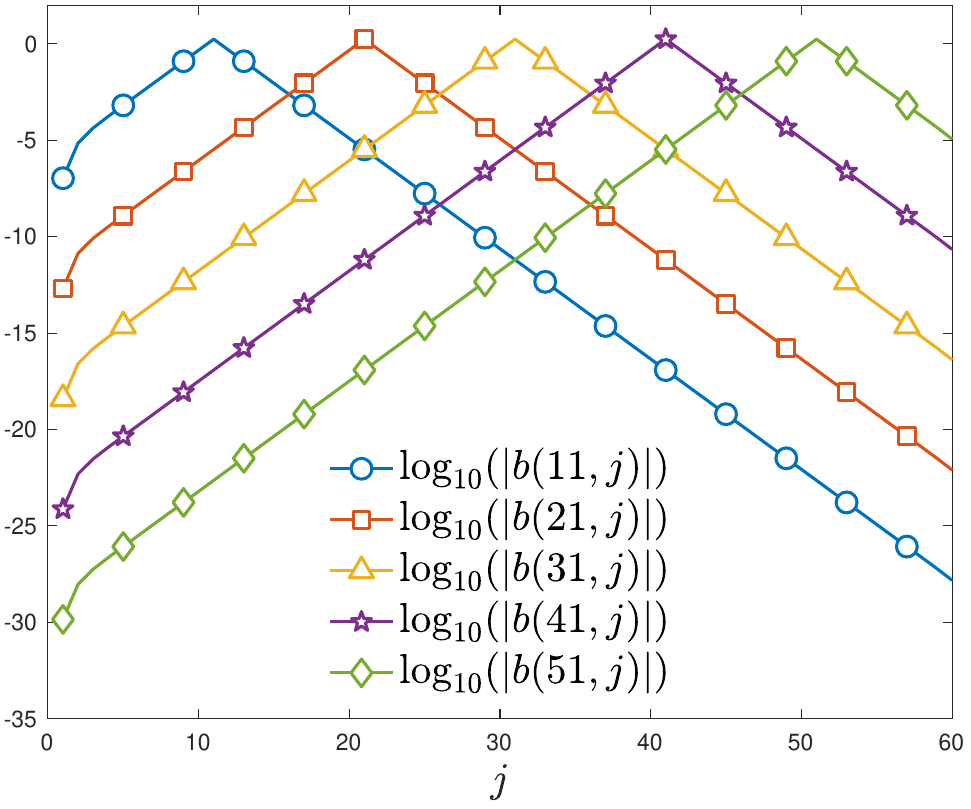}
\includegraphics[width=0.49\textwidth,height=0.24\textwidth]{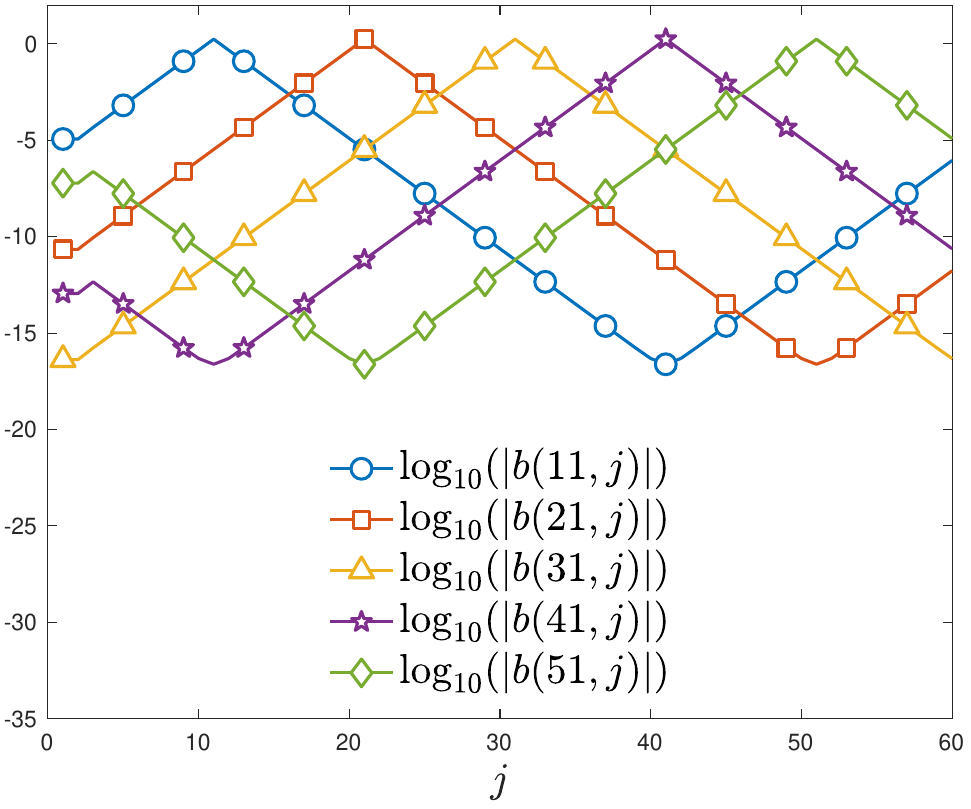}}
\caption{The non-diagonal elements $b_{ij}$ in $A^{-1}$ decay exponentially away from the main diagonal $b_{ii}$, and the coefficients ${\bm{\eta}} = A^{-1}(f(r_{-1}), \dots, f(r_{N_x+1}))^{T}$ can be recovered by $\widetilde{\bm{\eta}}^{(l)} =  (A_M^{(l)})^{-1}(\phi_L^{(l)}, f(r_0^{(l)}), \dots, f(r_M^{(l)}), \phi_R^{(l)})^T$. }
\end{figure}

Hereafter we focus on the distributed parallelization of B-spline interpolation, namely, the spline needs to be decomposed into several pieces and stored in multiple processors. In each direction, we can divide $N_x+1$ grid points on a line into $p$ uniform parts, with $M = N_x/p$,
\begin{align*}
\underbrace{r_0 < r_1 < \cdots < r_{M-1}}_{\textup{the first processor}} < \underbrace{r_{M}}_{\textup{shared}} < \cdots < \underbrace{r_{(p-1)M}}_{\textup{shared}} < \underbrace{r_{(p-1)M+1} < \cdots < r_{pM}}_{\textup{$p$-th processor}},
\end{align*}
The grid points $r_{M}, r_{2M}, \dots, r_{(p-1)M}$ are shared by the adjacent patches. The target is to recover the global B-spline by the local spline coefficients $\bm{\widetilde \eta}^{(l)}$ for $l$-th piece, that is, 
\begin{equation}\label{relation}
\bm{\widetilde\eta }^{(l)}= (\widetilde \eta_{-1}^{(l)}, \dots, \widetilde \eta_{M+1}^{(l)}) = ( \eta_{-1 +(l-1)M}, \dots,  \eta_{M+1 + (l-1)M}), \quad l = 1, \dots, p.
\end{equation}
This can be realized by imposing perfectly matched boundary conditions (PMBC) on two ends of local splines (see Figure \ref{domain_decomposition}) \cite{CrouseillesLatuSonnendrucker2009,XiongZhangShao2023},
\begin{equation}
-\frac{1}{2h} \widetilde \eta_{-1}^{(l)} + \frac{1}{2h} \widetilde \eta_{1}^{(l)}= \phi_L^{(l)}, \quad -\frac{1}{2h} \widetilde \eta_{M-1}^{(l)} + \frac{1}{2h} \widetilde \eta_{M+1}^{(l)}= \phi_R^{(l)},
\end{equation}
so that all the coefficients $ \widetilde{\bm{\eta}}^{(l)}= (\widetilde \eta_{-1}^{(l)}, \dots, \widetilde \eta_{M+1}^{(l)})$ in each patch can be obtained by solving the following equation independently,
\begin{equation}\label{spline_cofficient_eq_piece}
A_{M}^{(l)}
\begin{pmatrix}
\widetilde \eta^{(l)}_{-1} \\
\widetilde \eta^{(l)}_{0} \\
\widetilde \eta^{(l)}_{1} \\
\vdots \\
\widetilde \eta^{(l)}_{M} \\
\widetilde \eta^{(l)}_{M+1} \\
\end{pmatrix}
=
\frac{1}{6}
\begin{pmatrix}
-\frac{3}{h} & 0 & \frac{3}{h} & 0 & \cdots & 0 \\
1     & 4 & 1     & 0 &           & \vdots \\
0     & 1 & 4     & 1 &  & \vdots \\
\vdots & \vdots & \vdots & \vdots & \ddots & \vdots \\ 
\vdots &           & 0 & 1 & 4 & 1 \\
0         &  0       & 0 & -\frac{3}{h} & 0 & \frac{3}{h} \\
\end{pmatrix}
\begin{pmatrix}
\phi^{(l)}_L \\
f(r_{0 + (l-1)M}) \\
f(r_{1 + (l-1)M}) \\
\vdots \\
f(r_{M + (l-1)M}) \\
\phi^{(l)}_R \\
\end{pmatrix}.
\end{equation}

Now we need to specify $\phi_L^{(l)}$ and $\phi_R^{(l)}$ by mathcing $\widetilde \eta_{k}^{(l)}$ and $ \eta_{k+(l-1)M}$. The coefficients of the global spline can be solved by inverting $A$ in Eq.~\eqref{spline_cofficient_eq}. Denote by $(b_{ij}) = {A}^{-1}, -1\le i, j \le pM+1$, then the solution $\eta_i$ of Eq.~\eqref{spline_cofficient_eq} can be represented by
\begin{equation}\label{exact_solution}
\eta_i = b_{ii} f(r_i) + \sum_{j=-1}^{i-1} b_{ij} f(r_j) + \sum_{j = i+1}^{pM+1} b_{i j} f(r_j), \quad i = -1, \dots, pM+1.
\end{equation}

In order to avoid global communication, we utilize a key observation that the non-diagonal elements $b_{ij}$ decay exponentially away from the main diagonal $b_{ii}$ due to the rapid decay of the wavelet basis in its dual space \cite{MalevskyThomas1997}, which is clearly visualized in Figure \ref{element_decay}. Therefore, it allows us to truncate $b_{ij}$ when $|i - j| \ge n_{nb}$ for sufficiently large $n_{nb}$ and obtain
 \begin{equation}\label{truncated_solution}
\eta_i \approx b_{ii} f(r_i) + \sum_{j=i - (n_{nb}-1)}^{i-1} b_{ij} f(r_j) + \sum_{j = i+1}^{i + n_{nb} - 1} b_{i j} f(r_j),  \quad i = -1, \dots, pM+1.
\end{equation}
Now using the truncated stencils \eqref{truncated_solution} and further adding four more terms to complete the summations from $-n_{nb}$ to $n_{nb}$, it yields that
\begin{equation*}
\begin{split}
-\frac{1}{2h} \widetilde \eta^{(l+1)}_{-1} + \frac{1}{2h} \widetilde \eta^{(l+1)}_{1} \approx &\underbrace{\sum_{j=-n_{nb}}^{-1} \left(-\frac{1}{2h} b_{lM-1, lM+j} + \frac{1}{2h} b_{lM+1, lM+j}\right) f(r_{lM+j})}_{\textup{stored in left processor}} \\
&~+\underbrace{\sum_{j=1}^{n_{nb}} \left(-\frac{1}{2h} b_{lM-1, lM+j} + \frac{1}{2h} b_{lM+1, lM+j}\right) f(r_{lM+j})}_{\textup{stored in right processor}}. 
\end{split}
\end{equation*}
This gives the formulation of PMBC for $1 \le l \le p-1$,
\begin{equation}\label{PMBC}
\begin{split}
\phi_{R}^{(l)} =  \phi_{L}^{(l+1)} \approx & \underbrace{\frac{1}{2}c_{0,l} f(r_{lM}) +  \sum_{j = 1}^{n_{nb}} c_{j, l}^{-} f(r_{lM-j})}_{\textup{stored in left processor}} + \underbrace{\frac{1}{2}c_{0,l} f(r_{lM}) + \sum_{j = 1}^{n_{nb}} c_{j, l}^{+} f(r_{lM+j})}_{\textup{stored in right processor}},
\end{split}
\end{equation}
where
\begin{equation}\label{PMBC_coeffcients}
\begin{split}
&c_{j, l}^+ =  -\frac{b_{lM-1, lM+j}}{2h} + \frac{b_{lM+1, lM+j}}{2h}, \quad c_{j, l}^- =  -\frac{b_{lM-1, lM-j}}{2h} + \frac{b_{lM+1, lM-j}}{2h},
\end{split}
\end{equation}
and  $c_{0, l} = c_{0, l}^+ = c_{0, l}^- = -\frac{b_{lM-1, lM}}{2h} + \frac{b_{lM+1, lM}}{2h}$. 

Finally, the effective boundary conditions $\phi_{L}^{(1)}$ and $\phi_{R}^{(p)}$ on both ends should match the boundary conditions of the global cubic spline, namely, 
\begin{align}
&\phi_{L}^{(1)} = \frac{\eta_1-\eta_{-1}}{2h} \approx  c_{0, 0} f(r_0) + \sum_{j=1}^{n_{nb}} c_{j, p}^{-} f(r_{N_x-j}) + \sum_{j=1}^{n_{nb}} c_{j, 1}^{+} f(r_j),\\
&\phi_{R}^{(p)} = \frac{\eta_{N_x+1}-\eta_{N_x-1}}{2h} \approx c_{0, p} f(r_{N_x}) + \sum_{j=1}^{n_{nb}} c_{j, p}^{+}f(r_{N_x-j}) +  \sum_{j=1}^{n_{nb}} c_{j, 1}^{-}f(r_{j}). 
\end{align}

%\begin{align}
%\phi_{L}^{(1)} &=\frac{\eta_1-\eta_{-1}}{2h}\approx \sum_{j=-1}^{n_{nb}} c_{j, 0}^{-} f(r_j), \quad c_{j, 0}^{-} =  \frac{-{b}_{-1, j}  + {b}_{1, j}}{2h}, \\
%\phi_{R}^{(p)} &= \frac{\eta_{N_x+1}-\eta_{N_x-1}}{2h} \approx \sum_{j=-1}^{n_{nb}} c_{j, p}^{+}f(r_{N_x-j}), \quad c_{j, p}^{+} = \frac{-{b}_{N_x-1, N_x-j}  + {b}_{N_x+1, N_x-j}}{2h}.
%\end{align}

\section{Numerical experiments}
\label{sec.num}

This section begins to investigate the correction of dynamical XC functionals and two-body scattering in  quantum kinetic simulations and one-dimensional strongly correlated system. For convenience, we adopt the atomic units $\hbar = e = m = 1$.  Our numerical study is somehow preliminary as the XC functionals are greatly simplified. Nonetheless, it provides an intuitive picture of underlying physical mechanisms, which may contribute to a better understanding of the XC effects and many-body correction. We believe this makes the first step towards the practical usage of 2-RDM dynamics.

All the numerical experiments using our CHASM scheme are performed via our own Fortran implementation with the computing platform: 2*AMD EPYC 7773X (2.20 GHz, 768 MB cache, 64 cores, 128 threads) with 1024 GB memory $@$3200Mhz. The mesh size ranges from $1.66\times 10^8$ ($161^3\times 80^3$) to $1.08\times 10^9$ ($257^2\times 128^2$), and the domain is evenly divided into $8 \times 8$ patches and distributed by 64 processes via the Message Passing Interface (MPI), and each process manipulates 2 threads realized by the OpenMP library.

\subsection{Convergence test}
\label{sec_convergence}

As the first step, we need to investigate the convergence of the spectral approximations to $\pdo$, while the accuracy of the distributed semi-Lagrangian scheme and scalability of CHASM have already been thoroughly tested in \cite{XiongZhangShao2023} and thus omitted. 

 The relative $L^{\infty}$-error $\varepsilon_{\infty}(t)$ and the relative $L^2$-error $\varepsilon_{2}(t)$ are adopted as the performance metrics, 
 \begin{equation}
 \begin{split}
\varepsilon_{\infty}(t) &= \frac{\max_{(\bm{r},\bm{p})\in\Omega}\big |f_{12}^{\textup{ref}}\left(\bm{r},\bm{p},t\right)-f_{12}^{\textup{num}}\left(\bm{r},\bm{p},t\right) \big |}{{\max_{(\bm{r},\bm{p})\in\Omega}\big |f_{12}^{\textup{ref}}\left(\bm{r},\bm{p},t\right) \big |}}, \\
\varepsilon_{2}(t) &= \frac{\left[\iint_{\Omega} \left(f_{12}^{\textup{ref}}\left(\bm{r},\bm{p},t\right)-f_{12}^{\textup{num}}\left(\bm{r},\bm{p},t\right)\right)^{2}\textup{d}\bm{r}\textup{d} \bm{p}\right]^{\frac{1}{2}}}{\max_{(\bm{r},\bm{p})\in\Omega}\big |f_{12}^{\textup{ref}}\left(\bm{r},\bm{p},t\right) \big |},
\end{split}
\end{equation}
%\begin{align}
%\varepsilon_{2}(t) &= [\iint_{\Omega} \left(f^{\textup{ref}}\left(\bm{x},\bm{k},t\right)-f^{\textup{num}}\left(\bm{x},\bm{k},t\right)\right)^{2}\textup{d}\bm{x}\textup{d} \bm{k}]^{\frac{1}{2}},\label{eq:e2}
%\end{align}
%\begin{align}
%\varepsilon_{\infty}(t) &=\max_{(\bm{x},\bm{k})\in\Omega}\big |f^{\textup{ref}}\left(\bm{x},\bm{k},t\right)-f^{\textup{num}}\left(\bm{x},\bm{k},t\right) \big |, \label{eq:ef}
%\end{align}
%\begin{align}
%\varepsilon_{\textup{mass}}(t) &= \Big |\iint_{\Omega} f^{\textup{num}}\left(\bm{x},\bm{k},t\right)\textup{d}\bm{x}\textup{d} \bm{k}-\iint_{\Omega} f^{\textup{ref}}\left(\bm{x},\bm{k},t=0\right)\textup{d}\bm{x}\textup{d} \bm{k} \Big |,\label{eq:emass}
%\end{align}
with $f_{12}^{\textup{ref}}$ and $f_{12}^{\textup{num}}$ the reference and numerical solution, respectively, and $\Omega$ denotes the computational domain. The integral can be approximated by the average over all grid points.
%Besides, the relative maximal error and relative $L^2$-error are obtained by $\frac{\varepsilon_{\infty}(t)}{\max(|f(\bx, \bk, 0)|)}$ and $\varepsilon_{2}(t)/ \sqrt{\iint(|f(\bx, \bk, 0)|^2 \D \bx \D \bk)}$, respectively.  

\subsubsection{Convergence test of FSM and SEM for the nonlinear exchange potential}

Consider the exchange potential
\begin{equation}
V_{XC}(r_1, r_2) = \iiint_{\mathbb{R}^3} f(r_1, r_2, p_1, p_2) \D r_2 \D p_1 \D p_2 + \iiint_{\mathbb{R}^3} f(r_1, r_2, p_1, p_2) \D r_1 \D p_1 \D p_2,
\end{equation}
and let
\begin{equation}
f_{12}(r_1, r_2, p_1, p_2) = \pi^{-2} \me^{- \frac{(r_1 + r_1^0)^2}{2} - 2 |p_1|^2} \me^{- \frac{(r_2 + r_2^0)^2}{2} - 2 |p_2|^2},
\end{equation}
then the exact form the exchange potential is $V_{XC}(r_1, r_2) = V_{XC}[n_1](r_1) + V_{XC}[n_2](r_2)$, where
\begin{equation}
V_{XC}[n_1](r_1) = \frac{1}{\sqrt{2\pi}} \me^{-\frac{(r_1 + r_1^0)^2}{2}}, \quad  V_{XC}[n_2](r_2) = \frac{1}{\sqrt{2\pi}} \me^{-\frac{(r_2 + r_2^0)^2}{2}}.
\end{equation}

Under $\hbar = 1$, $p^{\prime} = \hbar k^{\prime} = k^{\prime} $,  the first Wigner kernel reads that
\begin{equation}
\begin{split}
V_W[n_1](r_1, p^{\prime}) &= \frac{1}{\mi\pi} \left[\me^{2\mi r_1 \cdot p^{\prime}} \mathcal{F}_{r_1 \to 2p^{\prime}}V_{XC}[n_1](r_1) -   \me^{-2 \mi r_1 \cdot p^{\prime}}  \mathcal{F}_{r_1 \to -2 p^{\prime}} V_{XC}[n_1](r_1)\right] \\
& =  \frac{1}{\mi\pi}  \left[ \me^{2 \mi (r_1 + r_1^0) \cdot p^{\prime}} e^{-2|p^{\prime}|^2} - \me^{-2 \mi (r_1 + r_1^0) \cdot p^{\prime}} e^{-2|p^{\prime}|^2}\right]
\end{split}
\end{equation}
and the exact form of the pseudo-differential operator reads that
\begin{equation}
\begin{split}
& \Theta_{V_{XC}[n_1]}[f_{12}](r_1, r_2, p_1, p_2) = \int_{\mathbb{R}} V_W[n_{1}](r_1,  p^{\prime}) f_{12}(r_1, r_2, p_1 - p^{\prime}, p_2) \D p^{\prime} \\
 & = \frac{1}{\mi\pi^3} \me^{- \frac{(r_1 + r_1^0)^2}{2}} \me^{- \frac{(r_2 + r_2^0)^2}{2} - 2 |p_2|^2} \int_{\mathbb{R}} \left[ \me^{2 \mi (r_1 + r_1^0) \cdot p^{\prime}} e^{-2|p^{\prime}|^2} - \me^{-2 \mi (r_1 + r_1^0) \cdot p^{\prime}} e^{-2|p^{\prime}|^2}\right] \me^{-2|p_1 - p^{\prime}|^2} \D p^{\prime} \\
  &  =\frac{1}{\mi\pi^3 } \me^{- \frac{(r_1 + r_1^0)^2}{2}} \me^{- \frac{(r_2 + r_2^0)^2}{2} - 2 |p_2|^2 - |p_1|^2} \int_{\mathbb{R}} \left[ \me^{2 \mi (r_1 + r_1^0) \cdot p^{\prime}}  - \me^{-2 \mi (r_1 + r_1^0) \cdot p^{\prime}}\right] \me^{-  4| p^{\prime} -  \frac{1}{2} p_1|^2} \D p^{\prime} \\
  & = \frac{1}{\pi^{5/2}}  \me^{- \frac{3(r_1 + r_1^0)^2}{4} - |p_1|^2}\me^{- \frac{(r_2 + r_2^0)^2}{2} - 2|p_2|^2}   \sin \left((r_1 + r_1^0) \cdot p_1\right).
   \end{split}
\end{equation}
Similarly, 
\begin{equation}
 \Theta_{V_{XC}[n_2]}[f_{12}](r_1, r_2, p_1, p_2)= \frac{1}{\pi^{5/2}}  \me^{- \frac{3(r_2 + r_2^0)^2}{4} - |p_2|^2}\me^{- \frac{(r_1 + r_1^0)^2}{2} - 2 |p_1|^2}  \sin \left((r_2 + r_2^0) \cdot p_2\right),
 \end{equation}
so that
\begin{equation}
\begin{split}
&\iint_{\mathbb{R}^2} \Theta_{V_{XC}[n_1]}[f_{12}](r, r_2, p, p_2) \D r_2 \D p_2 +  \iint_{\mathbb{R}^2} \Theta_{V_{XC}[n_2]}[f_{12}](r_1, r, p_1, p) \D r_1 \D p_1 \\
& = \frac{1}{\pi^{3/2}}  \me^{- \frac{3(r + r_1^0)^2}{4}} \me^{- |p|^2} \sin \left((r + r_1^0) \cdot p\right) + \frac{1}{\pi^{3/2}}  \me^{- \frac{3(r + r_2^0)^2}{4}} \me^{- |p|^2} \sin \left((r + r_2^0) \cdot p\right).
\end{split}
\end{equation} 

Tables \ref{XC_FSM_convergence_data} and \ref{XC_SEM_convergence_data} present the numerical errors and the computational time  (without  parallelization) of FSM and SEM, respectively,  under the computational domain $\Omega = [-10, 10]^2 \times [-\frac{5\pi}{2}, \frac{5\pi}{2}]^2$. The spectral convergence of both methods is confirmed. Although it is difficult to compare the computational time due to different mesh strategies, it is still seen that FSM and SEM are  comparably efficient for the mean-field $\pdo$.  

\begin{table}[!h]
  \centering
  \caption{\small XC potential: Numerical errors and computational time of FSM.}
  \label{XC_FSM_convergence_data}
   \begin{tabular}{c|cc|c|c|c}
    \hline\hline
   Convergence & $N_x$ &  $N_p$ & $\varepsilon_\infty$ & $\varepsilon_2$ & Time(s)\\
   \hline
    \multirow{6}{*}{\includegraphics[scale=0.18]{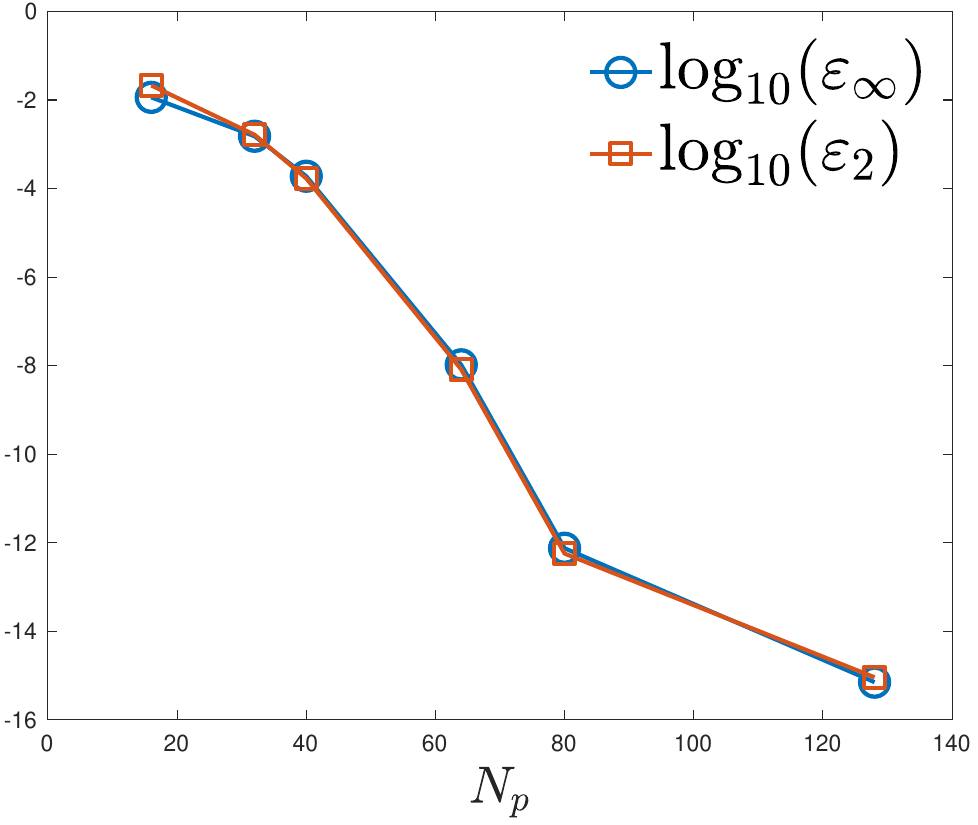}}  
&	33	&	16	&	1.148$\times10^{-2}$	&	2.136$\times10^{-2}$	&	0.017 	\\
&	65	&	32	&	1.519$\times10^{-3}$	&	1.636$\times10^{-3}$	&	0.448 	\\
&	81	&	40	&	1.898$\times10^{-4}$	&	1.695$\times10^{-4}$	&	1.348 	\\
&	129	&	64	&	1.042$\times10^{-8}$	&	8.215$\times10^{-9}$	&	16.000 	\\
&	161	&	80	&	7.503$\times10^{-13}$	&	5.680$\times10^{-13}$	&	42.377 	\\
&	257	&	128	&	7.078$\times10^{-16}$	&	9.111$\times10^{-16}$	&	701.480 	\\
   \hline\hline
 \end{tabular}  
  \end{table}

\begin{table}[!h]
  \centering
  \caption{\small XC potential: Numerical errors and computational time of SEM.}
  \label{XC_SEM_convergence_data}
   \begin{tabular}{c|cccc|c|c|c}
    \hline\hline
   Convergence & $N_x$	&$N_p$	& $M$ &  $N$ & $\varepsilon_\infty$ & $\varepsilon_2$ & Time(s)\\
   \hline
    \multirow{6}{*}{\includegraphics[scale=0.18]{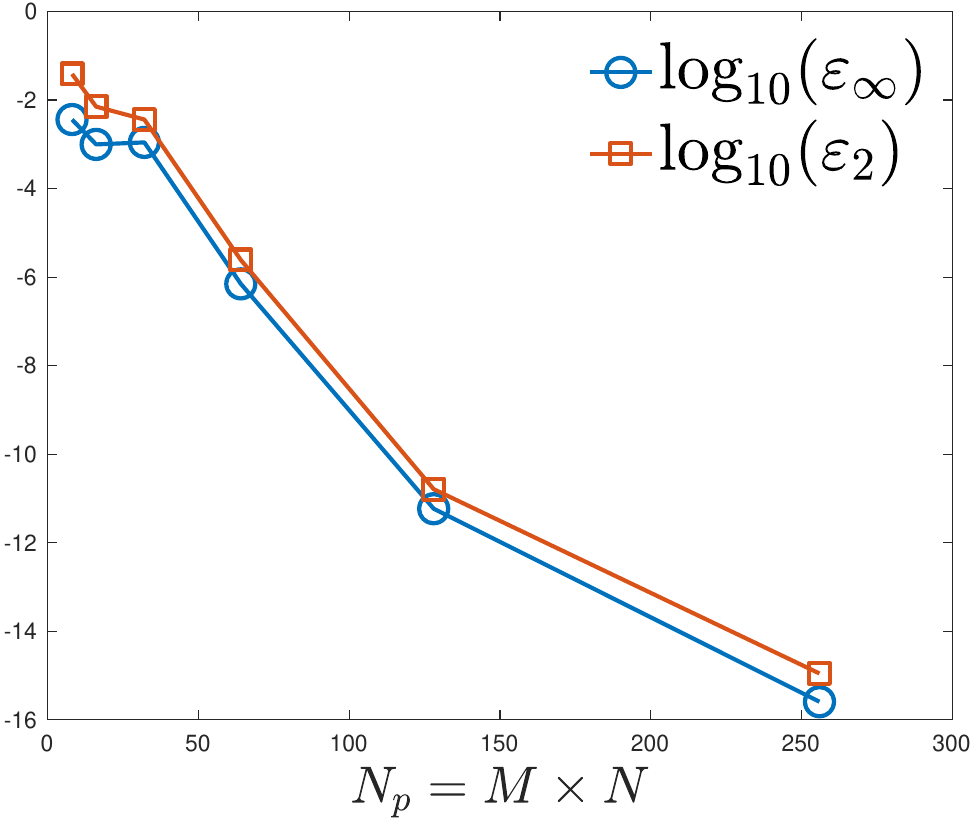}}  
&	101	&	8	&	4	&	2	&	3.618$\times10^{-3}$	&	3.872$\times10^{-2}$	&	0.380 	\\
&	101	&	16	&	4	&	4	&	9.930$\times10^{-4}$	&	7.083$\times10^{-3}$	&	1.585 	\\
&	101	&	32	&	8	&	4	&	1.102$\times10^{-3}$	&	3.622$\times10^{-3}$	&	6.399 	\\
&	101	&	64	&	8	&	8	&	7.033$\times10^{-7}$	&	2.439$\times10^{-6}$	&	26.321 	\\
&	101	&	128	&	8	&	16	&	5.858$\times10^{-12}$	&	1.618$\times10^{-11}$	&	117.419 	\\
&	101	&	256	&	8	&	32	&	2.560$\times10^{-16}$	&	1.112$\times10^{-15}$	&	757.915 	\\
   \hline\hline
 \end{tabular}  
  \end{table}

\subsubsection{Convergence test of FSM and SEM for the two-body scattering}

 Consider the two-body interaction of the Gaussian type
\begin{equation}
V_{12}(r_1, r_2) = V_{ee}(|r_1 - r_2|) = H_B e^{-|r_1 - r_2|^2/2}
\end{equation}
and
\begin{equation}
f_{12}(r_1, r_2, p_1, p_2) = \pi^{-2} \me^{- (r_1 + r_1^0)^2 -  |p_1|^2} \me^{- (r_2 + r_2^0)^2 - |p_2|^2}.
\end{equation}
Then under $\hbar=1$, $\pdo$ reads that
\begin{equation*}
\begin{split}
 &\Theta_{V_{12}}[f_{12}](r_1, r_2, p_1, p_2) \\
& =  \frac{2 H_B}{\mi \sqrt{2\pi}} \int_{\mathbb{R}} \D p^{\prime} \me^{-2 \mi p^{\prime} \cdot (r_1 - r_2)} \me^{-2(p^{\prime})^2} \left\{ f_{12}(r_1, r_2, p_1 - p^{\prime}, p_2+ p^{\prime}) -  f_{12}(r_1, r_2, p_1 + p^{\prime}, p_2 - p^{\prime})\right\} \\
& = \frac{ \sqrt{2} H_B}{\mi\pi^{5/2} } \me^{- (r_1 + r_1^0)^2 - (r_2 + r_2^0)^2} \me^{- p_1^2 - p_2^2 + \frac{(p_1 - p_2)^2}{4}}  \int_{\mathbb{R}} \D p^{\prime}~\me^{-2 \mi p^{\prime} \cdot (r_1 - r_2)}\left[\me^{-4(p^{\prime} - \frac{p_1 - p_2}{4})^2} - \me^{-4(p^{\prime} + \frac{p_1 - p_2}{4})^2} \right] \\
& =  \frac{\sqrt{2} H_B}{\pi^2} \me^{- (r_1 + r_1^0)^2 - (r_2 + r_2^0)^2 -\frac{(r_1 - r_2)^2}{4}} \me^{- p_1^2 - p_2^2 + \frac{(p_1 - p_2)^2}{4}} \sin\frac{(r_1- r_2)(p_1 - p_2)}{2}.
\end{split}
\end{equation*}

Tables \ref{2B_FSM_convergence_data} and \ref{2B_SEM_convergence_data} present the numerical errors and the computational time  (without parallelization) of FSM and SEM, respectively,  under the computational domain $\Omega = [-10, 10]^2 \times [-\frac{5\pi}{2}, \frac{5\pi}{2}]^2$. Again, both methods are able to achieve spectral convergence. Here it is observed that FSM seems to be more efficient than SEM  for two-body scattering part,  as double integrals are greatly simplified in the pseudo-difference approach.
\begin{table}[!h]
  \centering
  \caption{\small Two-body scattering: Numerical errors and computational time of FSM.}
  \label{2B_FSM_convergence_data}
   \begin{tabular}{c|cc|c|c|c}
    \hline\hline
   Convergence & $N_x$ &  $N_p$ & $\varepsilon_\infty$ & $\varepsilon_2$ & Time(s)\\
   \hline
    \multirow{6}{*}{\includegraphics[scale=0.18]{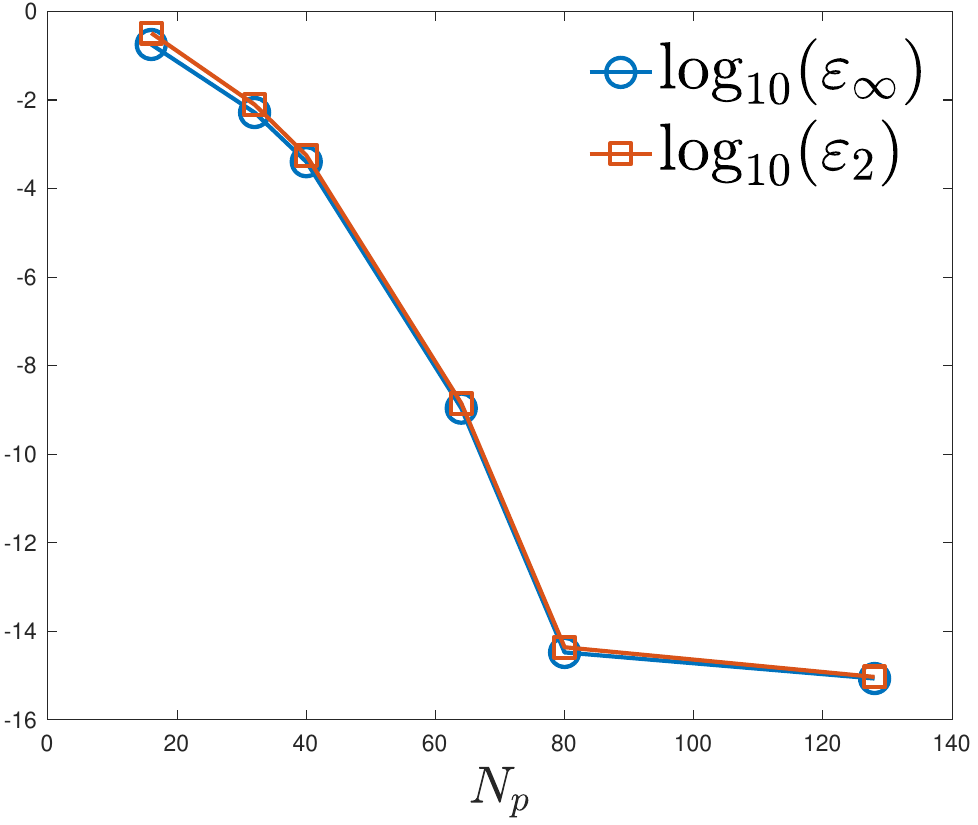}}  
&	33	&	16	&	1.795$\times10^{-1}$	&	3.257$\times10^{-1}$	&	0.011 	\\
&	65	&	32	&	5.210$\times10^{-3}$	&	7.877$\times10^{-3}$	&	0.265 	\\
&	81	&	40	&	4.037$\times10^{-4}$	&	5.646$\times10^{-4}$	&	0.771 	\\
&	129	&	64	&	1.094$\times10^{-9}$	&	1.418$\times10^{-9}$	&	7.610 	\\
&	161	&	80	&	3.324$\times10^{-15}$	&	4.356$\times10^{-15}$	&	23.916 	\\
&	257	&	128	&	8.552$\times10^{-16}$	&	9.289$\times10^{-16}$	&	274.068 	\\
   \hline\hline
 \end{tabular}  
  \end{table}

\begin{table}[!h]
  \centering
  \caption{\small Two-body scattering: Numerical errors and computational time of SEM.}
  \label{2B_SEM_convergence_data}
   \begin{tabular}{c|cccc|c|c|c}
    \hline\hline
   Convergence & $N_x$	&$N_p$	& $M$ &  $N$ & $\varepsilon_\infty$ & $\varepsilon_2$ & Time(s)\\
   \hline
    \multirow{6}{*}{\includegraphics[scale=0.18]{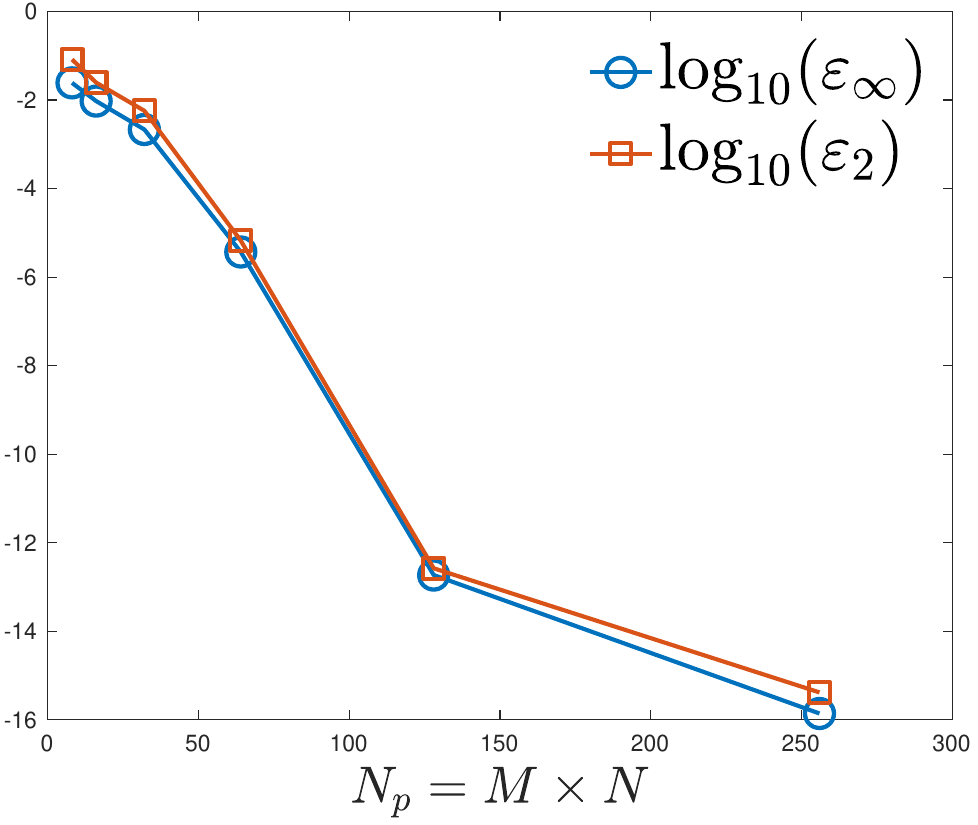}}  
&	101	&	8	&	4	&	2	&	2.439$\times10^{-2}$	&	8.241$\times10^{-2}$	&	25.7192	\\
&	101	&	16	&	4	&	4	&	9.453$\times10^{-3}$	&	2.472$\times10^{-2}$	&	29.0108	\\
&	101	&	32	&	8	&	4	&	2.158$\times10^{-3}$	&	5.707$\times10^{-3}$	&	126.4249	\\
&	101	&	64	&	8	&	8	&	3.687$\times10^{-6}$	&	6.646$\times10^{-6}$	&	139.5739	\\
&	101	&	128	&	8	&	16	&	1.859$\times10^{-13}$	&	2.644$\times10^{-13}$	&	174.3935	\\
&	101	&	256	&	8	&	32	&	1.388$\times10^{-16}$	&	4.155$\times10^{-16}$	&	250.9817	\\
   \hline\hline
 \end{tabular}  
  \end{table}

\subsubsection{Effect of zero-padding for the Hartree potential}

\begin{figure}[!h]
\subfigure[Approximated Hartree potential under different $\tilde N$ and convergence with respect to $\tilde N/N_x$.\label{zero_padding}]{
\includegraphics[width=0.48\textwidth,height=0.27\textwidth]{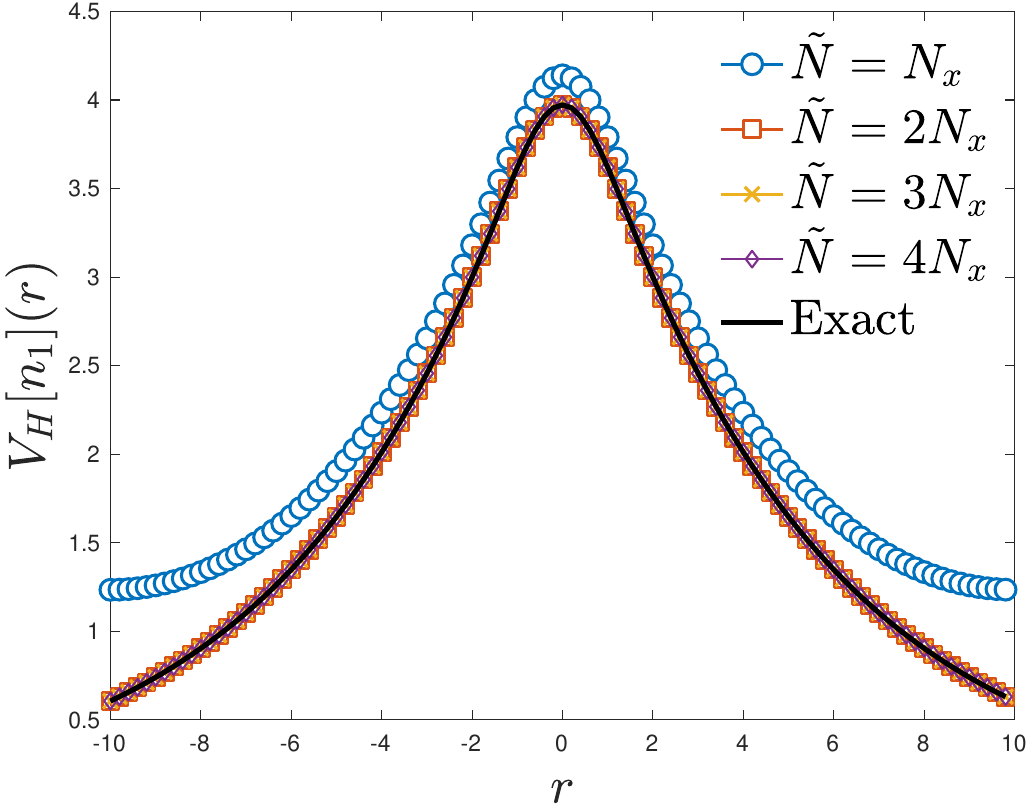}
\includegraphics[width=0.48\textwidth,height=0.27\textwidth]{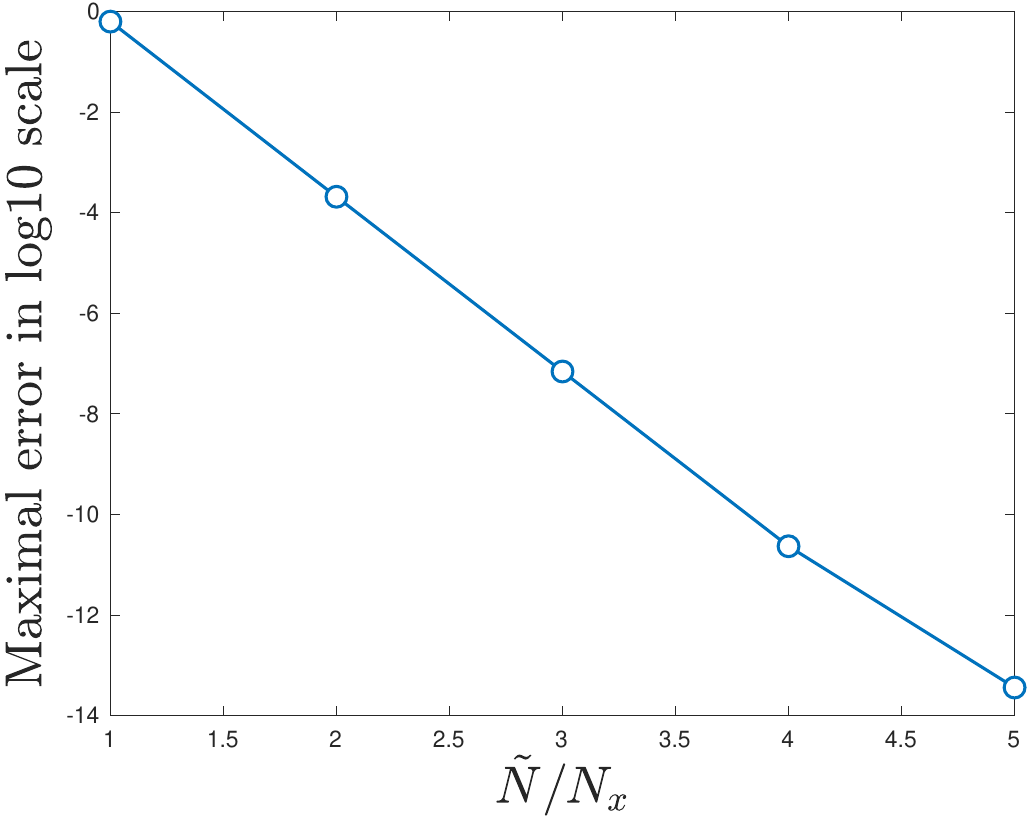}}
\\
\centering
\subfigure[$\pdo$ for the Hartree potential (left: FSM, middle SEM, right: their difference approaches machine precision).\label{Hartree_comparsion}]{
\includegraphics[width=0.32\textwidth,height=0.22\textwidth]{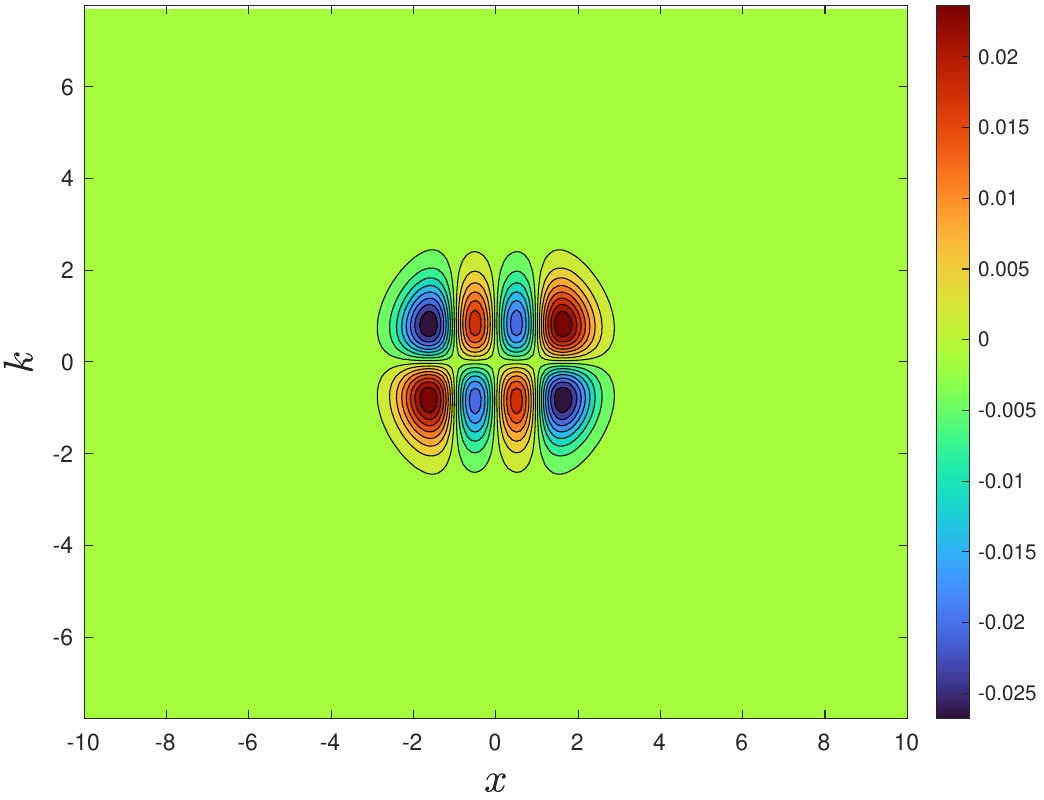}
\includegraphics[width=0.32\textwidth,height=0.22\textwidth]{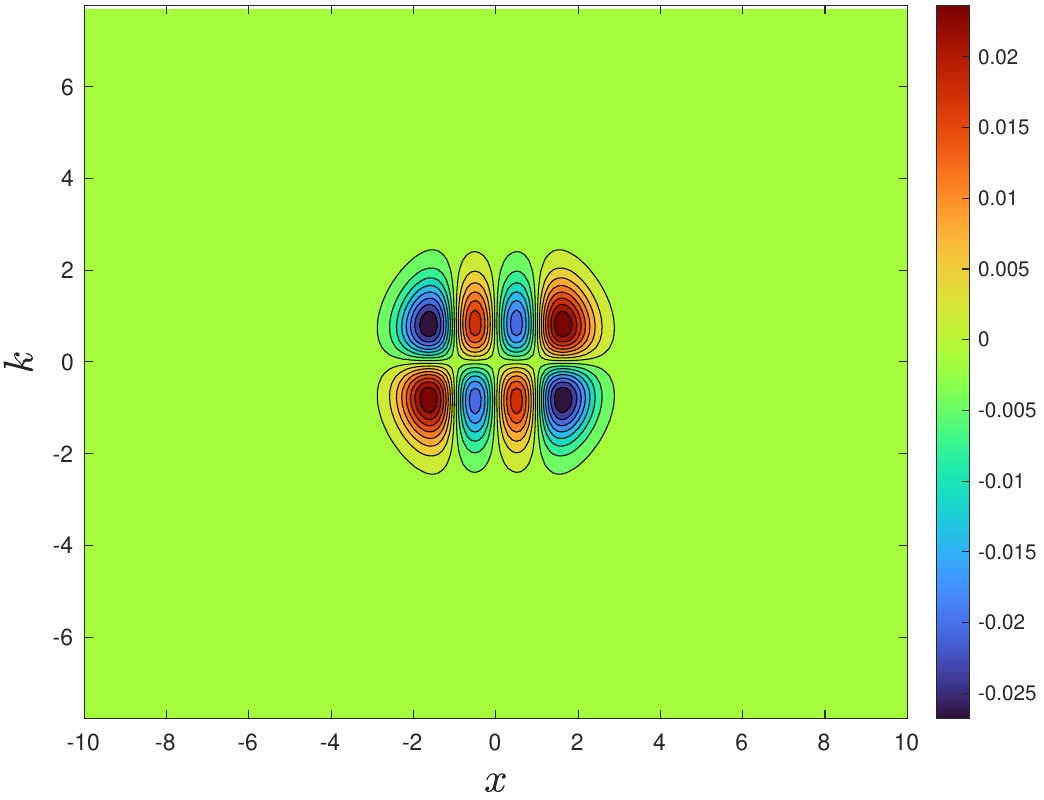}
\includegraphics[width=0.32\textwidth,height=0.22\textwidth]{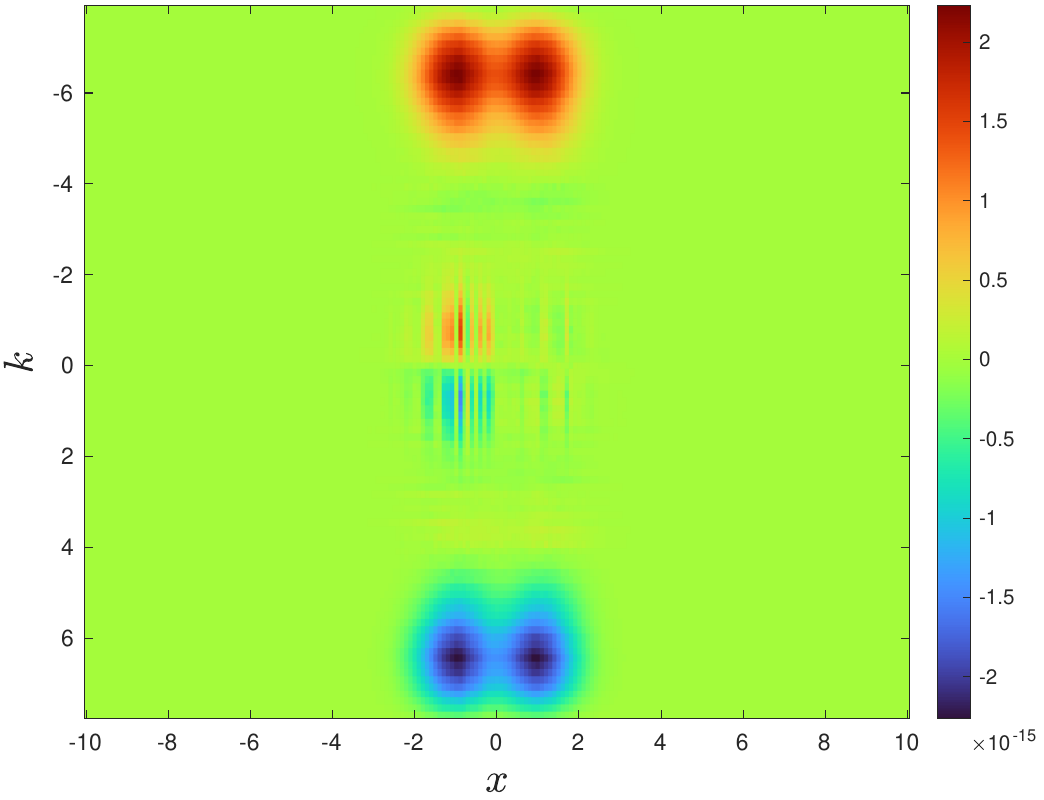}}
\caption{Three-fold or four-fold zero-padding is necessary in evaluating the Hartree potential.}
\end{figure}

Now we would like to show that zero-padding technique is indispensable in Eq.~\eqref{Hartree_approximation} for slow-decay potential. Here we choose $n_1(r_1) = \me^{-r_1^2}$ and obtain the reference solution 
\begin{equation}
\begin{split}
V_H[n_1](r_1, t)  &= \int_{\mathbb{R}} \frac{1}{\sqrt{ (r_1 - s)^2 + \epsilon}} \me^{-s^2}\D s  =  \int_{0}^{\infty} \left[\frac{1}{\sqrt{ (r_1 - s)^2 + \epsilon}} + \frac{1}{\sqrt{ (r_1 + s)^2 + \epsilon}}\right] \me^{-s^2}\D s\\
& =\frac{1}{2} \int_{0}^{\infty} \left[\frac{1}{\sqrt{ (r_1 - \sqrt{s})^2 + \epsilon}} + \frac{1}{\sqrt{ (r_1 + \sqrt{s})^2 + \epsilon}}\right] s^{-1/2} \me^{-s}\D s,
\end{split}
\end{equation}
which can be calculated by the Laguerre-Gauss quadrature. As shown in Figure \ref{zero_padding}, when zero padding is not imposed, the overlap of two neighbor images brings in large errors near the boundary. This can be alleviated by three-fold or four-fold zero-padding, coinciding with assertions in \cite{LiuTangZhangZhang2024}.

Figure \ref{Hartree_comparsion} compares pseudo-difference approximations and the Chebyshev spectral approximation to $\pdo$ for the Hartree part. It is shown that both methods are capable to resolve the high oscillations, and their  difference approaches machine precision.

%Table \ref{TKM_convergence_data} presents the spectral convergence of SEM under different $M$ and $N$, with $N_{p_1} = N_{p_2} = M N$.
%\begin{table}[!h]
%  \centering
%  \caption{\small Numerical errors and computational time of SEM.}
%  \label{TKM_convergence_data}
%   \begin{tabular}{c|c|c|c|c|c}
%    \hline\hline
%   Convergence & $M$ &  $N$ & $l^\infty$-error & $l^2$-error & Time(s)\\
%   \hline
%    \multirow{6}{*}{\includegraphics[scale=0.18]{./SEM_convergence.pdf}}  
%&	2	&	4	&	1.160		&	3.096	&	0.074 	\\
%&	4	&	4	&	5.044$\times10^{-1}$		&	8.647$\times10^{-1}$		&	0.278 	\\
%&	8	&	4	&	1.062$\times10^{-1}$		&	1.654$\times10^{-1}$		&	1.080 	\\
%&	8	&	8	&	8.542$\times10^{-4}$		&	1.865$\times10^{-3}$		&	4.386 	\\
%&	8	&	16	&	2.333$\times10^{-9}$		&	4.321$\times10^{-9}$		&	18.764 	\\
%&	8	&	32	&	1.983$\times10^{-9}$		&	4.560$\times10^{-14}$		&	83.140 	\\
%   \hline\hline
% \end{tabular}  
%  \end{table}
%

\subsection{Dynamical XC potential and two-body corrections to quantum Landau damping}

The Coulomb interaction offers a crucial advantage for a correct physical analysis of the system. To study this, we first examine the quantum Landau damping phenomenon, the damping of a collective mode of oscillations in plasmas without collisions of charged particle. Our interest focus on the building up of correlations via two-body collisions and correlations of higher order \cite{WangCassing1985}. To this end, we choose the initial uncorrelated condition as
\begin{equation}
f^{init}_{12}(r_1, r_2, p_1, p_2) = \frac{1}{2\pi}(1 + \varepsilon_1 \cos(k_1 r_1)) (1 + \varepsilon_2 \cos(k_2 r_2)) \me^{- \frac{p_1^2 + p_2^2}{2}},
\end{equation}
with parameters: $k_1 = 0.4$, $k_2 = 0.4$. The computational domain $\mathcal{X} \times \mathcal{P} = [-5\pi, 5\pi]^2 \times [-6.4, 6.4]^2$ is represented by a uniform mesh with $257^2\times 128^2 \approx 1.08\times10^9$ grid points and a fixed spatial spacing $\Delta x = \frac{5\pi}{128}$, $\Delta p = 0.1$. The time step is $\Delta t = 0.01$ and the final time is $T = 200$. 

The Hartree part is set to be the neutralized Poisson equation subject to a periodic boundary condition (i.e., the Wigner-Poisson system). The H-L potential is chosen to model the exchange and correlation effect. The two-body interaction is the smoothed Coulomb potential $V_{ee}(r_1, r_2) = \frac{\gamma}{\sqrt{|r_1 - r_2|^2 + \epsilon}}$ with  $\epsilon = 0.01, 1$, $\gamma = 1$. In absence of two-body scattering term, since the initial condition is uncorrelated, the two-body Wigner equation \eqref{two_body_truncated_Wigner} reduces to the mean-field equation by separation of variables. We compare the electrostatic field $E(x_1, t) = \frac{\partial }{\partial x_1} V_H(x_1, t)$ in $x_1$-direction, and number density $n_{12}(r_1, r_2, t)$ and the reduced Wigner function $W^{red}(r_1, p_1, t) = \iint_{\mathbb{R}^2}  f_{12}(r_1, r_2, p_1, p_2, t) \D r_2 \D p_2$ under the Hartree approximation, Hartree with H-L potential and Hartree with two-body correction.

% \begin{figure}[!h]
% \centering
% \subfigure[Linear Landau damping.\label{LD}]{
% \includegraphics[width=0.32\textwidth,height=0.22\textwidth]{./LD_comp_short}}
% \subfigure[$n_{12}(r_1, r_2, 5)$ under the two-body scattering (left: $\epsilon=1$, right: $\epsilon = 0.01$).\label{LD_xdist}]{
% \includegraphics[width=0.32\textwidth,height=0.22\textwidth]{./LD_2B_eps1_T5}
% \includegraphics[width=0.32\textwidth,height=0.22\textwidth]{./LD_2B_eps001_T5}}
% \caption{Linear quantum Landau damping ($\varepsilon_1 = \varepsilon_2 = 0.002$): The correction of H-L potential  seems to be negligible, whereas the two-body scattering impedes the electric field to be damped away. }
% \end{figure} 

\begin{figure}[h]
\centering
\subfigure[Linear regime.\label{NLD}]{
\includegraphics[width=0.48\textwidth,height=0.27\textwidth]{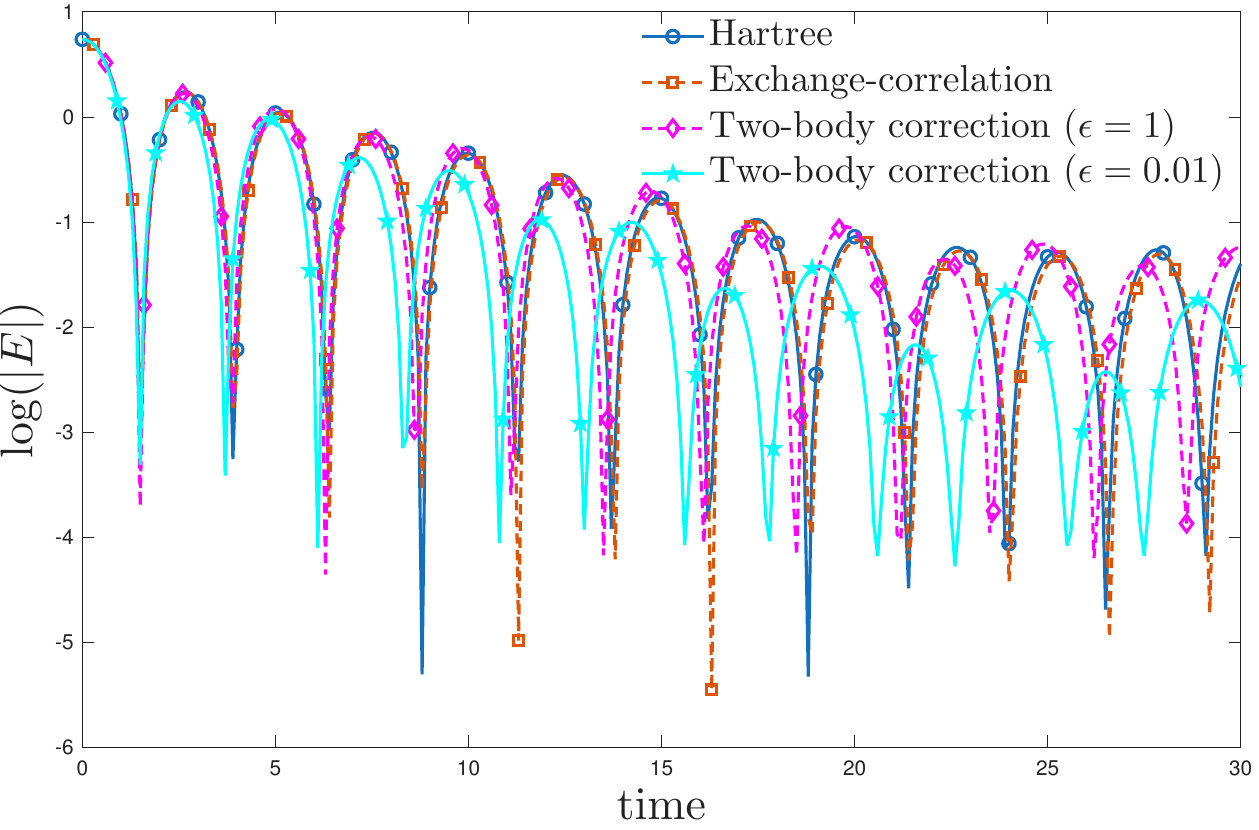}}
\\
\centering
\subfigure[Nonlinear Landau damping up to $T =200$ (left: with XC potential, right: with two-body interaction). \label{two_stream_long}]{
\includegraphics[width=0.48\textwidth,height=0.27\textwidth]{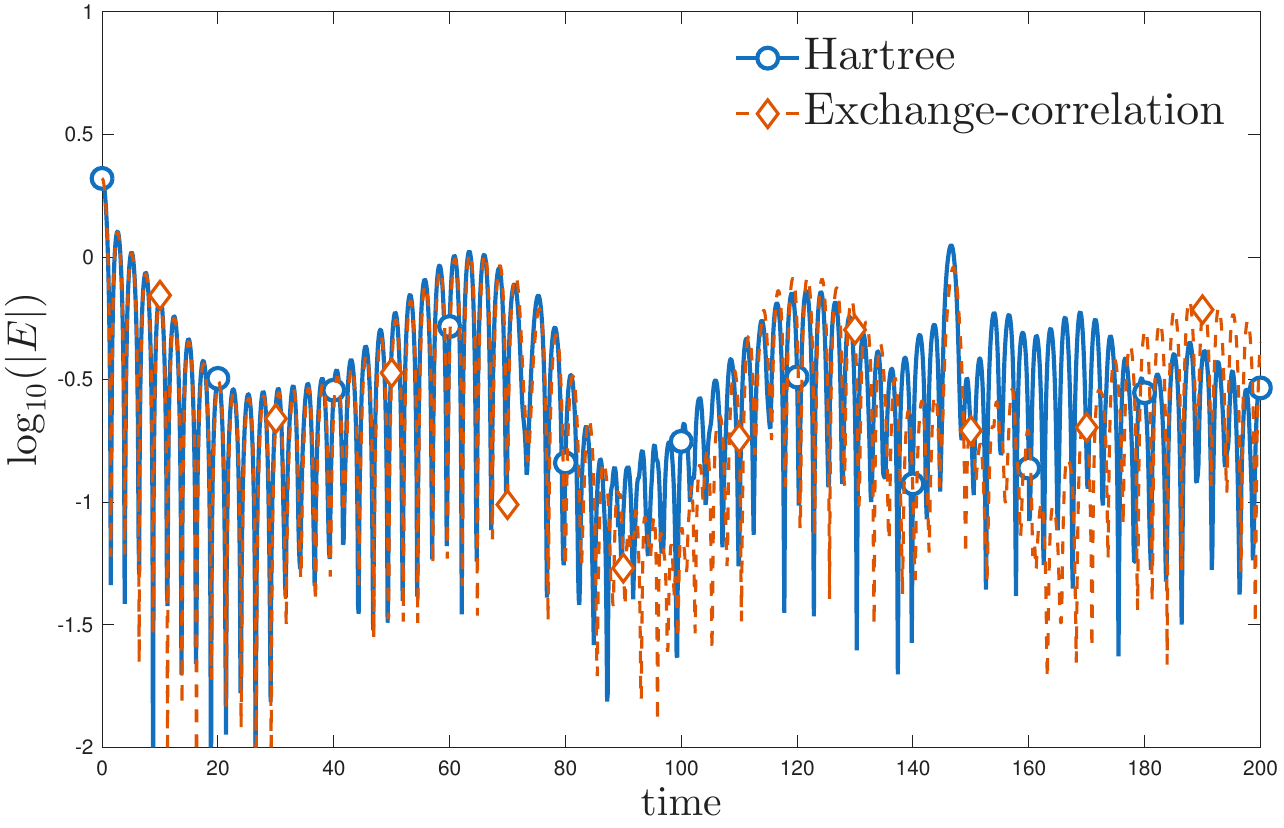}
\includegraphics[width=0.48\textwidth,height=0.27\textwidth]{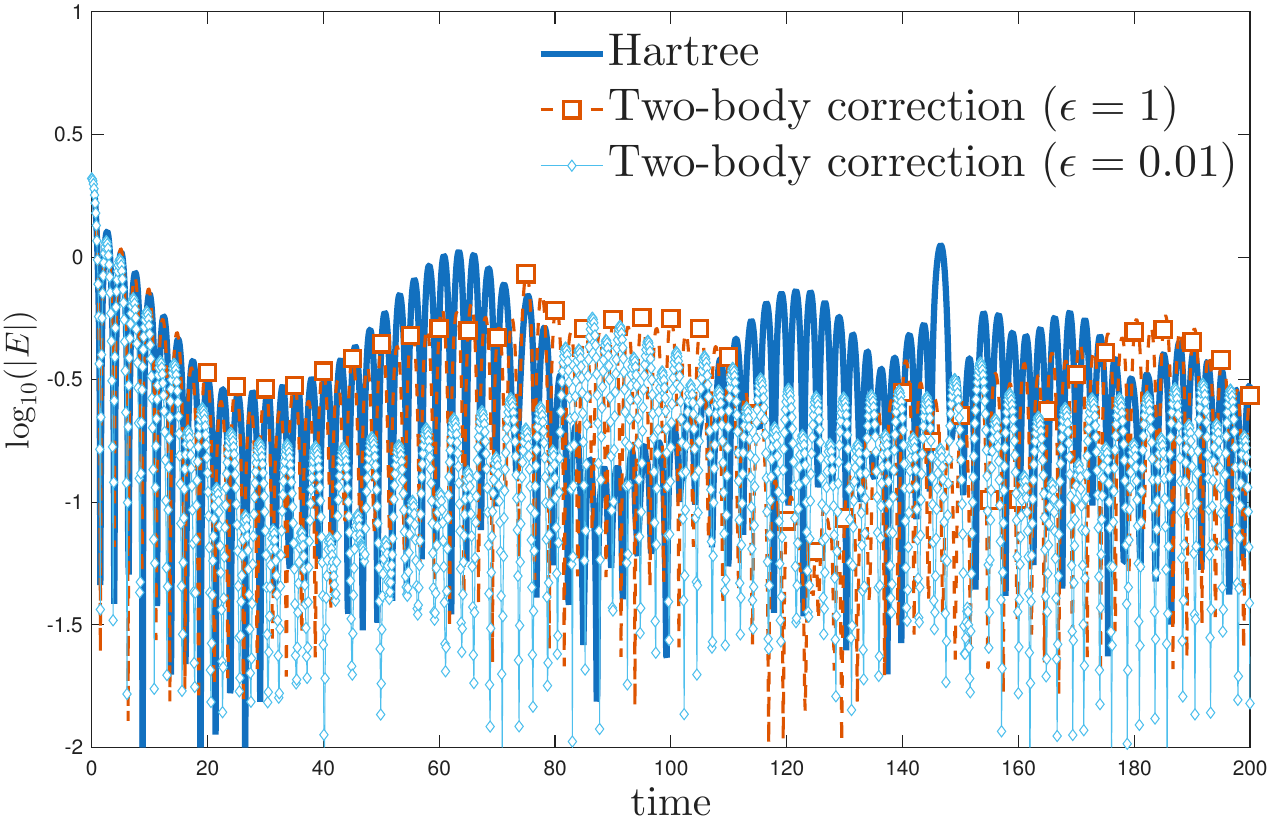}}
\caption{Nonlinear quantum Landau damping up to $T =200$ (left: with XC potential, right: with two-body interaction). In the linear stage ($t = 0$-$15$), the purely local exchange–correlation effect does not alter the Landau damping rate of the plasma, whereas the two-body interactions significantly modify the linear Landau damping rate and can even change the nonlinear coupling processes.  \label{wNLD_long}}
\end{figure}

 \begin{figure}[!h]
    \centering
    \subfigure[$t=12$.]{
    \includegraphics[width=1.4in,height=1.0in]{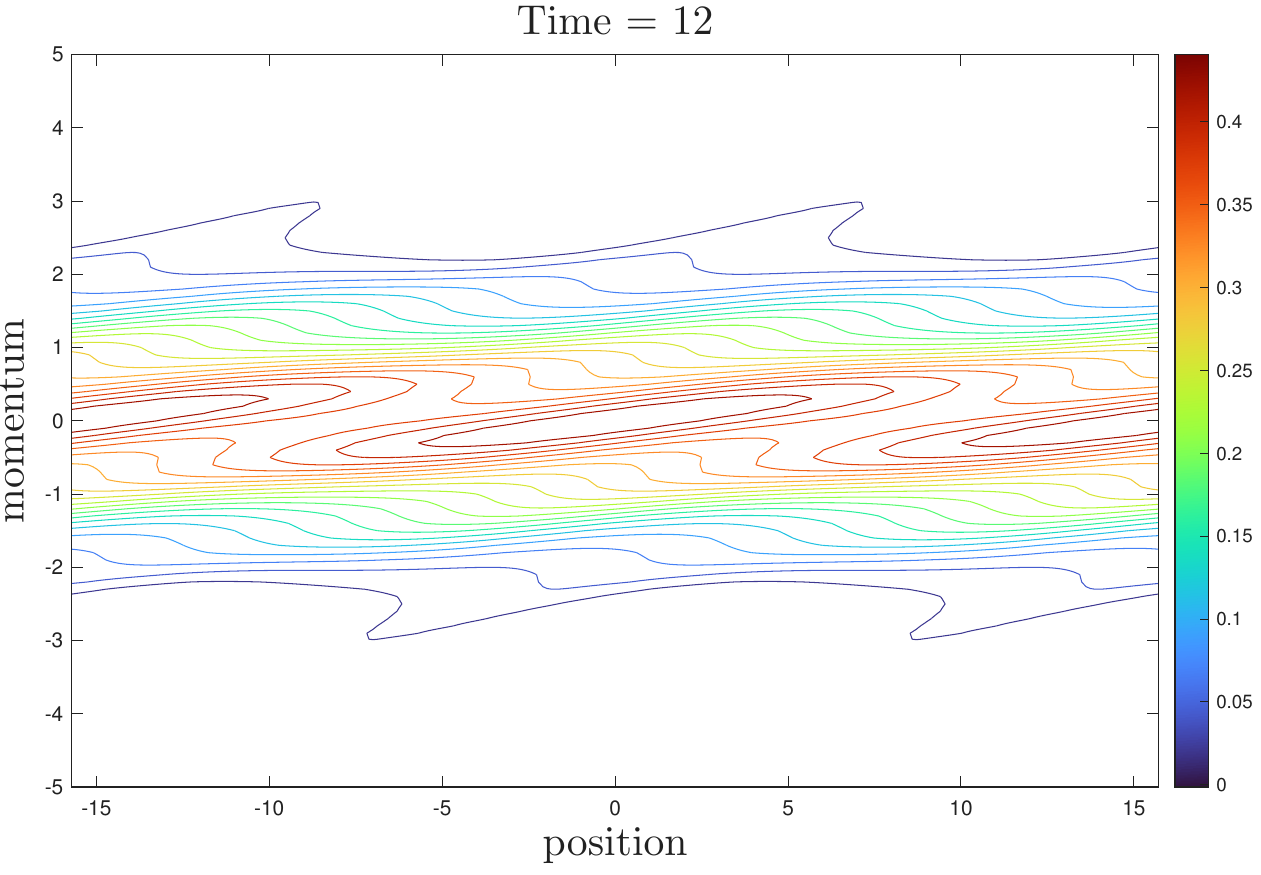}
    \includegraphics[width=1.4in,height=1.0in]{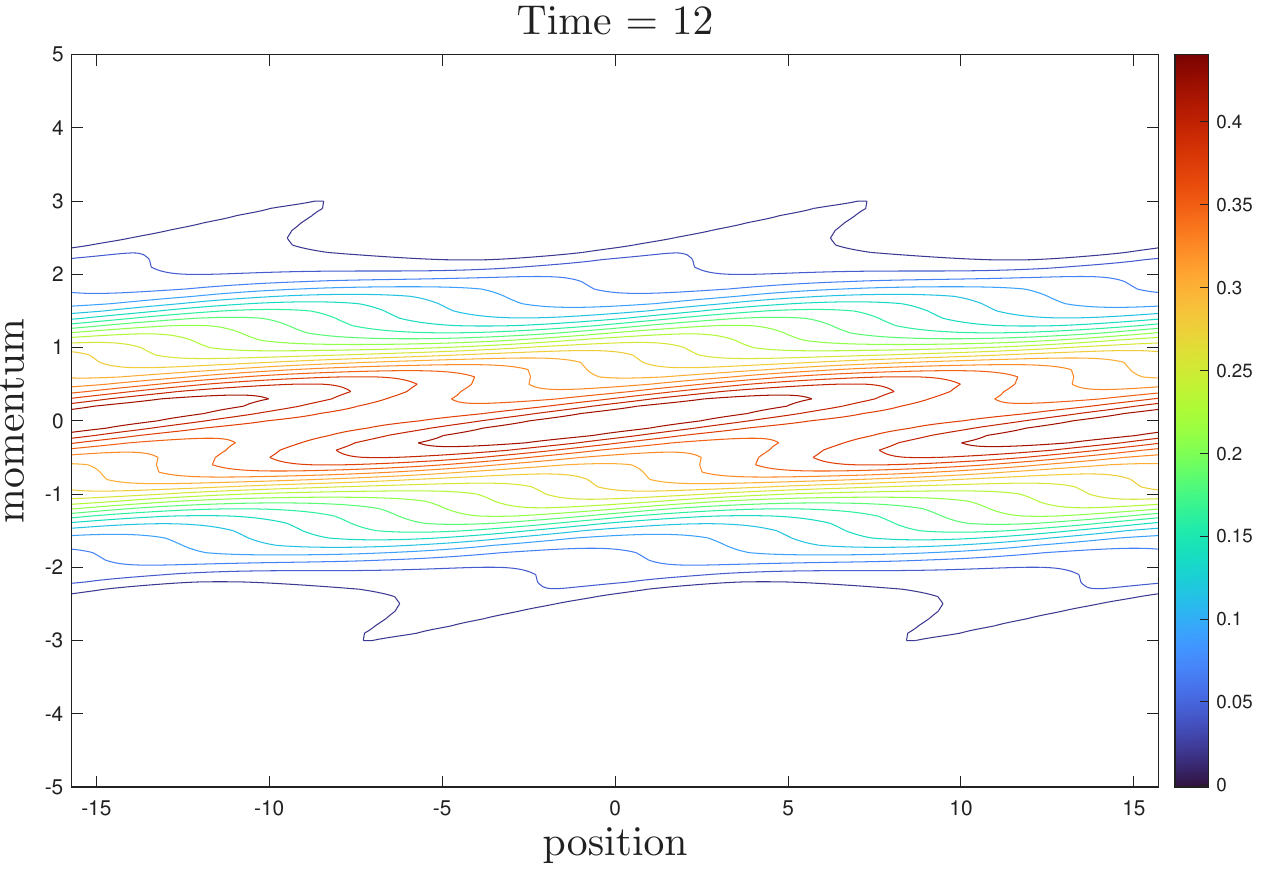}
    \includegraphics[width=1.4in,height=1.0in]{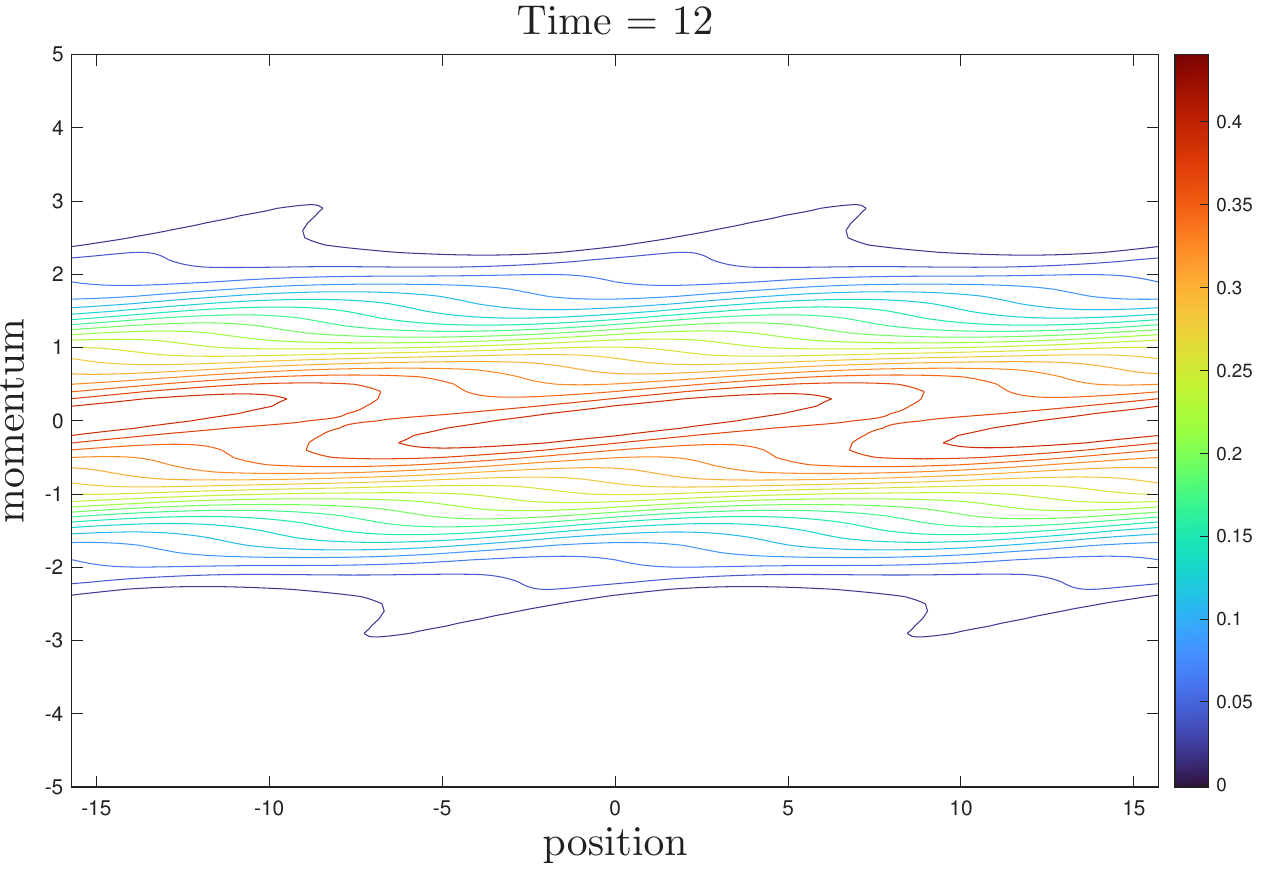}
    \includegraphics[width=1.4in,height=1.0in]{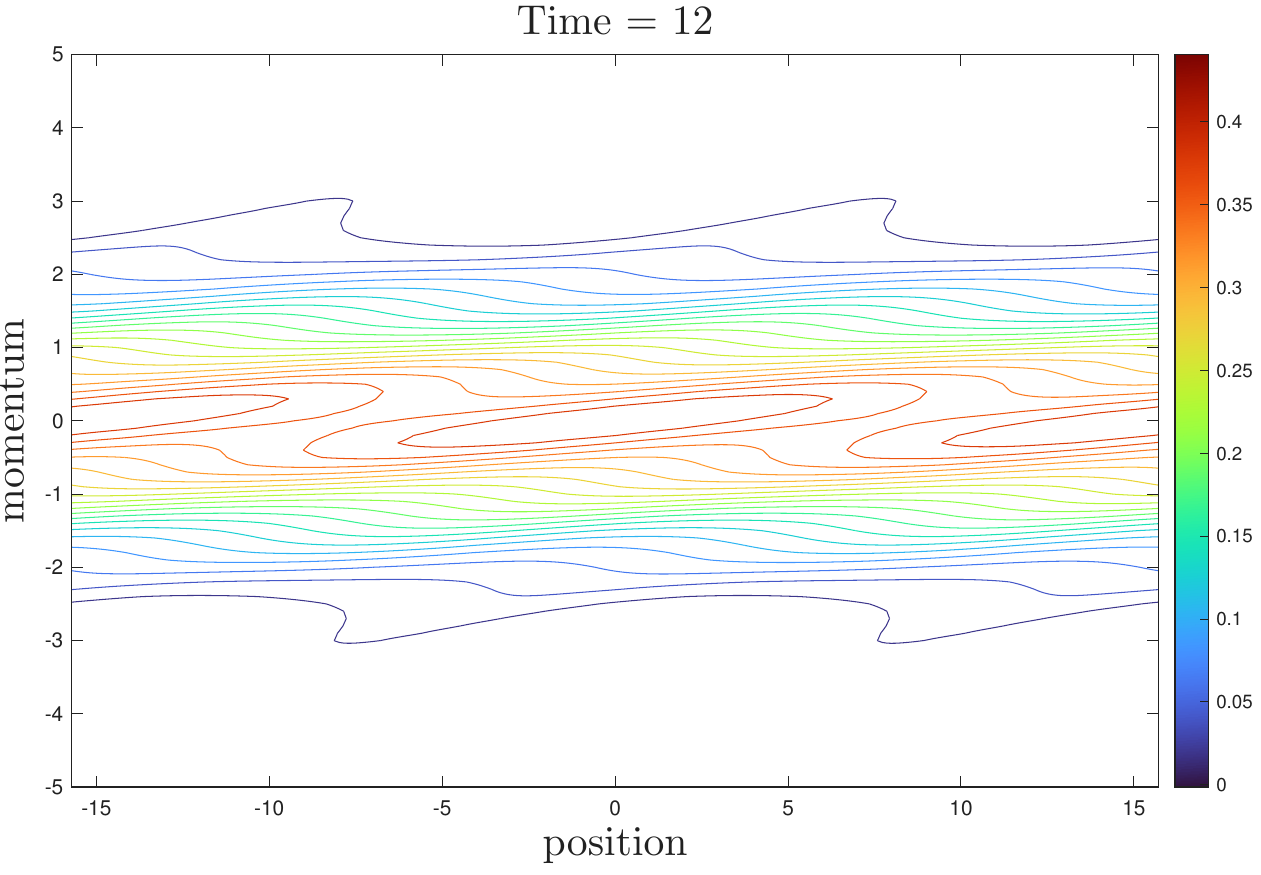}}
    \\
    \centering
    \subfigure[$t=24$.]{
    \includegraphics[width=1.4in,height=1.0in]{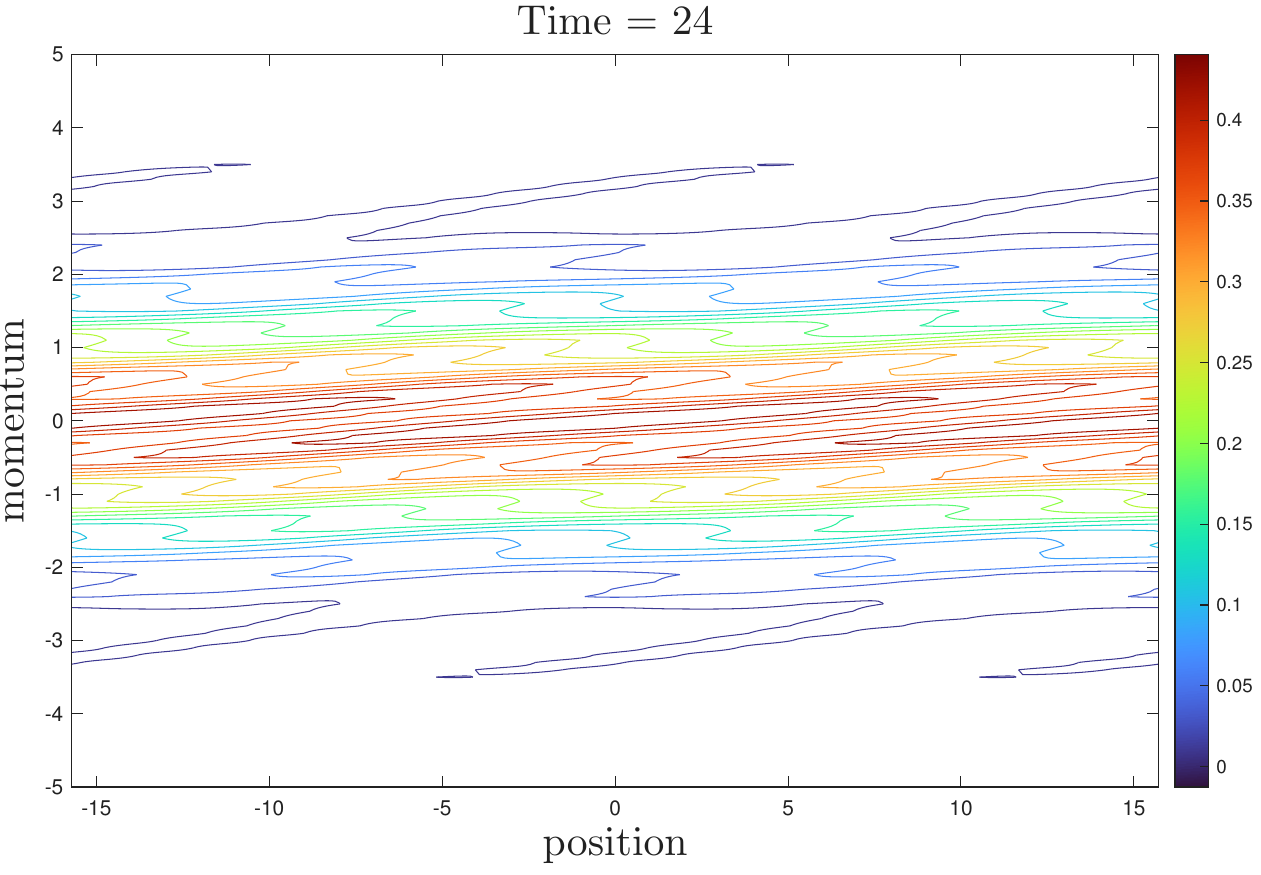}
    \includegraphics[width=1.4in,height=1.0in]{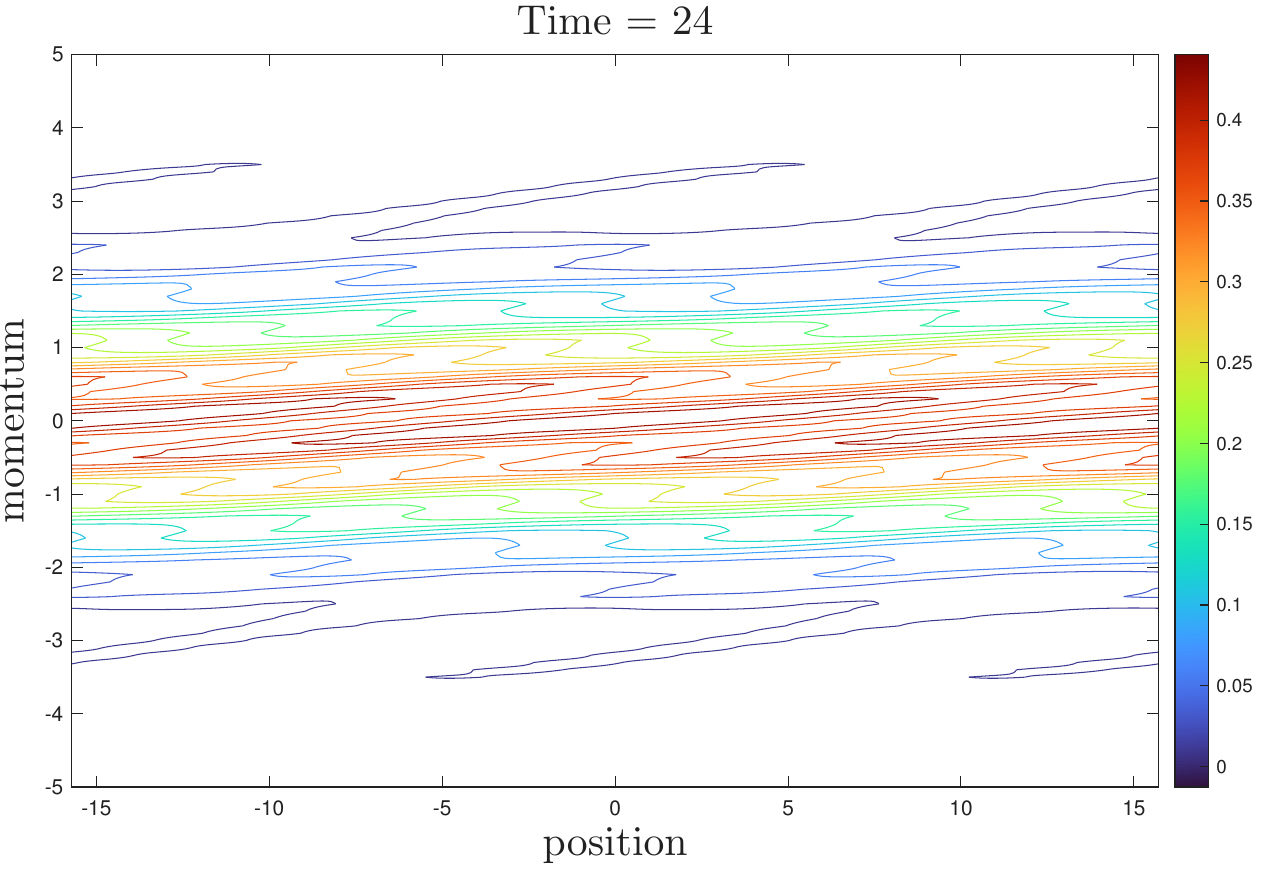}
    \includegraphics[width=1.4in,height=1.0in]{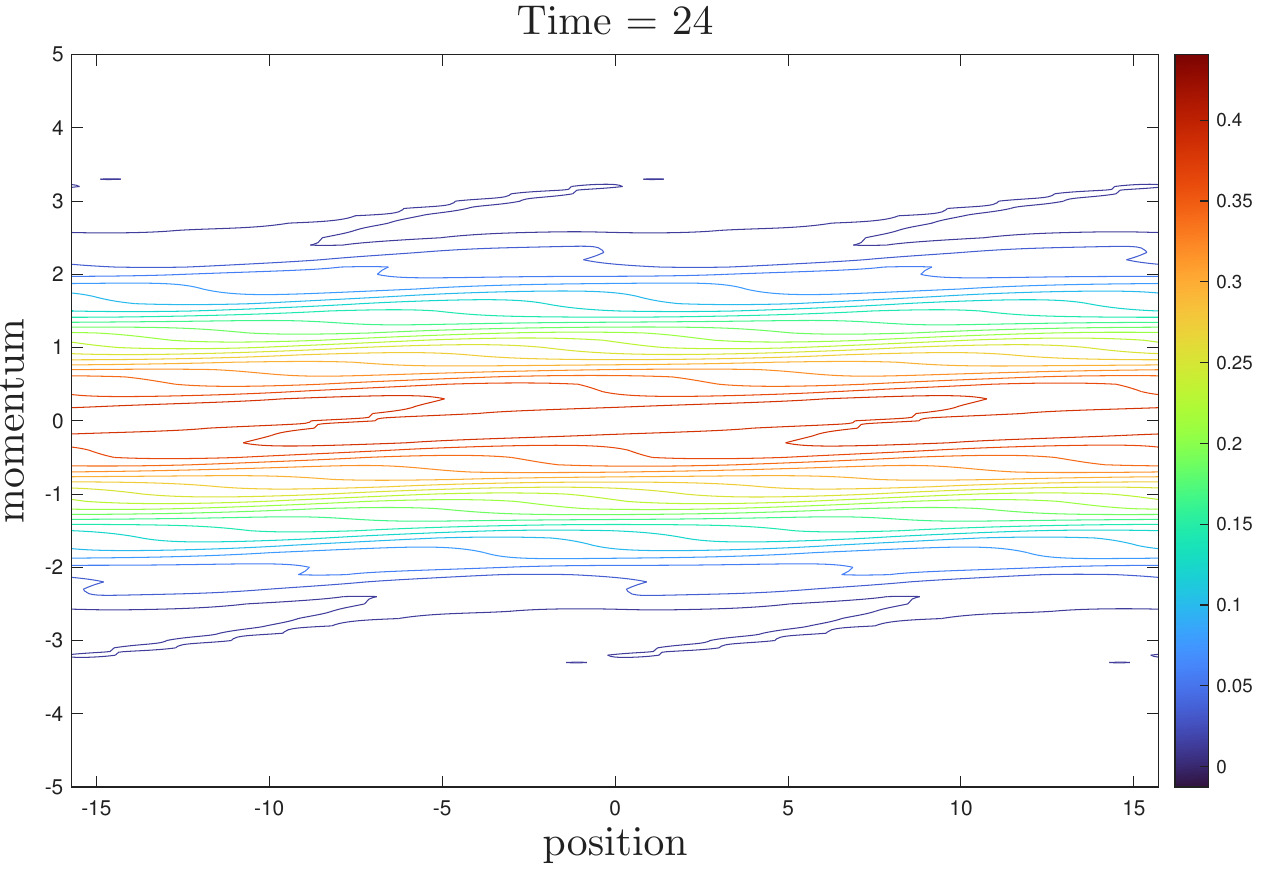}
    \includegraphics[width=1.4in,height=1.0in]{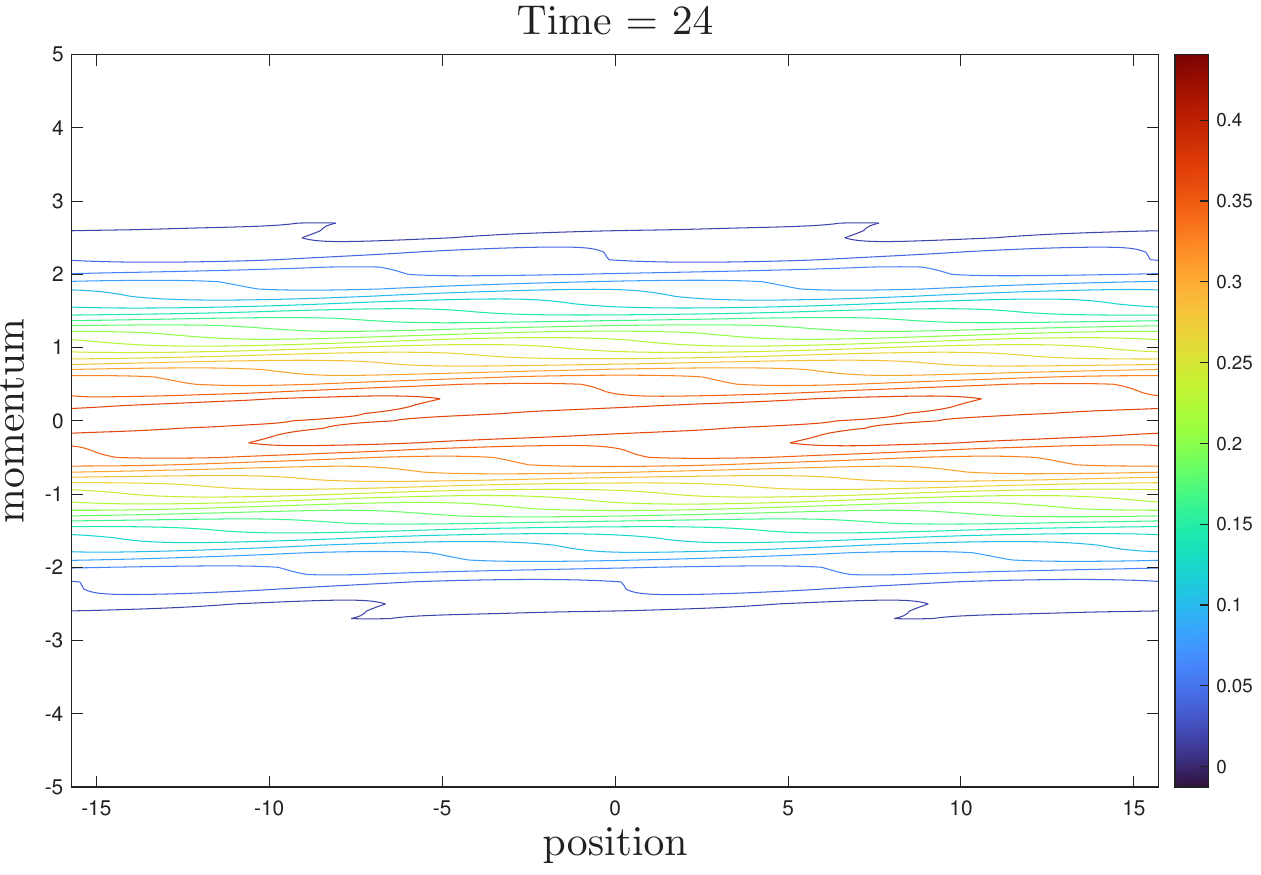}}
   \\
    \centering
    \subfigure[$t=36$.]{
    \includegraphics[width=1.4in,height=1.0in]{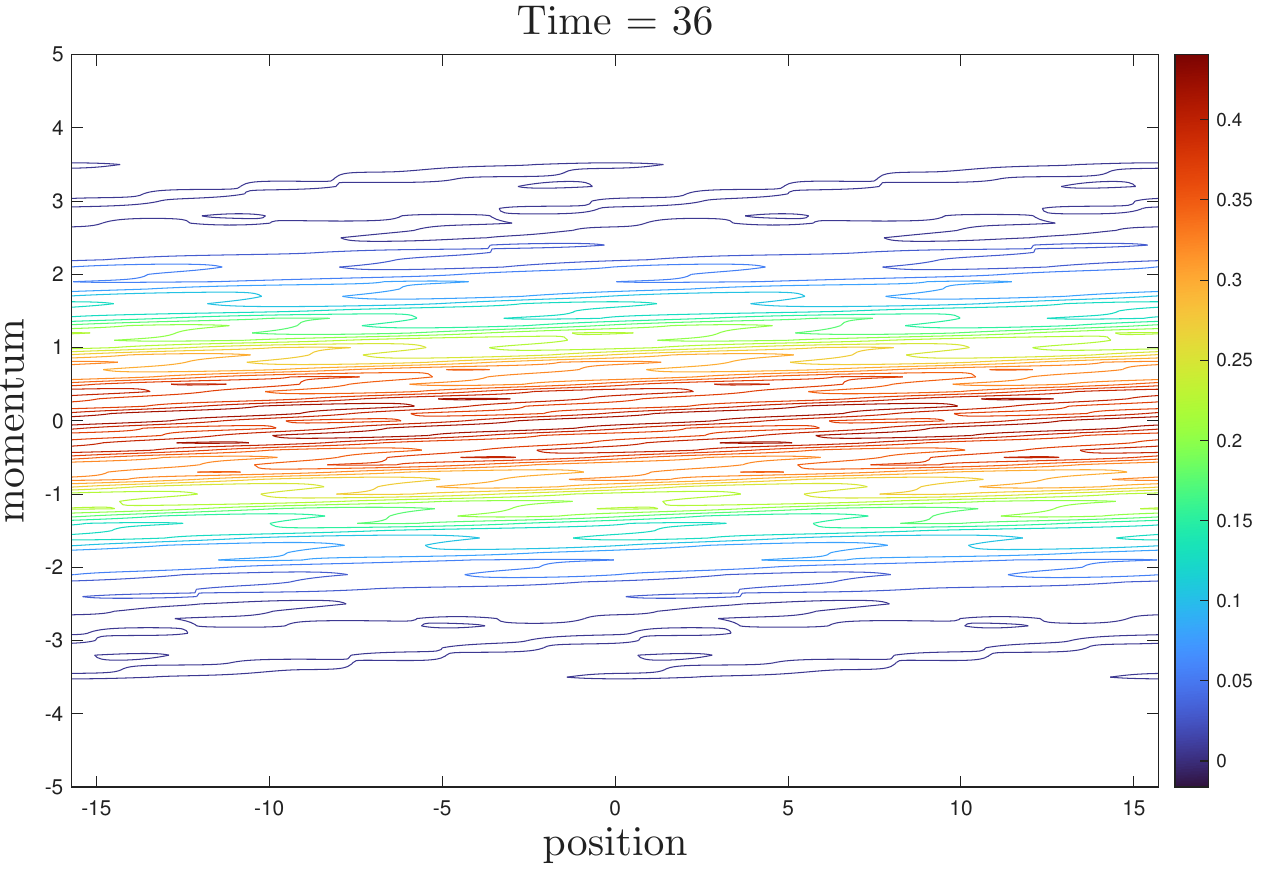}
    \includegraphics[width=1.4in,height=1.0in]{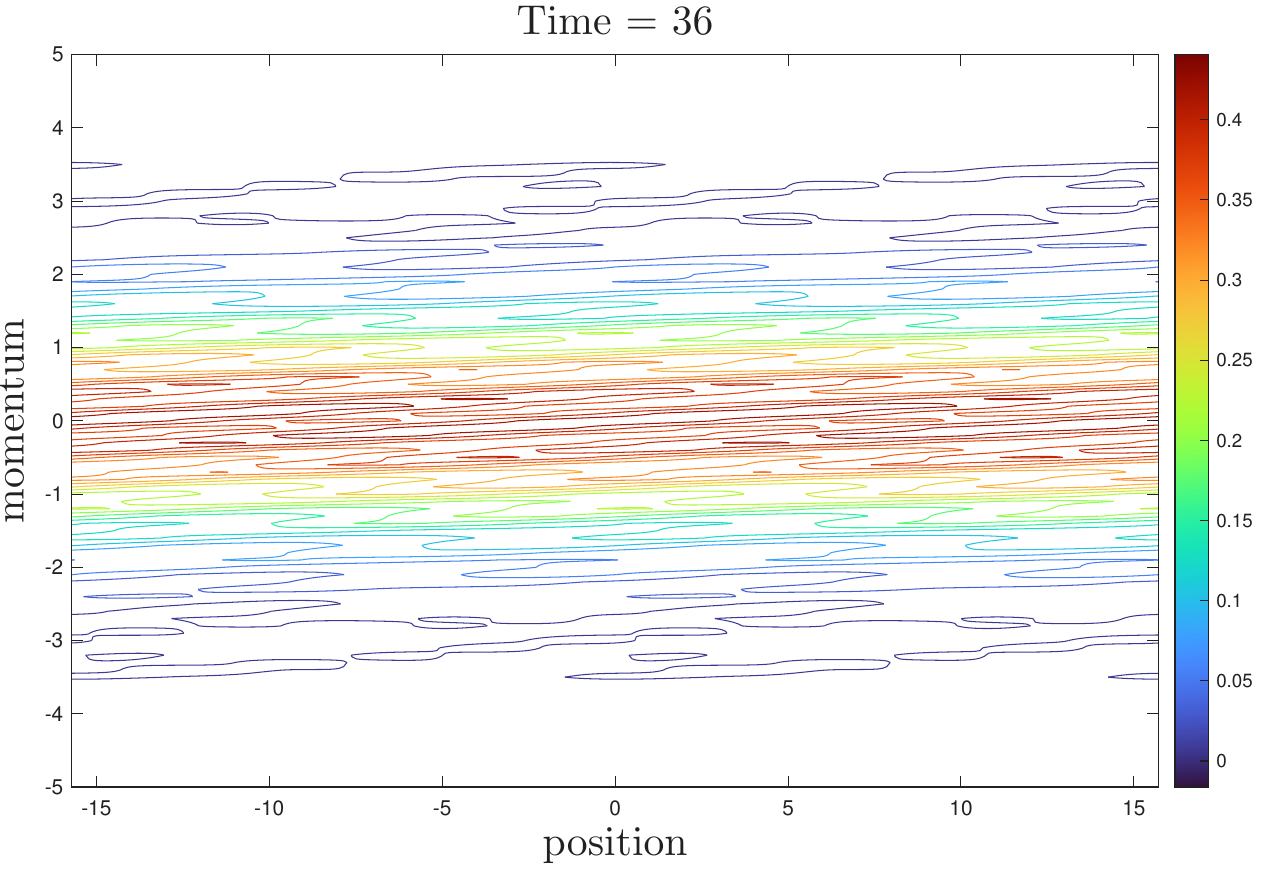}
    \includegraphics[width=1.4in,height=1.0in]{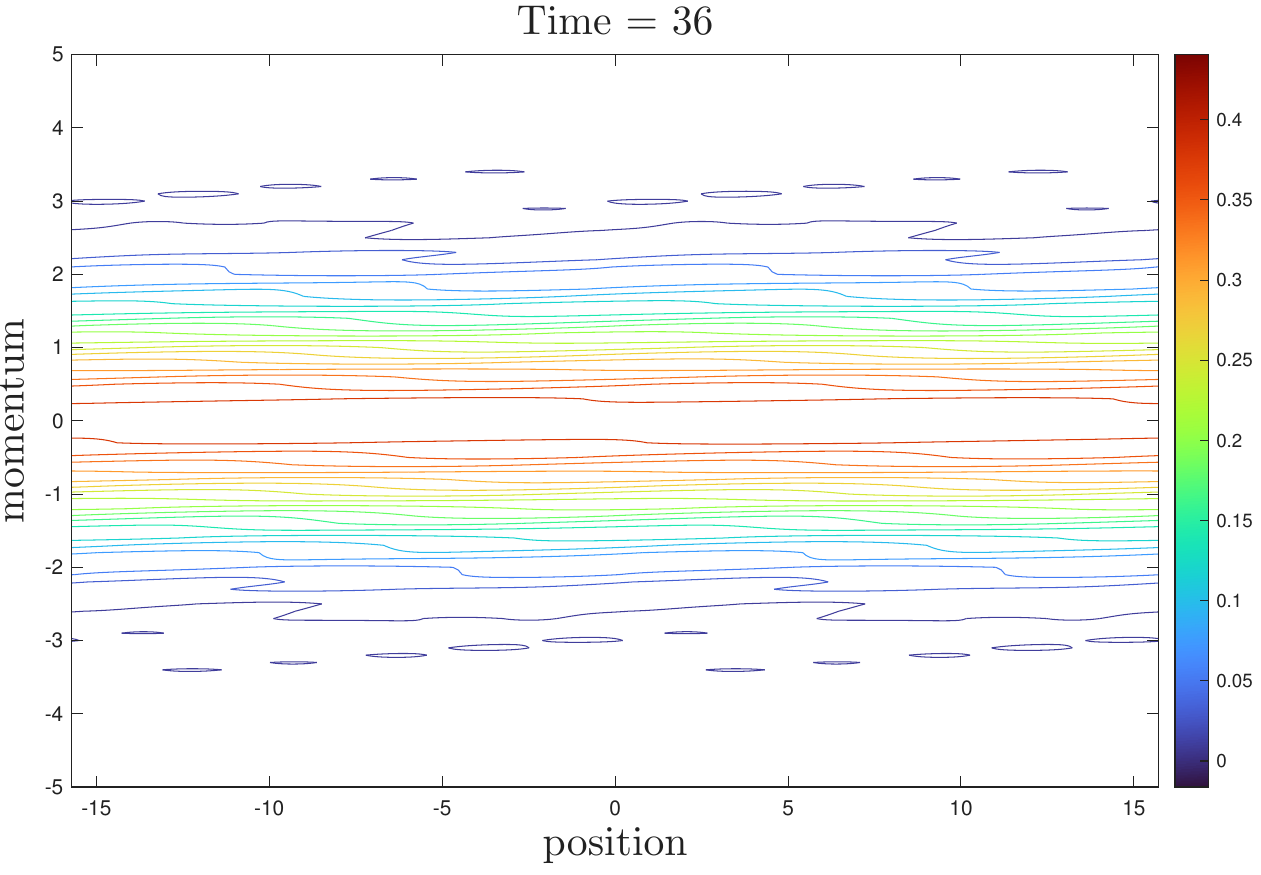}
    \includegraphics[width=1.4in,height=1.0in]{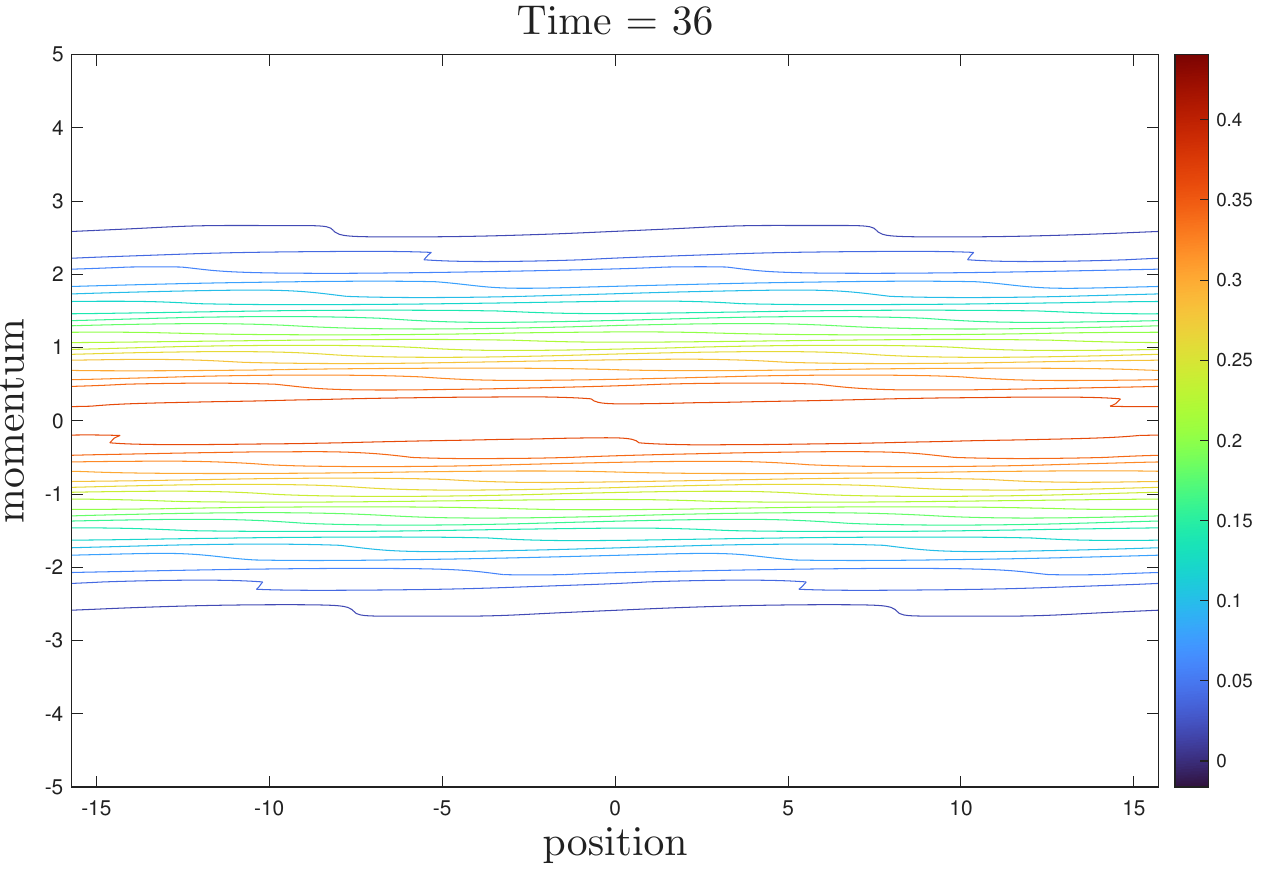}}
   \\
     \centering
    \subfigure[$t=48$.]{
    \includegraphics[width=1.4in,height=1.0in]{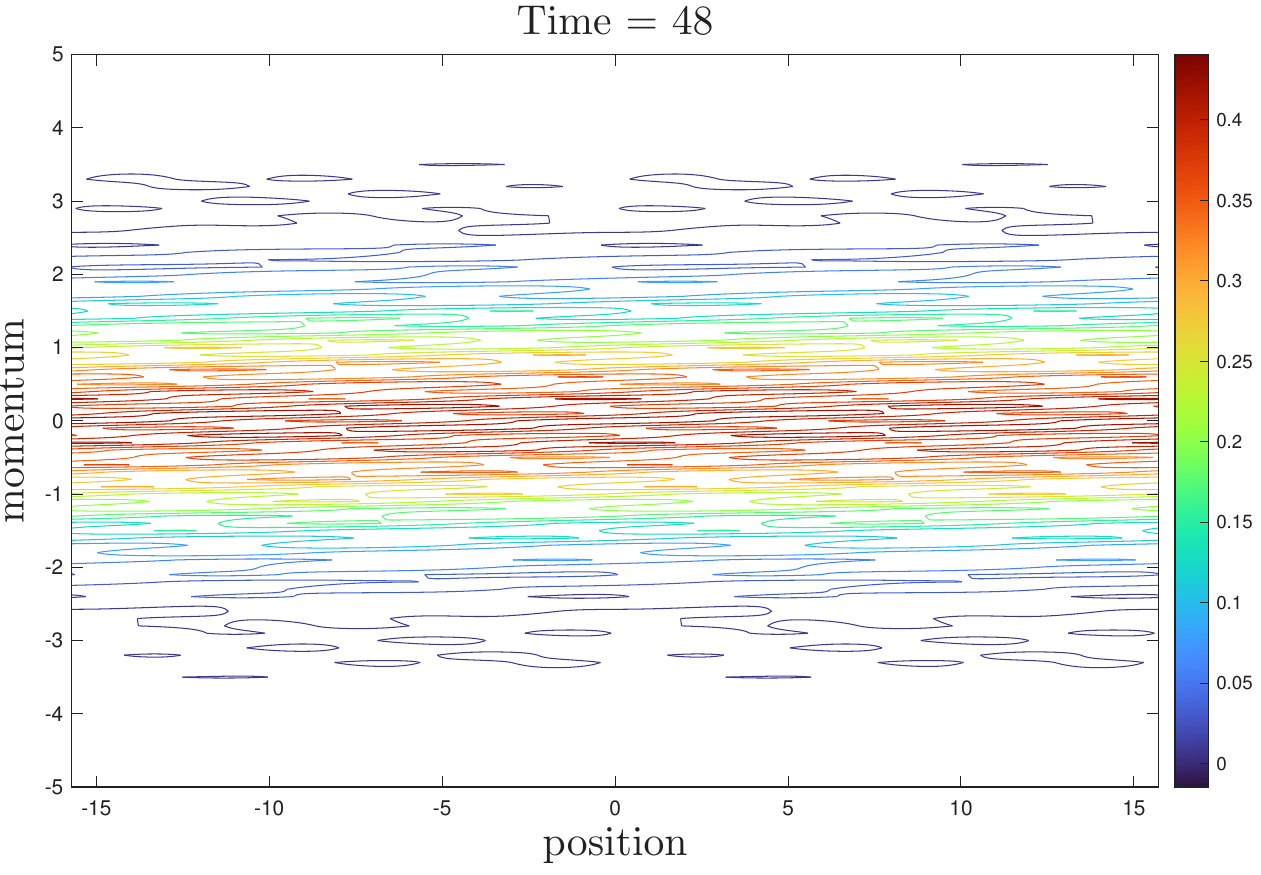}
    \includegraphics[width=1.4in,height=1.0in]{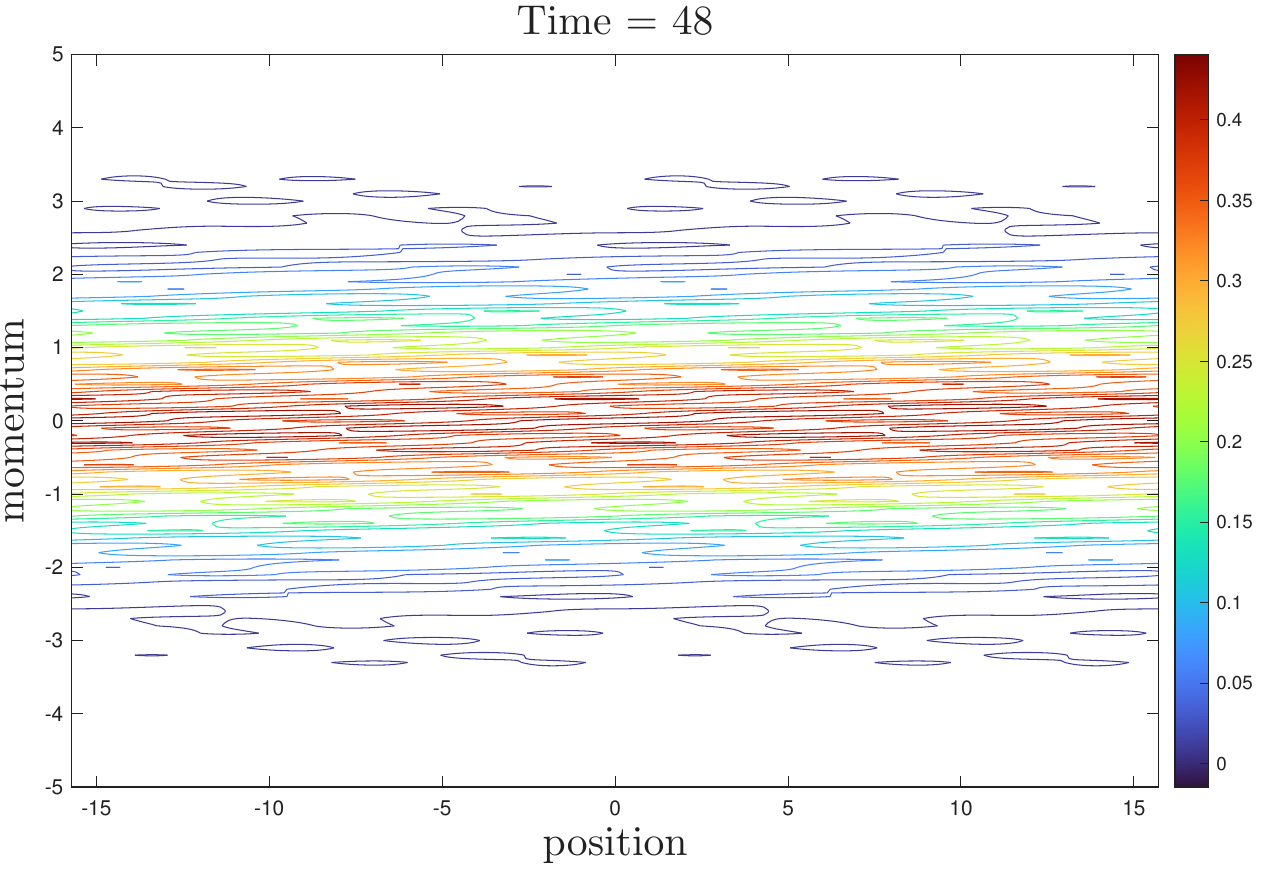}
    \includegraphics[width=1.4in,height=1.0in]{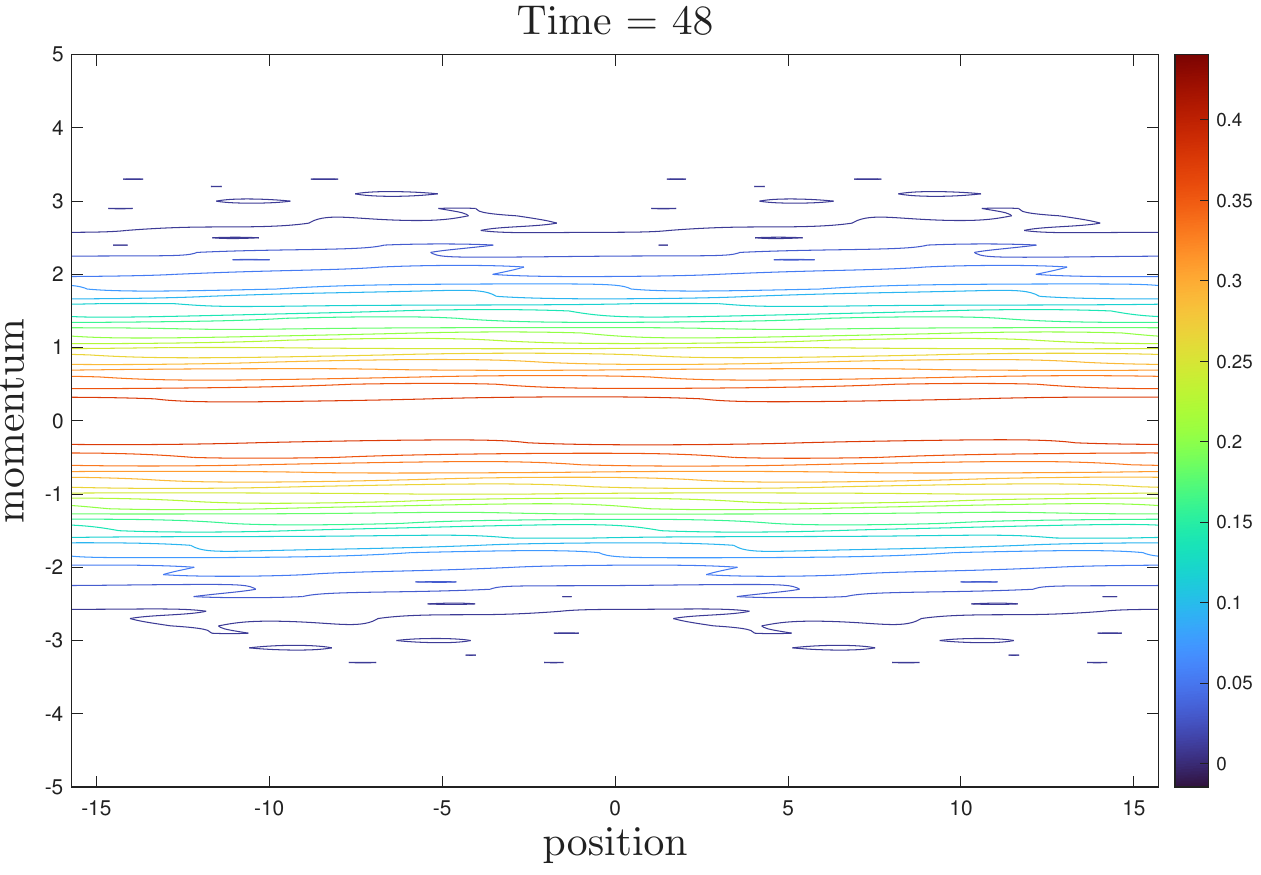}
    \includegraphics[width=1.4in,height=1.0in]{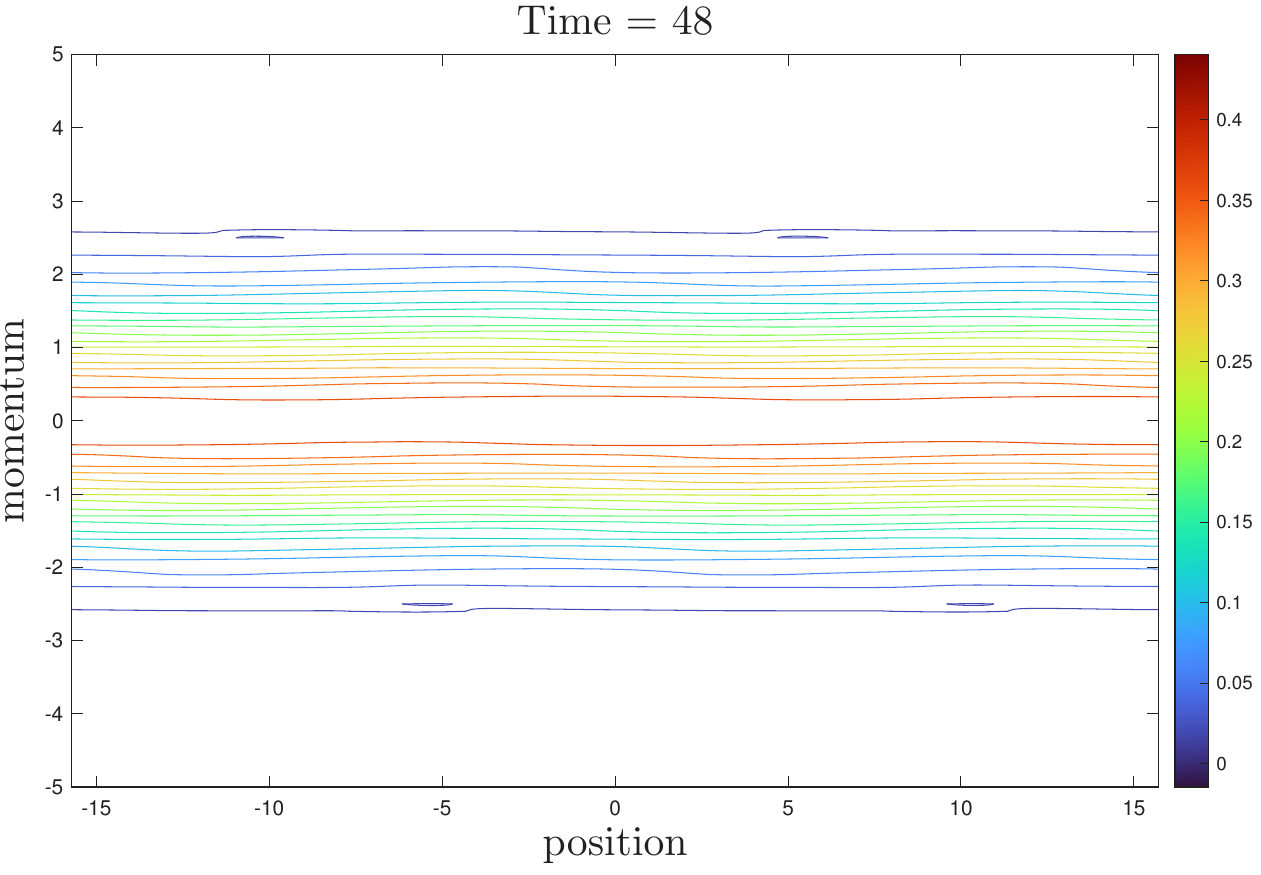}}
    \\
    \centering
    \subfigure[$t=60$.]{
    \includegraphics[width=1.4in,height=1.0in]{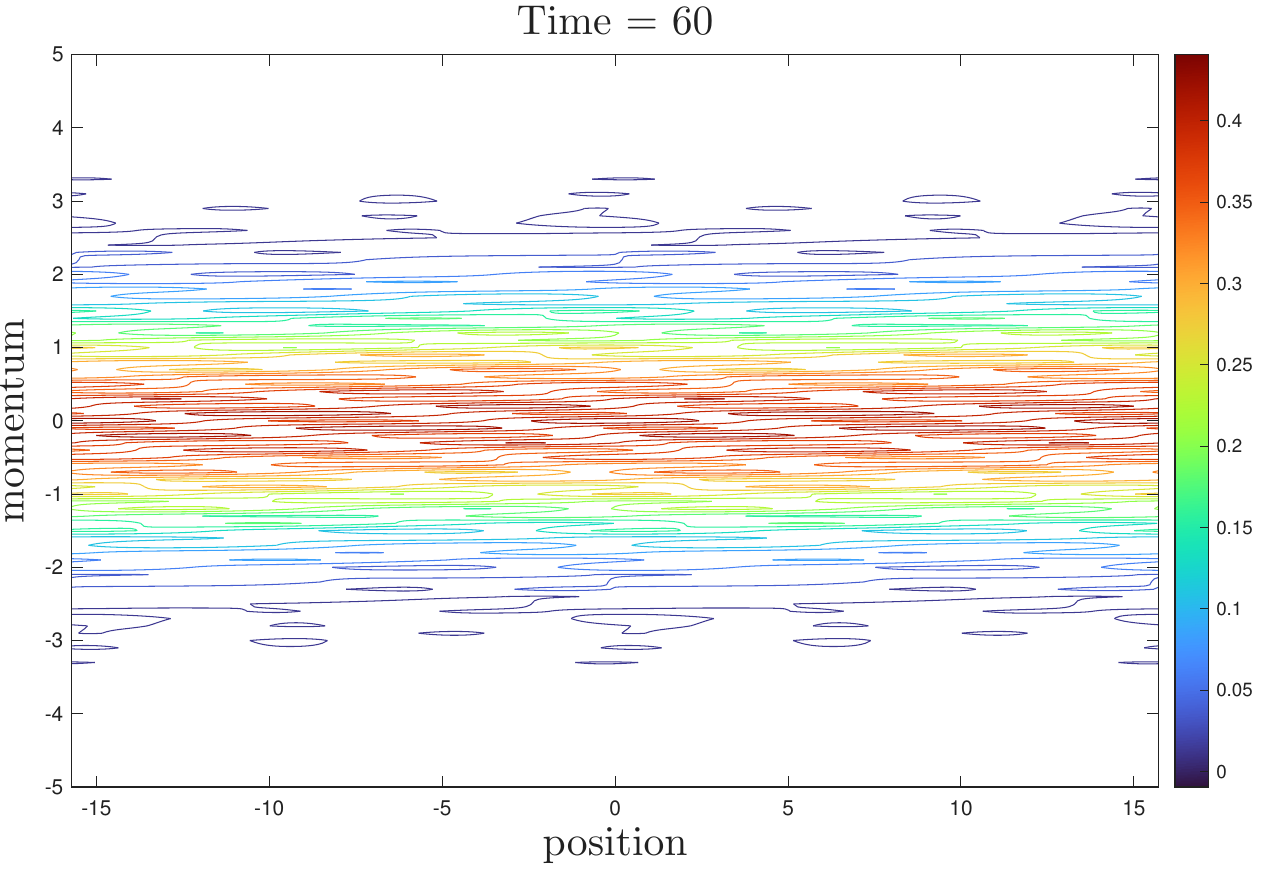}
    \includegraphics[width=1.4in,height=1.0in]{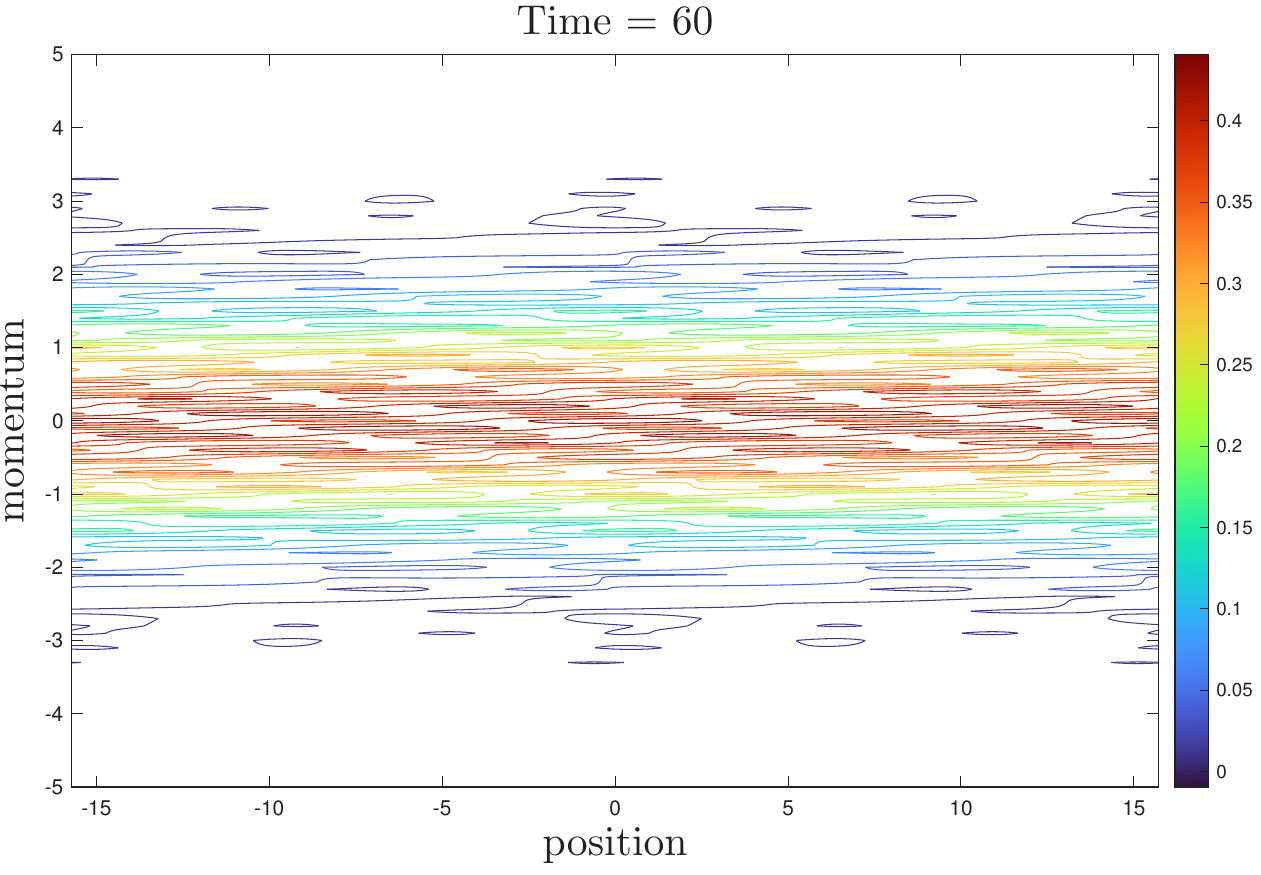}
    \includegraphics[width=1.4in,height=1.0in]{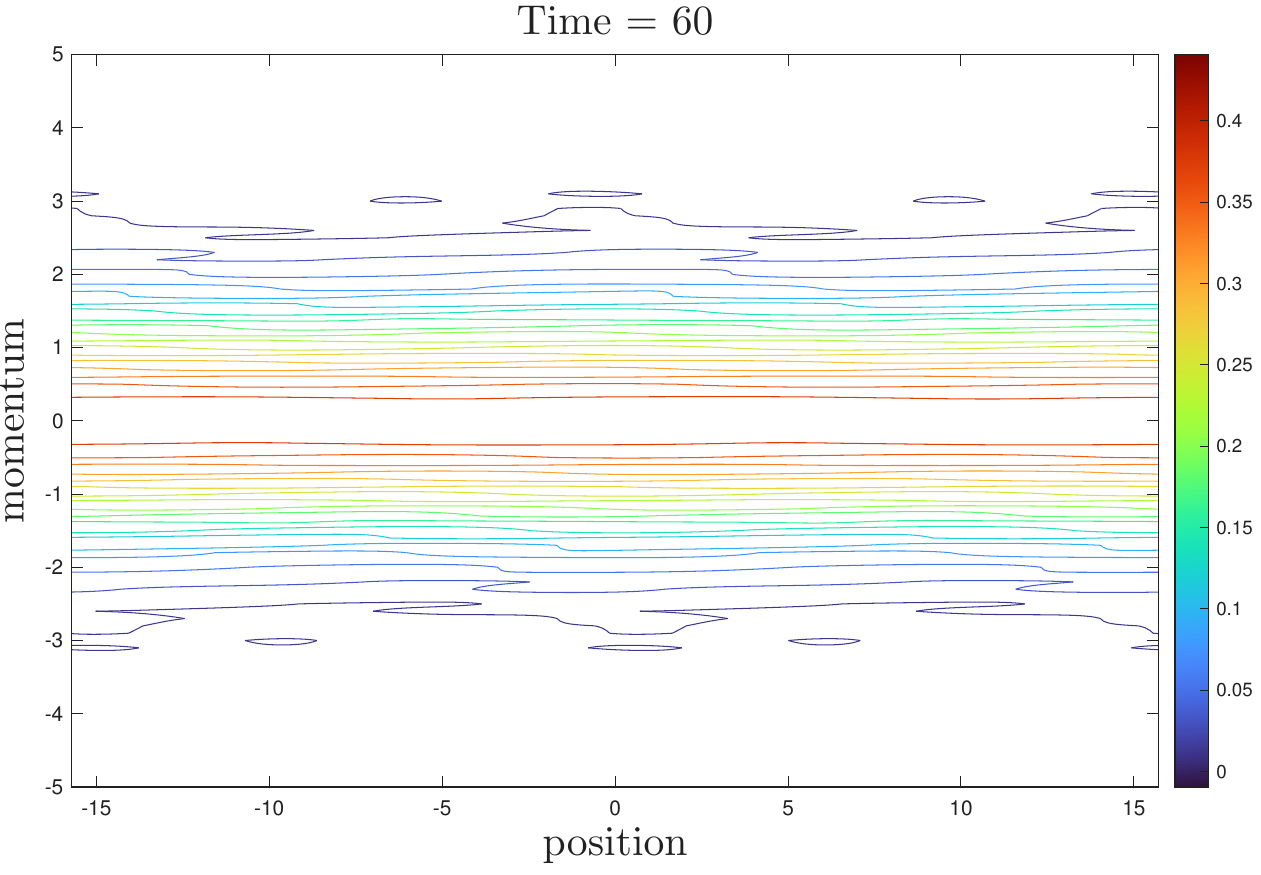}
    \includegraphics[width=1.4in,height=1.0in]{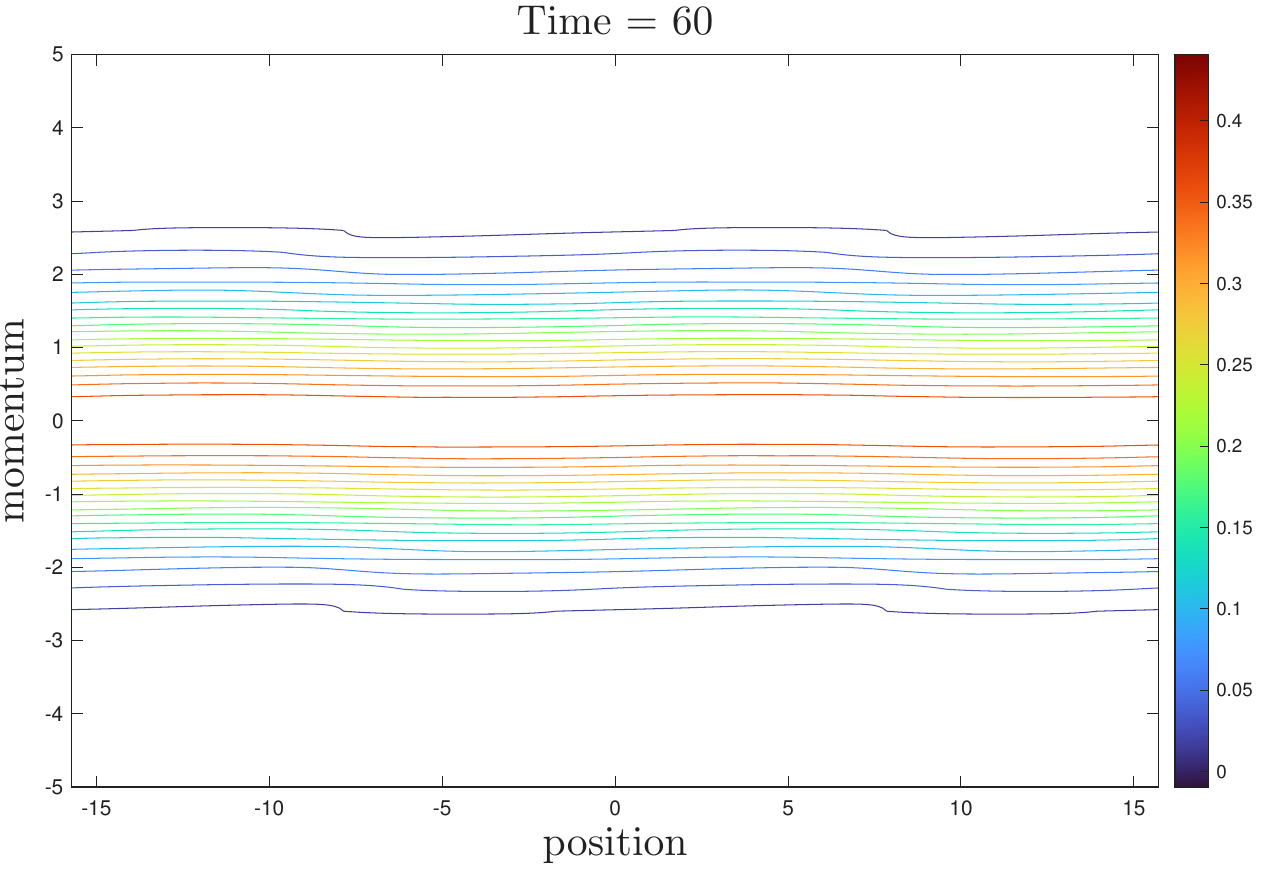}}   
    \\
    \centering
\subfigure[The slice of reduced Wigner function $W^{red}(0, p, t)$ at $r =0$, $t = 20, 40, 60$. \label{slice_Wigner_NLD}]{
\includegraphics[width=0.32\textwidth,height=0.22\textwidth]{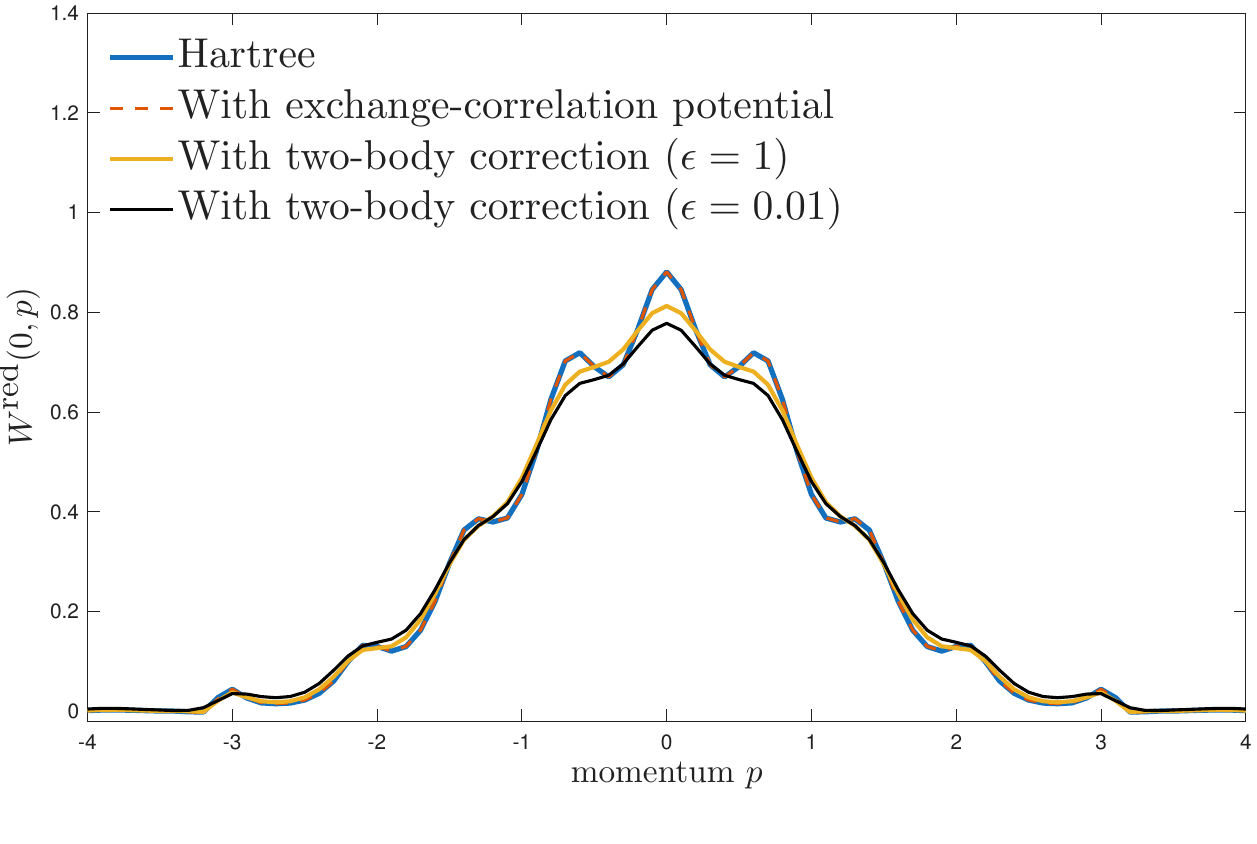}
\includegraphics[width=0.32\textwidth,height=0.22\textwidth]{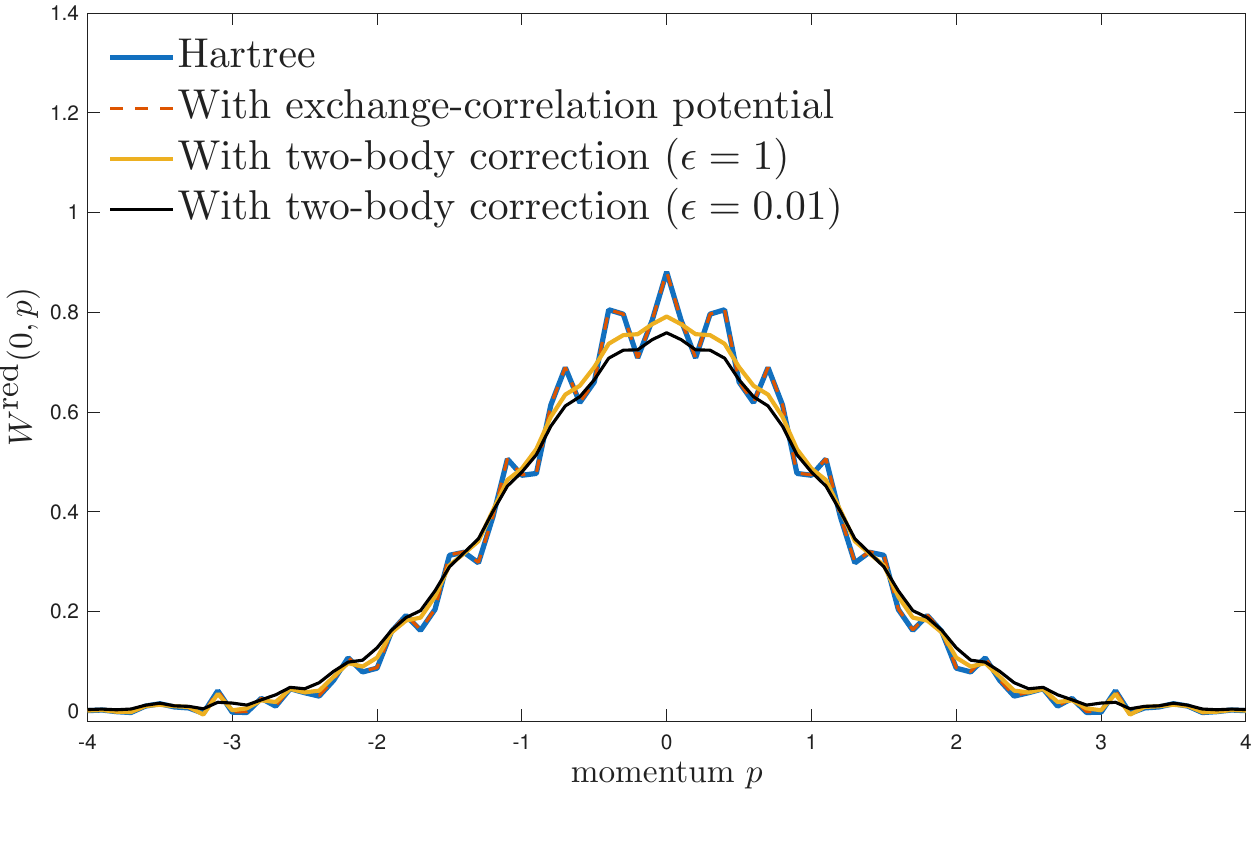}
\includegraphics[width=0.32\textwidth,height=0.22\textwidth]{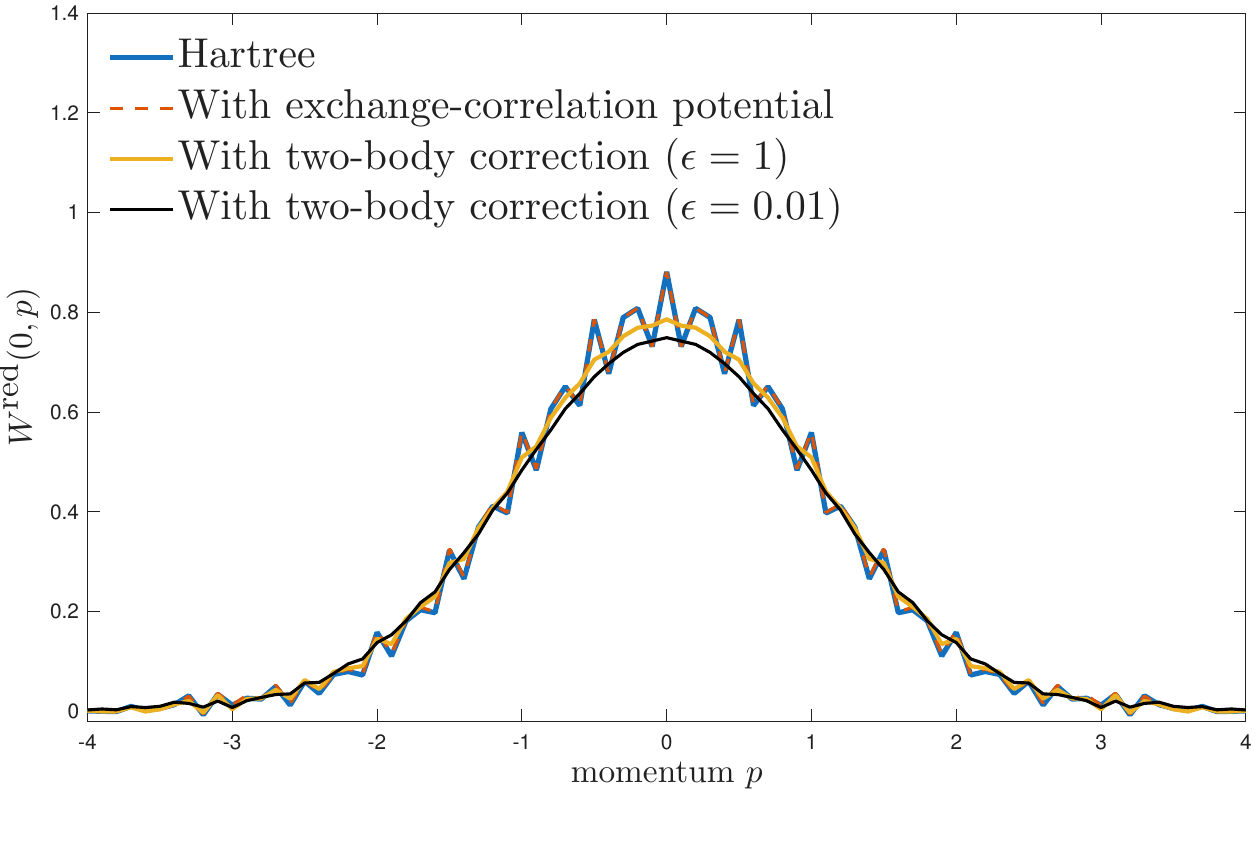}}            
     \caption{Nonlinear quantum Landau damping: Evolution of the reduced Wigner function. The cases from left side to right side are: Hartree, Hartree with XC correction, Hartree with two-body correction ($\epsilon = 1$ or $\epsilon = 0.01$). \label{Wigner_NLD}}    
\end{figure}

  \begin{figure}[!h]
    \centering
    \subfigure[$t=12$.]{
    \includegraphics[width=1.4in,height=1.0in]{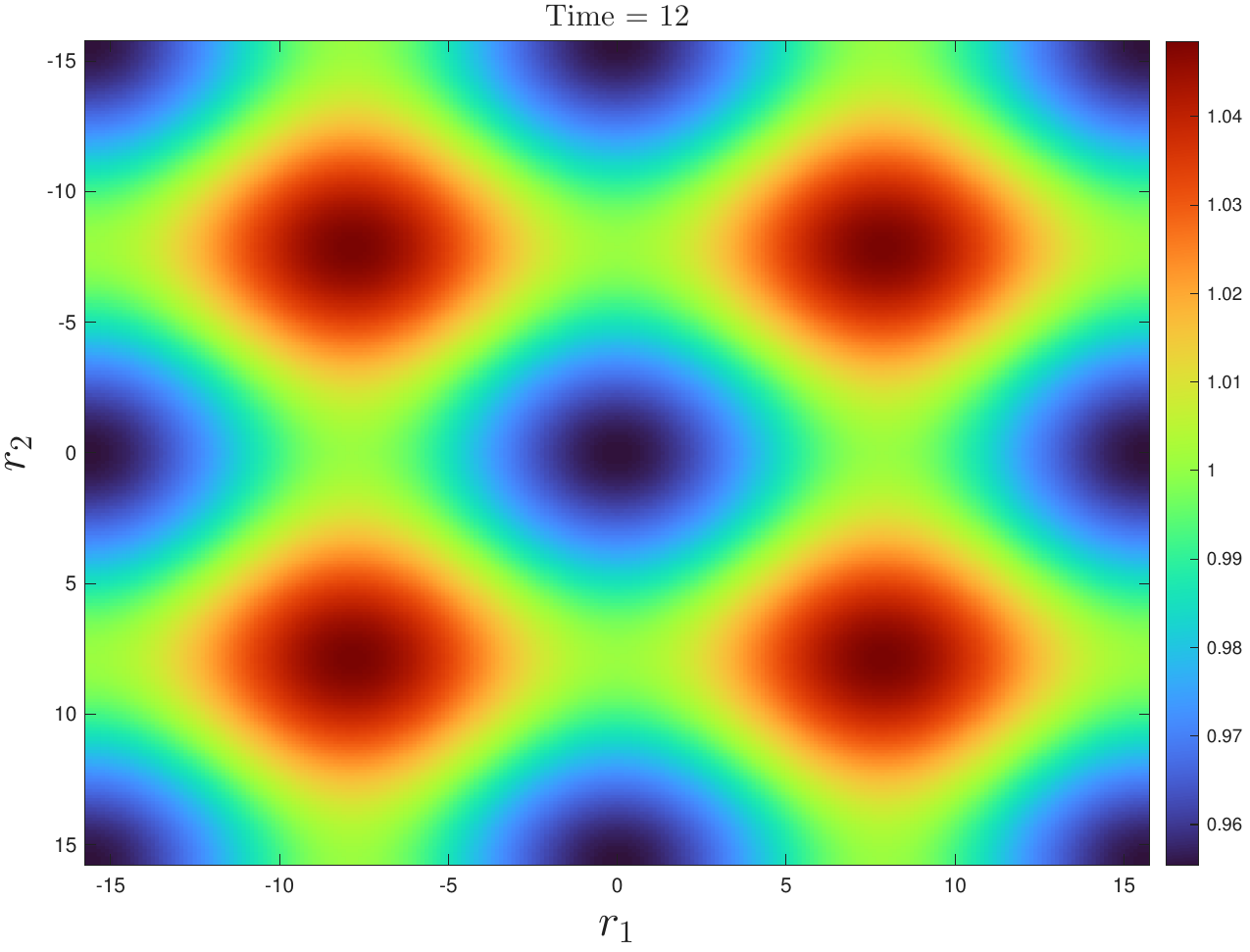}
    \includegraphics[width=1.4in,height=1.0in]{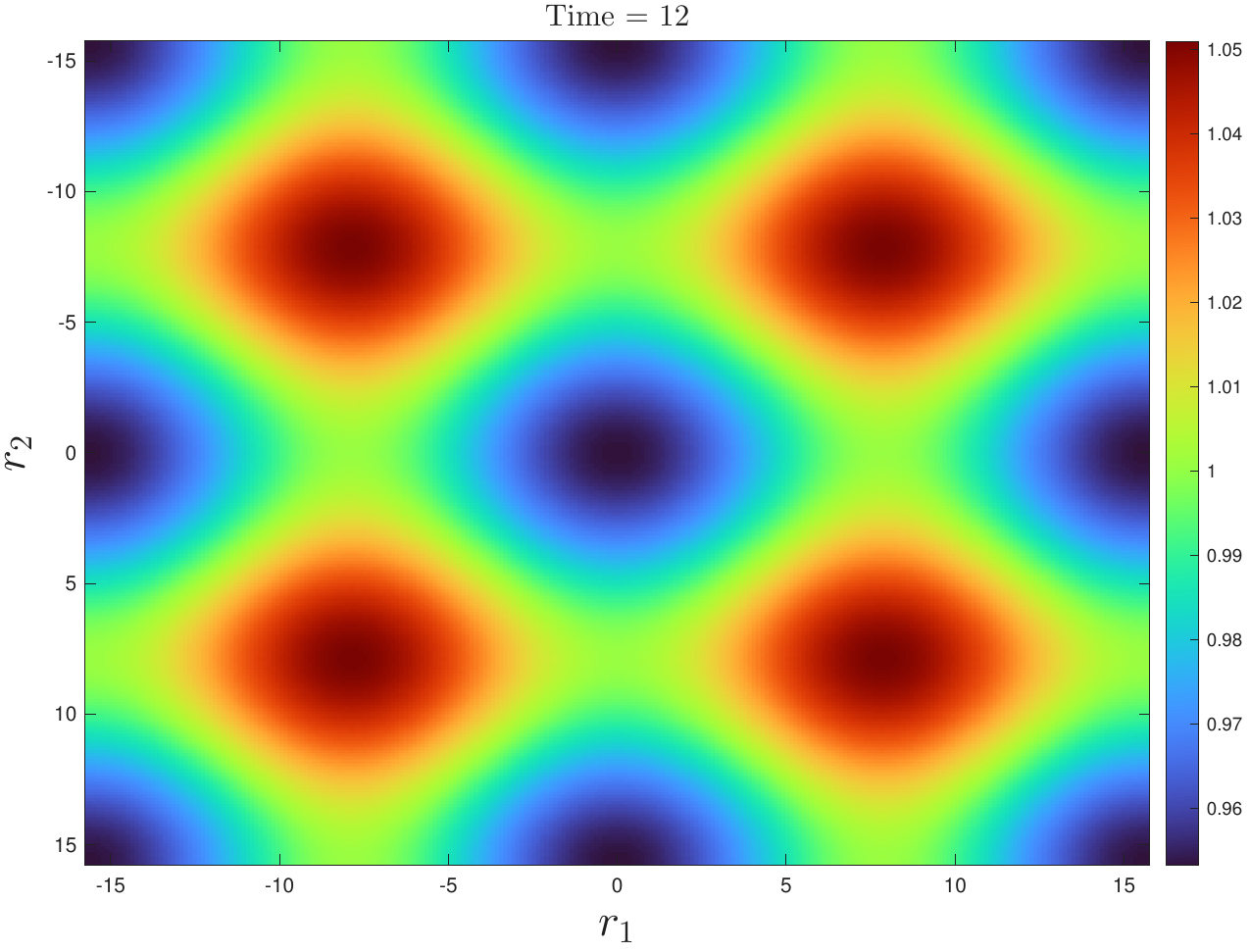}
    \includegraphics[width=1.4in,height=1.0in]{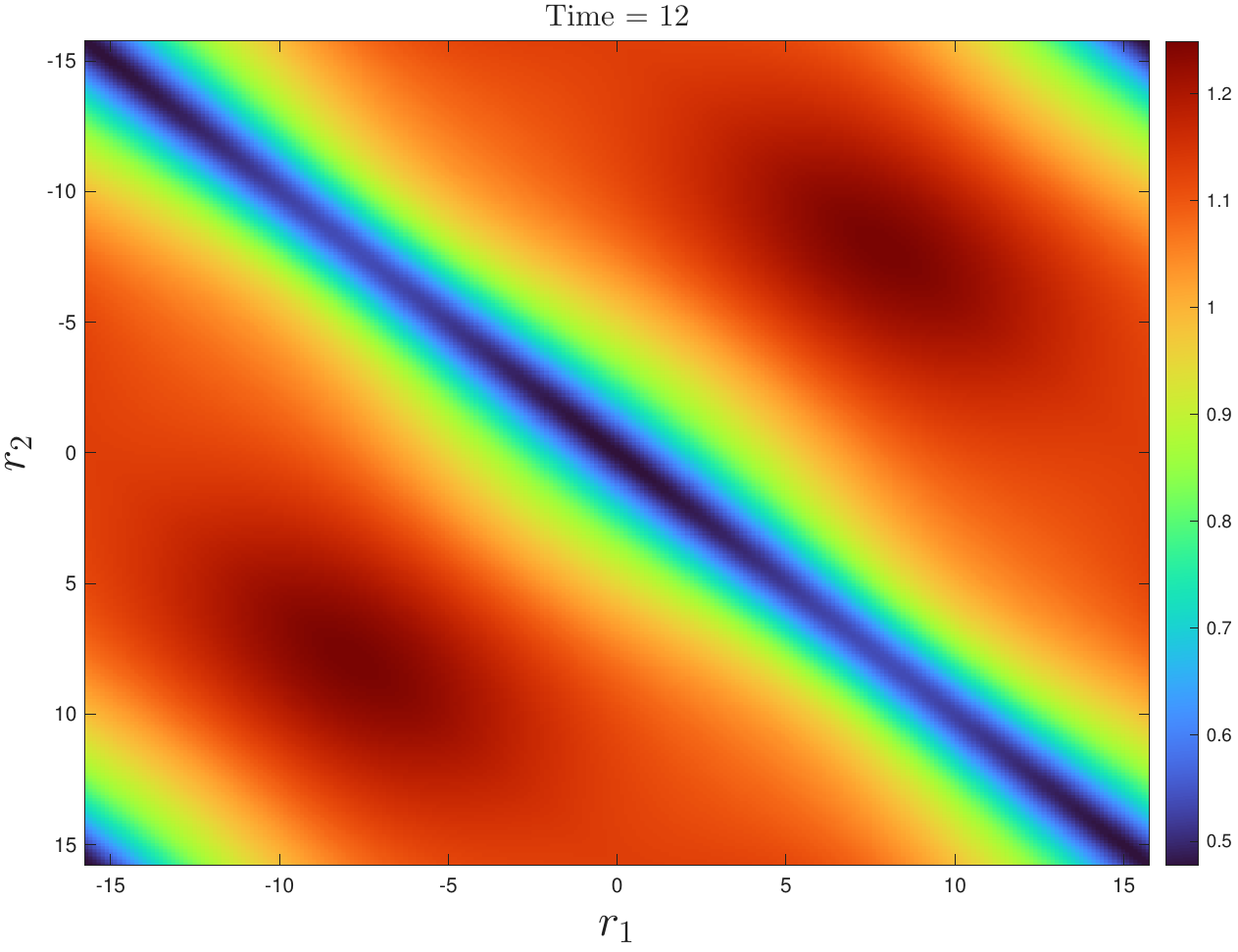}
    \includegraphics[width=1.4in,height=1.0in]{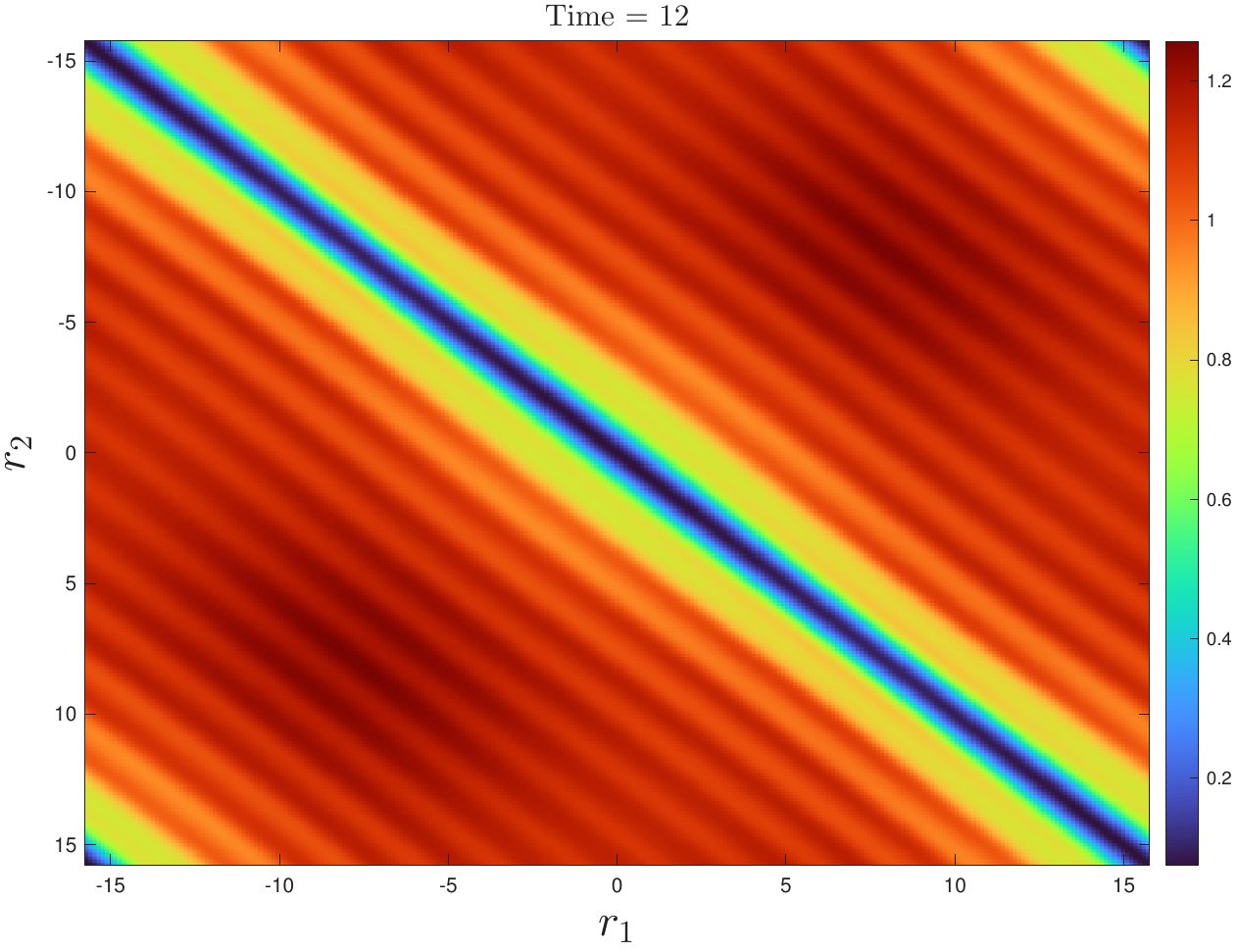}}
    \\
    \centering
    \subfigure[$t=24$.]{
    \includegraphics[width=1.4in,height=1.0in]{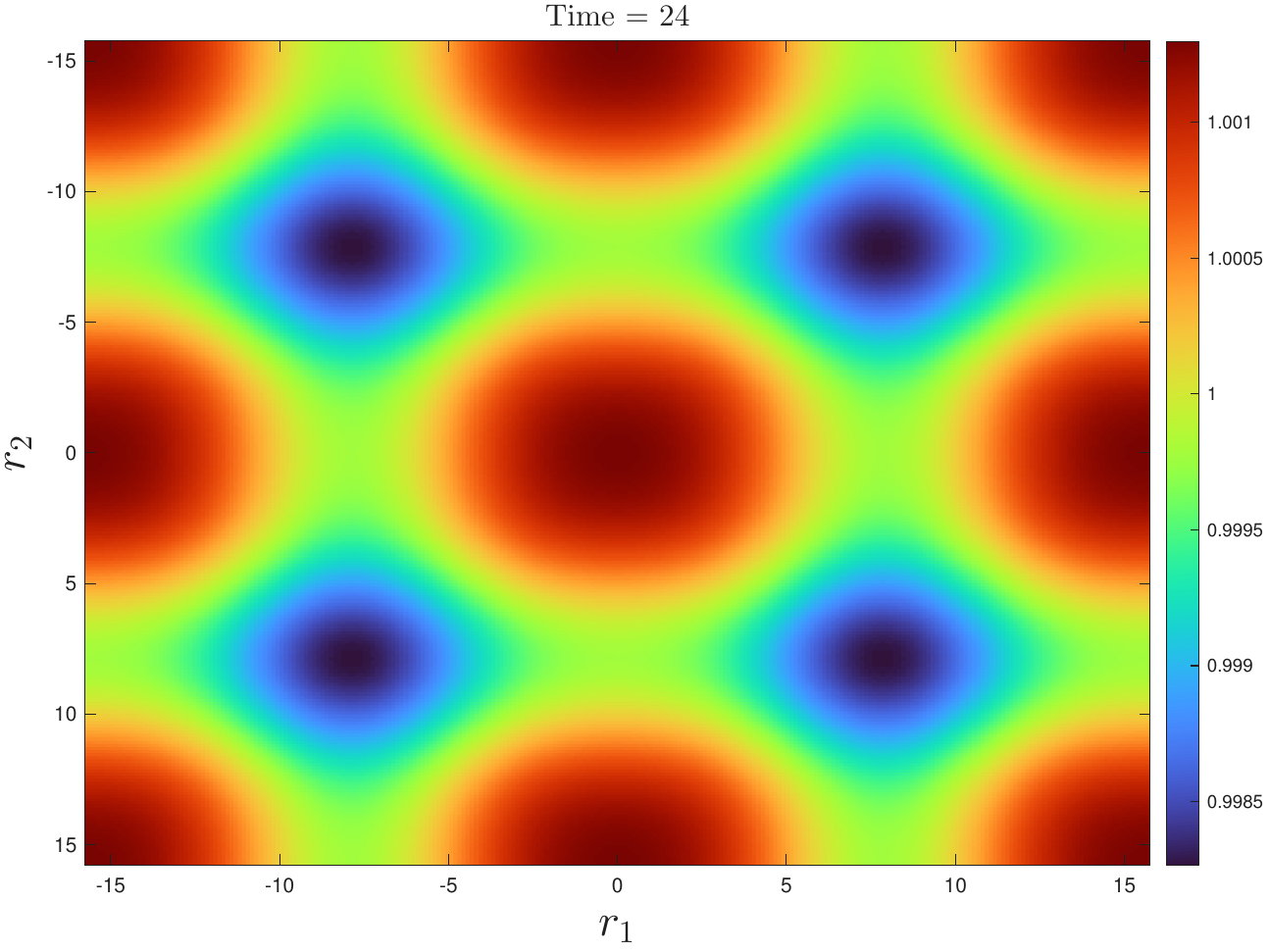}
    \includegraphics[width=1.4in,height=1.0in]{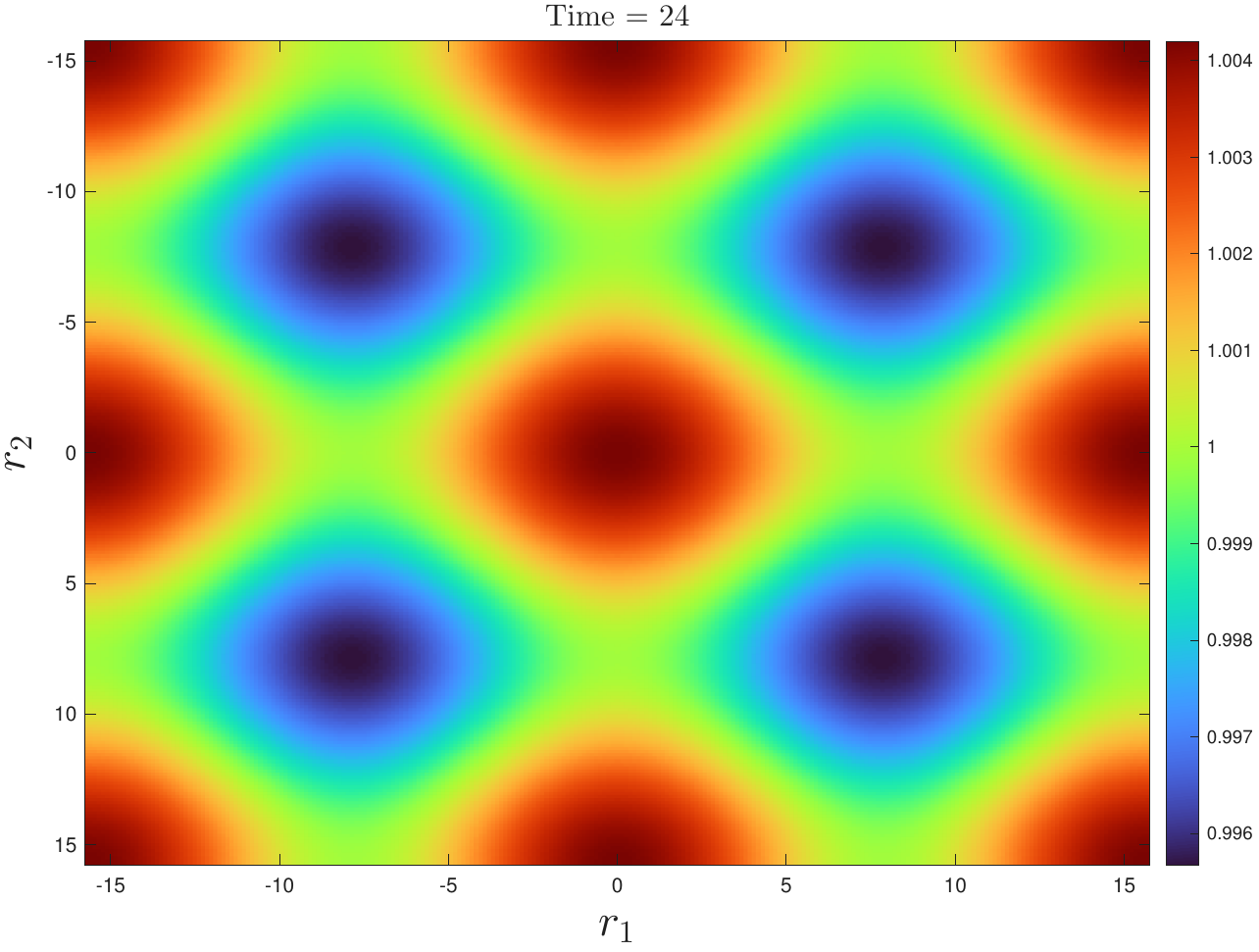}
    \includegraphics[width=1.4in,height=1.0in]{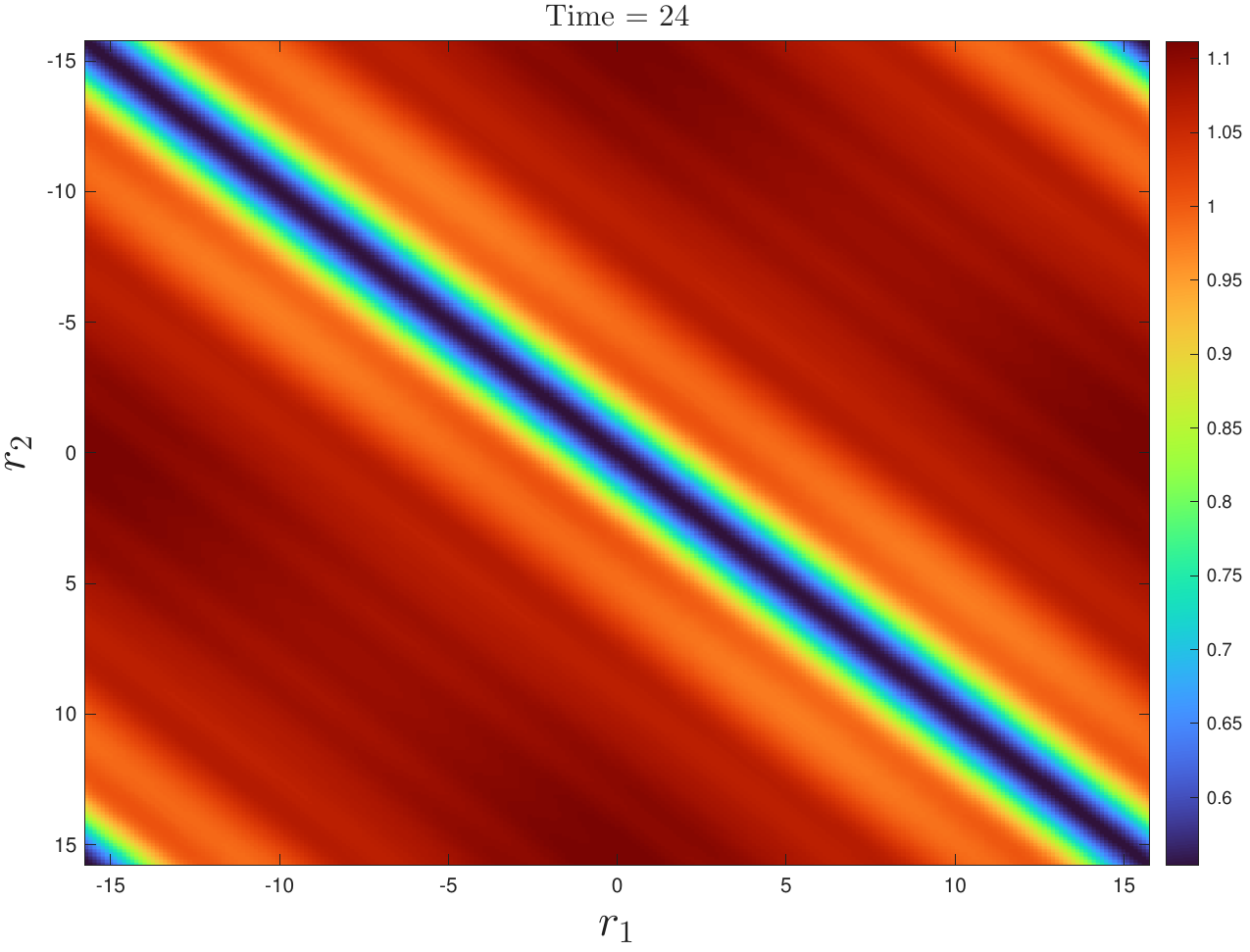}
    \includegraphics[width=1.4in,height=1.0in]{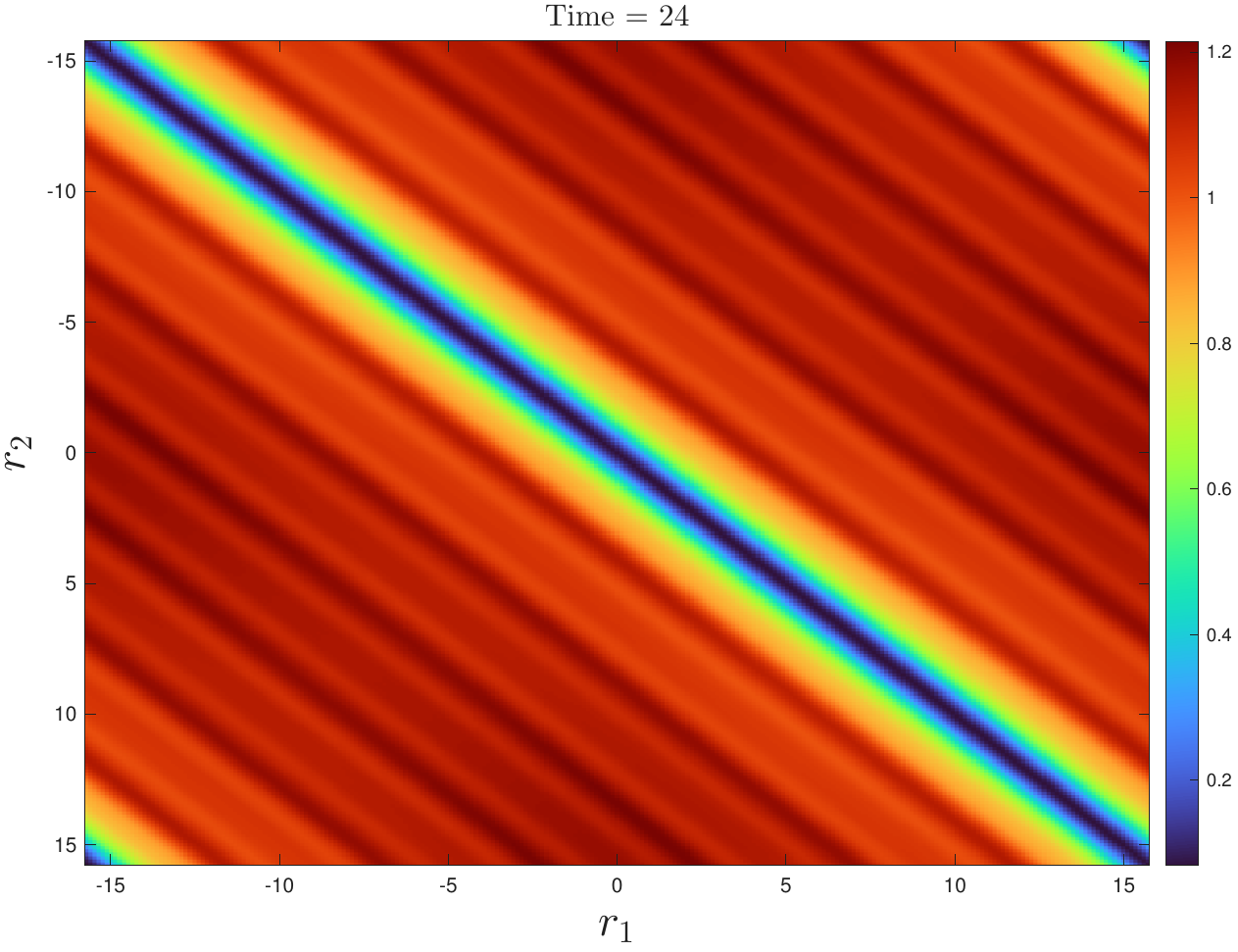}}
   \\
    \centering
    \subfigure[$t=36$.]{
    \includegraphics[width=1.4in,height=1.0in]{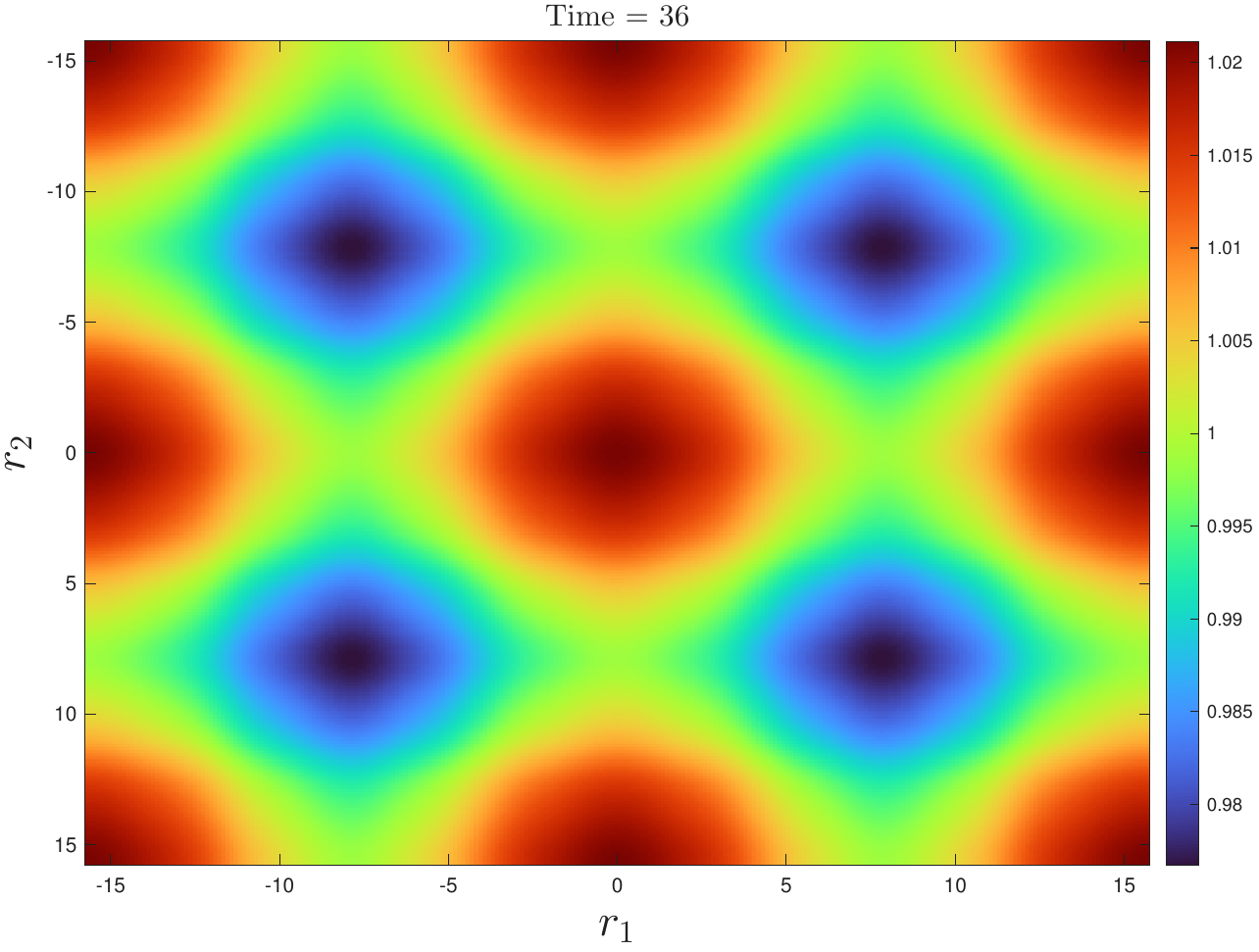}
    \includegraphics[width=1.4in,height=1.0in]{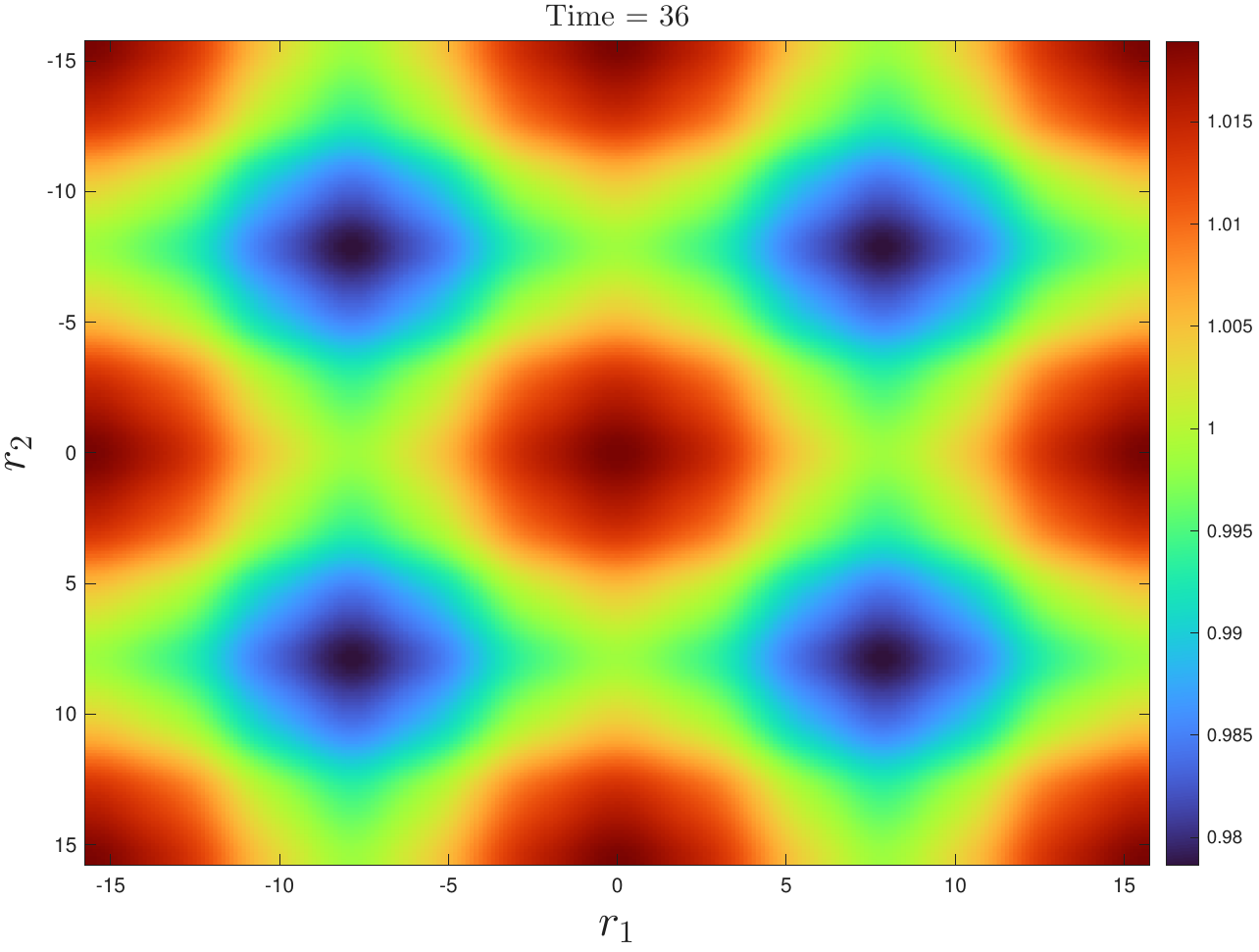}
    \includegraphics[width=1.4in,height=1.0in]{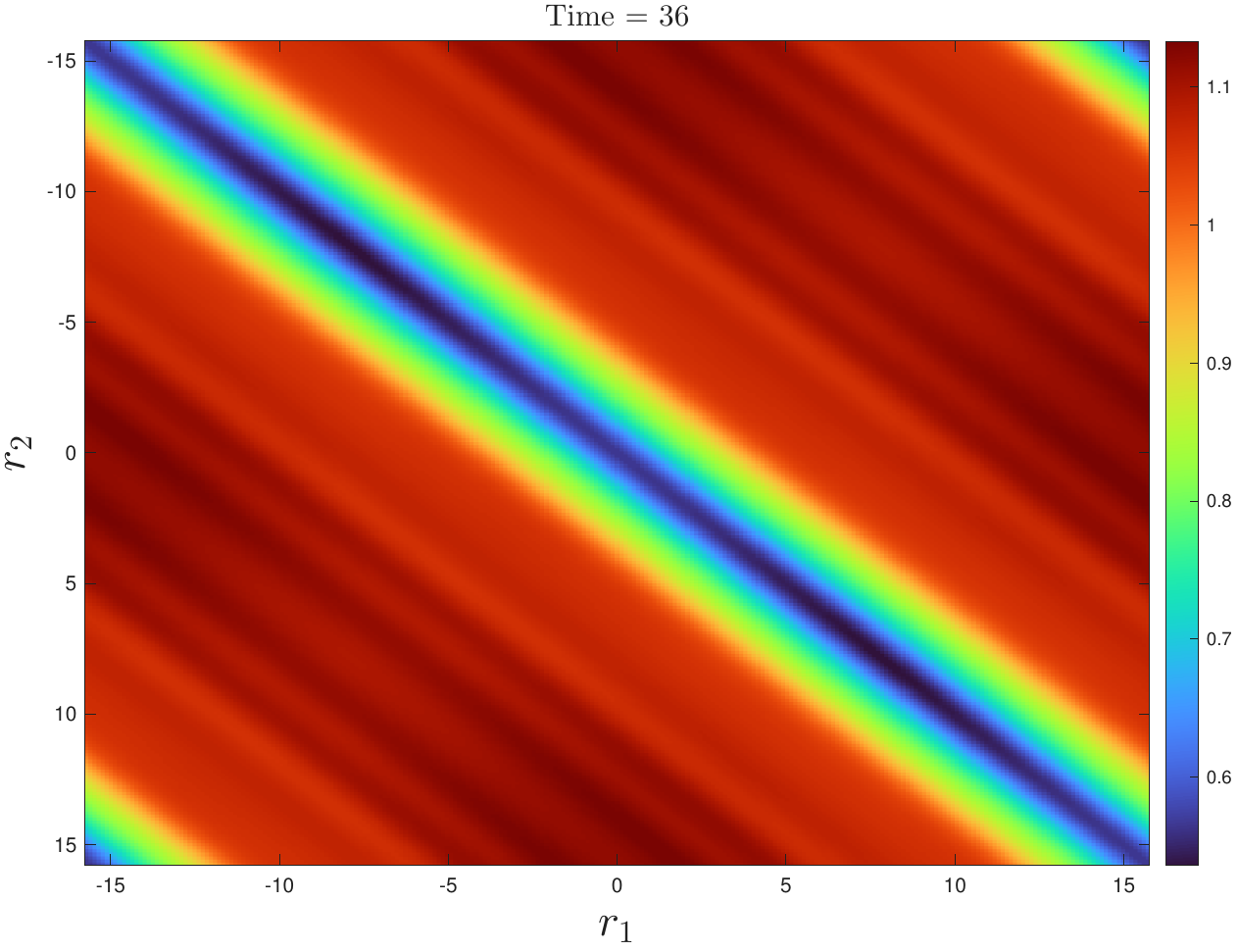}
    \includegraphics[width=1.4in,height=1.0in]{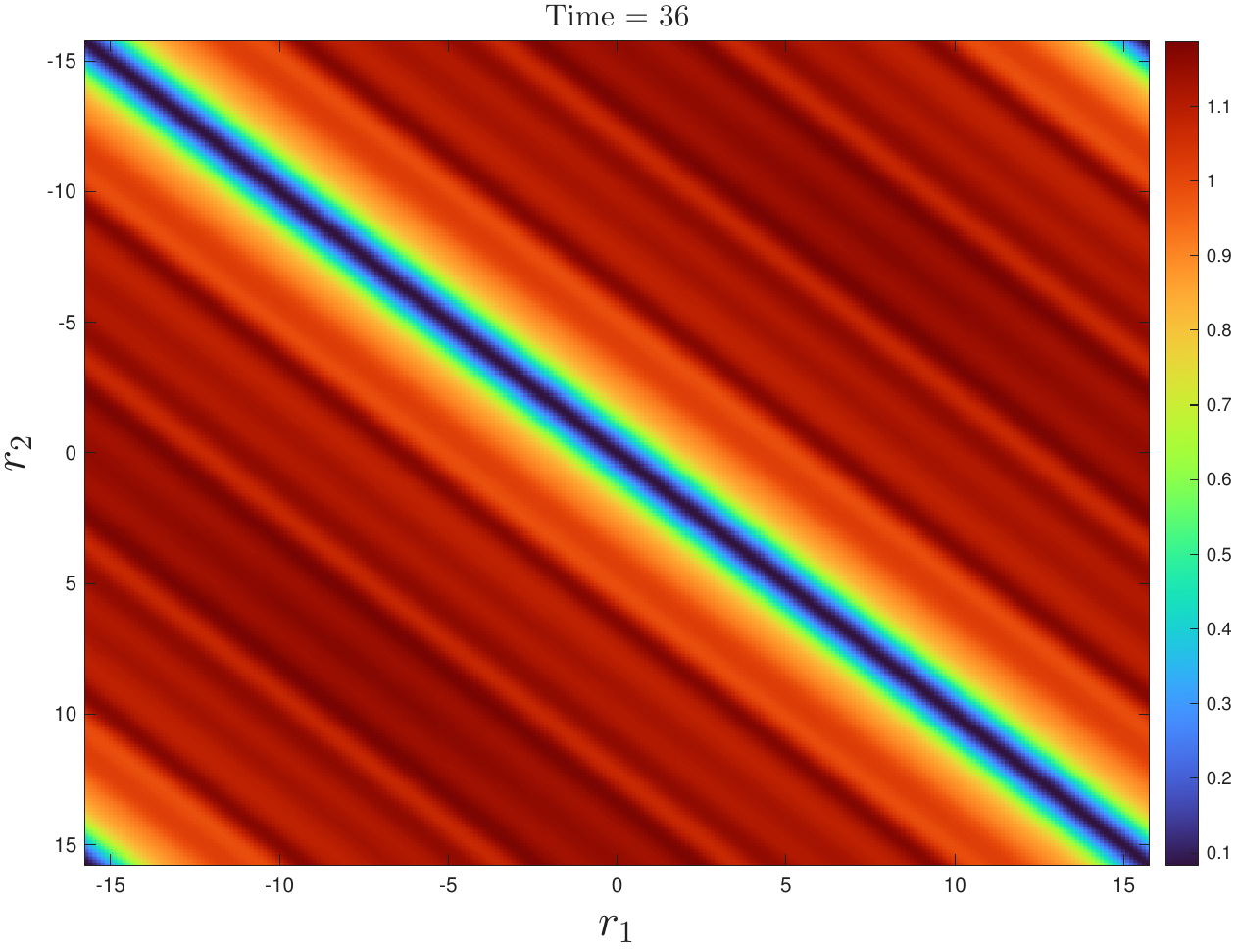}}
   \\
     \centering
    \subfigure[$t=48$.]{
    \includegraphics[width=1.4in,height=1.0in]{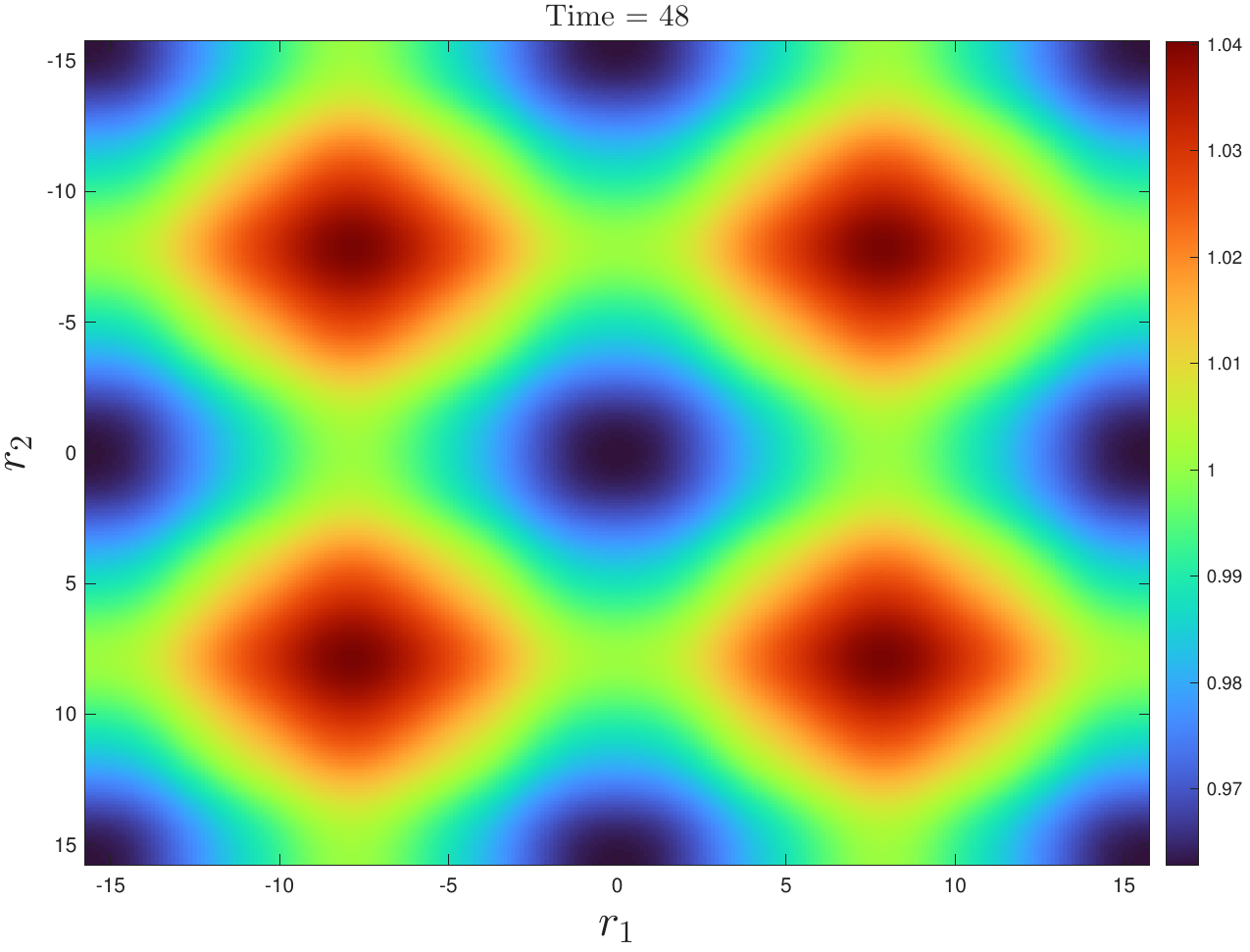}
    \includegraphics[width=1.4in,height=1.0in]{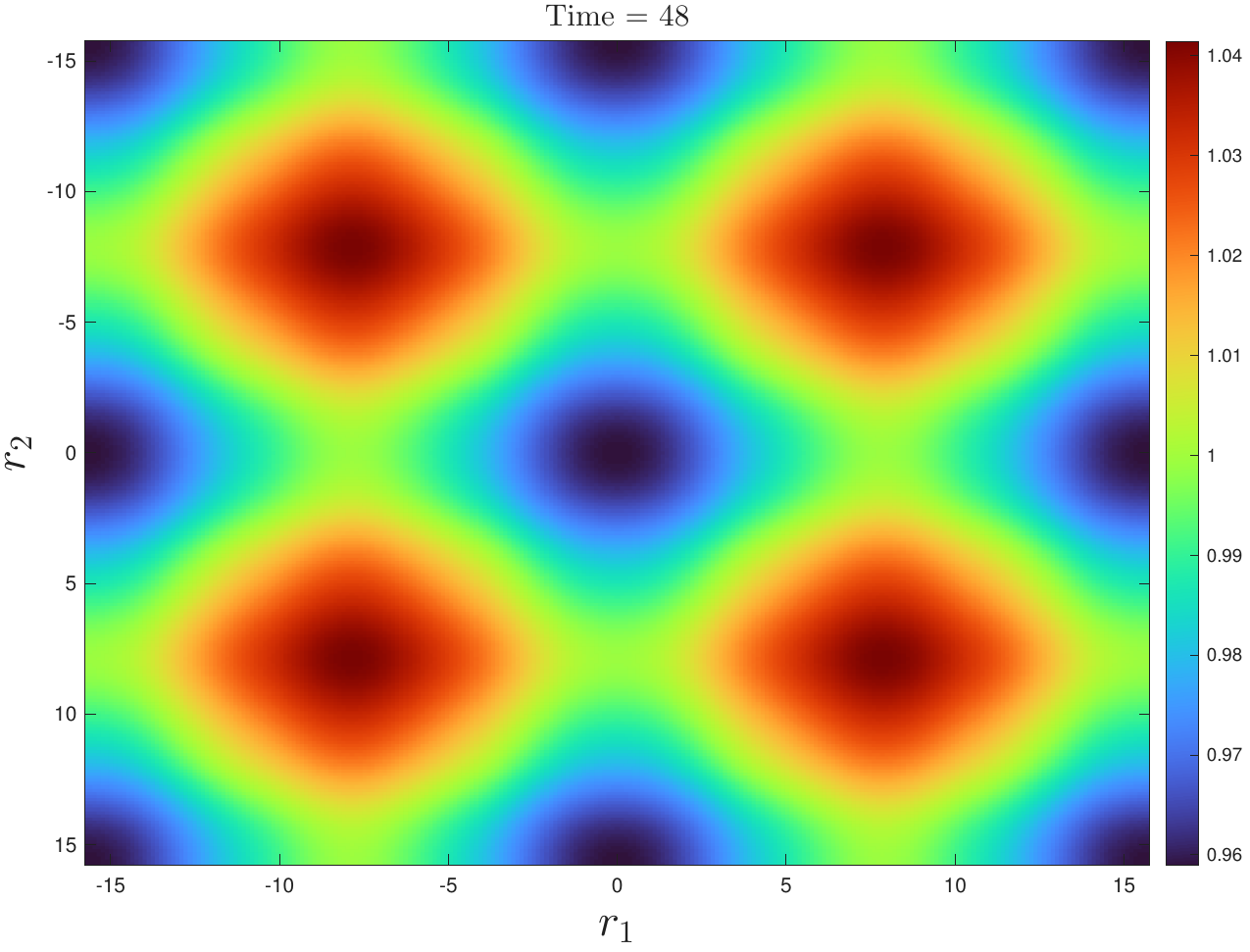}
    \includegraphics[width=1.4in,height=1.0in]{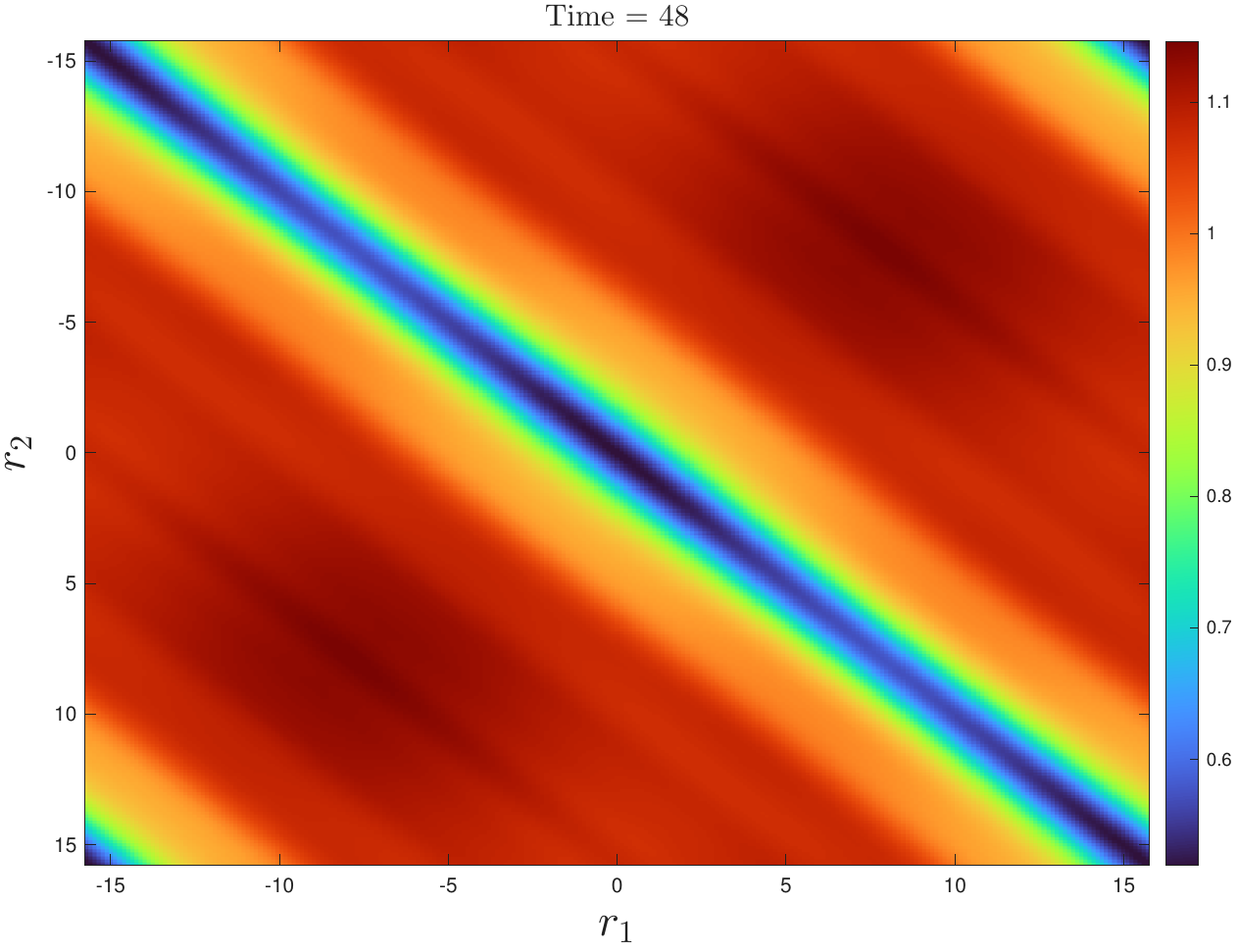}
    \includegraphics[width=1.4in,height=1.0in]{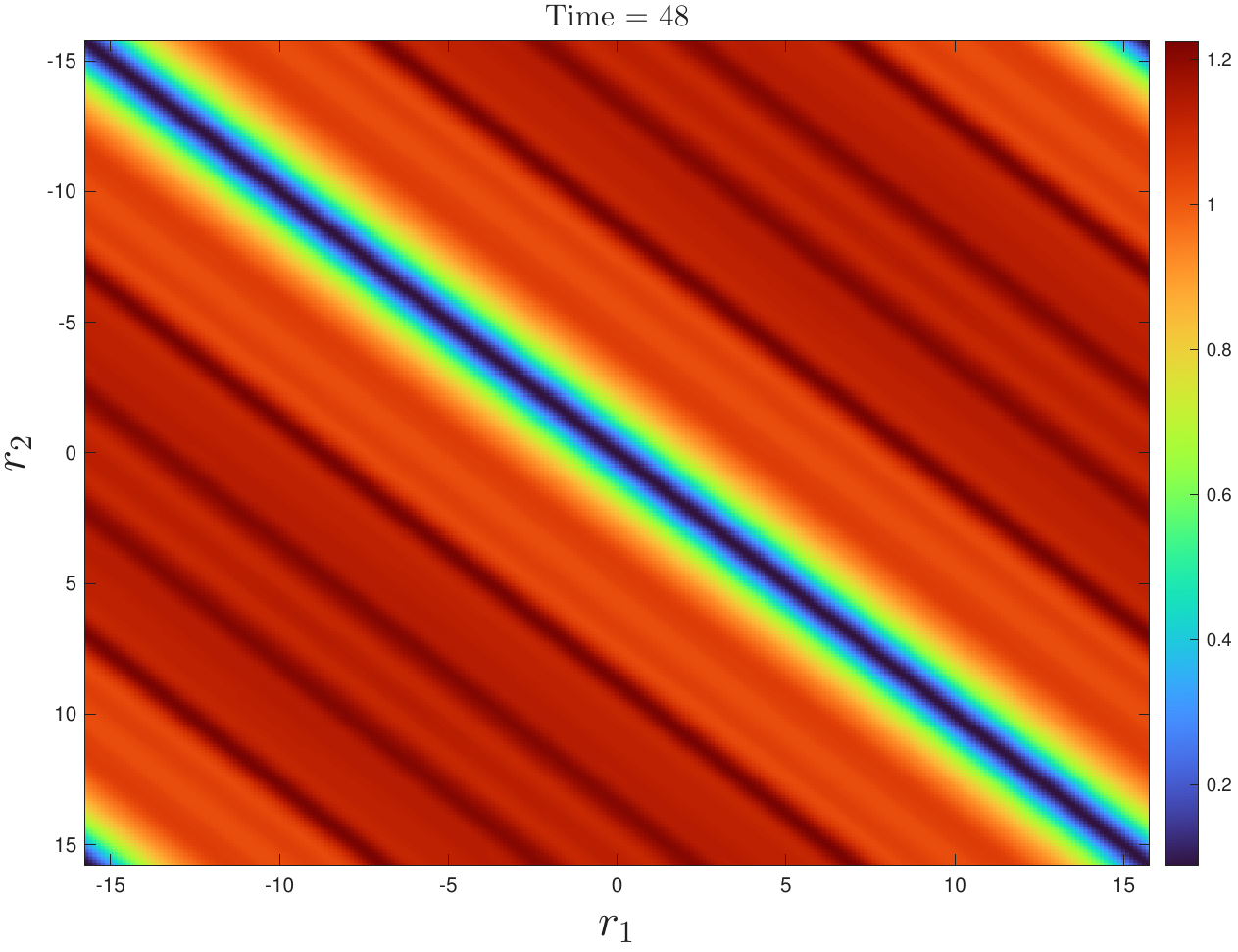}}
    \\
    \centering
    \subfigure[$t=60$.]{
    \includegraphics[width=1.4in,height=1.0in]{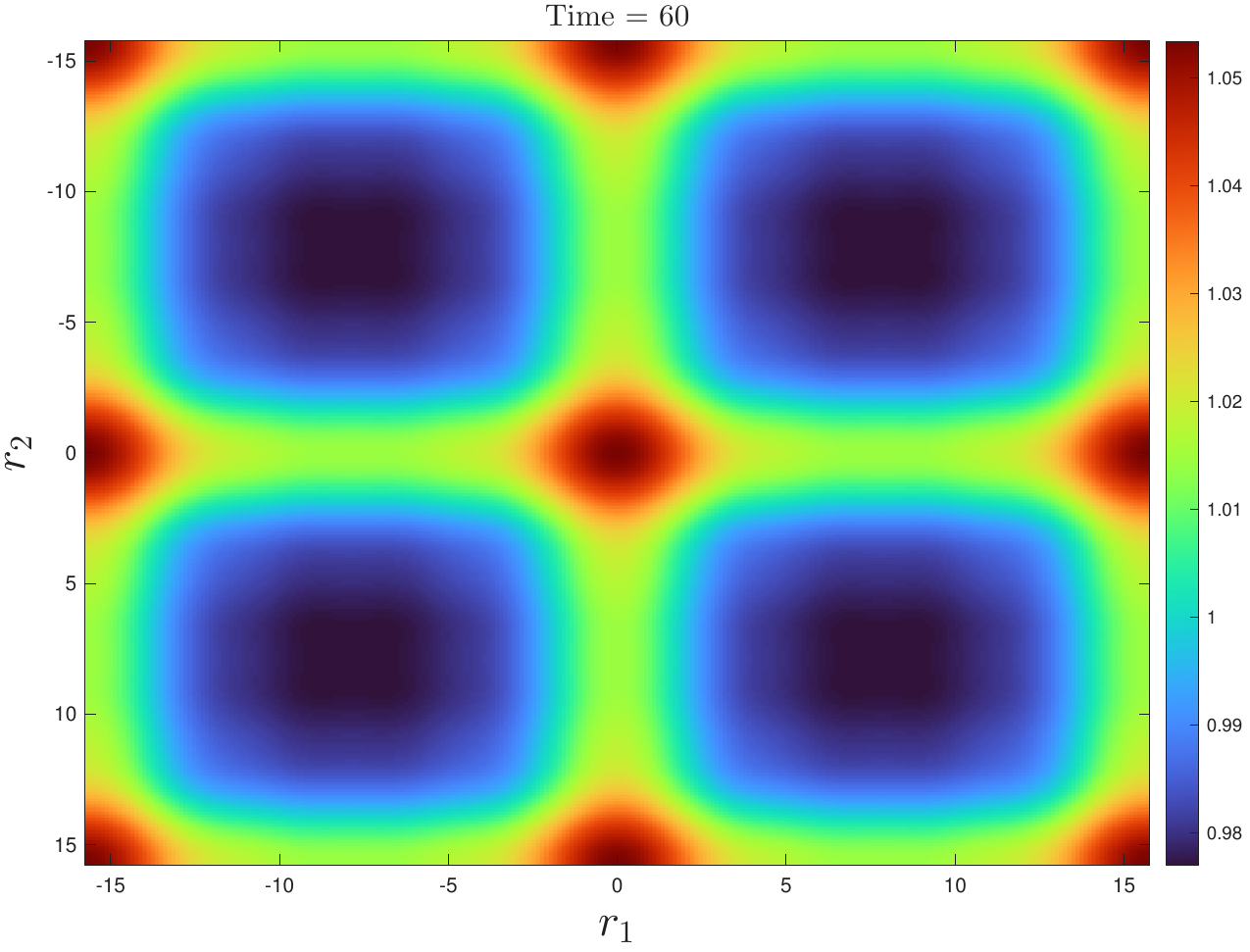}
    \includegraphics[width=1.4in,height=1.0in]{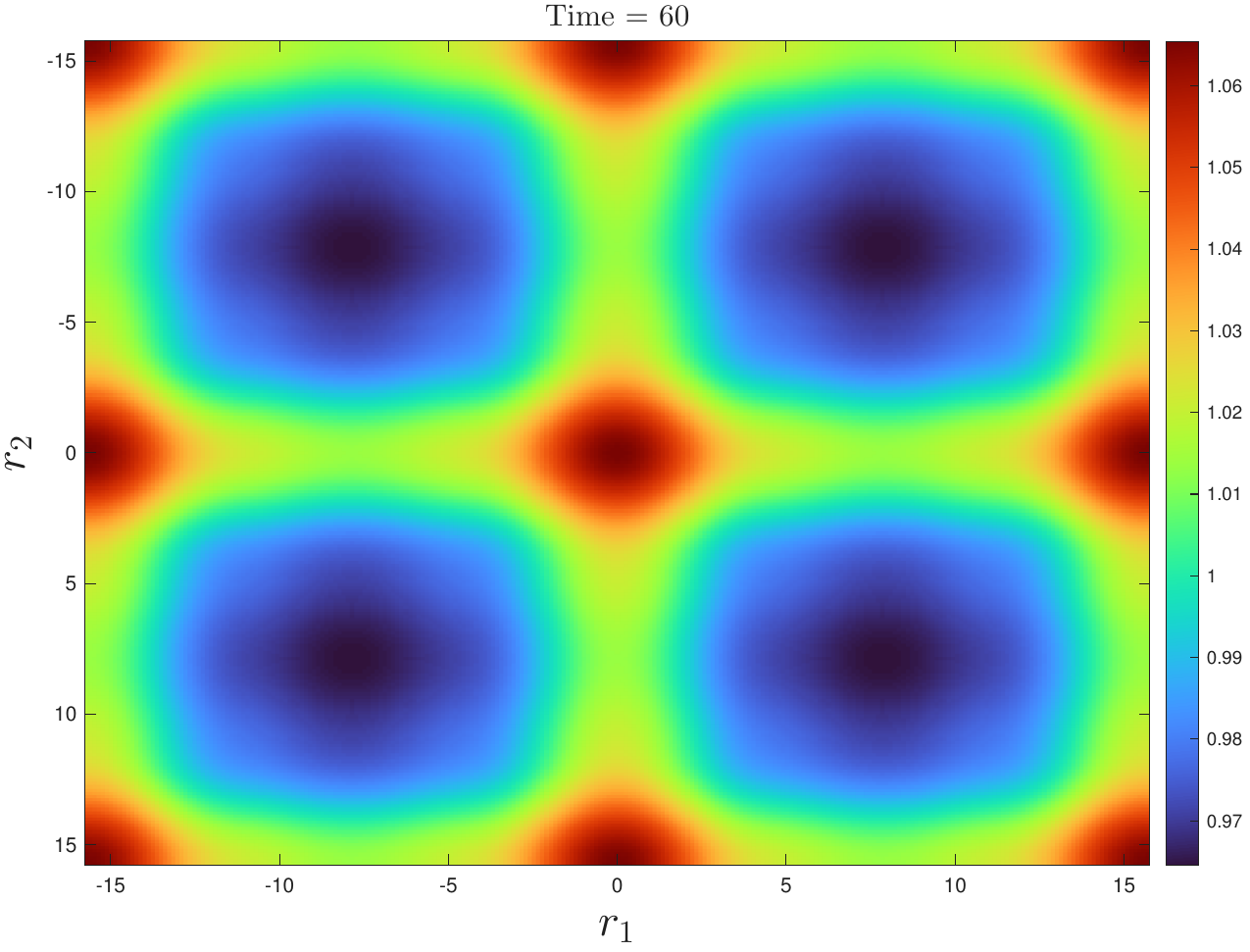}
    \includegraphics[width=1.4in,height=1.0in]{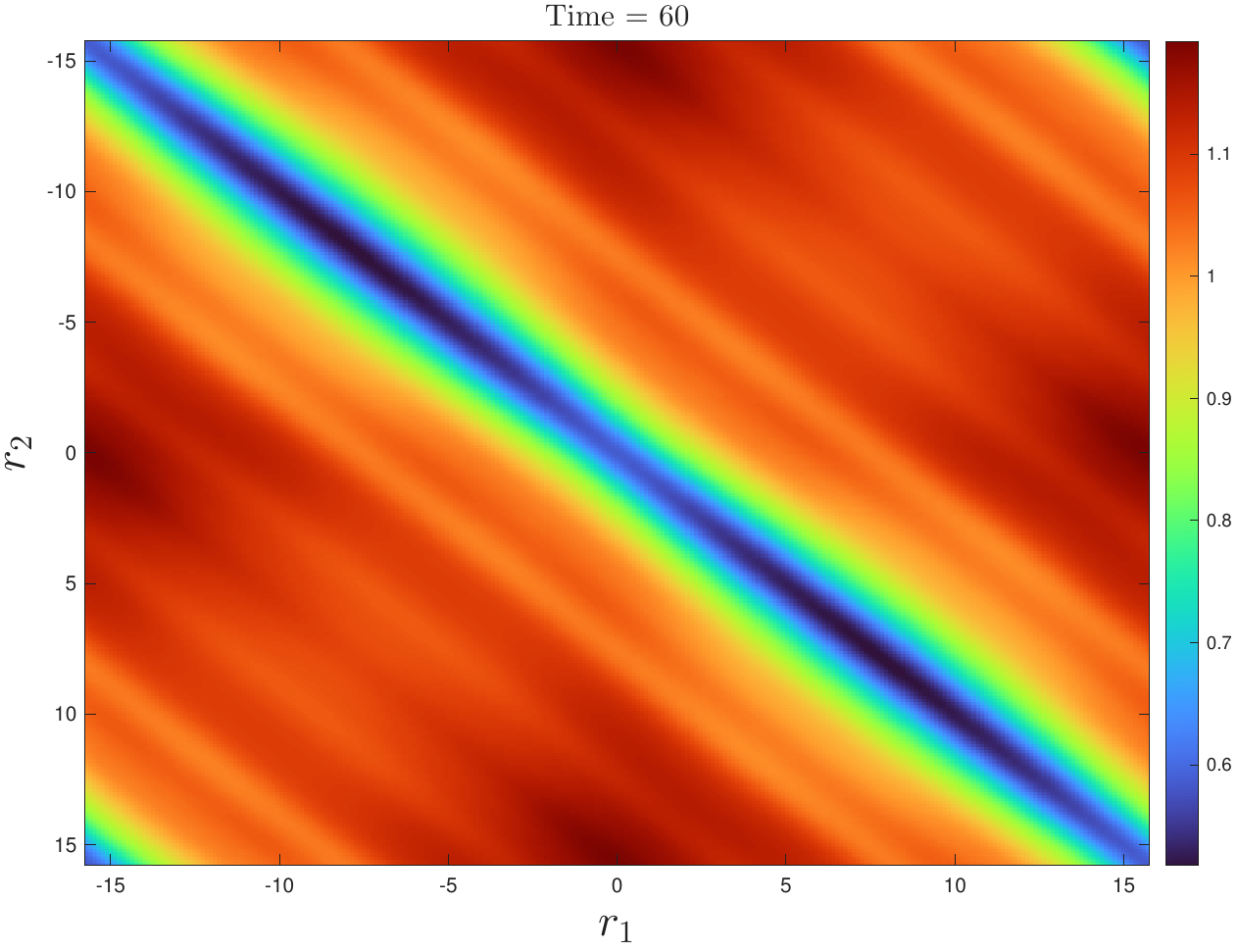}
    \includegraphics[width=1.4in,height=1.0in]{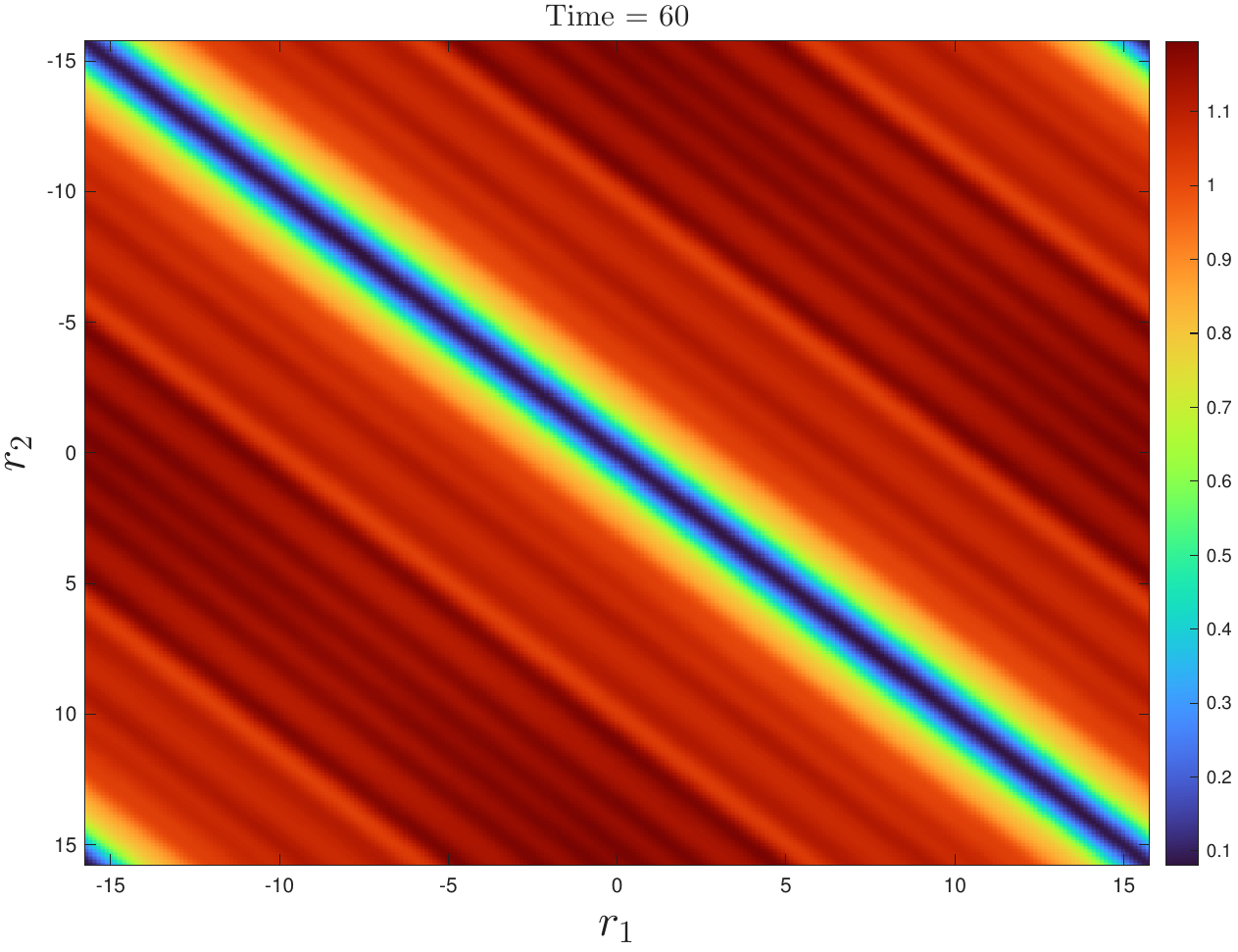}}               
     \caption{Nonlinear quantum Landau damping: Evolution of the number density $n_{12}(r_1, r_2, t)$. The cases from left side to right side are: Hartree, Hartree with XC correction, Hartree with two-body correction ($\epsilon = 1$ or $\epsilon = 0.01$).  Few particles can occupy the same position ($r_1=r_2$) in the two-particle configuration space when two-body interaction is included. \label{xdist_NLD}}  
\end{figure}

% First, we add a small perturbation on the initial Gaussian distribution, say,  $\varepsilon_1 = \varepsilon_2 = 0.002$, and study the evolution of the electrostatic field $E(x_1, t)$. From Figure \ref{LD}, the electrostatic field damps away, which recovers the linear Landau damping in the Wigner-Poisson system. The correction of H-L potential \eqref{mean_field_approximation} seems to be negligible, whereas the two-body scattering impedes the electric field to be damped away. This results from the polarization of electrons, as clearly visualized by the number density in Figure \ref{LD_xdist}, since the repulsive smoothed Coulomb interaction forbids particles to be located at the same location $x_1 = x_2$.

When we add a large perturbation on the initial Gaussian ($\varepsilon = 0.1$), as seen in Figure \ref{wNLD_long}, Landau
damping of the plasma vanishes after a short time because the nonlinear interaction of a plasma wave with resonant electrons results in a plateau in the
electron distribution function close to the phase velocity of the plasma wave. In the linear stage ($t = 0$-$15$), the purely local exchange–correlation effect does not alter the Landau damping rate of the plasma, whereas the two-body interactions significantly modify the linear Landau damping rate and can even change the nonlinear coupling processes. From both the amplitude of electric field and the plots of number densities in Figure \ref{xdist_NLD}, the correction of H-L potential seems to be negligible, but the two-body interaction directly gives rises to the strong two-particle correlation. Due to the smoothed Coulomb interaction, few particles can occupy the same position ($r_1=r_2$) in the two-particle configuration space. Such correlation has an evident influence on the linear damping rate (see Figure \ref{wNLD_long}, $\epsilon=0.01$), which may provide an evidence on the necessity to incorporate the two-body interaction for an accurate description of dense plasma.  

The nonlinear Landau damping comes with filamentation and oscillations of the distribution function \cite{MouhotVillani2011}. This holds true for the quantum system as visualized in Figure \ref{Wigner_NLD}, say, the phase-space distribution becomes more and more oscillating, whereas the hole structure near the phase velocity in classical Vlasov-Poisson system is replaced by the negative region reflecting the uncertainty. An interesting finding is that two-body interaction can suppress the filamentation, as seen in the sliced plot \ref{slice_Wigner_NLD}, so that the Wigner distribution function relaxes to a smoother state. This coincides with our intuition that the collisions of two particles exhibit typical dissipation and thus results in irreversible smoothing effects \cite{WenBartonRiosStevensen2018}.

%The initial 1-D Fermi-Dirac distribution 
%\begin{equation}
%f(v; \Theta) = \frac{3}{4} \frac{n_0}{v_F}\Theta \ln \left(1 + \exp\left[ \Theta^{-1} \left(1 - \frac{v^2}{v_F^2}\right)\right]\right),
%\end{equation}
%with the $\Theta = \frac{k_B T}{\varepsilon_F}$ the degeneracy degree, the Fermi energy $\varepsilon_F = \frac{\hbar^2 (3\pi^2 n_0)^{2/3}}{2m_e}$, and $v_F = \sqrt{\frac{2\varepsilon_F}{m_e}}$.

\subsection{Dynamical XC potential and two-body corrections to quantum two-stream instability}

The two-stream instability is one of the most classical examples of reactive-type instability that occurs in plasma physics, and has a quantum analog \cite{SuhFeixBertrand1991,Haas2019,HuLiangShengWu2022}. In the following, we use the Wigner-Poisson system under the periodic boundary condition as the baseline and investigate the corrections provided by XC potential and two-body interaction. The H-L potential is chosen to model the exchange and correlation effect. The two-body interaction is the smoothed Coulomb potential $V_{ee}(r_1, r_2) = \frac{\gamma}{\sqrt{|r_1 - r_2|^2 + \epsilon}}$ with $\epsilon = 0.001, 0.01, 1$, $\gamma = 1$.

The initial condition is chosen as
\begin{equation}
f_{12}^{init}(r_1, r_2, p_1, p_2) =  \frac{(1 + \varepsilon_1 \cos(k_1 r_1)) (1 + \varepsilon_2 \cos(k_2 r_2)) }{(6\sqrt{2\pi})^2}(1 + 5p_1^2) (1 + 5p_2^2) \me^{- \frac{p_1^2 + p_2^2}{2}}
\end{equation}
with parameters: $k_1 = 0.4$, $k_2 = 0.4$. The computational domain $\mathcal{X} \times \mathcal{P} = [-5\pi, 5\pi]^2 \times [-6.4, 6.4]^2$ is represented by a uniform mesh with $257^2\times 128^2 \approx 1.08\times10^9$ grid points and a fixed spatial spacing $\Delta x = \frac{5\pi}{128}$, $\Delta p = 0.1$. The time step is $\Delta t = 0.01$ and the final time is $T = 200$. 

\begin{figure}[!h]
\centering
\subfigure[Two-stream instability up to $T = 60$. \label{two_stream_short}]{
\includegraphics[width=0.48\textwidth,height=0.27\textwidth]{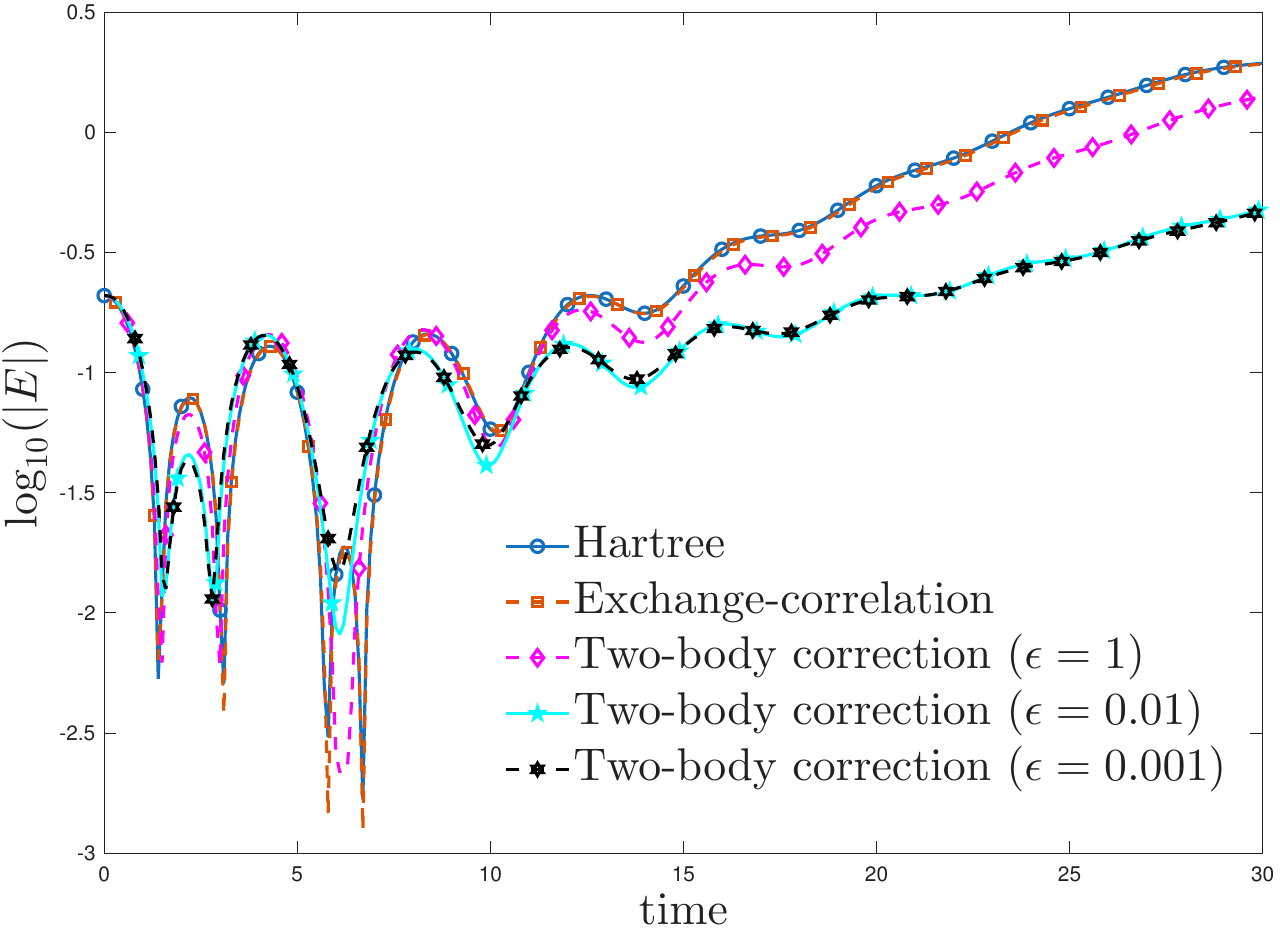}}
\\
\centering
\subfigure[Two-stream instability up to $T =200$. (left: with XC potential, right: with two-body interaction) \label{two_stream_long}]{
\includegraphics[width=0.48\textwidth,height=0.27\textwidth]{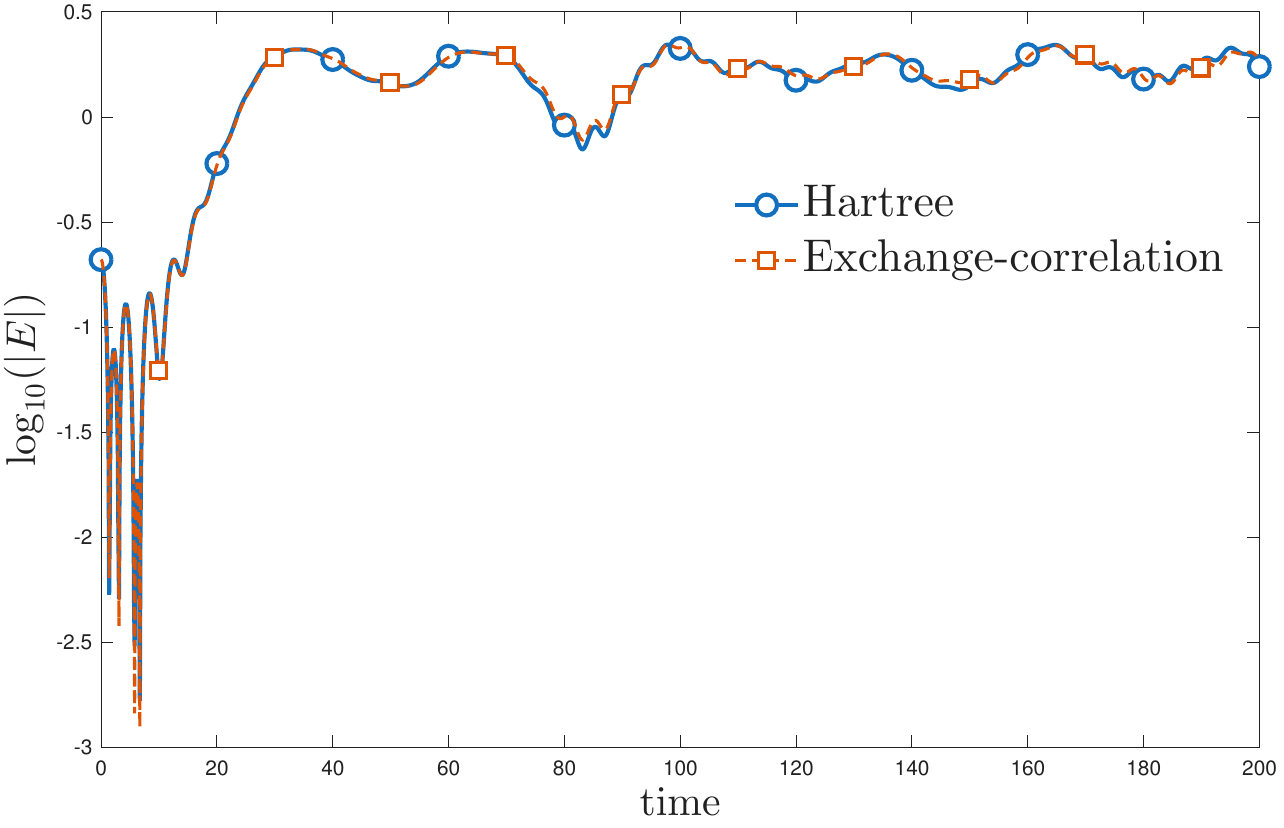}
\includegraphics[width=0.48\textwidth,height=0.27\textwidth]{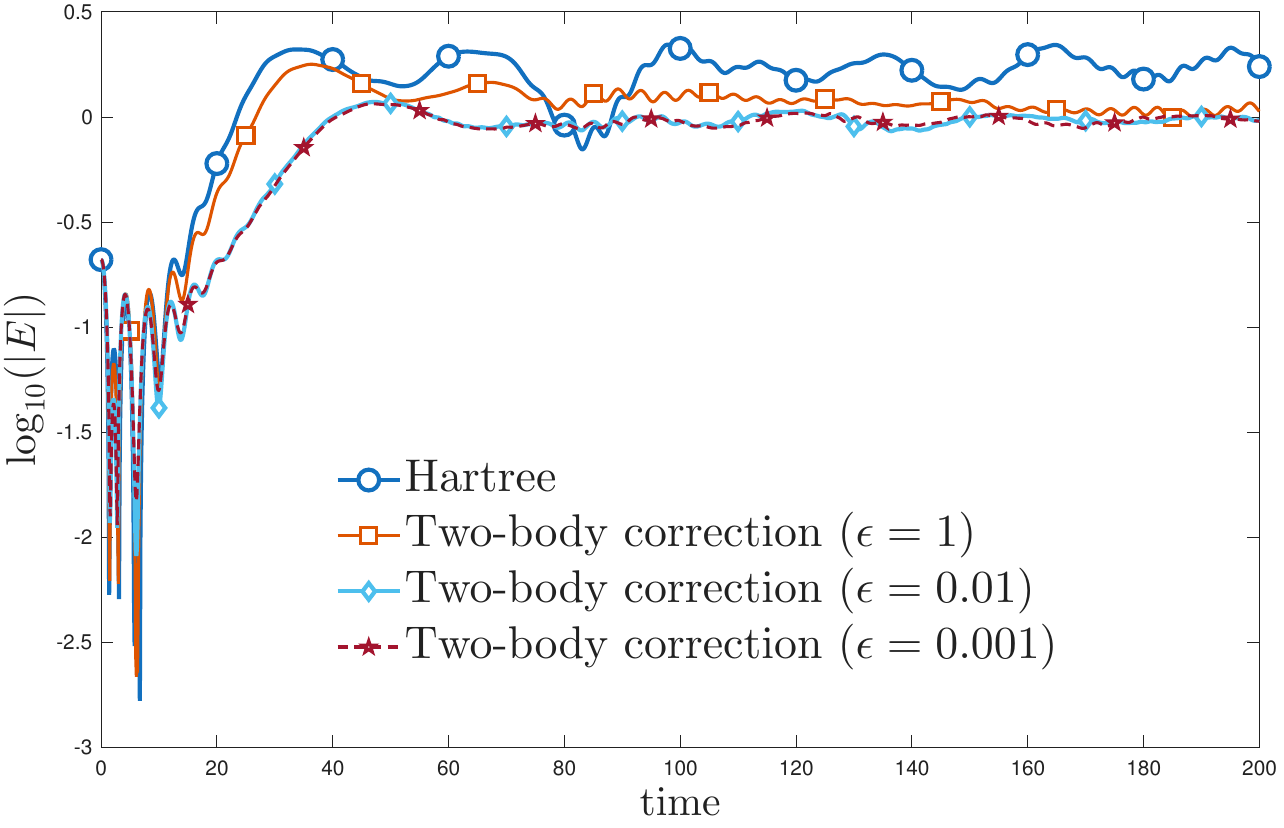}}
\caption{Quantum two-stream instability up to $T =200$. The correction of H-L potential  seems to be negligible, whereas the two-body scattering may suppress the instability. }
\end{figure}

\begin{figure}[!h]
    \centering
    \subfigure[$t=12$.]{
    \includegraphics[width=1.4in,height=1.0in]{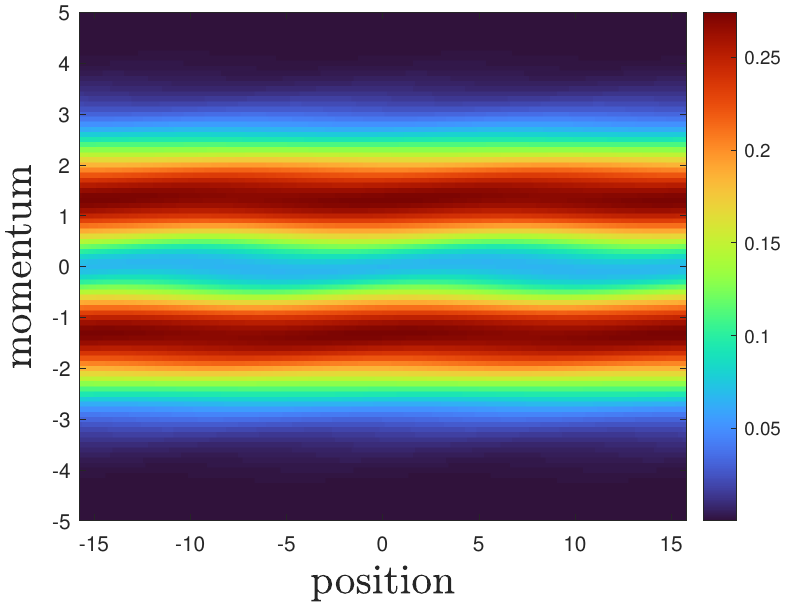}
    \includegraphics[width=1.4in,height=1.0in]{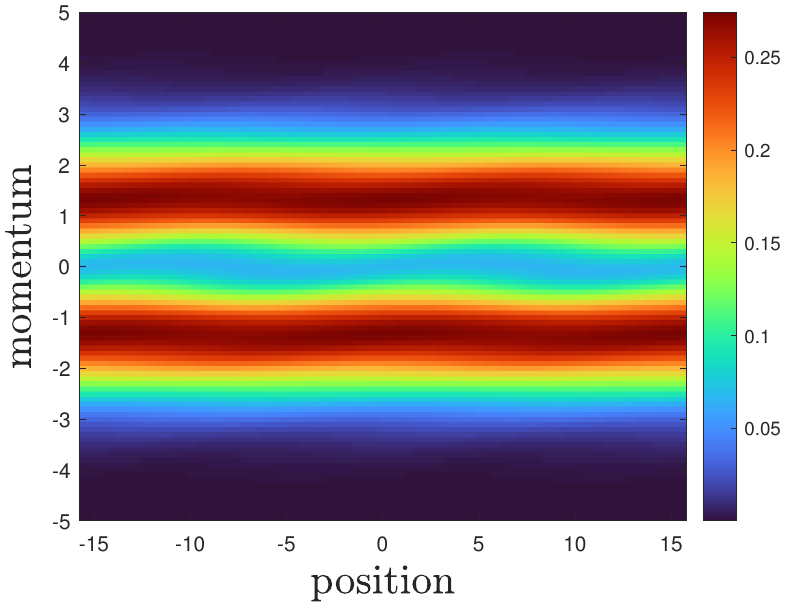}
    \includegraphics[width=1.4in,height=1.0in]{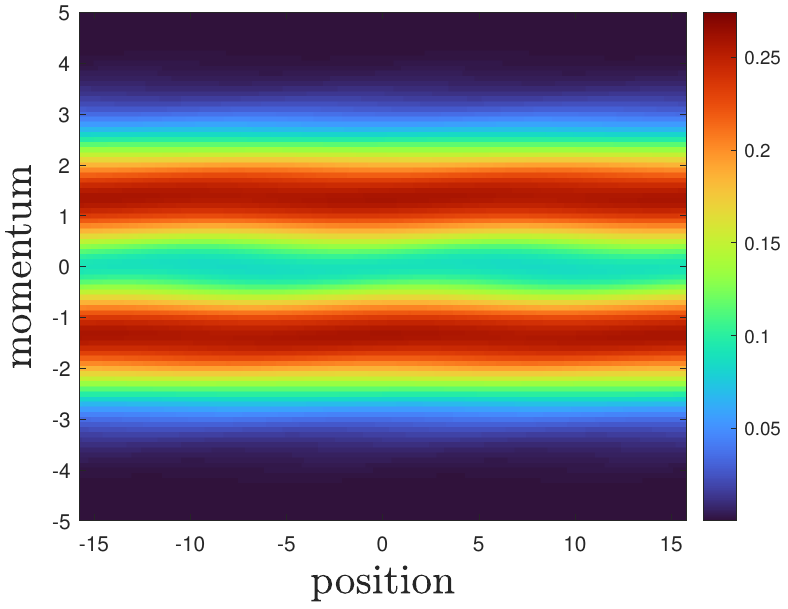}
    \includegraphics[width=1.4in,height=1.0in]{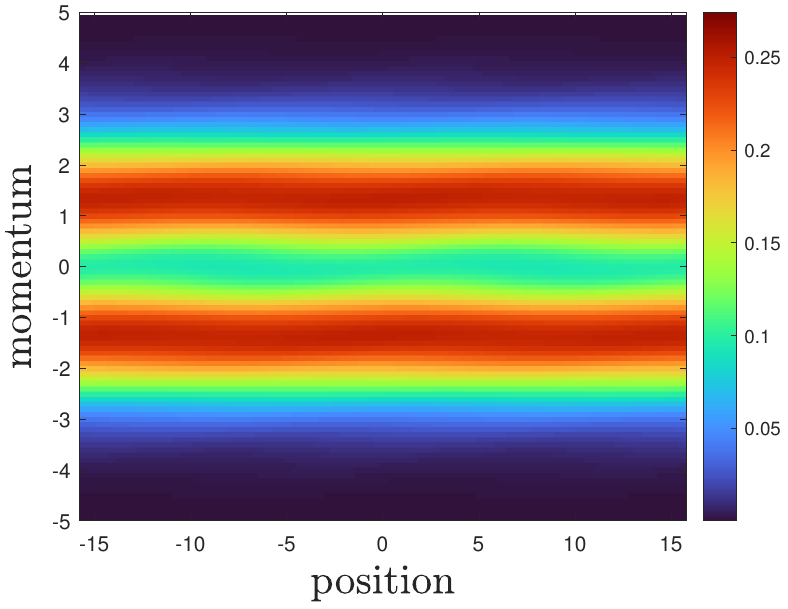}}
    \\
    \centering
    \subfigure[$t=24$.]{
    \includegraphics[width=1.4in,height=1.0in]{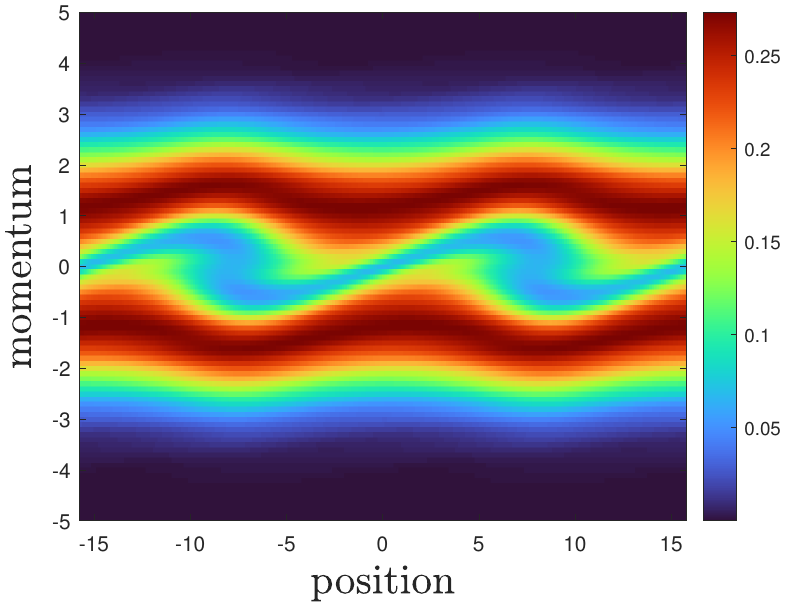}
    \includegraphics[width=1.4in,height=1.0in]{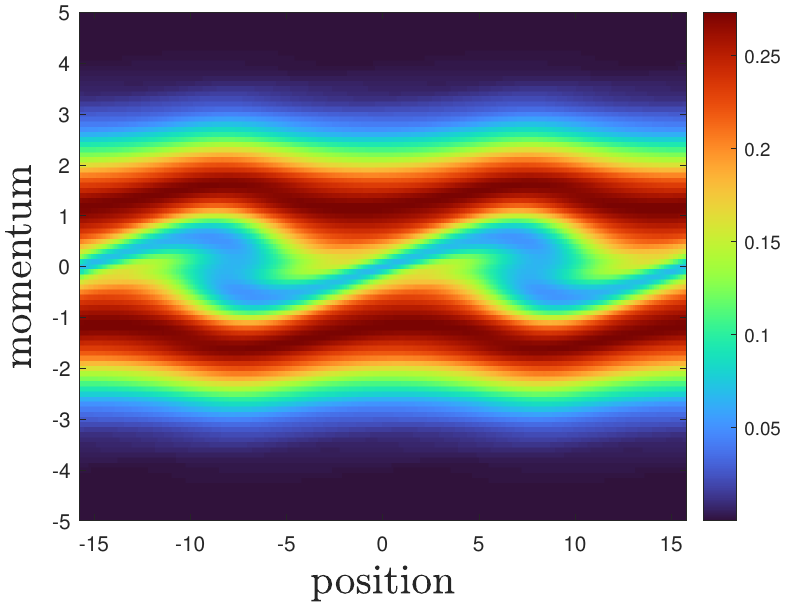}
    \includegraphics[width=1.4in,height=1.0in]{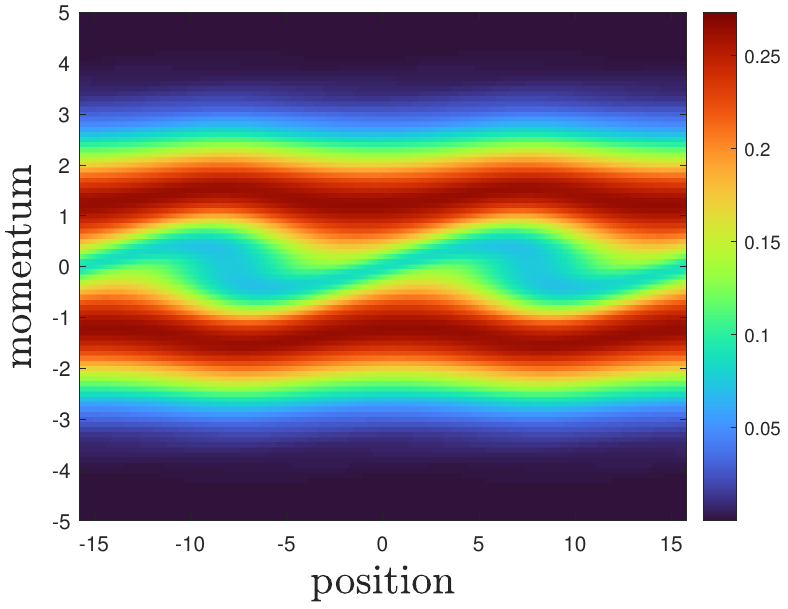}
    \includegraphics[width=1.4in,height=1.0in]{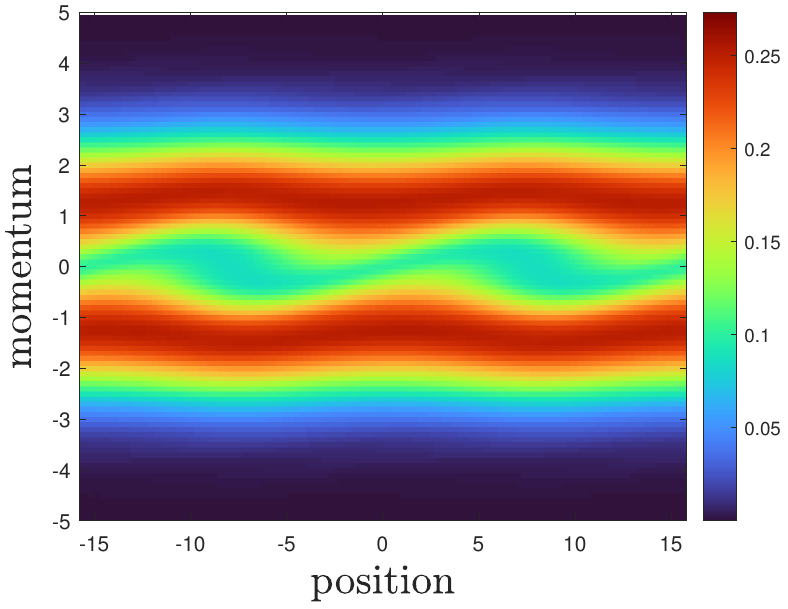}}    
     \\
    \centering
    \subfigure[$t=36$.]{
    \includegraphics[width=1.4in,height=1.0in]{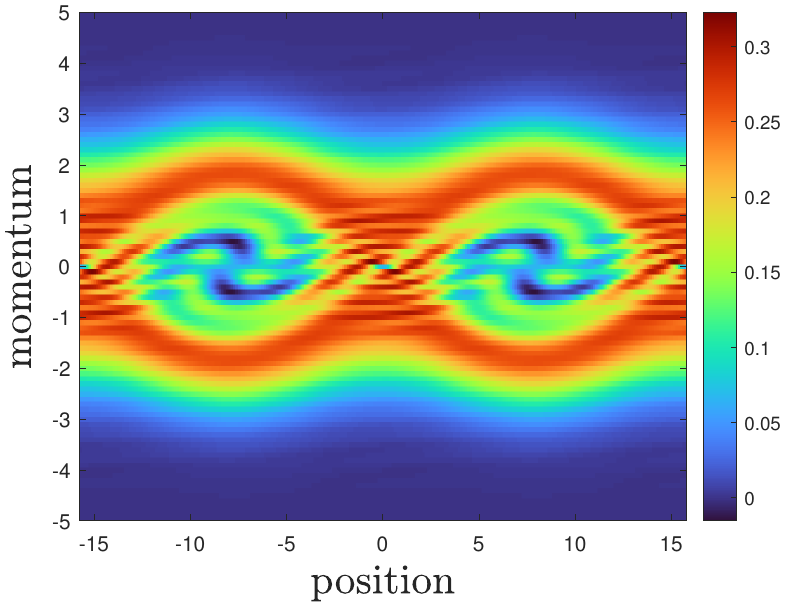}
    \includegraphics[width=1.4in,height=1.0in]{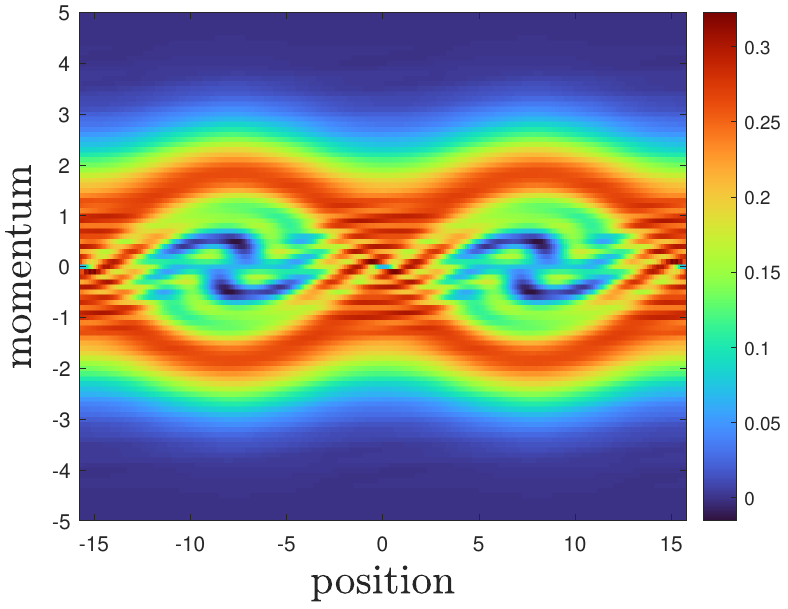}
    \includegraphics[width=1.4in,height=1.0in]{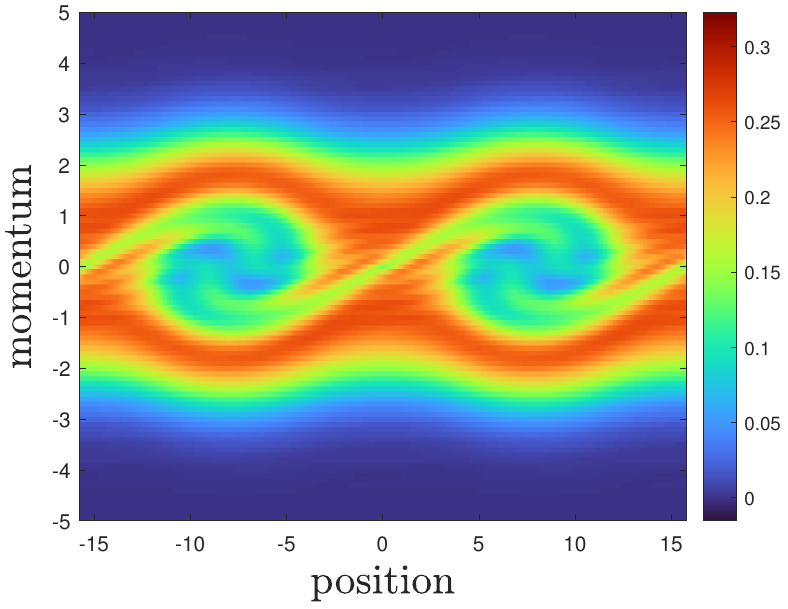}
    \includegraphics[width=1.4in,height=1.0in]{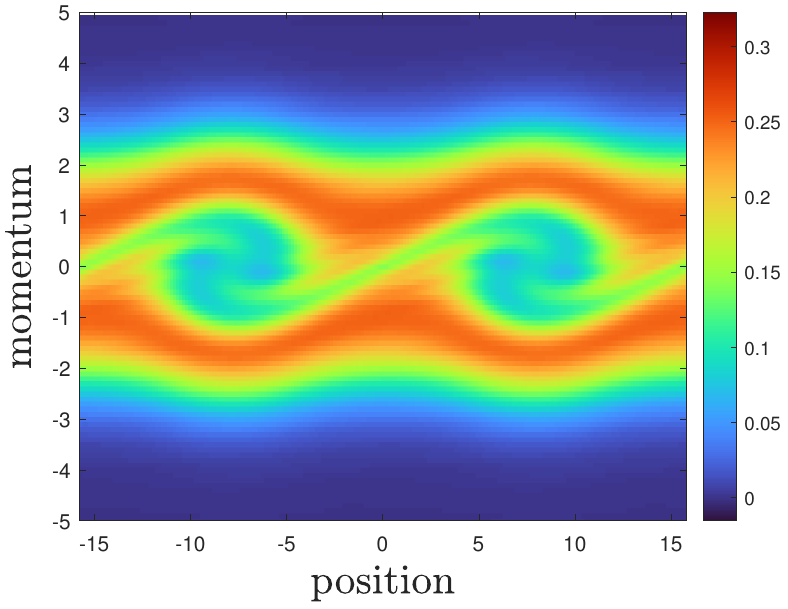}}      
     \\
    \centering
    \subfigure[$t=48$.]{
    \includegraphics[width=1.4in,height=1.0in]{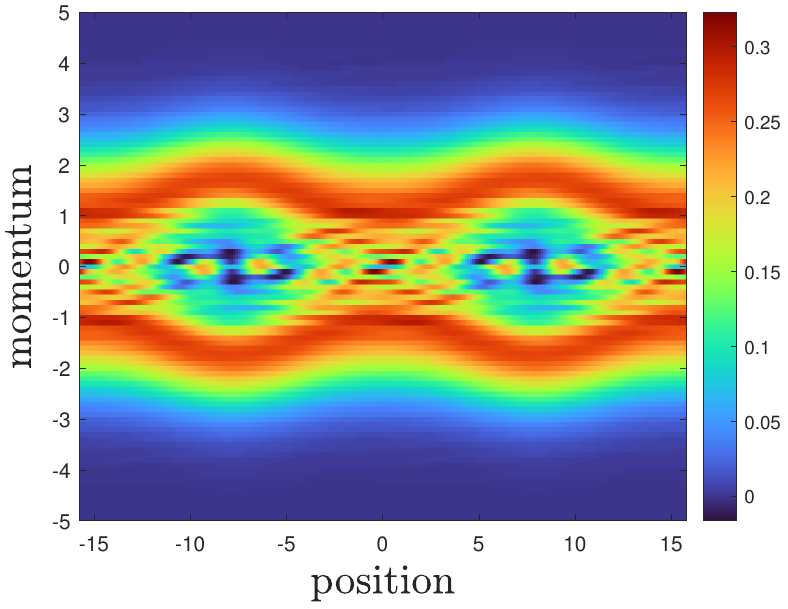}
    \includegraphics[width=1.4in,height=1.0in]{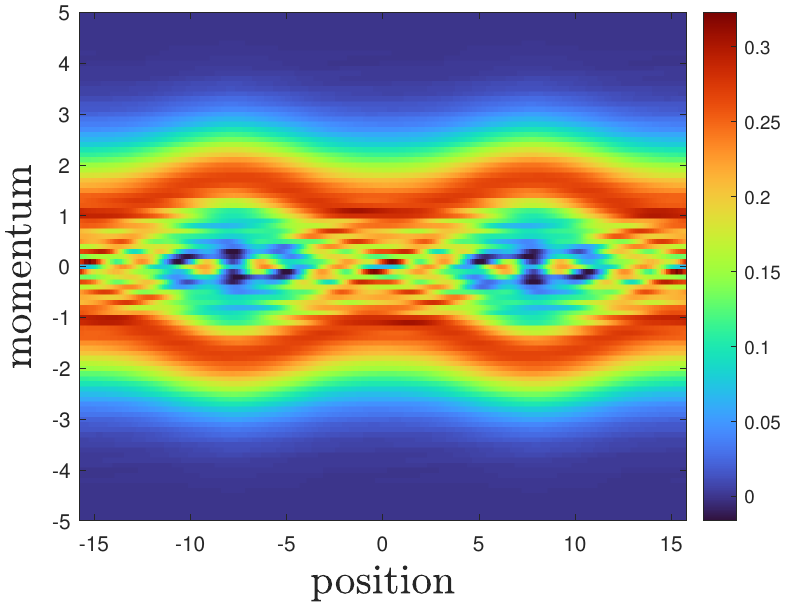}
    \includegraphics[width=1.4in,height=1.0in]{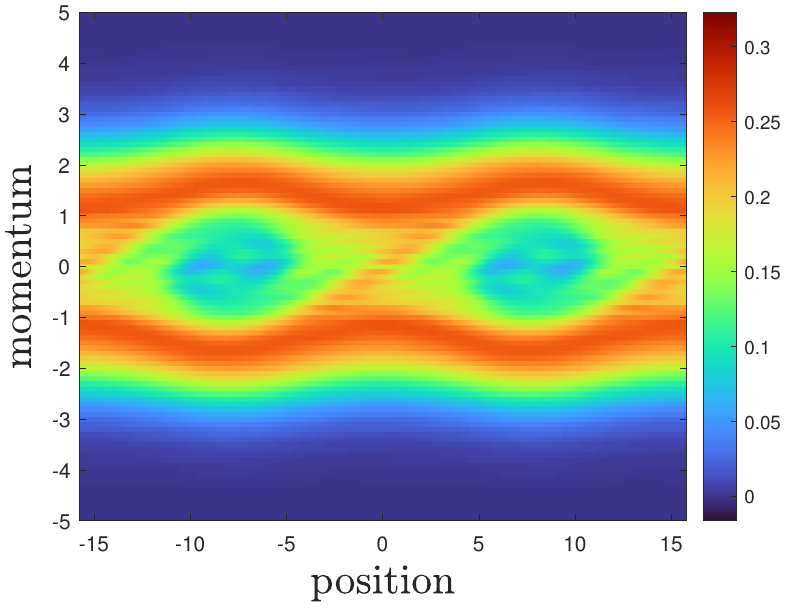}
    \includegraphics[width=1.4in,height=1.0in]{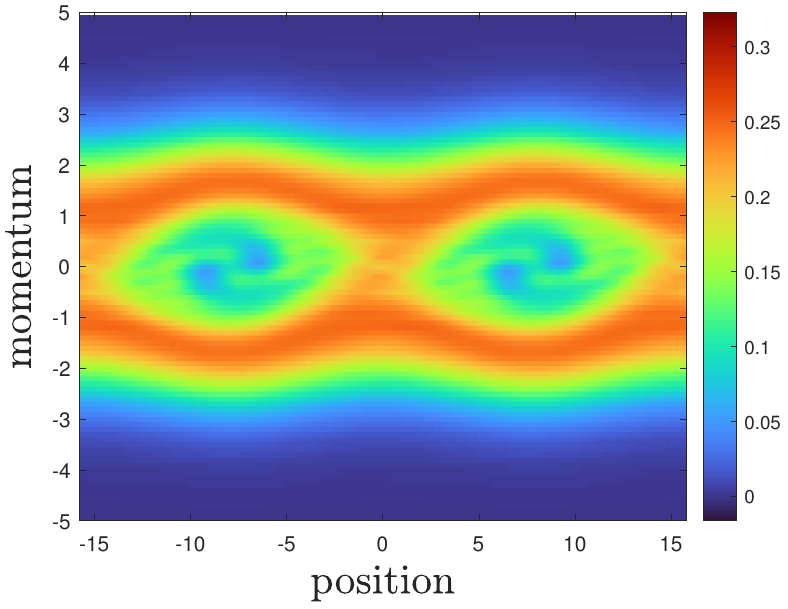}}          
     \\
    \centering
    \subfigure[$t=60$.]{
    \includegraphics[width=1.4in,height=1.0in]{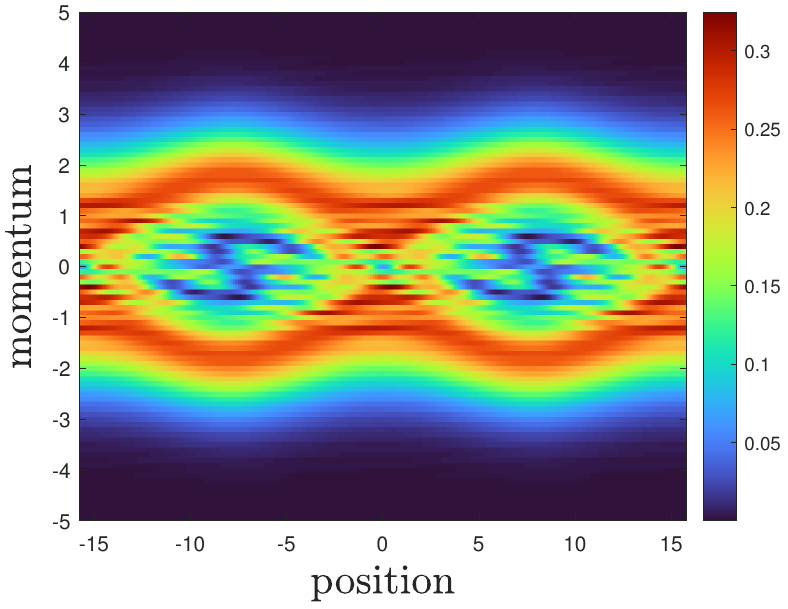}
    \includegraphics[width=1.4in,height=1.0in]{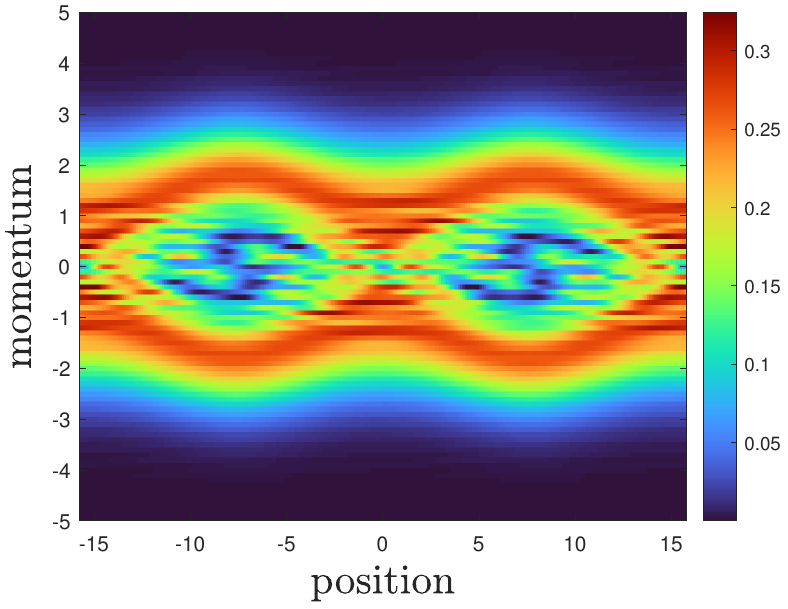}
    \includegraphics[width=1.4in,height=1.0in]{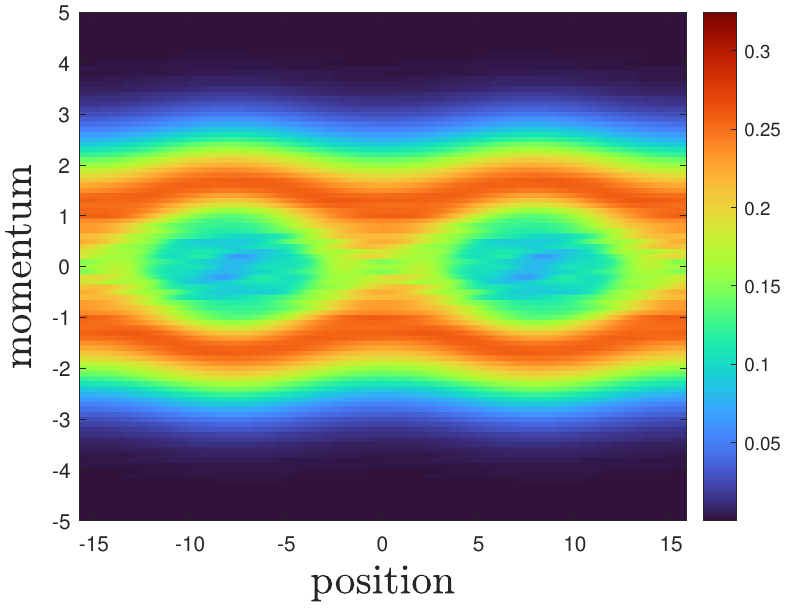}
    \includegraphics[width=1.4in,height=1.0in]{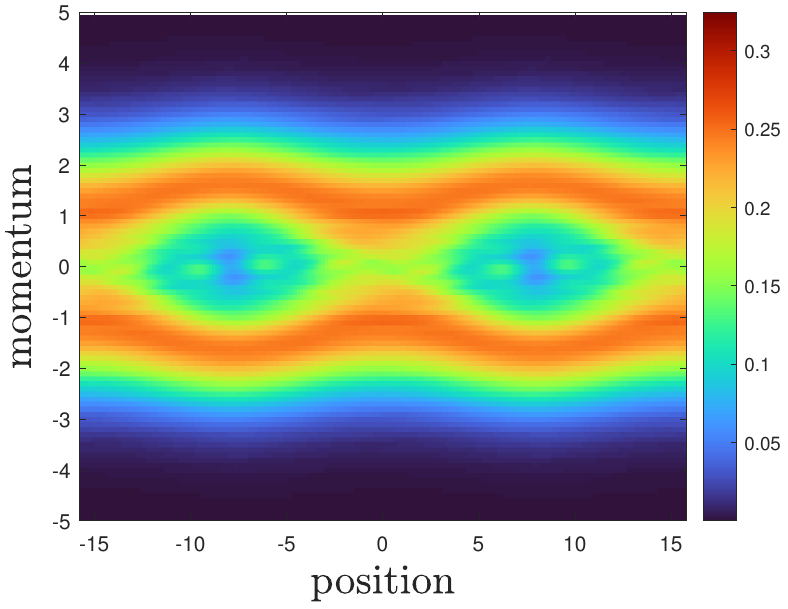}}      
  \\
  \centering
\subfigure[The slice of reduced Wigner function $W^{red}(0, p, t)$ at $r =0$, $t = 20, 40, 60$.  \label{slice_Wigner_two_stream}]{
\includegraphics[width=0.32\textwidth,height=0.22\textwidth]{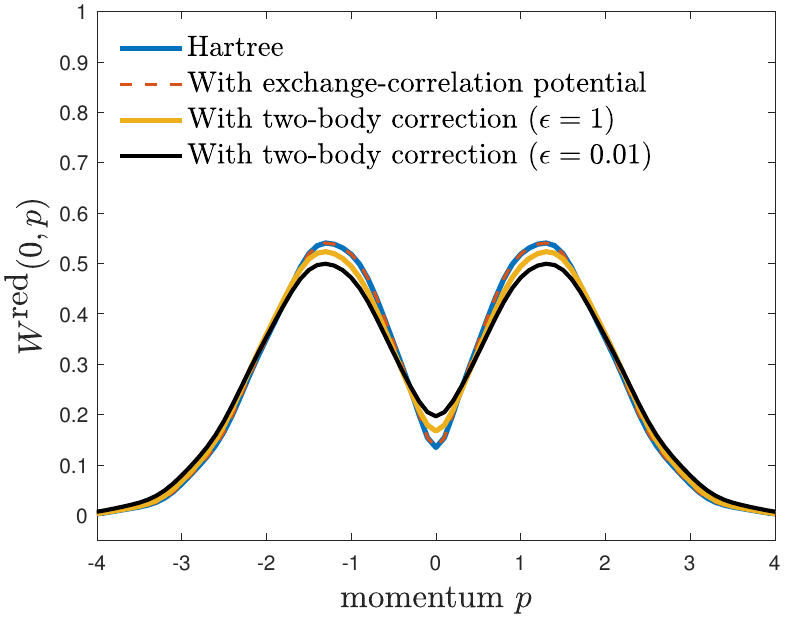}
\includegraphics[width=0.32\textwidth,height=0.22\textwidth]{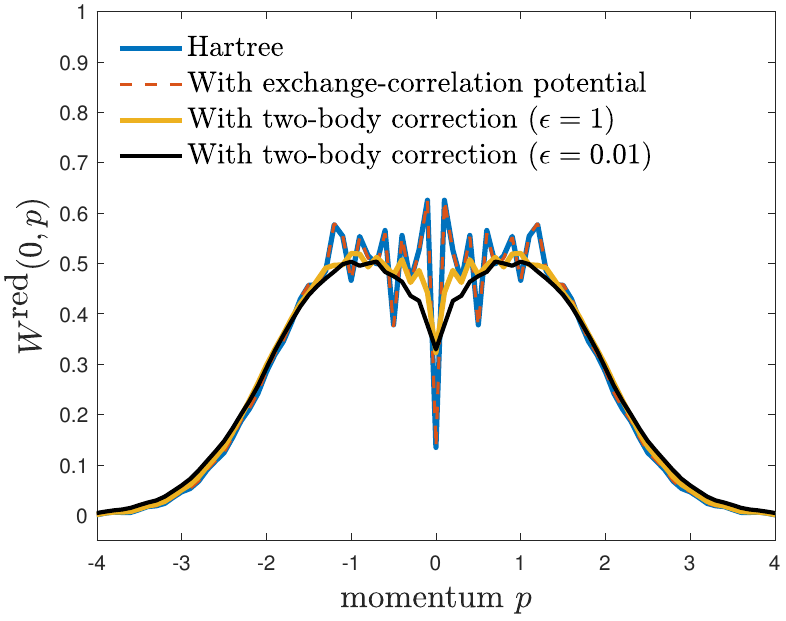}
\includegraphics[width=0.32\textwidth,height=0.22\textwidth]{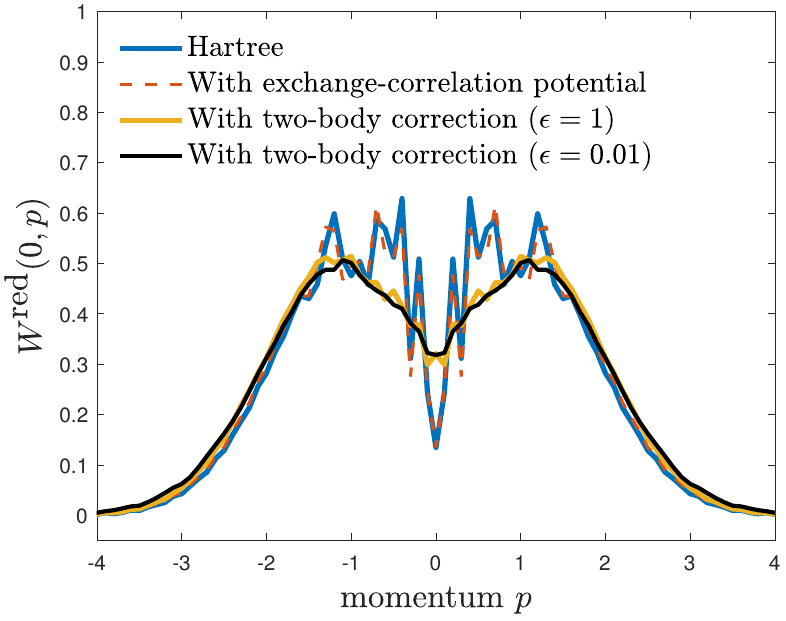}}          
     \caption{Quantum two-stream instability: The evolution of the reduced Wigner function. The cases from left side to right side are: Hartree, Hartree with XC correction, Hartree with two-body correction $\epsilon = 1$ or $\epsilon = 0.01$. \label{Wigner_two_stream}}    
\end{figure}

Under the perturbation level $\varepsilon_1 = \varepsilon_2 = 0.02$,  the instability of electric field energy starts at $t = 12$ (see Figure \ref{two_stream_short}), which is a consequence of the coalescence of the double peaks in the phase space (see Figure \ref{Wigner_two_stream}). After a long-time evolution, the Wigner function relaxes to  a nonlinear equilibrium mode, namely the BGK
mode, in an electron electrostatic plasma, which provides a nonlinear saturation mechanism for
plasma instabilities. This coincides with the early simulation results in \cite{SuhFeixBertrand1991}. Moreover, in absence of XC potential and two-body interaction, our simulations recover the spirals in the central region and the abnormal oscillation resulting from the energy exchange between electrons and the plasmon  as in \cite{HuLiangShengWu2022}. 

  \begin{figure}[!h]
    \centering
    \subfigure[$t=12$.]{
    \includegraphics[width=1.4in,height=1.0in]{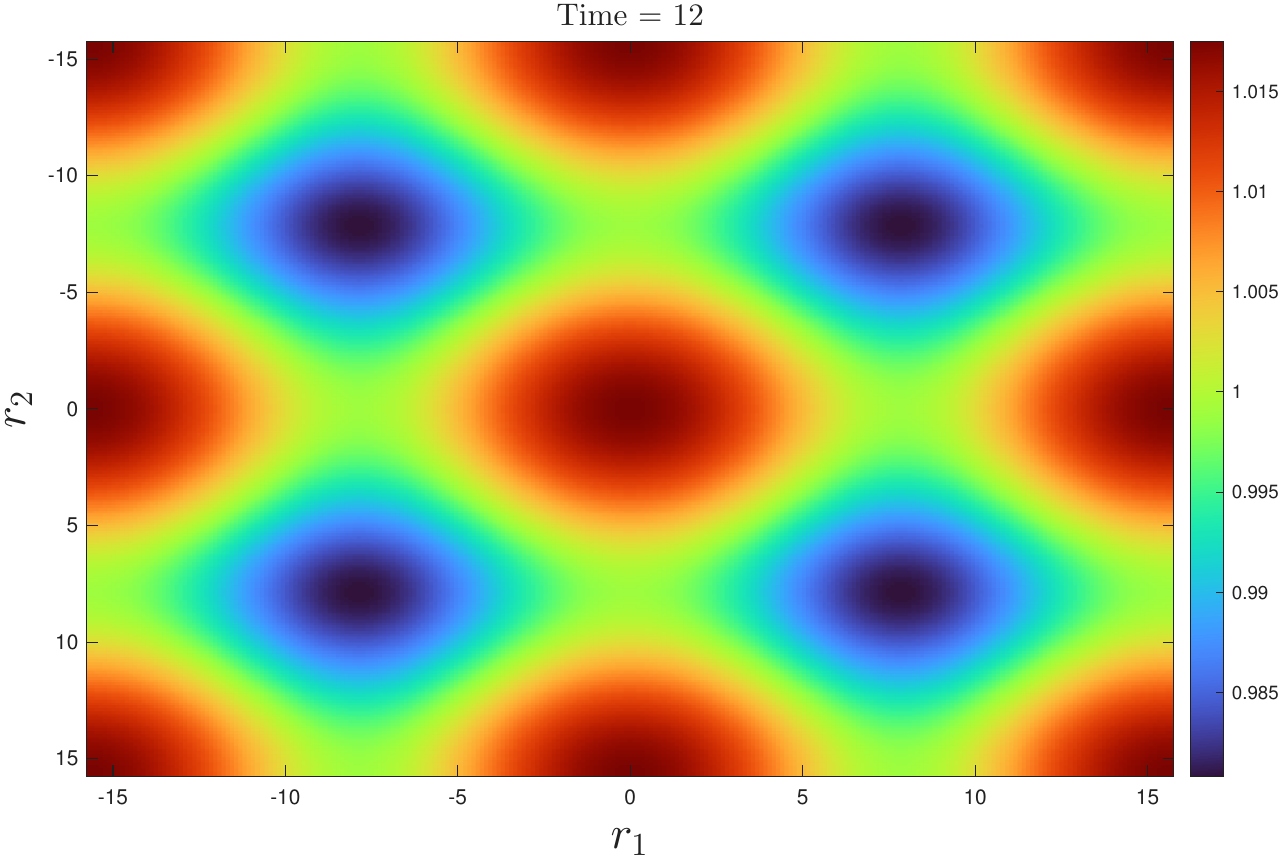}
    \includegraphics[width=1.4in,height=1.0in]{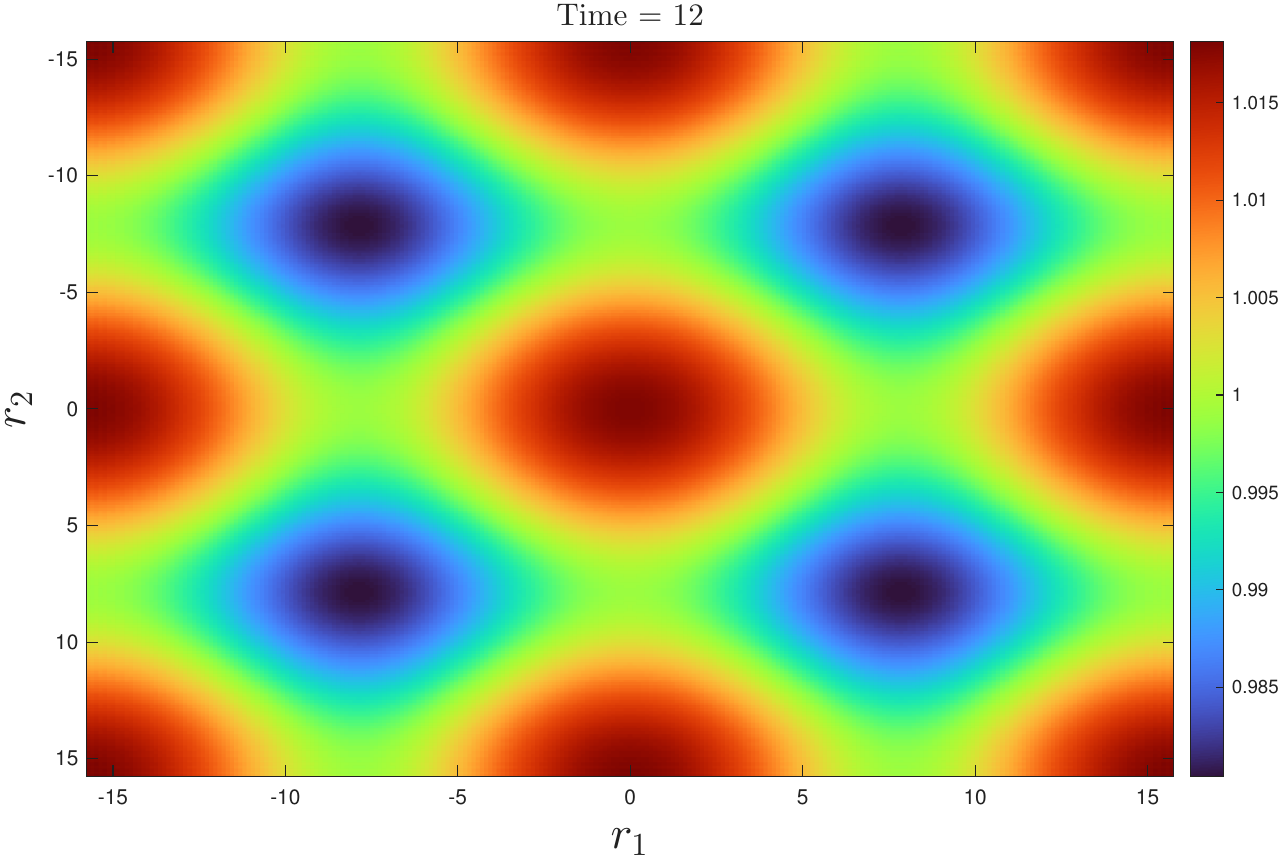}
    \includegraphics[width=1.4in,height=1.0in]{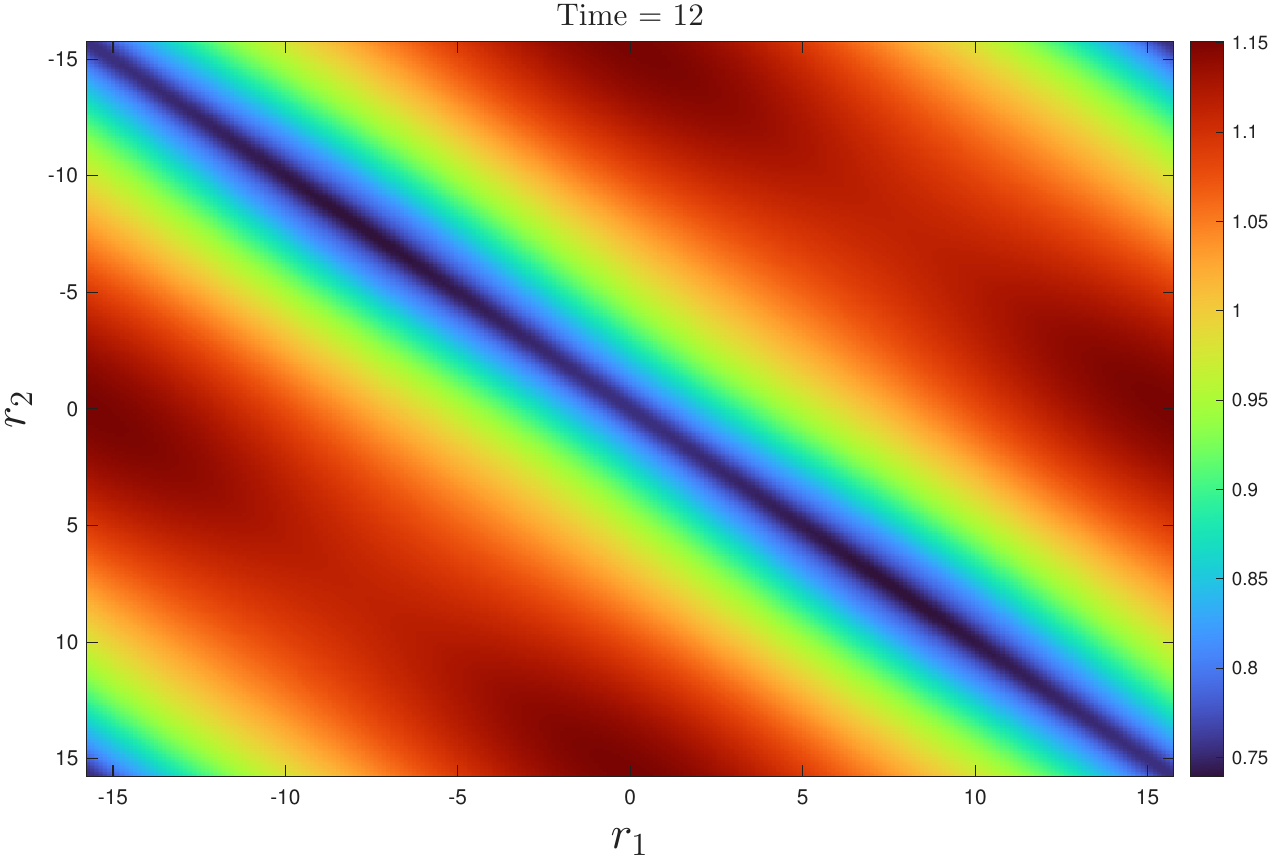}
    \includegraphics[width=1.4in,height=1.0in]{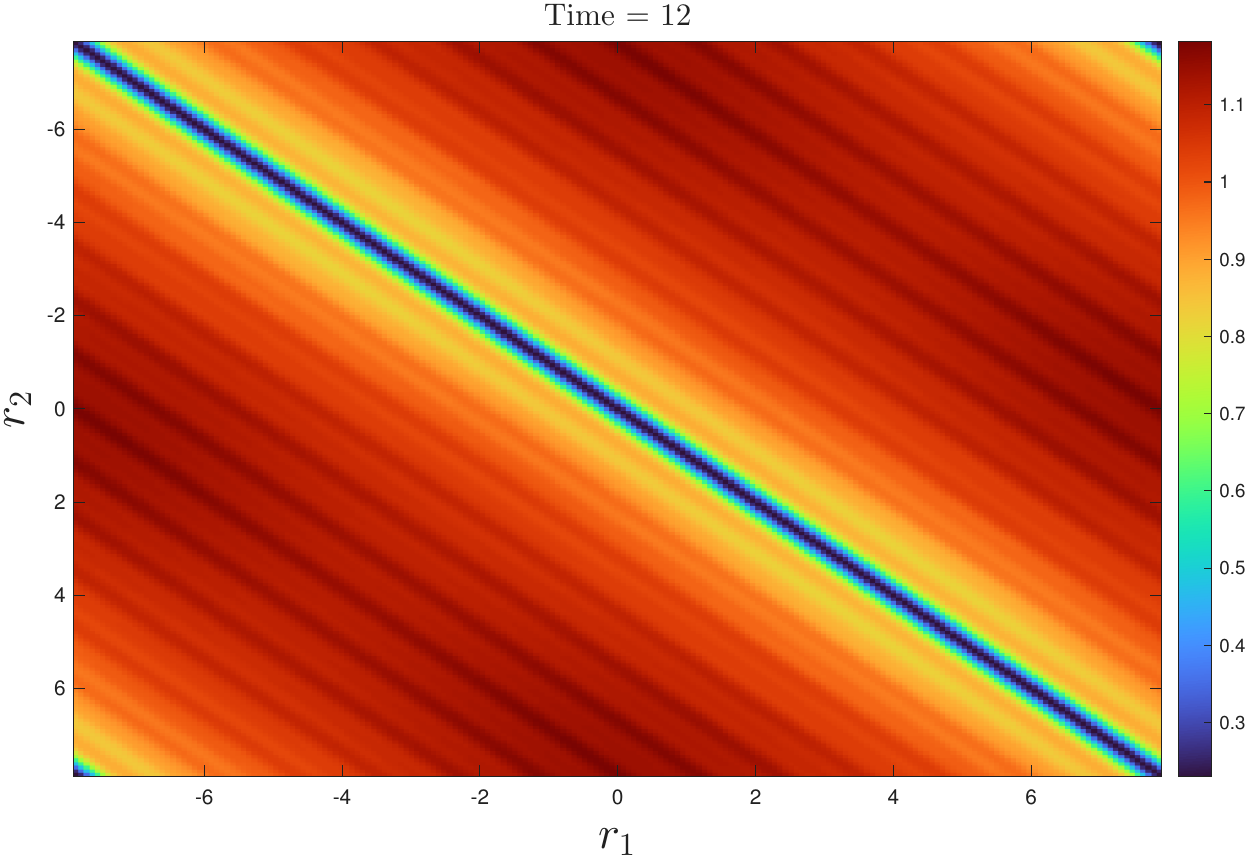}}
    \\
    \centering
    \subfigure[$t=24$.]{
    \includegraphics[width=1.4in,height=1.0in]{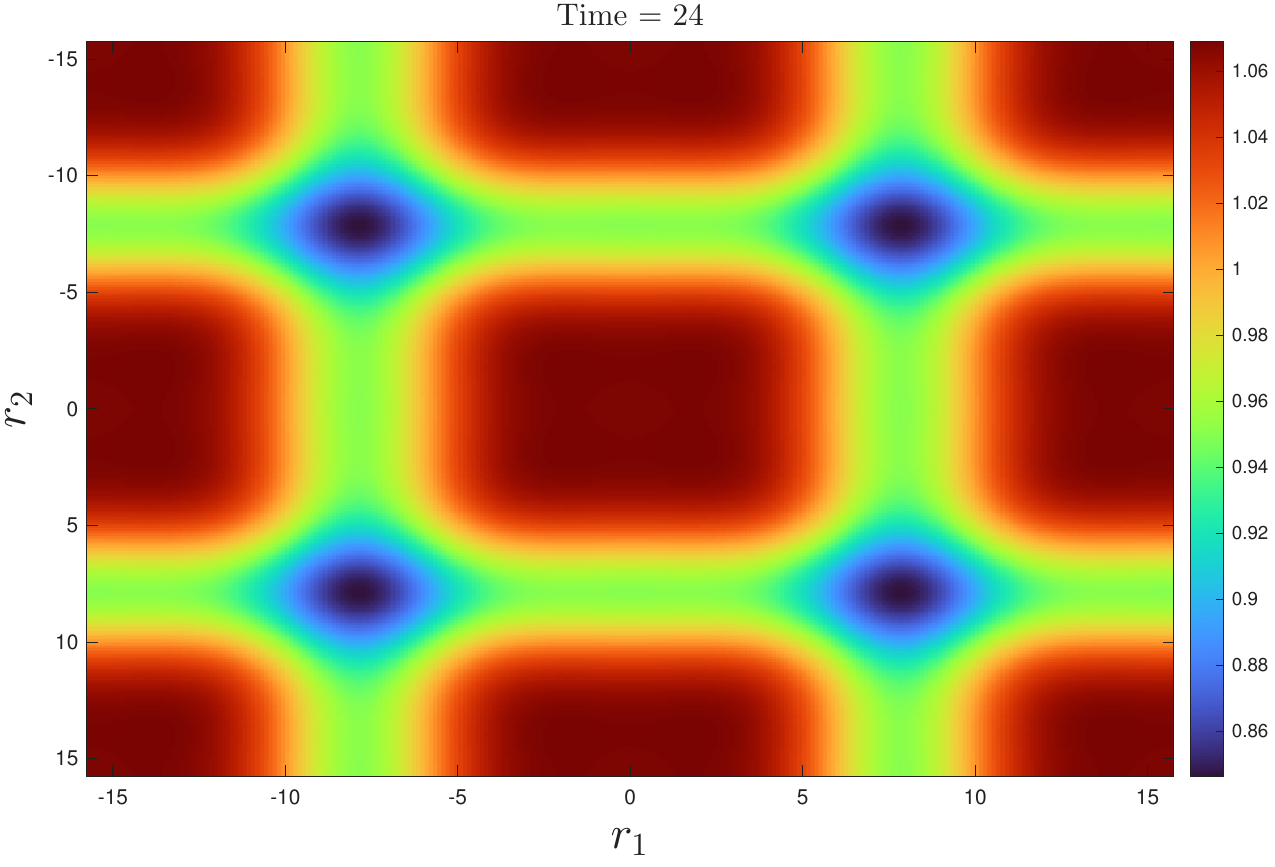}
    \includegraphics[width=1.4in,height=1.0in]{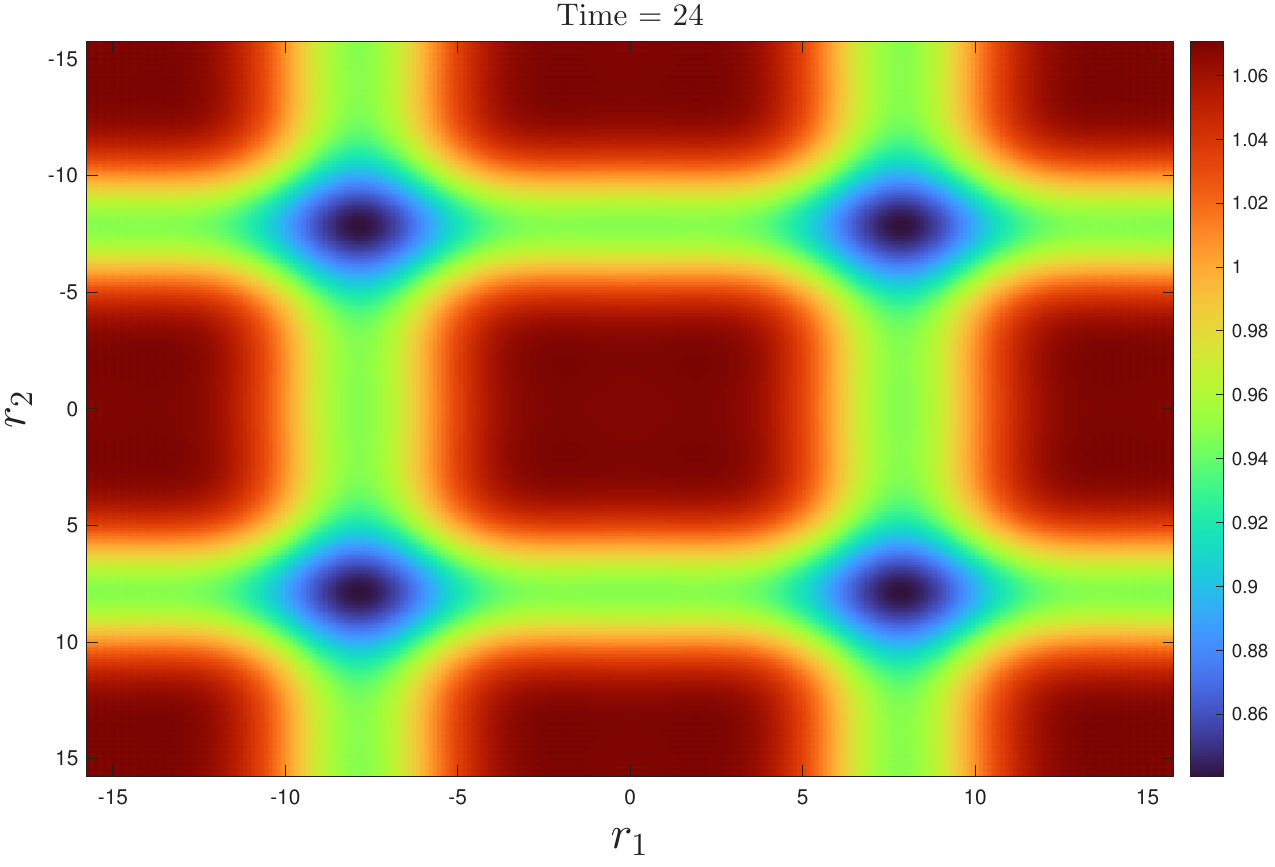}
    \includegraphics[width=1.4in,height=1.0in]{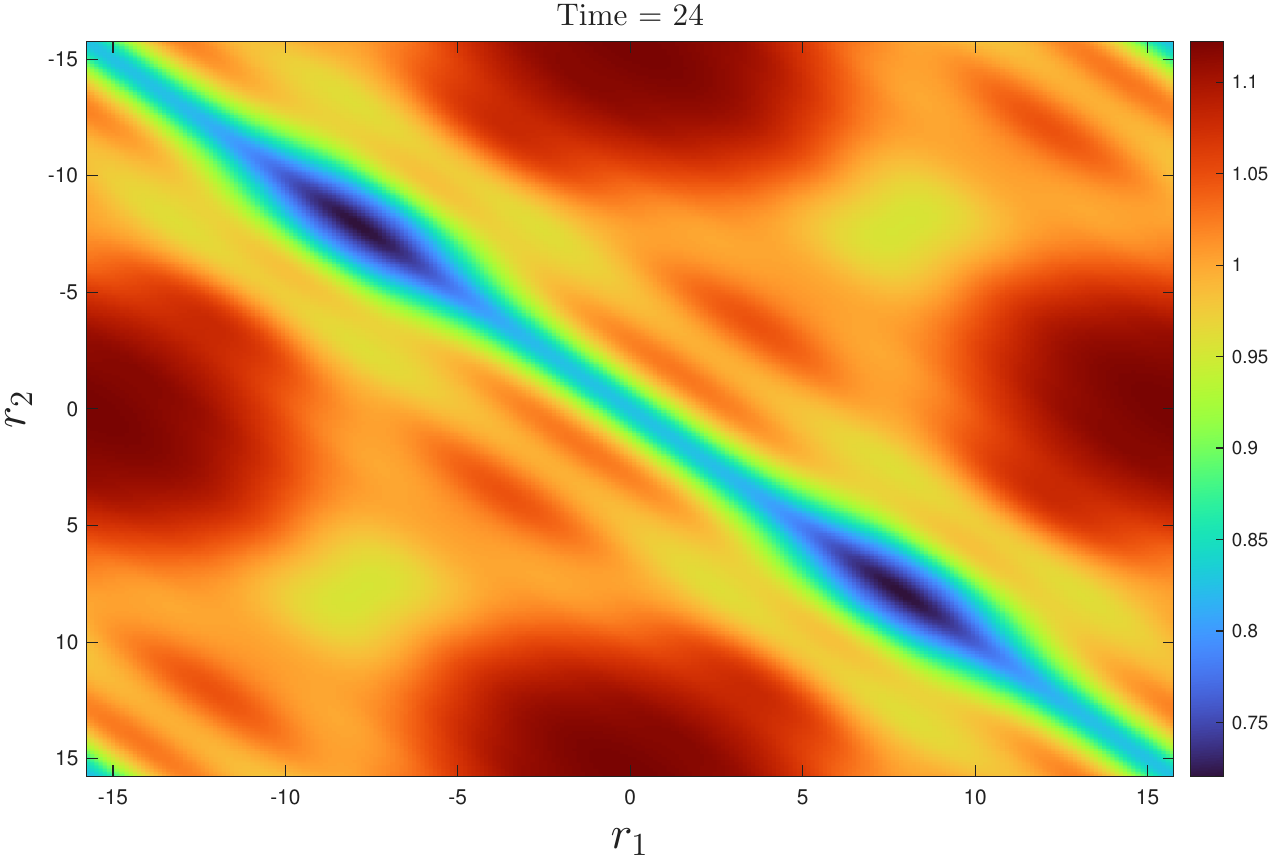}
    \includegraphics[width=1.4in,height=1.0in]{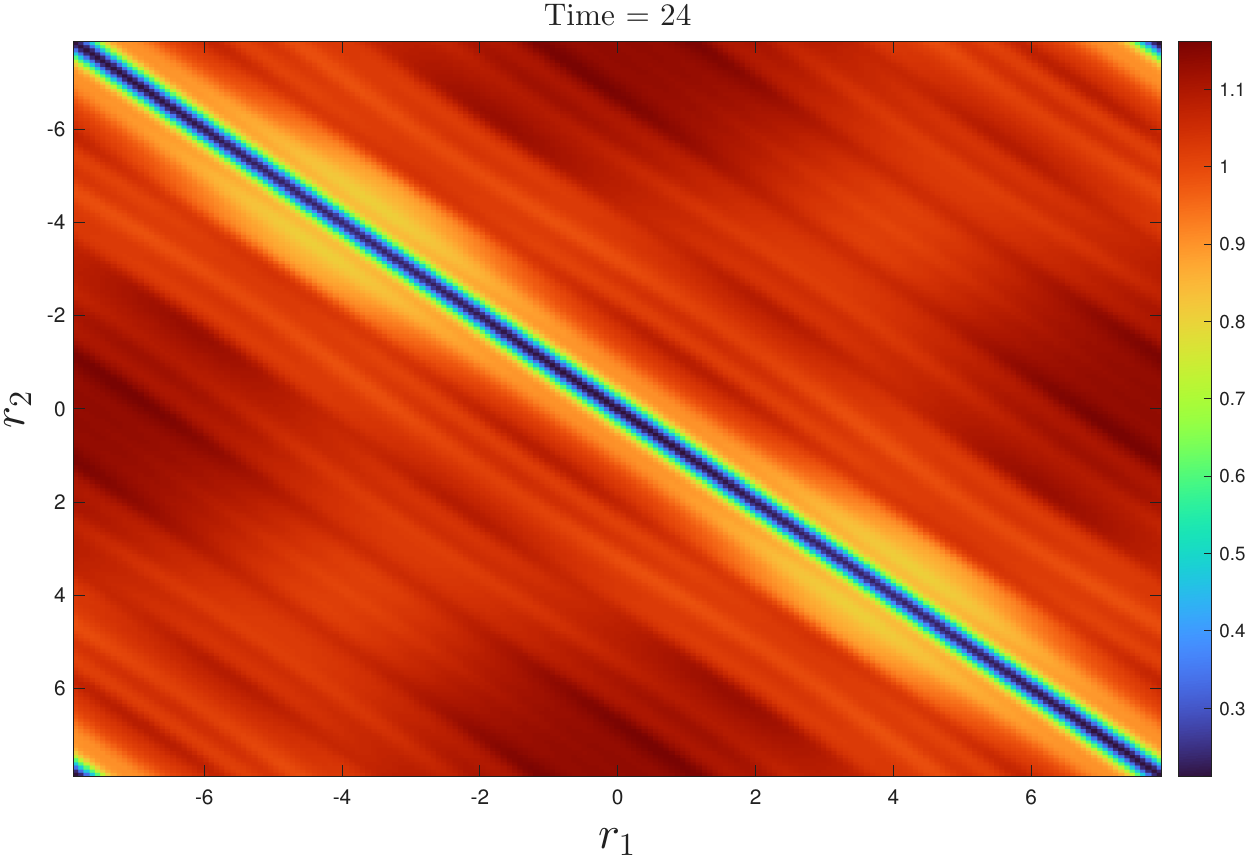}}    
     \\
    \centering
    \subfigure[$t=36$.]{
    \includegraphics[width=1.4in,height=1.0in]{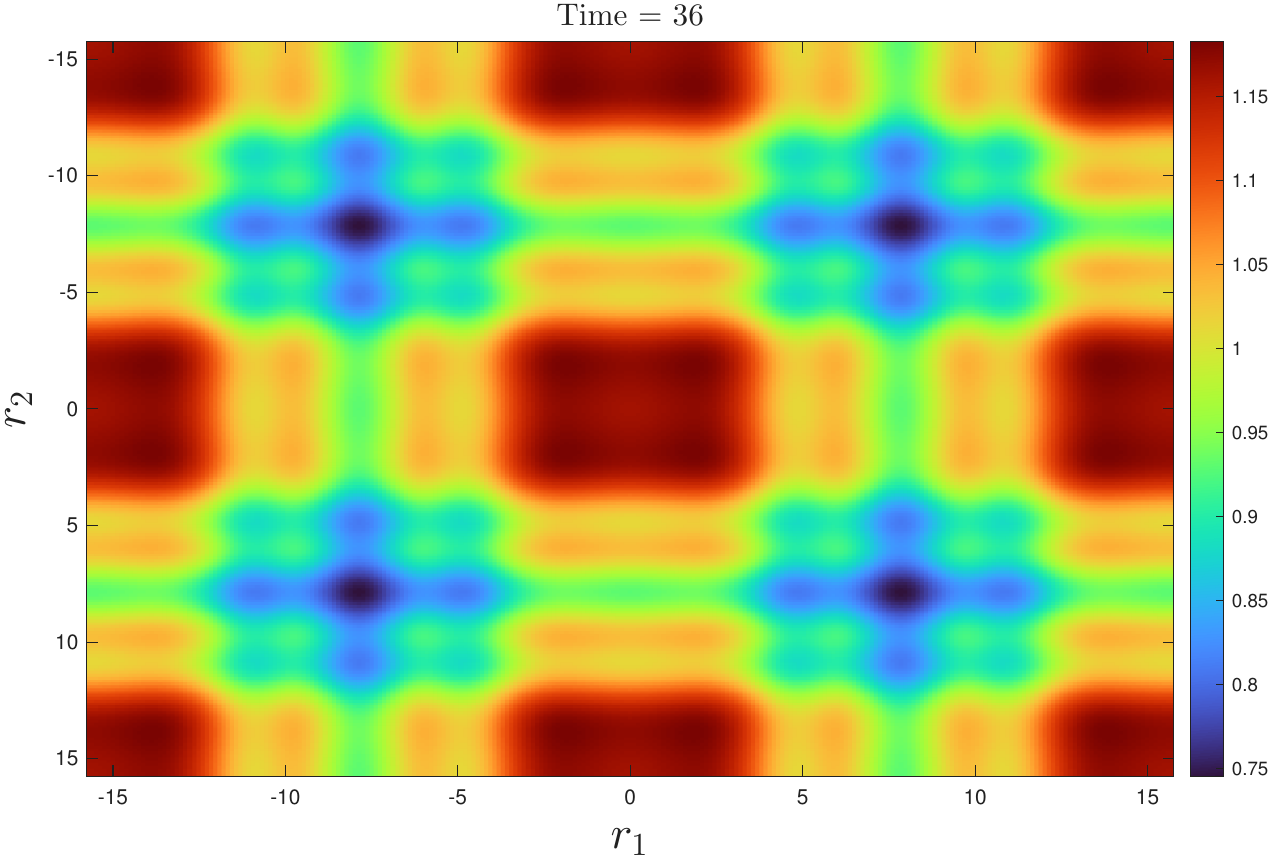}
    \includegraphics[width=1.4in,height=1.0in]{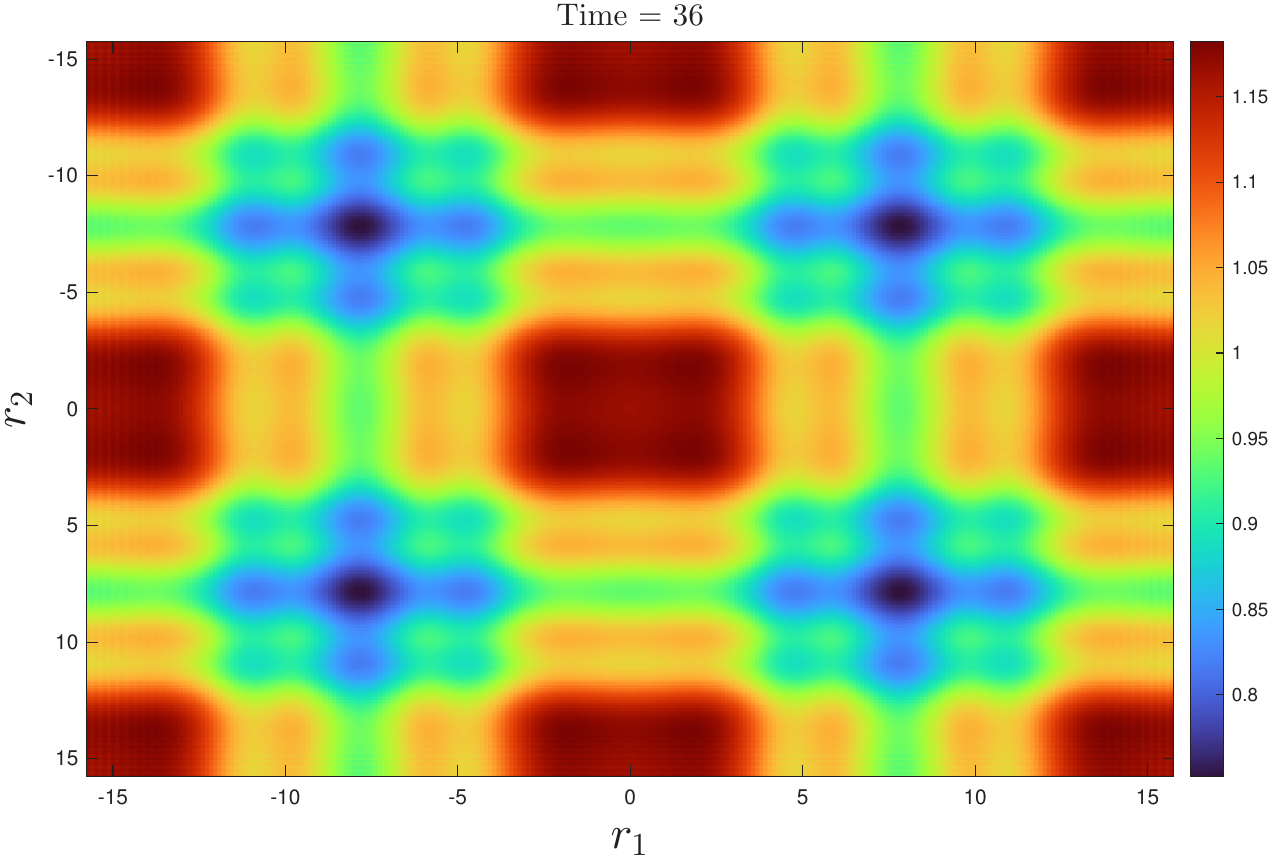}
    \includegraphics[width=1.4in,height=1.0in]{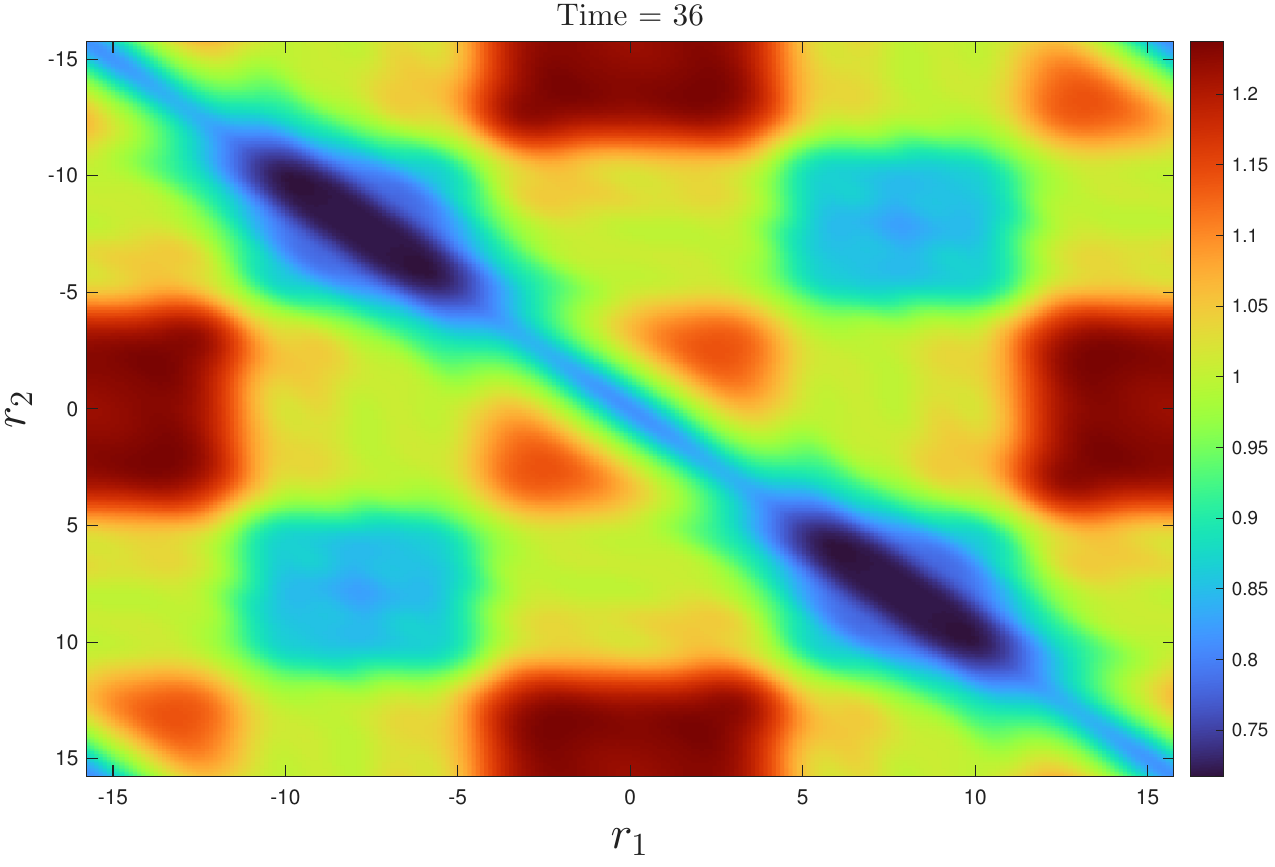}
    \includegraphics[width=1.4in,height=1.0in]{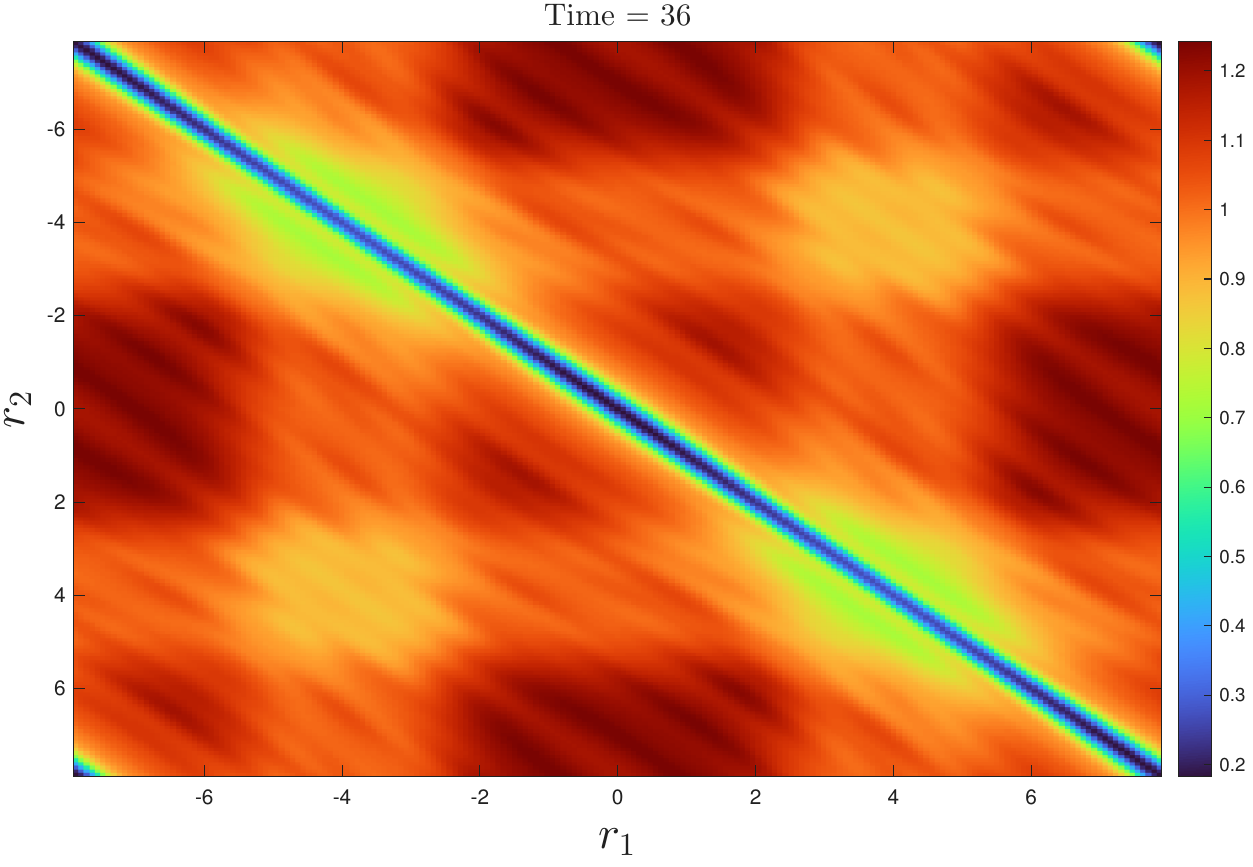}}      
     \\
    \centering
    \subfigure[$t=48$.]{
    \includegraphics[width=1.4in,height=1.0in]{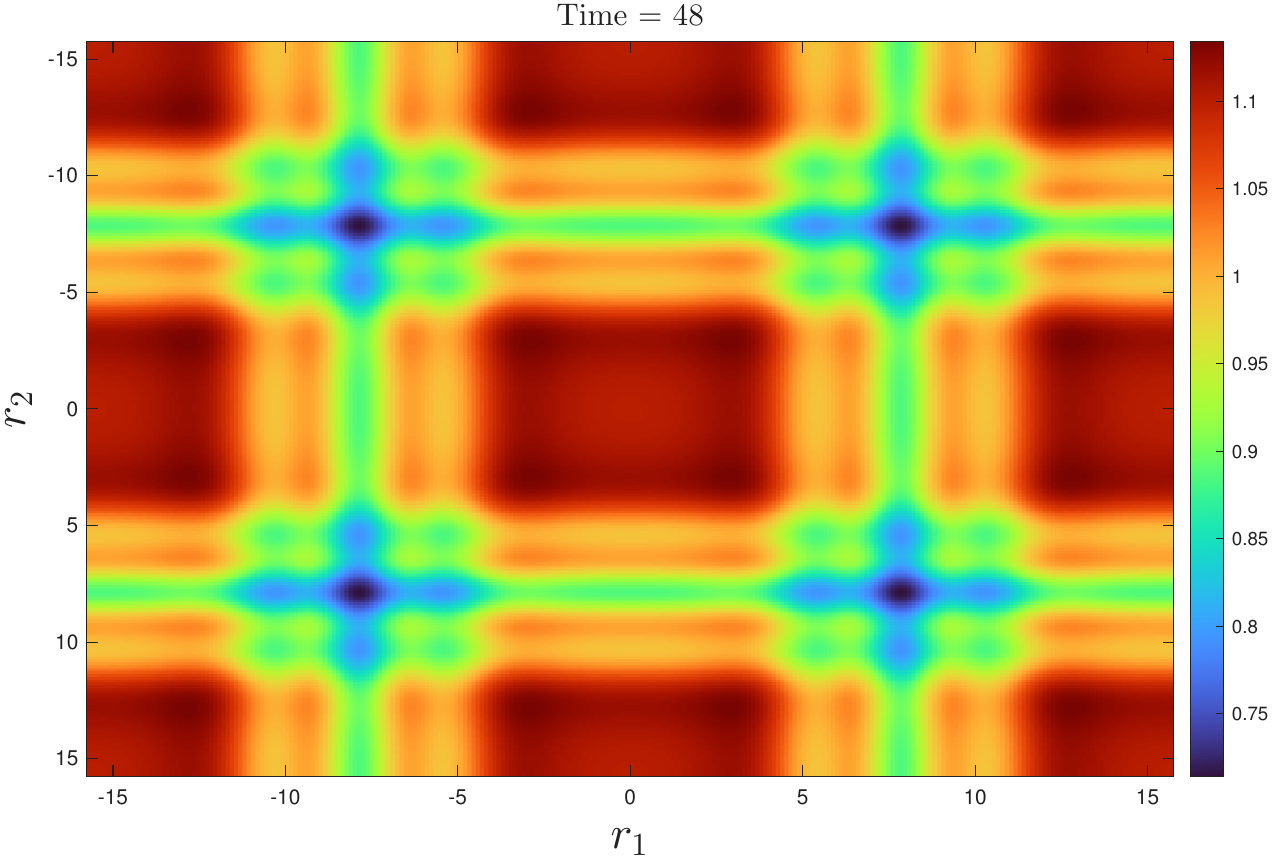}
    \includegraphics[width=1.4in,height=1.0in]{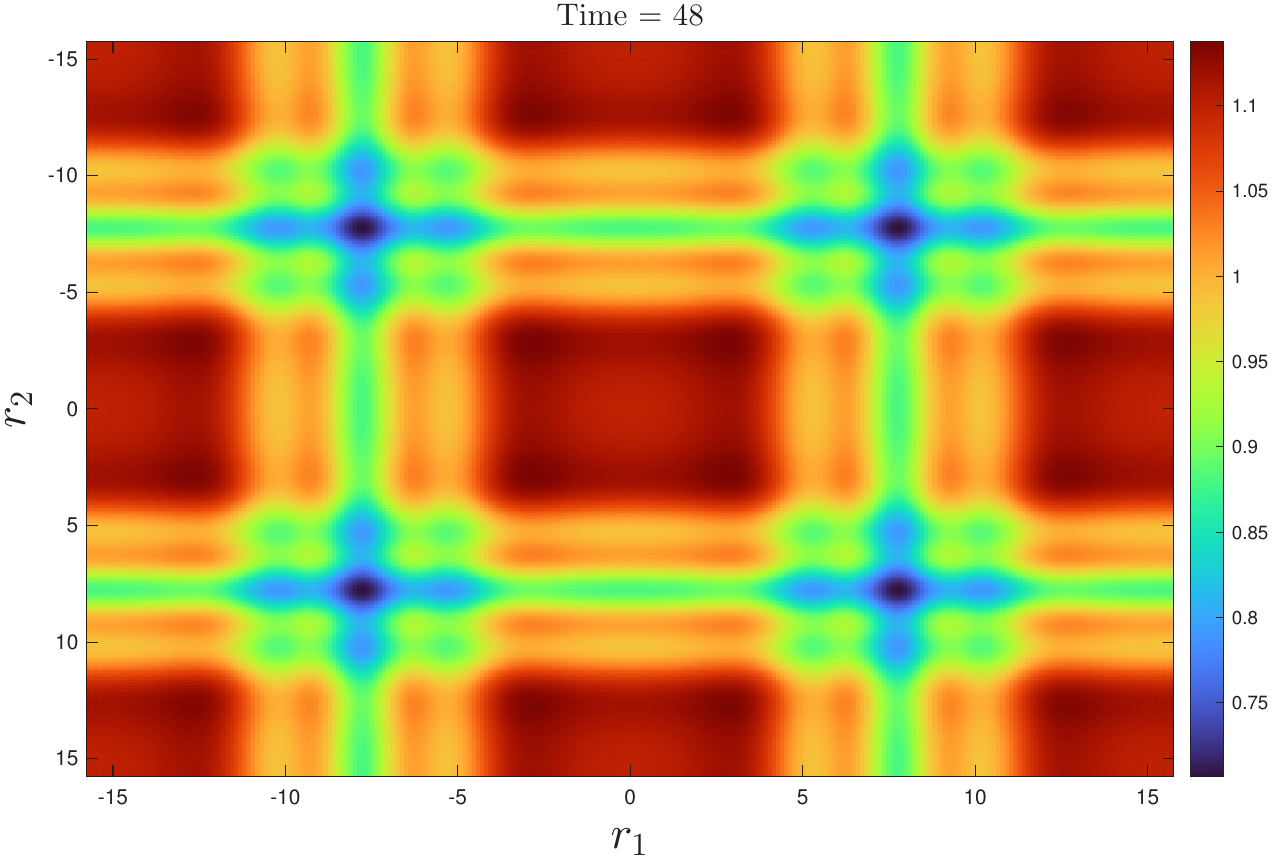}
    \includegraphics[width=1.4in,height=1.0in]{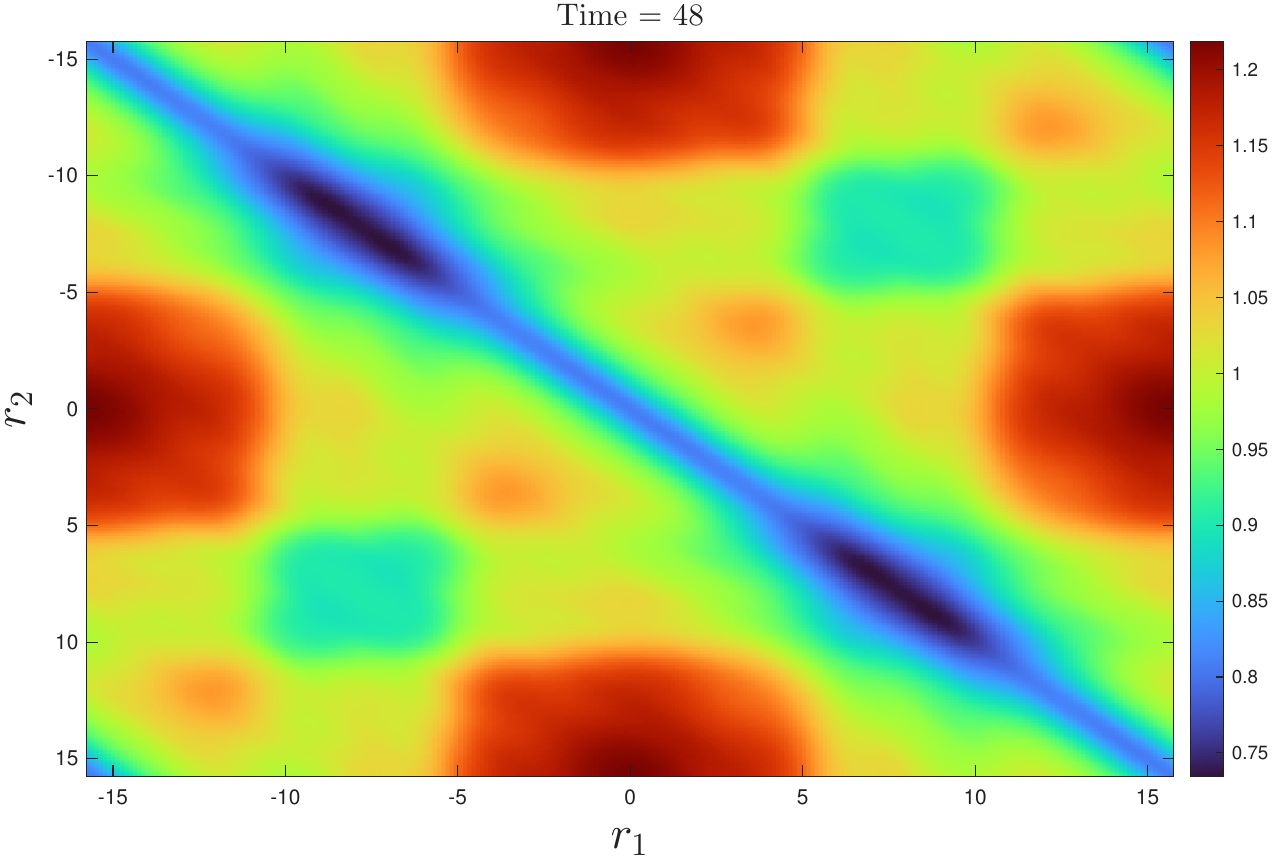}
    \includegraphics[width=1.4in,height=1.0in]{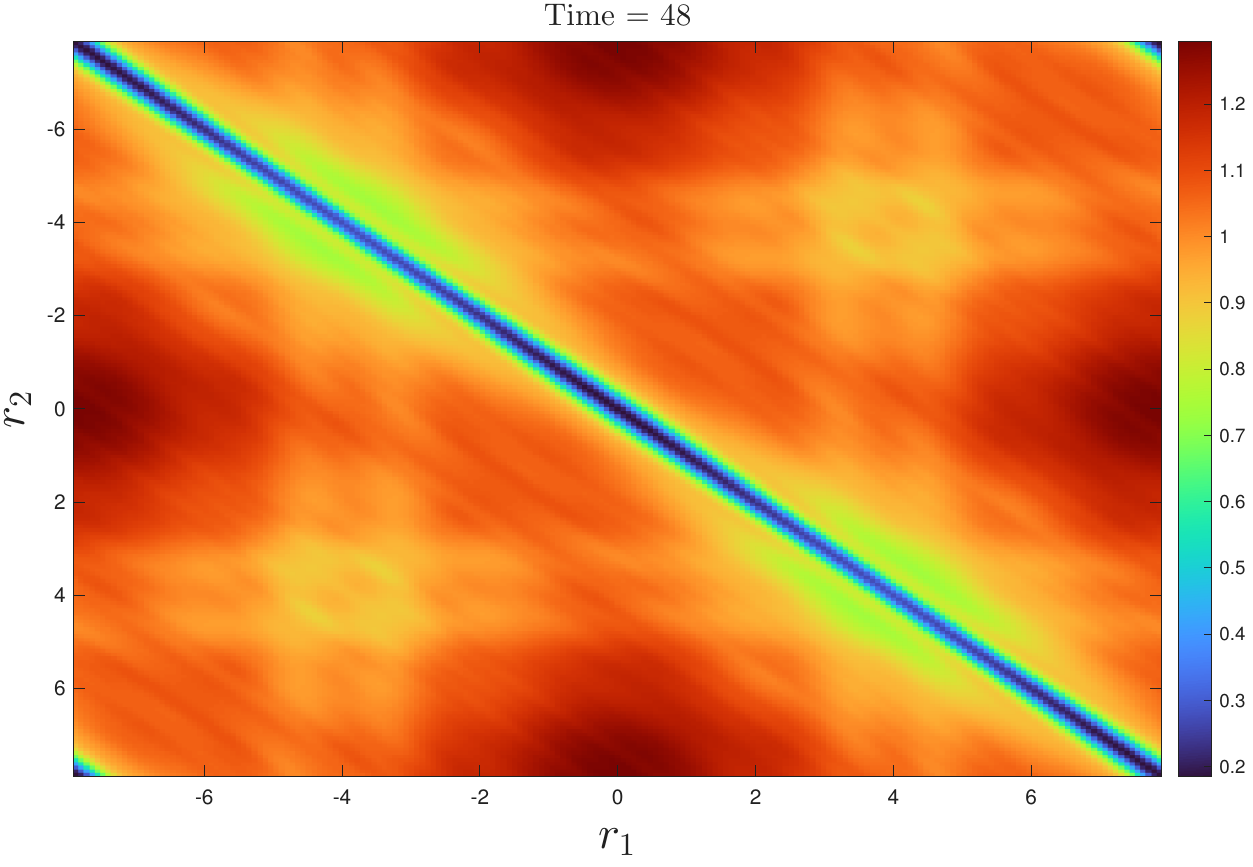}}          
     \\
    \centering
    \subfigure[$t=60$.]{
    \includegraphics[width=1.4in,height=1.0in]{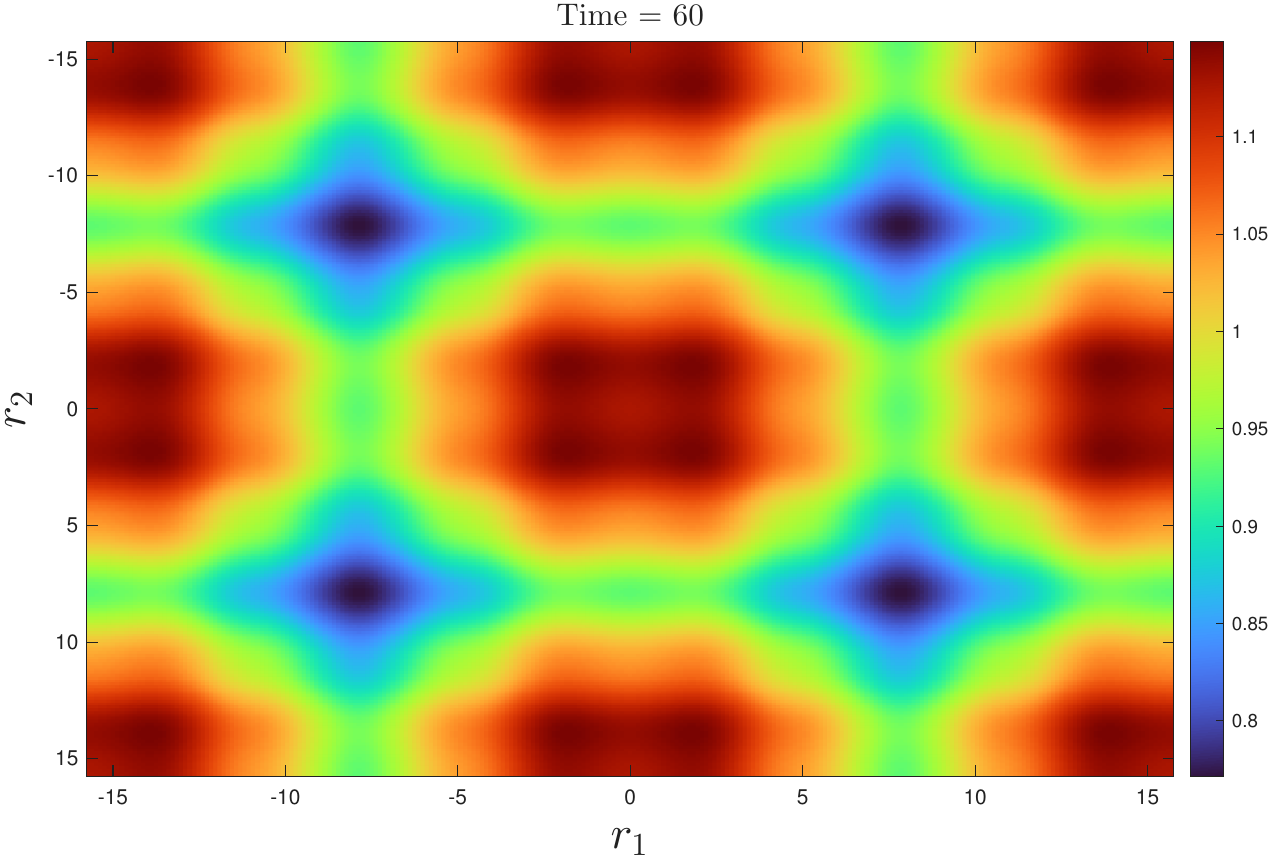}
    \includegraphics[width=1.4in,height=1.0in]{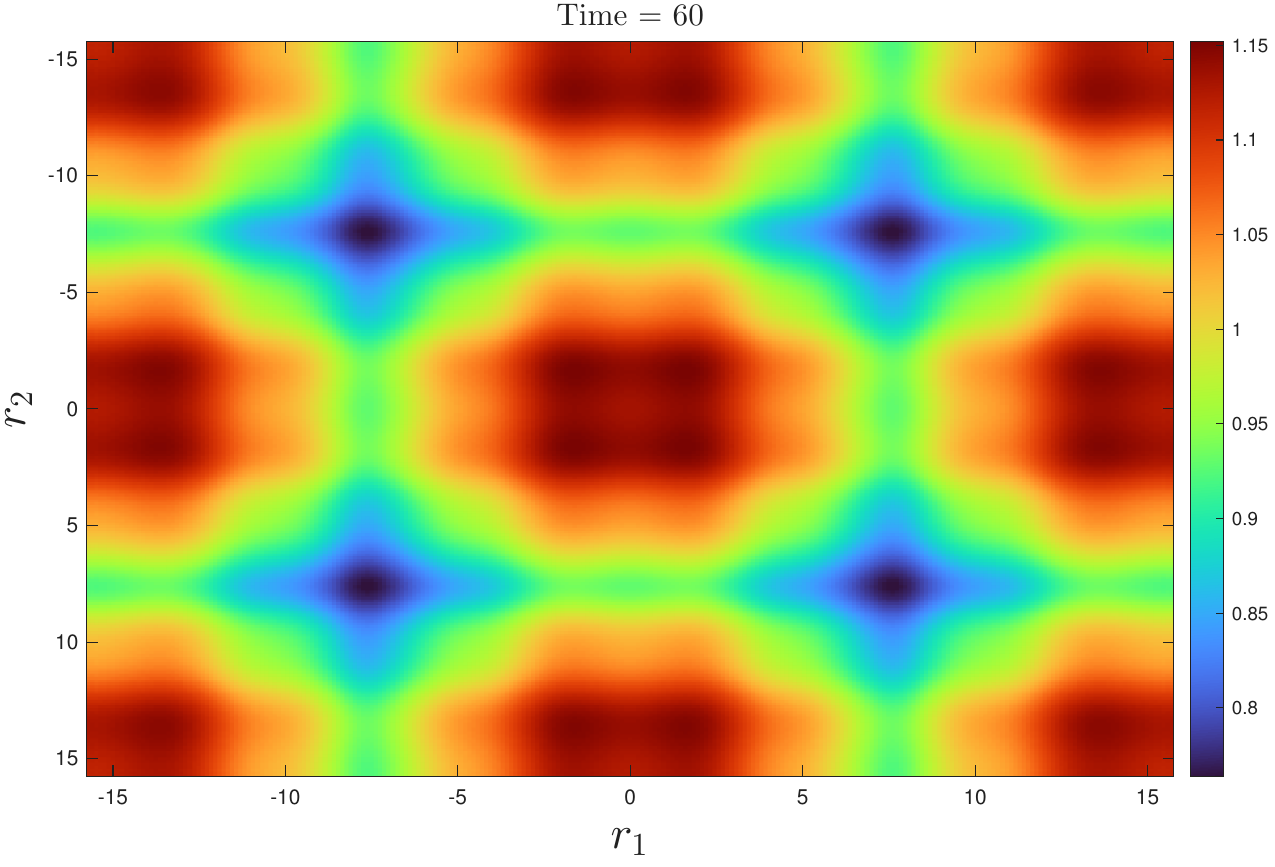}
    \includegraphics[width=1.4in,height=1.0in]{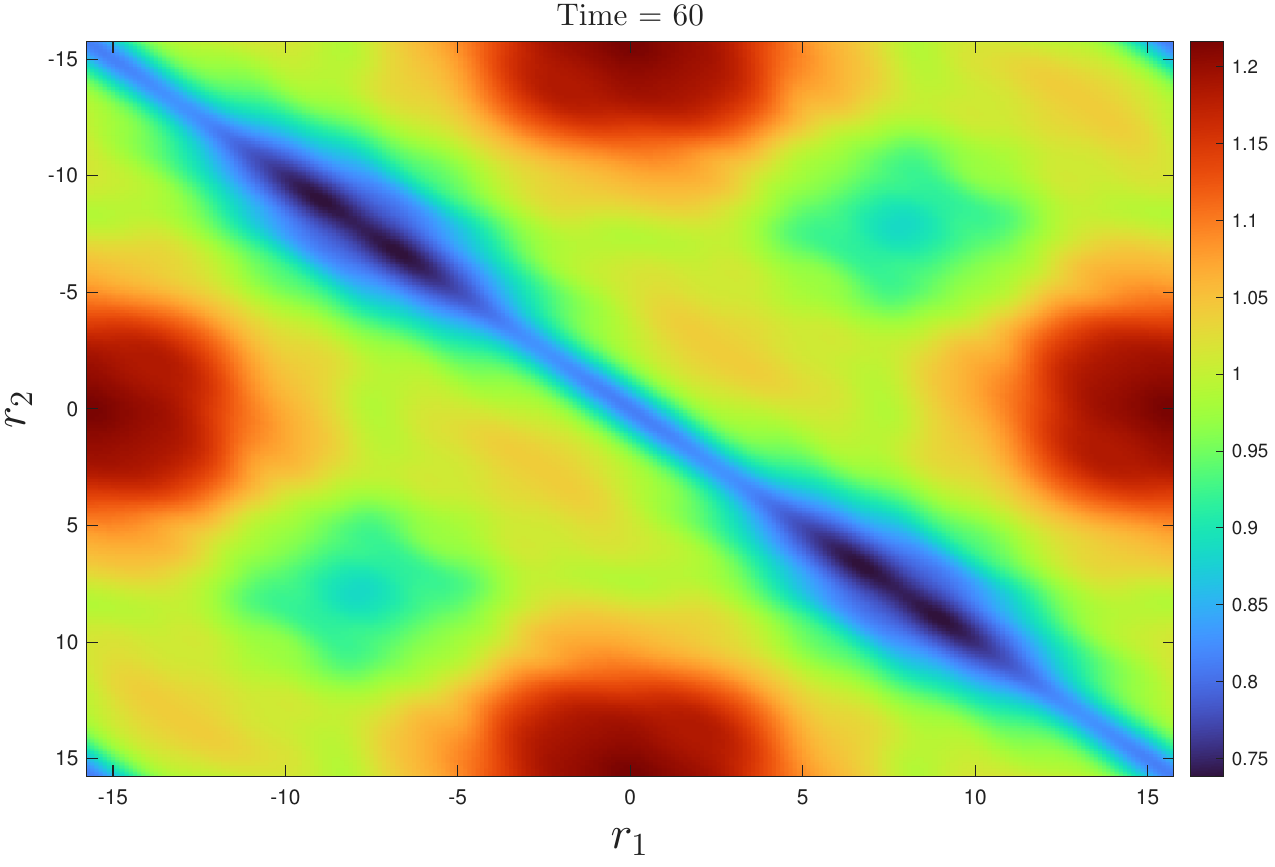}
    \includegraphics[width=1.4in,height=1.0in]{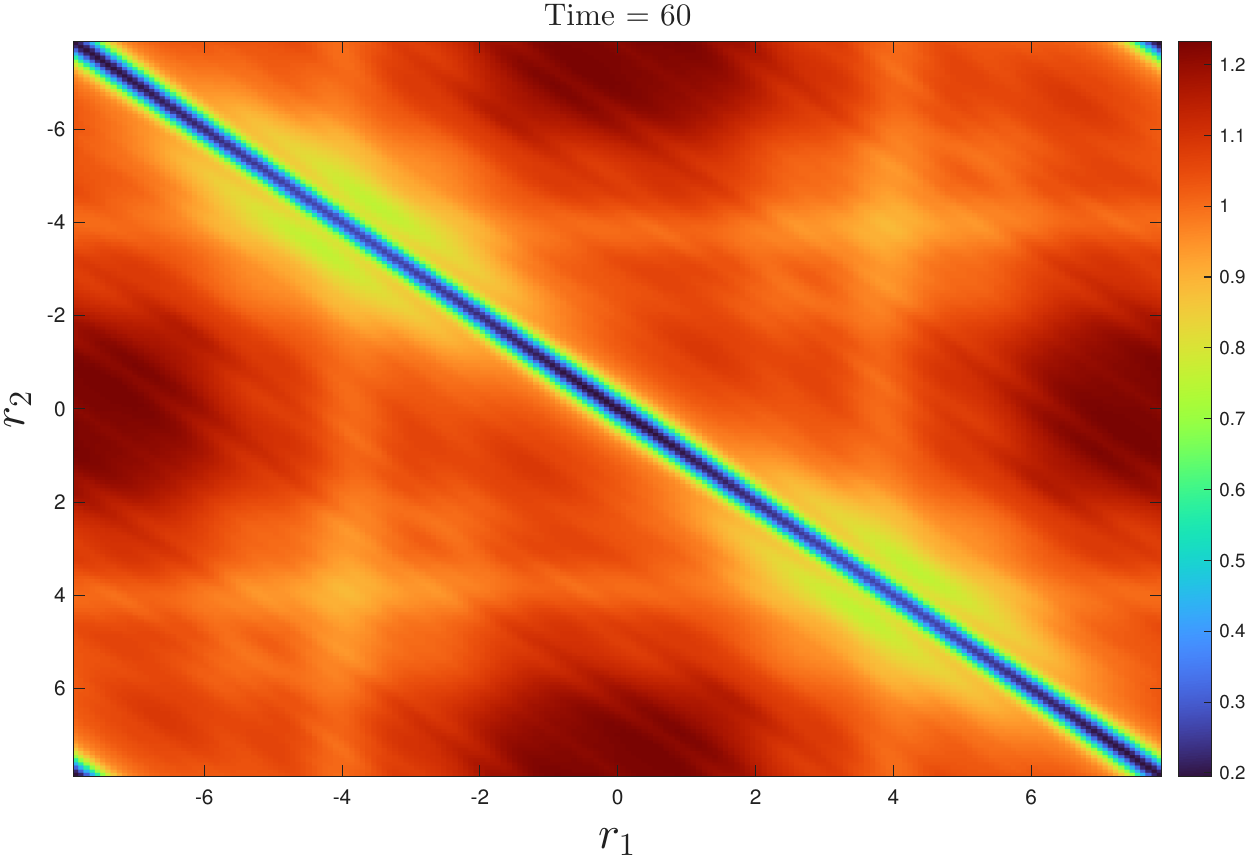}}              
     \caption{Quantum two-stream instability: Evolution of the number density $n_{12}(r_1, r_2, t)$. The cases from left side to right side are: Hartree, Hartree with XC correction, Hartree with two-body correction $\epsilon = 1$ or $\epsilon = 0.01$. Few particles can occupy the same position ($r_1=r_2$) in the two-particle configuration space when two-body interaction is included. \label{xdist_two_stream}}  
\end{figure}

When the two-body interaction is incorporated, the explicit correlation emerges in the number densities (see Figure \ref{xdist_two_stream}). The dissipation by the two-body scattering smoothes out the spiral structure and negative values in the BGK mode (see Figure \ref{Wigner_two_stream}), as well as the filamentation in the momentum space (see Figure \ref{slice_Wigner_two_stream}), and consequently has a significant suppression on the instability (see Figure \ref{two_stream_long}). The suppression effect is more evident as $\epsilon$ decreases.  Apparently, the Coulomb repulsion is much more dominant than the exchange energy, providing a numerical evidence on the assertion in \cite{MoldabekovShaoBellenbaum2025_arXiv}.

\subsection{Dynamical XC potential and two-body effects in a non-periodic, strongly correlated system}

Finally, we study a non-periodic, strongly correlated quantum system under four kinds of Hamiltonians: 
\begin{equation}\label{Hamiltonian}
\begin{split}
\textup{(I)}: ~H(p_1, p_2, r_1,r_2)= &\frac{p_1^2}{2m}  + \frac{p_2^2}{2m} + V_H[n_1](r_1, t) + V_H[n_2](r_2, t),\\
\textup{(II)}: ~H(p_1, p_2, r_1,r_2)= &\frac{p_1^2}{2m}  + \frac{p_2^2}{2m} + V_{HXC}[n_1](r_1, t) + V_{HXC}[n_2](r_2, t), \\
\textup{(III)}: ~H(p_1, p_2, r_1,r_2)= &\frac{p_1^2}{2m}  + \frac{p_2^2}{2m} + V_{HXC}[n_1](r_1, t) + V_{HXC}[n_2](r_2, t) + \frac{\gamma}{\sqrt{|r_1 - r_2|^2 + \epsilon}}, \\
\textup{(IV)}: ~H(p_1, p_2, r_1,r_2)= &\frac{p_1^2}{2m}  + \frac{p_2^2}{2m} + V_{HXC}[n_1](r_1, t) + V_{HXC}[n_2](r_2, t) + \frac{\gamma}{\sqrt{|r_1 - r_2|^2 + \epsilon}} \\
&- \frac{2\gamma}{\sqrt{|r_1|^2 + \epsilon}} - \frac{2\gamma}{\sqrt{|r_2|^2 + \epsilon}}.
\end{split}
\end{equation}
Here $V_H[n_1]$ is defined in Eq.~\eqref{Hartree_smoothed_Coulomb} and $V_{XC}[n_1]$ is chosen as the H-L potential \eqref{mean_field_approximation}. The parameters $\epsilon$ is $0.01$ or $1$ and $\gamma = 1$. The corresponding $\pdo$ is resolved by SEM. 

Since  the Pauli principle should be assured, we start from the 4-dimensional 2-reduced Wigner function \cite{SchroedterBonitz2024}
\begin{equation}
f_{12}^{init}(r_1, r_2, p_1, p_2) = f_1^{init}(r_1, p_2) f_1^{init}(r_2, p_2) g_{12}(|r_1 - r_2|),
\end{equation}
where
\begin{equation}
\begin{split}
f_1^{init}(r, p) = \frac{1}{2\pi \hbar} \left[\me^{-2a(r - r_0^{(1)})^2 -\frac{1}{2a\hbar^2}(p - p_0^{(1)})^2}  + \me^{-2a(r - r_0^{(2)})^2 -\frac{1}{2a\hbar^2}(p - p_0^{(2)})^2}\right],
\end{split}
\end{equation}
and the correlation term $g_{12}$ reads that
\begin{equation}
g_{12}(r) = \exp(-\beta \exp(-\alpha r)/r),
\end{equation}
where $\beta$ is the inverse temperature, and $\alpha$ controls the intensity of screening effect.

We use the correlation entropy to measure the quantum correlation induced by the two-body interaction. For the density matrix for the quantum mixed state
\begin{equation}
\rho_1(r_1, r_1^{\prime}, t) = \sum_{k=1}^{\infty} n_k(t) \psi_k^\ast (r_1, t) \psi_k(r_1^{\prime}, t).
\end{equation}
the correlation entropy $s(t)$ is defined as 
\begin{equation}
s(t)  = \frac{1}{N} \sum_{k=1}^N n_k(t) \ln n_k(t).
\end{equation}
Details on its calculation are put in \ref{sec.correlated}.

The parameters are set as follows: $\alpha = \beta = 0.1$, $r_0^{(1)} = -2$, $r_0^{(2)} = 2$, $p_0^{(1)} = p_0^{(2)} = 0$. The computational domain $\mathcal{X} \times \mathcal{P} = [-15, 15]^2 \times [-\frac{10\pi}{3}, \frac{10\pi}{3}]^2$ is represented by a non-uniform mesh with $201^2\times 128^2 \approx 6.62\times10^8$ grid points and a fixed spatial spacing $\Delta x = 0.15$. In each direction, the momentum space is divided into $8$ cells and each cell contains $16$ collocation points.  The time step is $\Delta t = 0.01$ and the final time is $T = 4$.

\begin{figure}[!h]
    \centering
    \subfigure[$t=0.5$a.u.]{
    \includegraphics[width=1.4in,height=1.0in]{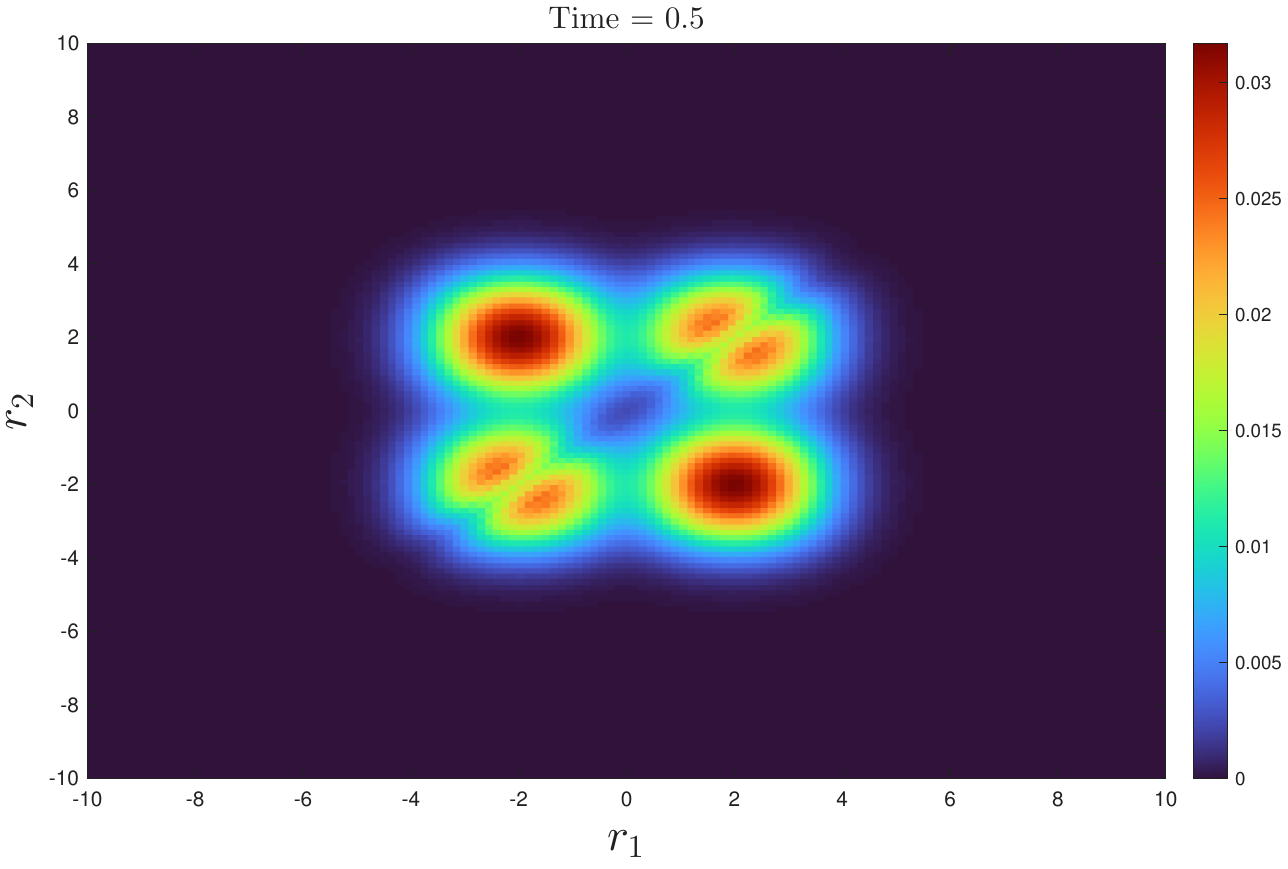}
    \includegraphics[width=1.4in,height=1.0in]{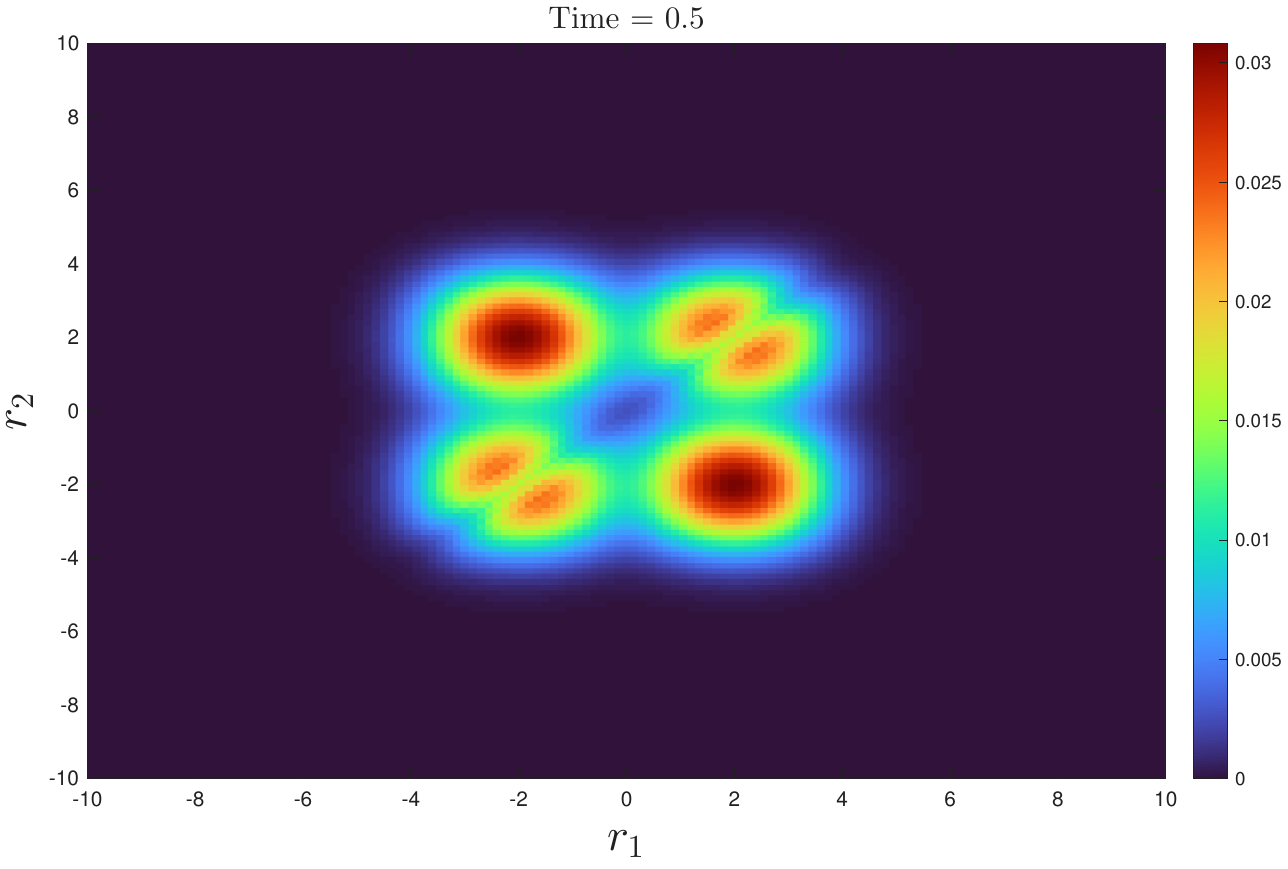}
    \includegraphics[width=1.4in,height=1.0in]{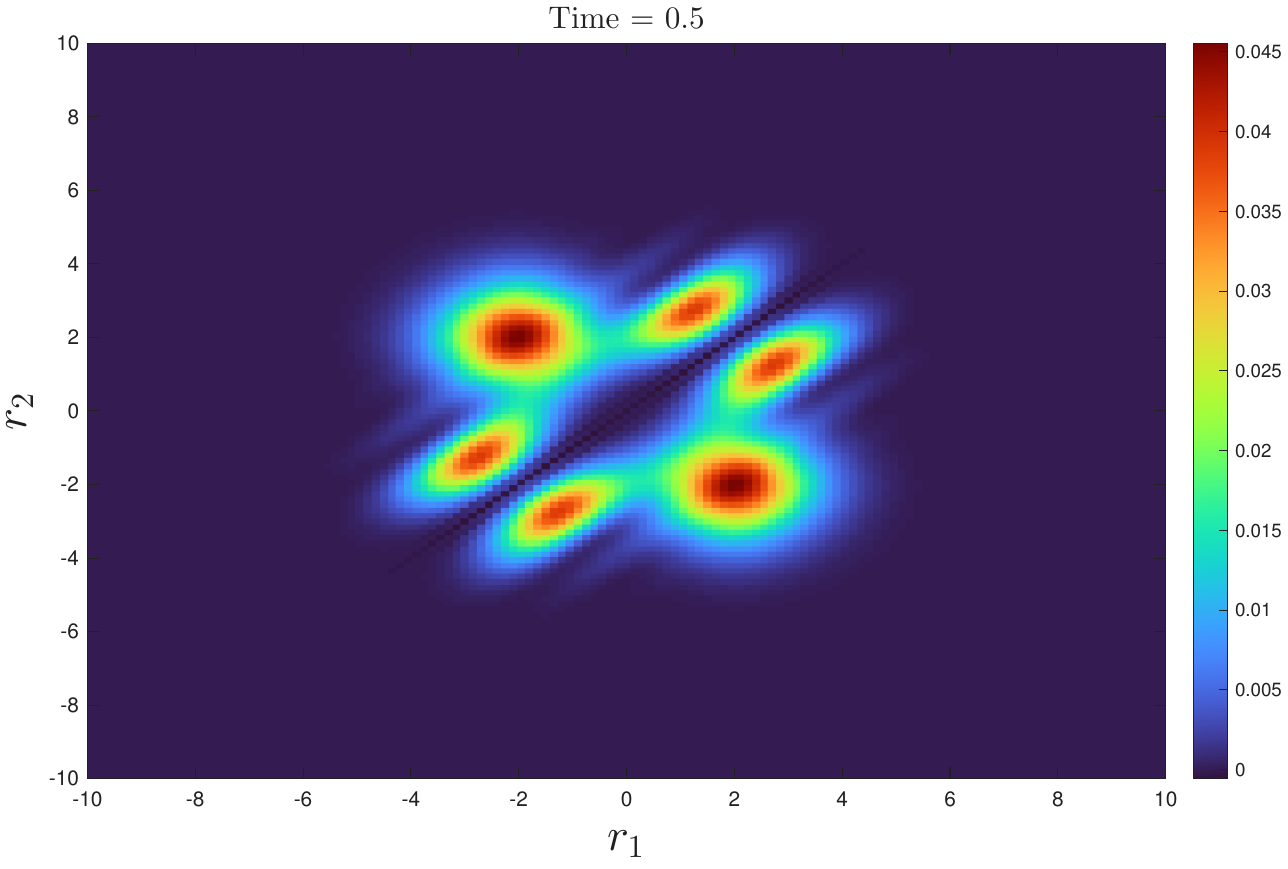}
    \includegraphics[width=1.4in,height=1.0in]{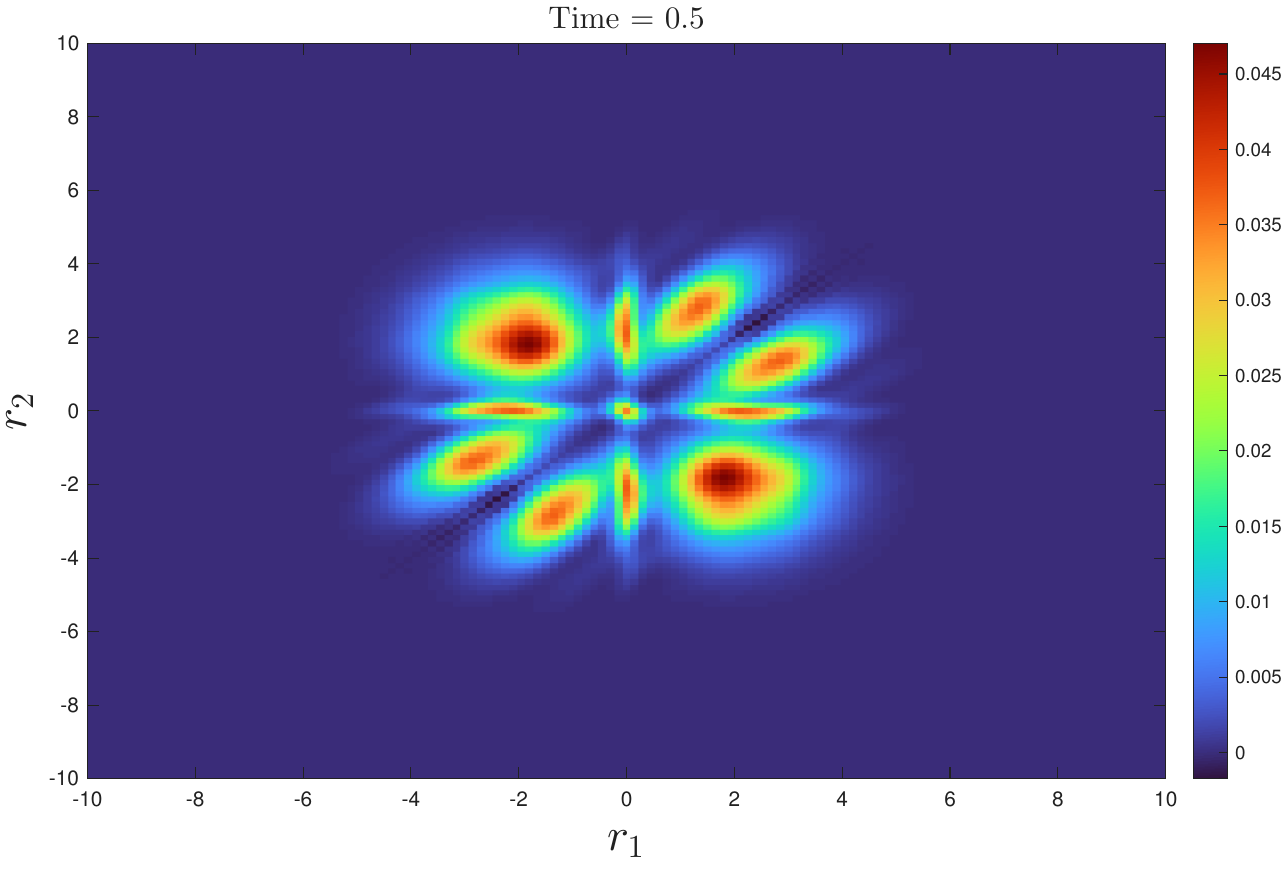}}
    \\
    \centering
    \subfigure[$t=1$a.u.]{
    \includegraphics[width=1.4in,height=1.0in]{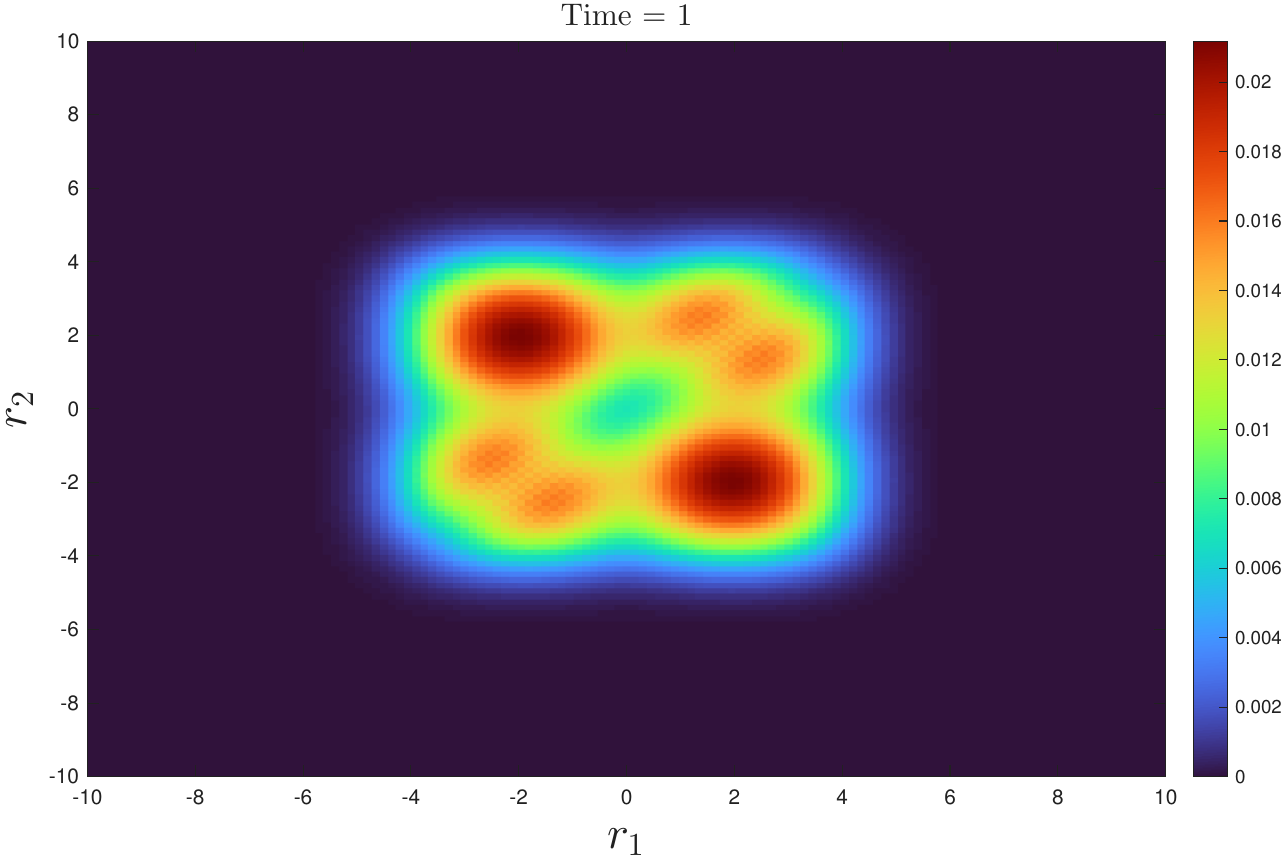}
    \includegraphics[width=1.4in,height=1.0in]{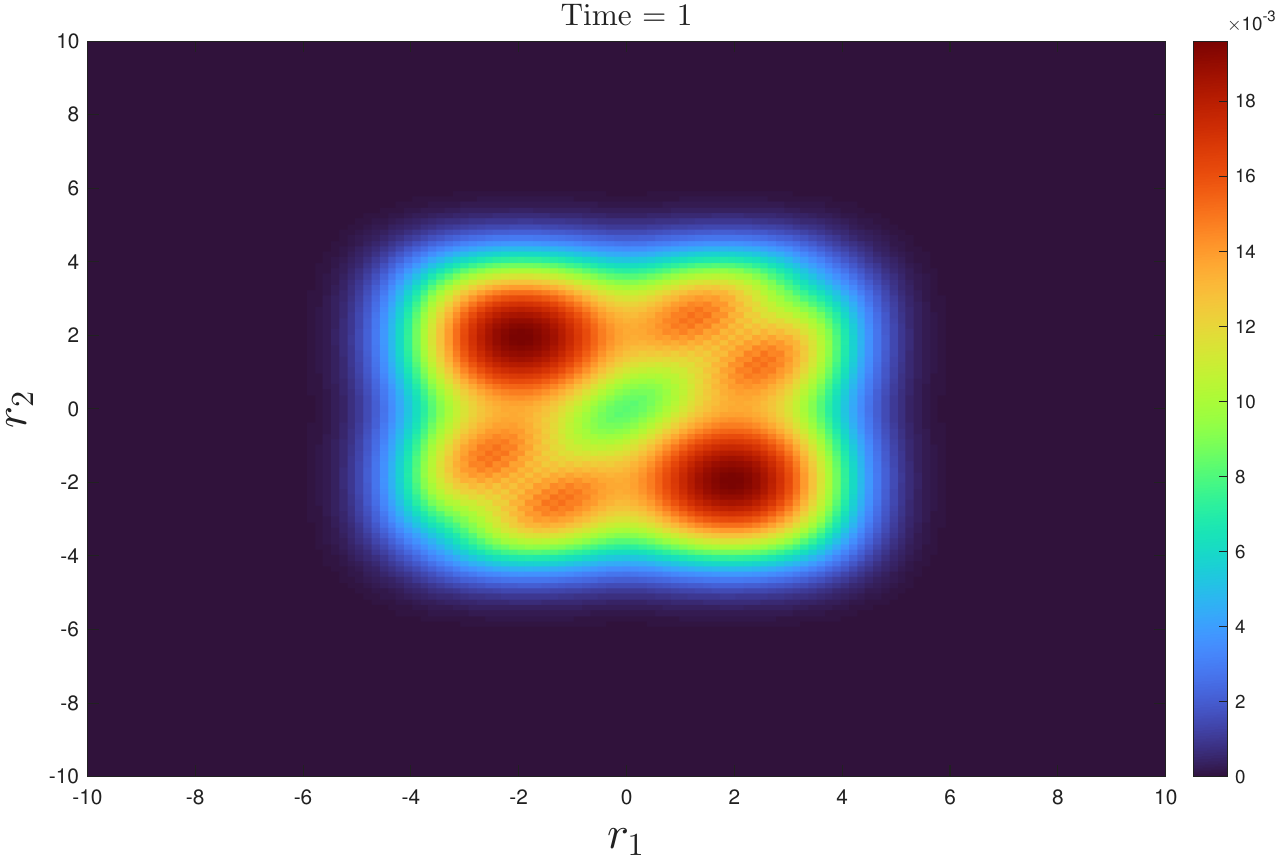}
    \includegraphics[width=1.4in,height=1.0in]{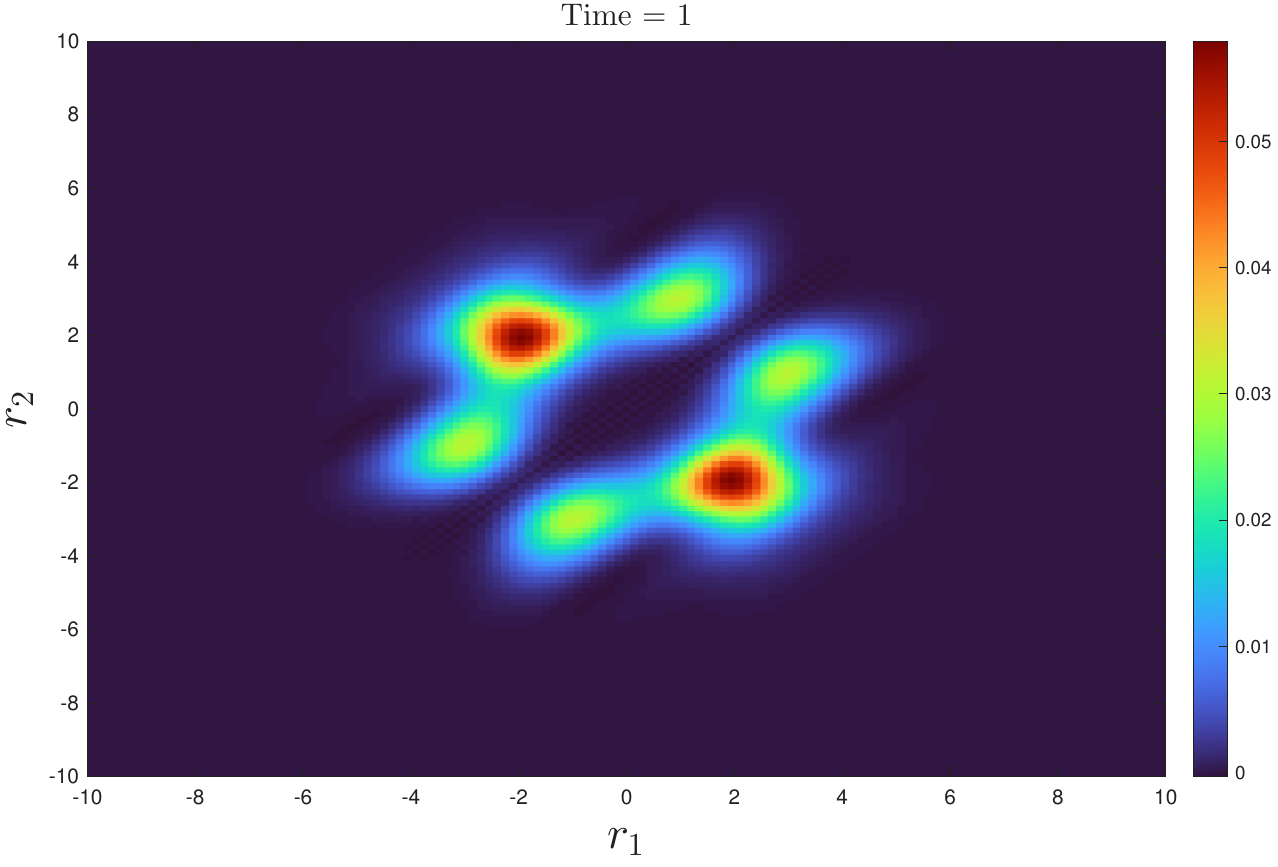}   
    \includegraphics[width=1.4in,height=1.0in]{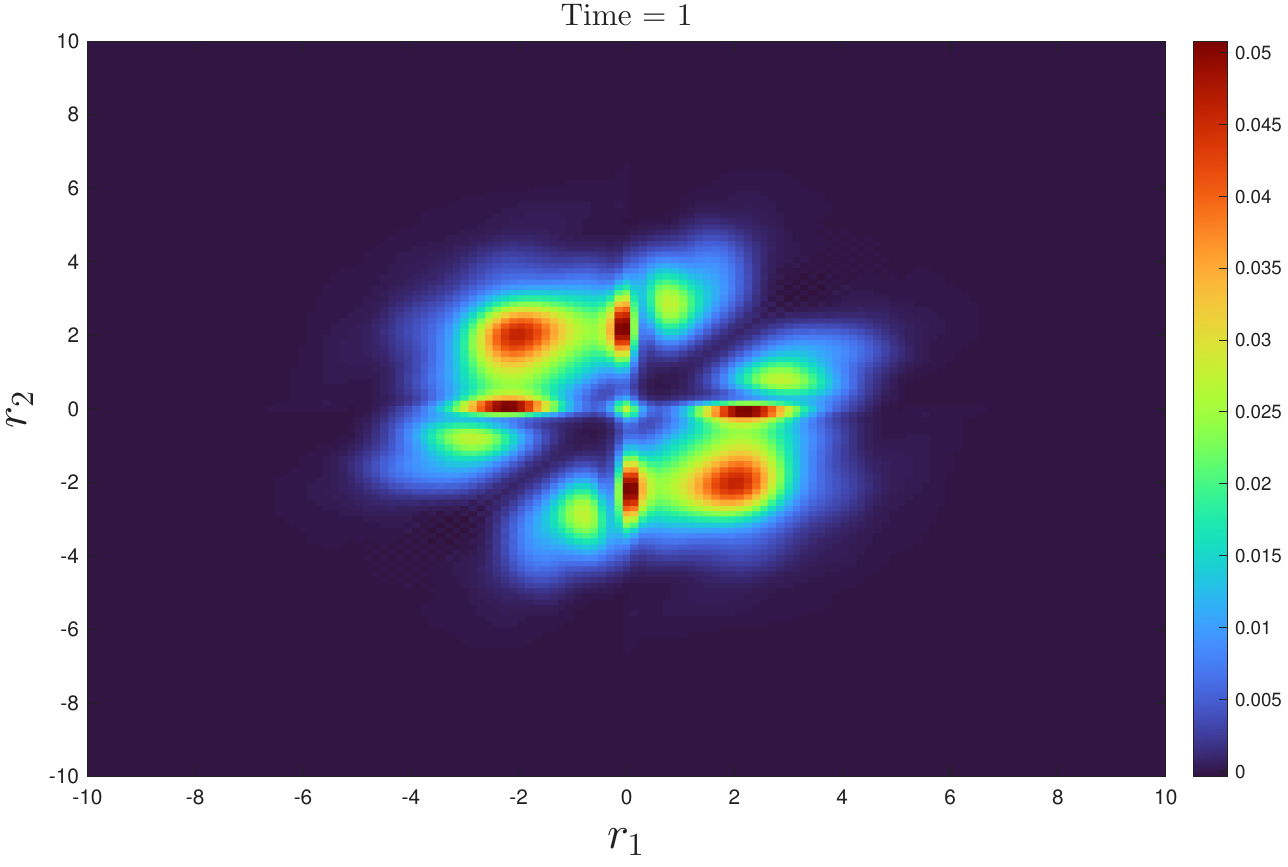}}
     \\
    \centering
    \subfigure[$t=2$a.u.]{
    \includegraphics[width=1.4in,height=1.0in]{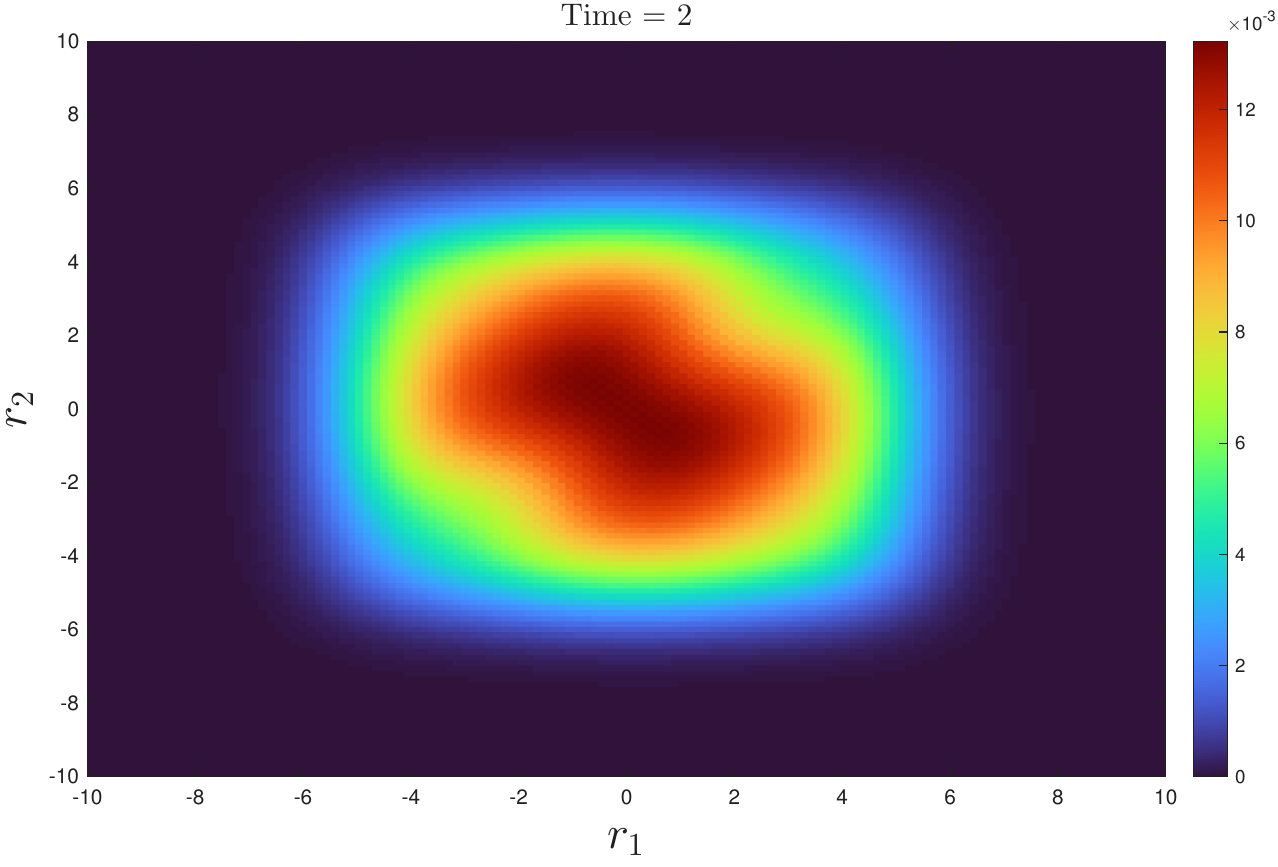}
    \includegraphics[width=1.4in,height=1.0in]{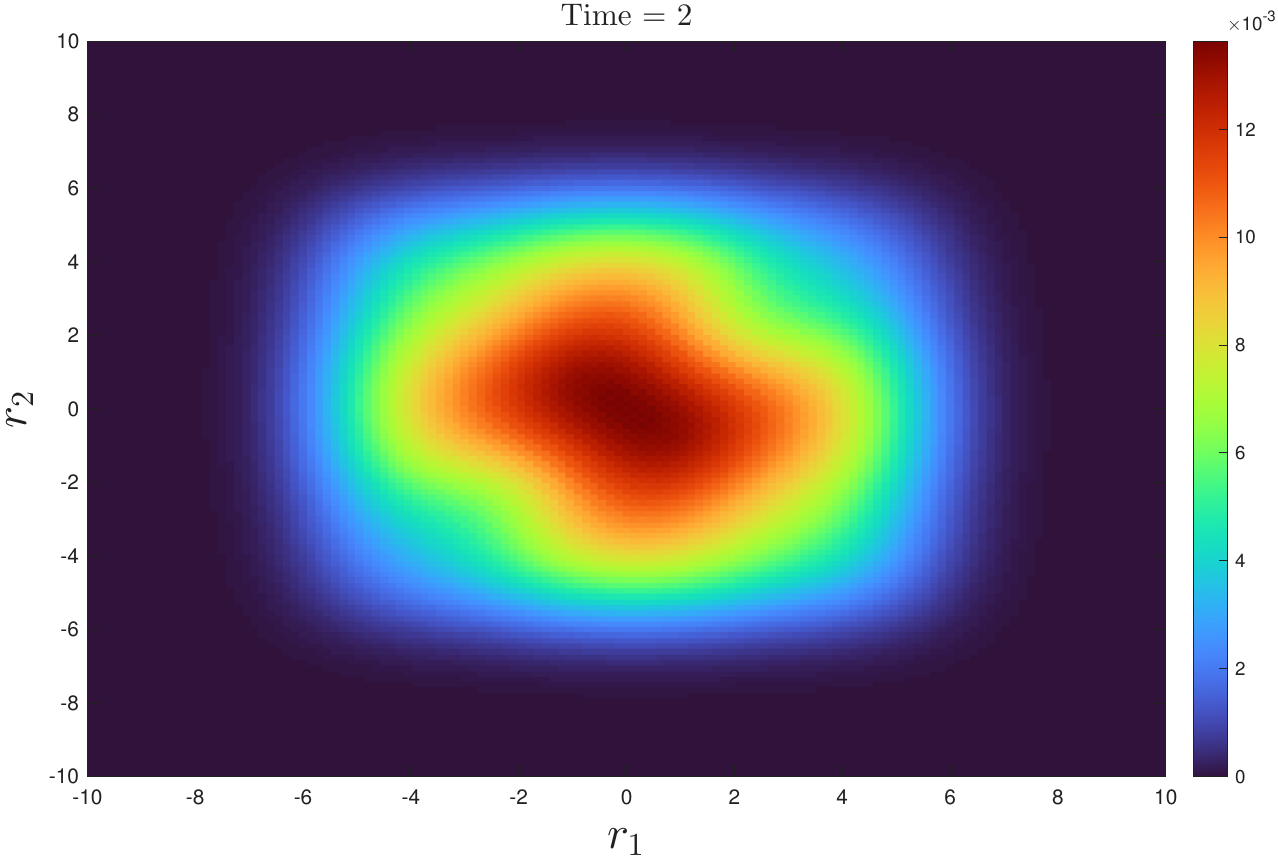}
    \includegraphics[width=1.4in,height=1.0in]{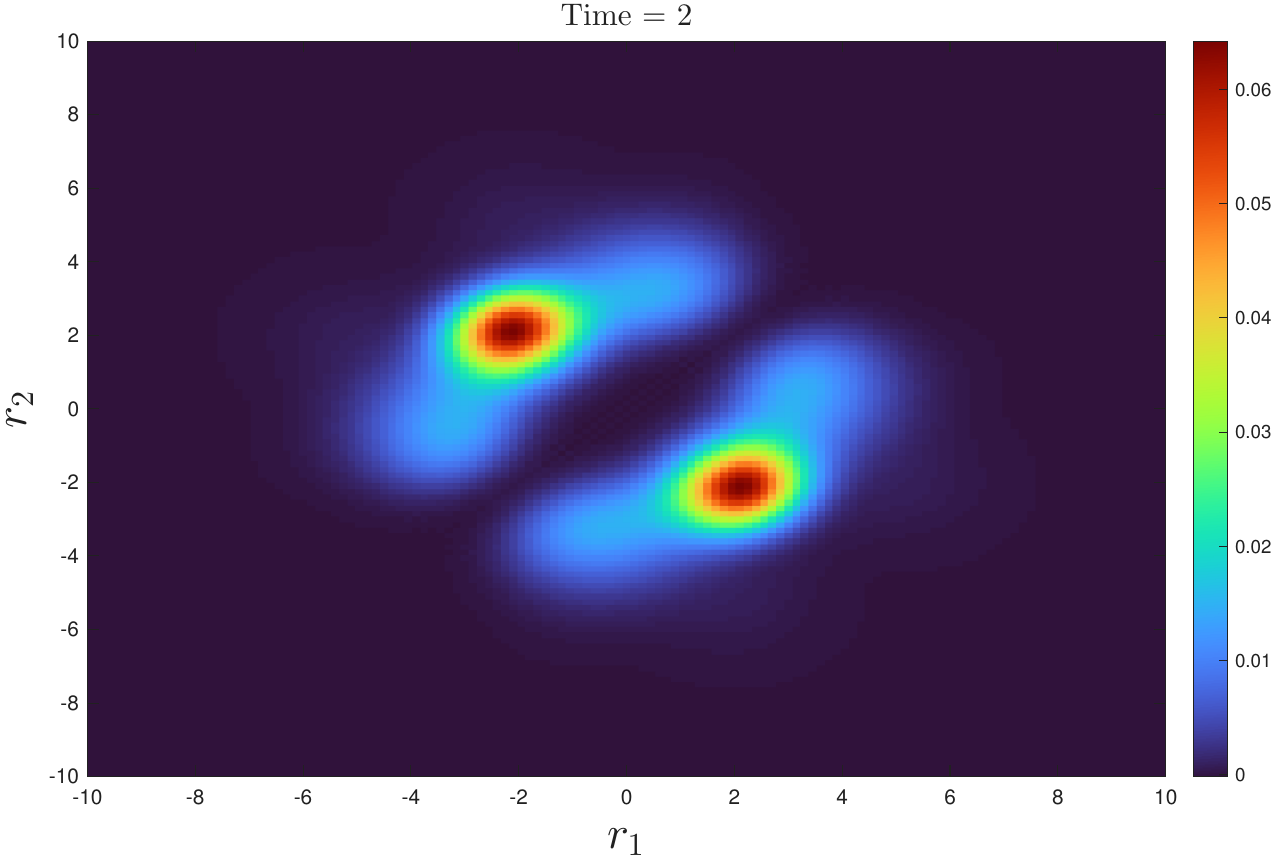}      
    \includegraphics[width=1.4in,height=1.0in]{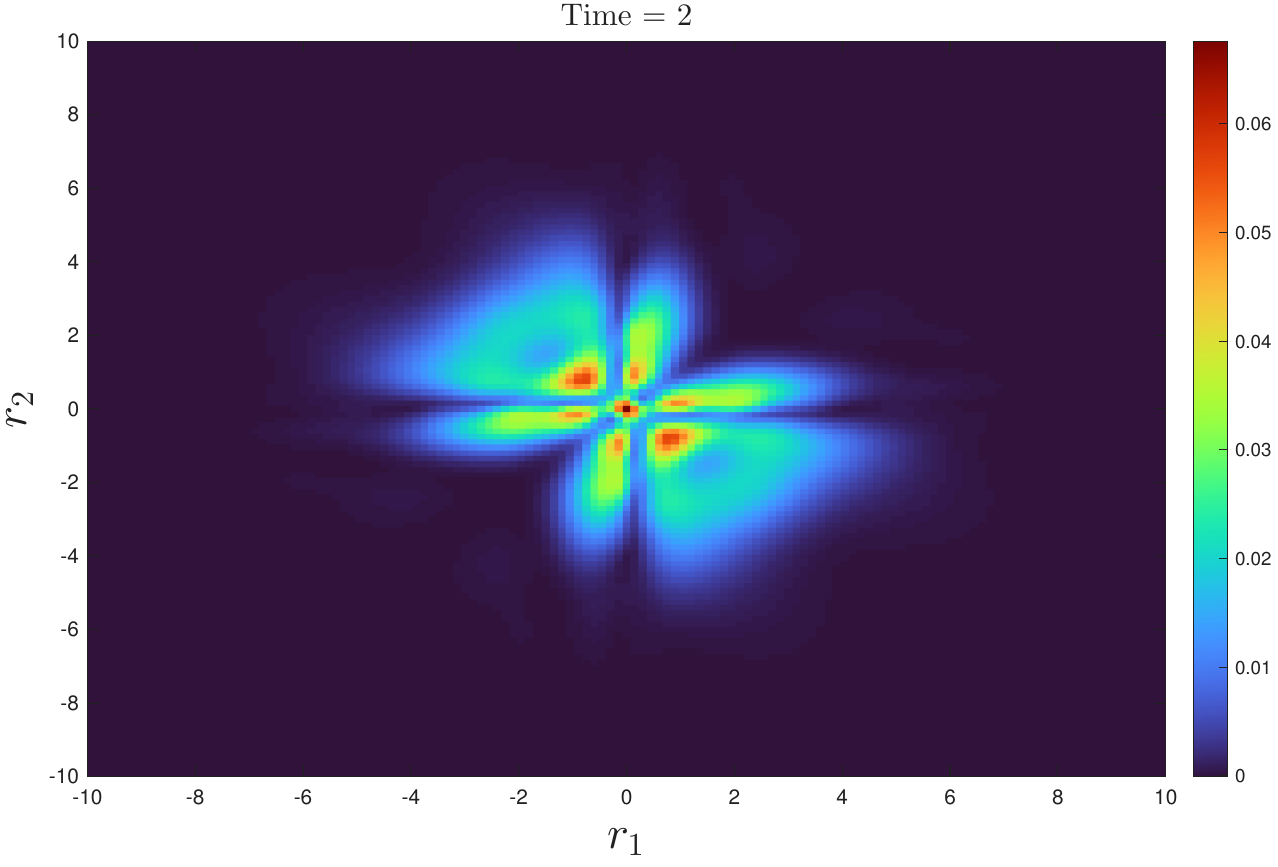}}    
     \\
    \centering
    \subfigure[$t=3$a.u.]{
    \includegraphics[width=1.4in,height=1.0in]{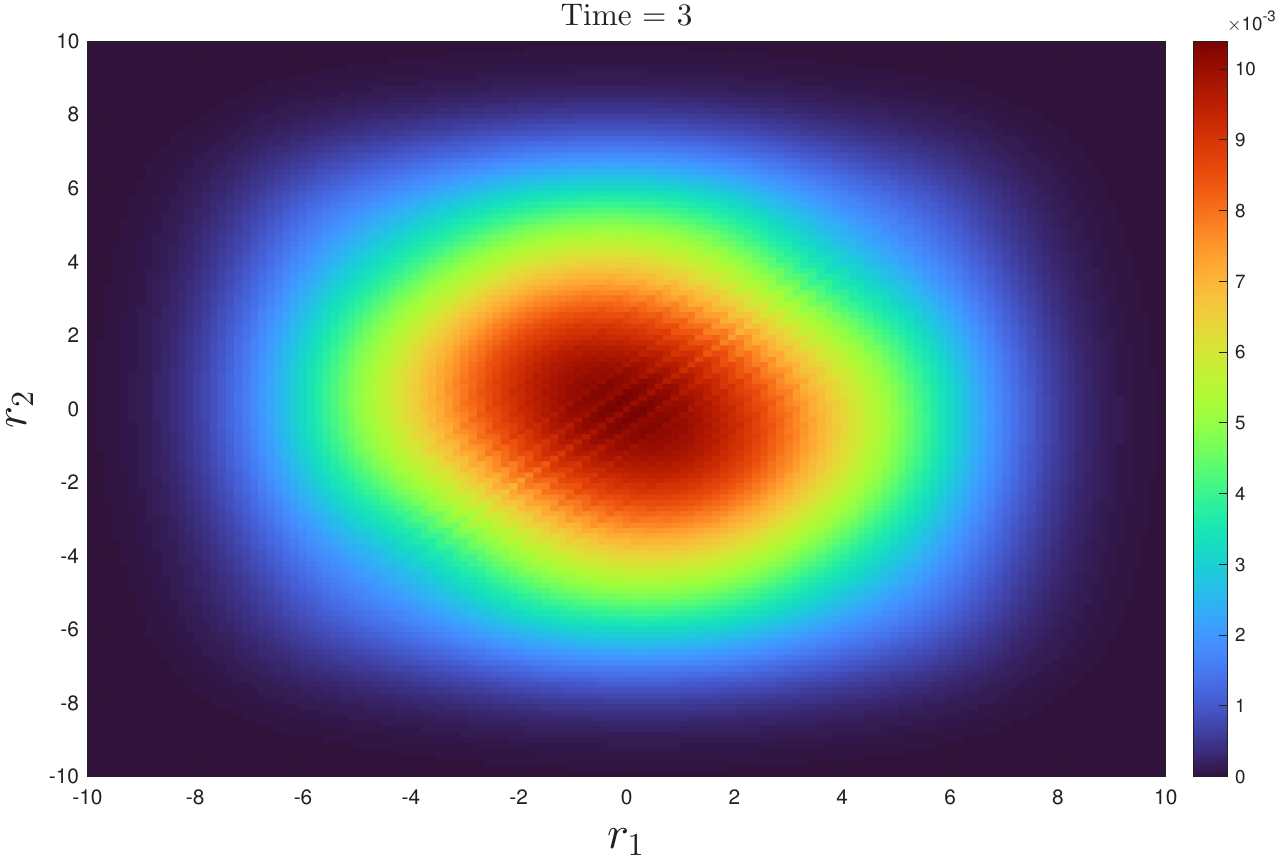}
    \includegraphics[width=1.4in,height=1.0in]{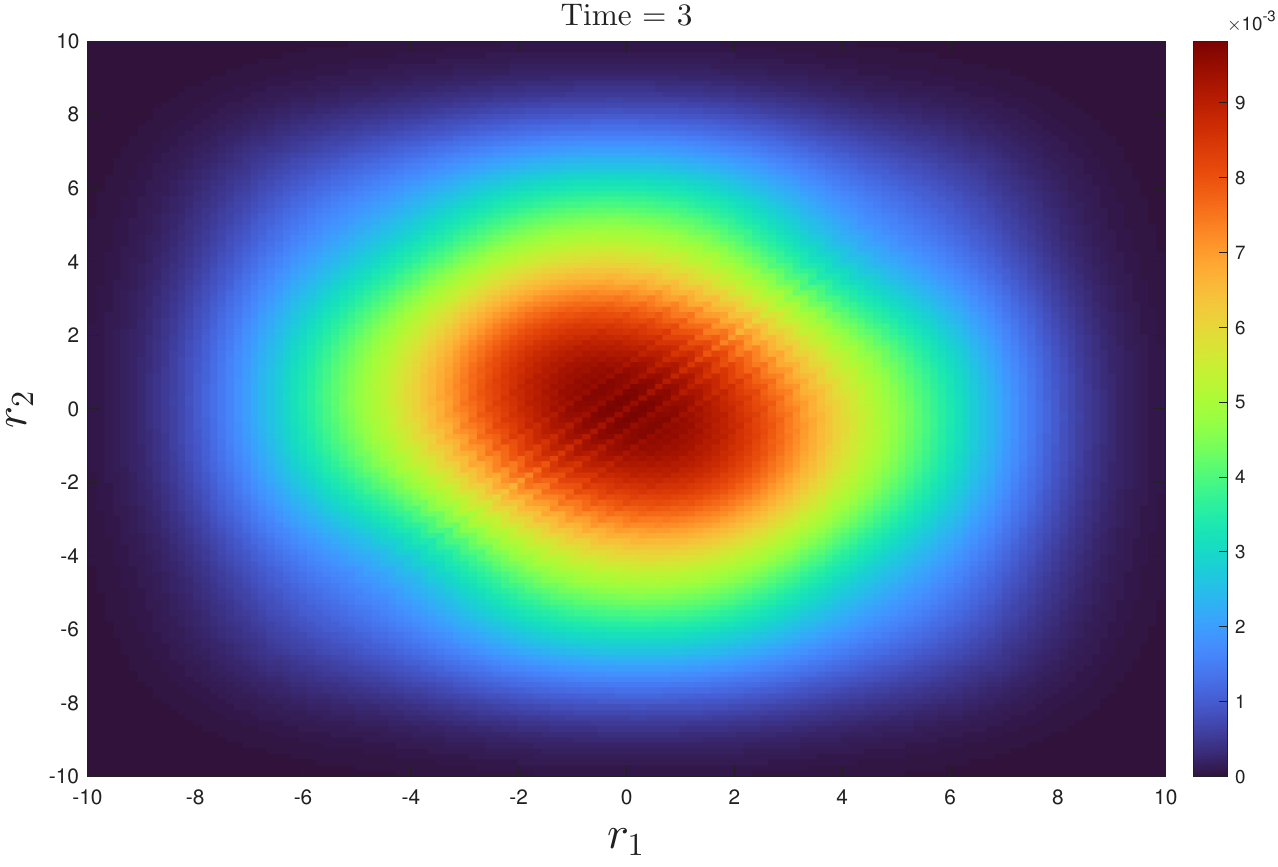}
    \includegraphics[width=1.4in,height=1.0in]{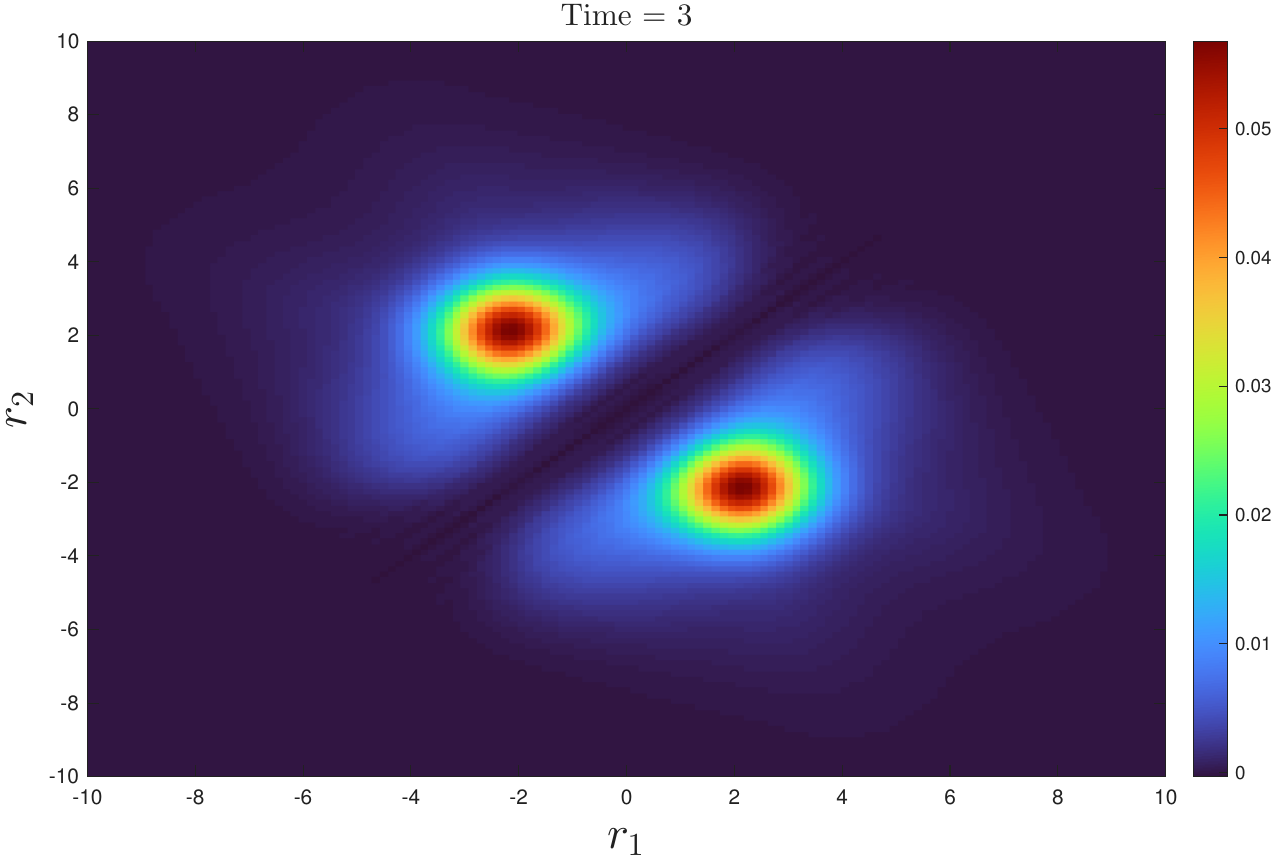}       
    \includegraphics[width=1.4in,height=1.0in]{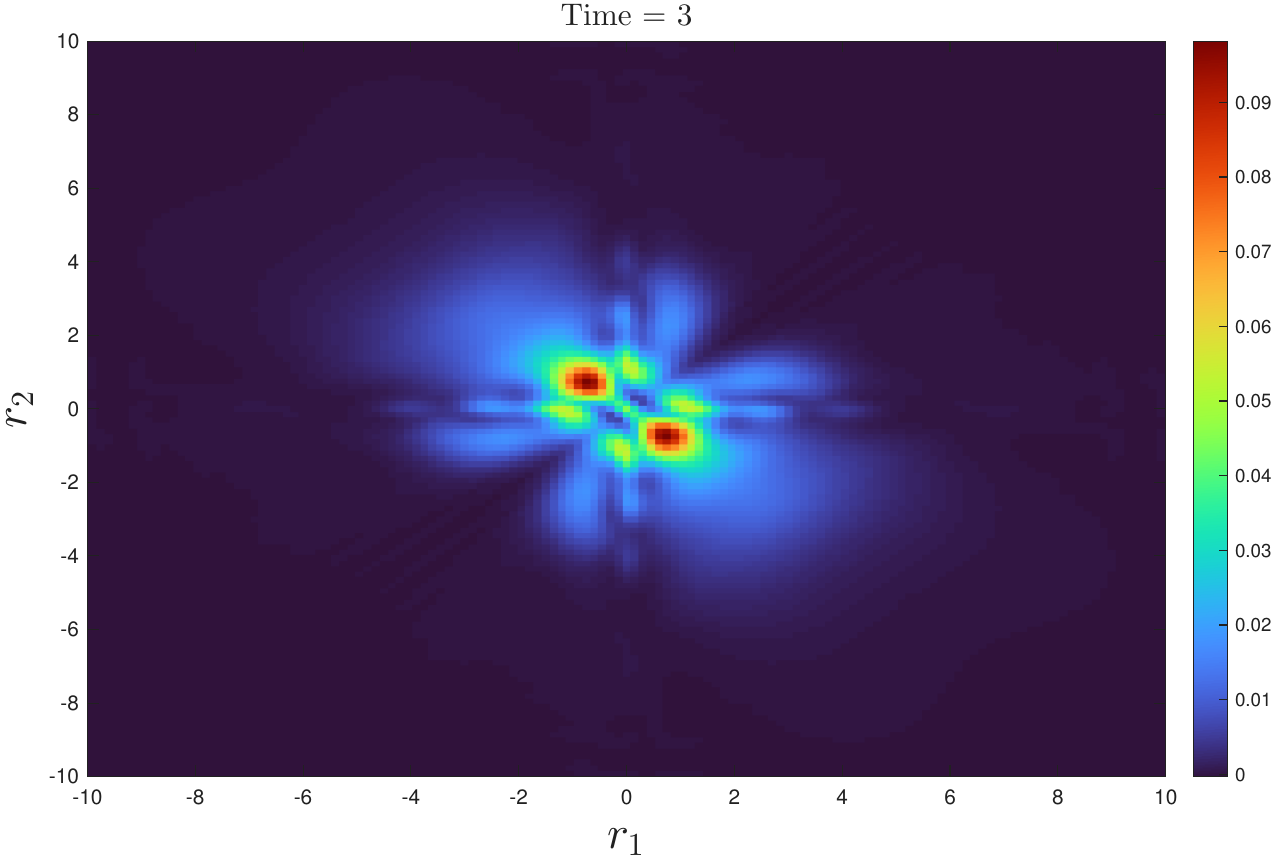}}       
     \\
    \centering
    \subfigure[$t=4$a.u.]{
    \includegraphics[width=1.4in,height=1.0in]{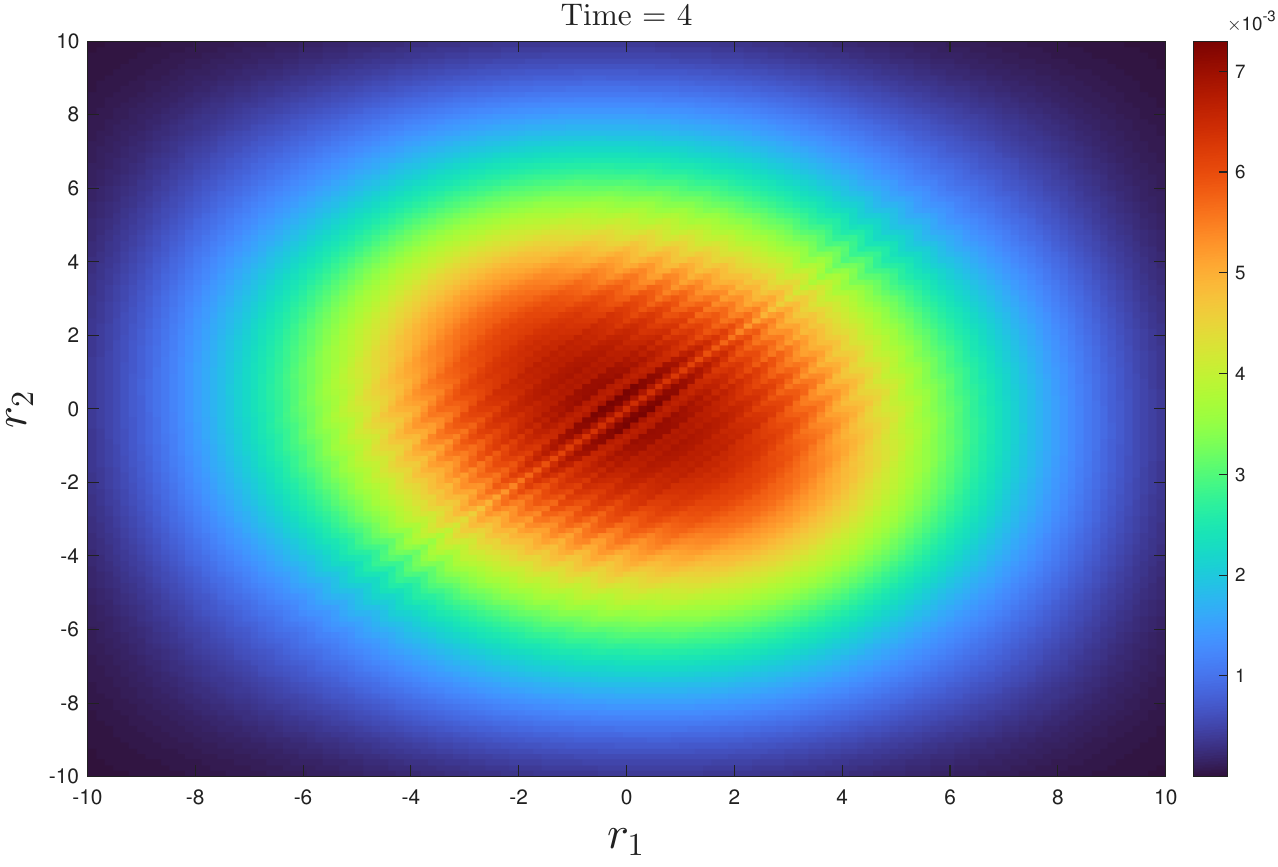}
    \includegraphics[width=1.4in,height=1.0in]{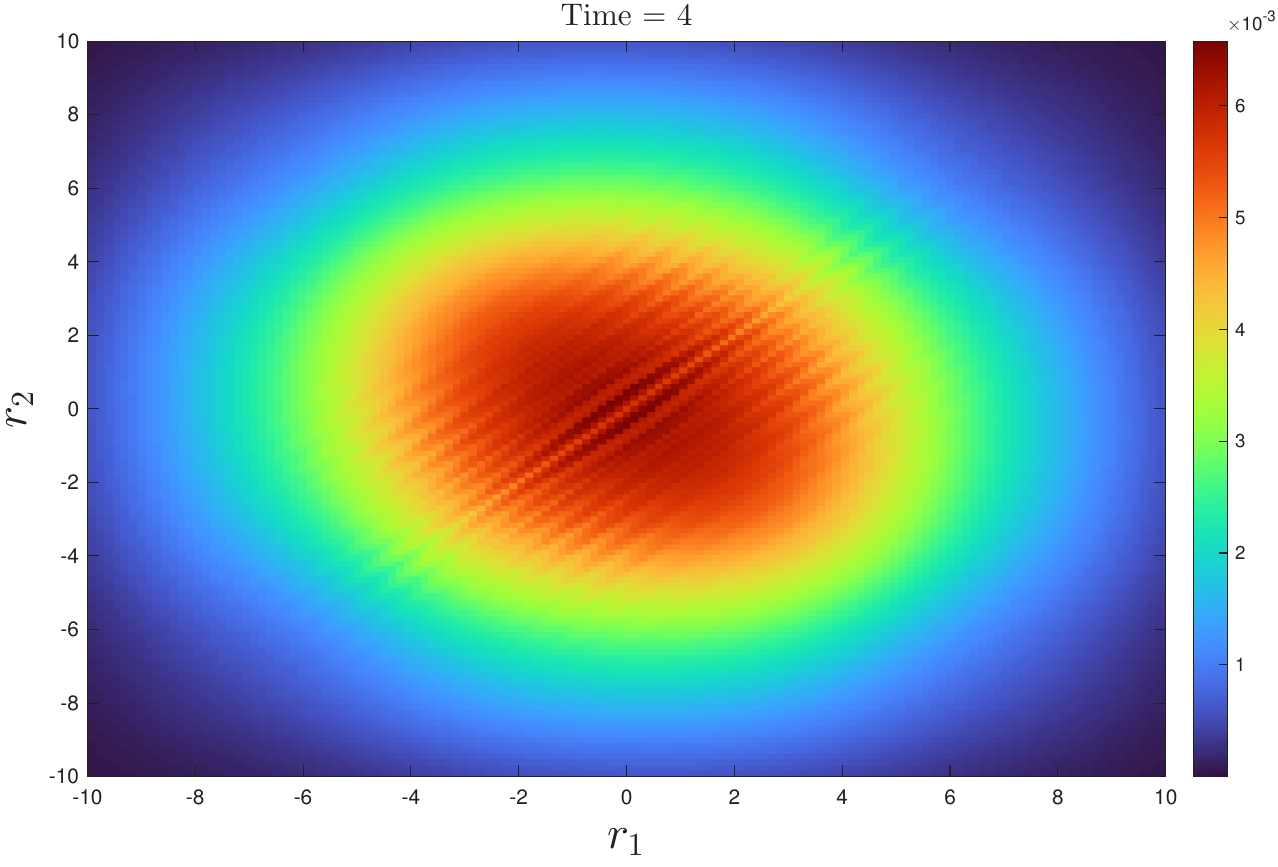}
    \includegraphics[width=1.4in,height=1.0in]{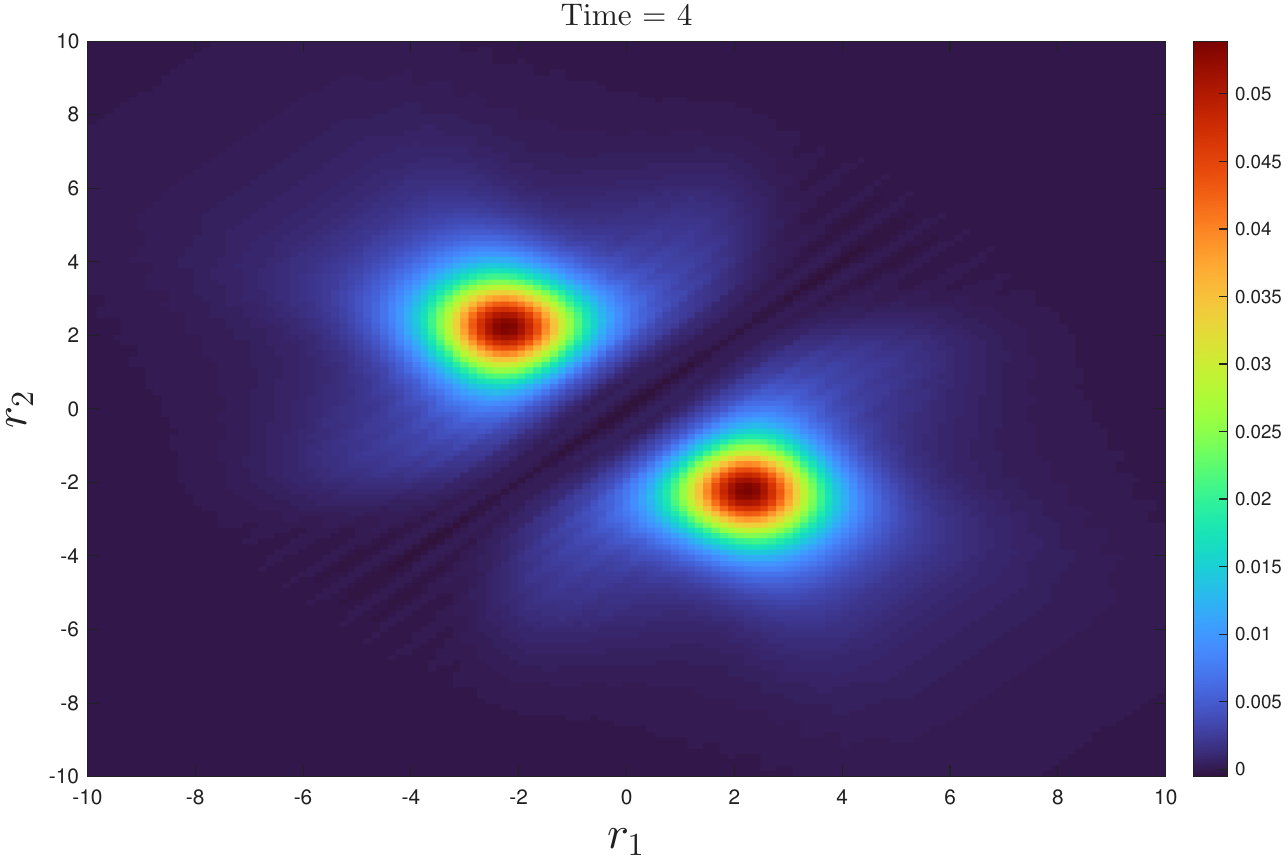}     
    \includegraphics[width=1.4in,height=1.0in]{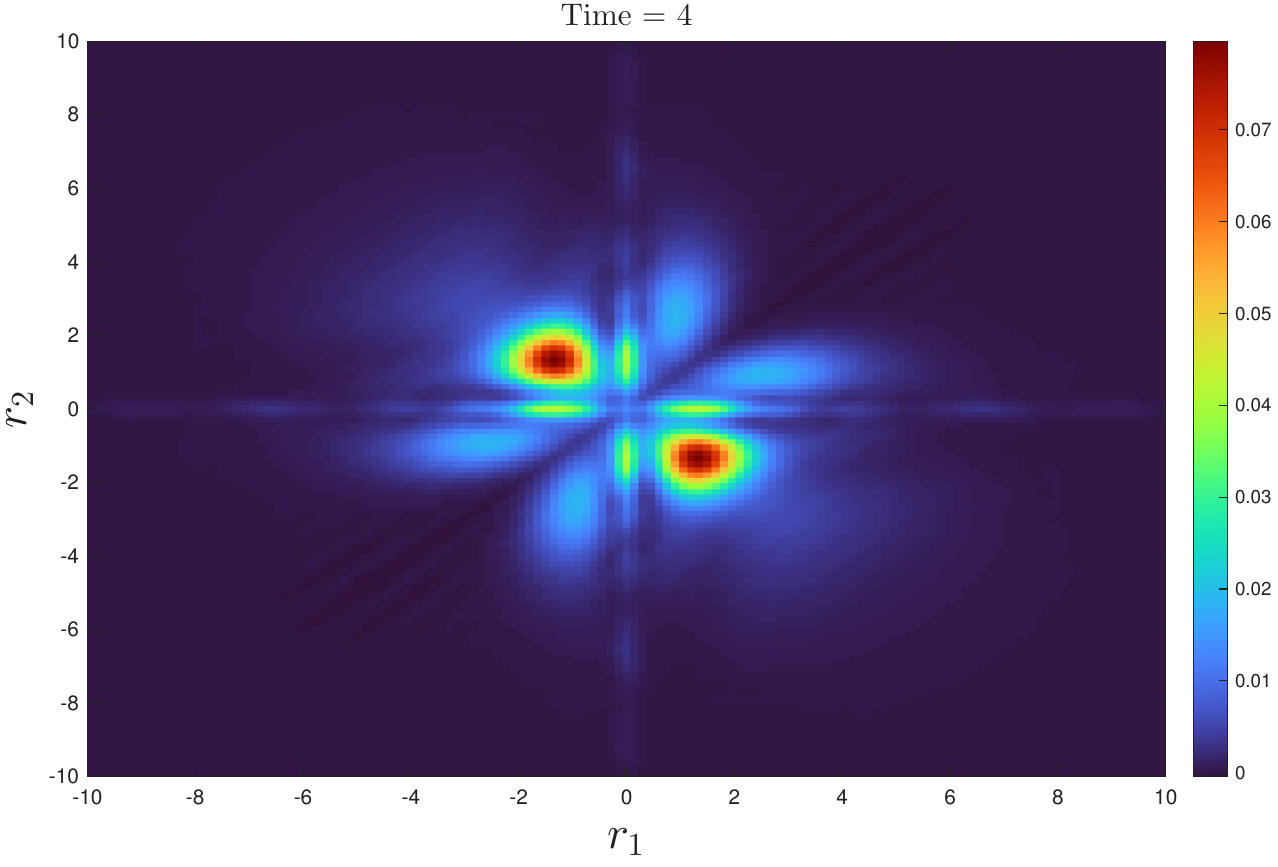}}     
     \\
    \centering
    \subfigure[The projection $n_1(r, t) + n_2(r, t)$.]{
    \includegraphics[width=1.4in,height=1.0in]{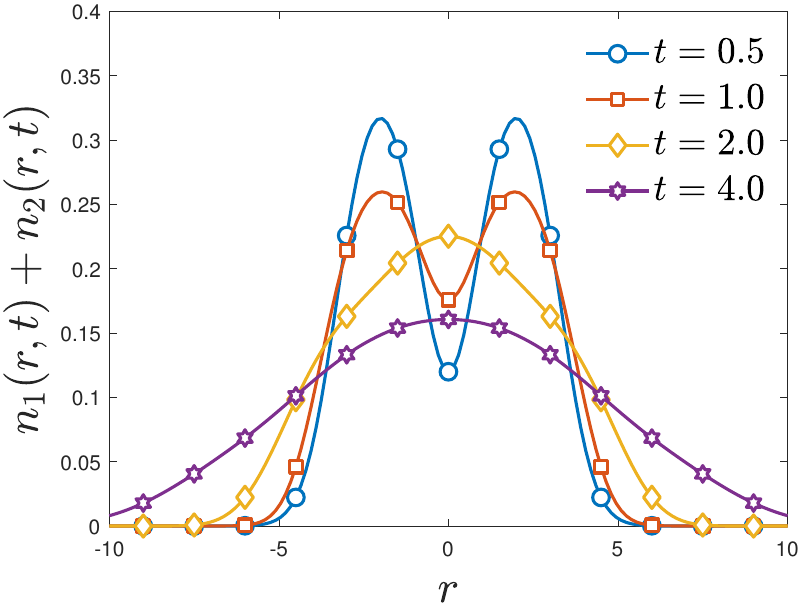}
    \includegraphics[width=1.4in,height=1.0in]{./SC_xdist_Hartree_eps001}
    \includegraphics[width=1.4in,height=1.0in]{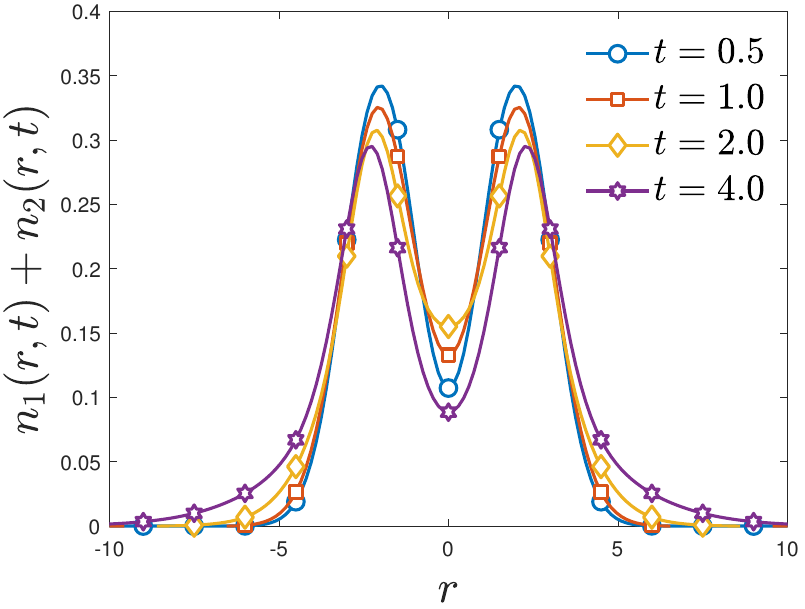}     
    \includegraphics[width=1.4in,height=1.0in]{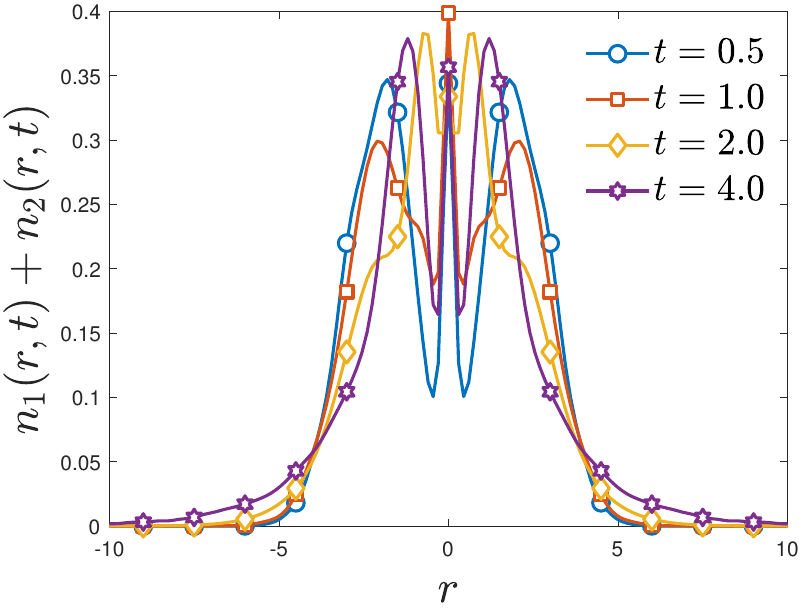}}                    
     \caption{Non-periodic, strongly correlated system: Evolution of the number density $n_{12}(r_1, r_2, t)$. The cases from left side to right side correspond to the Hamiltonians (I) to (IV) in Eq.~\eqref{Hamiltonian}. Here the smoothing parameter is $\epsilon = 0.01$.\label{xdist_correlated}}  
\end{figure}

From the plots of number densities in Figure \ref{xdist_correlated}, the double peaks soon coalesce and the initial correlation seems to disappear when only the Hartree part is involved. The same picture holds when the XC potential is included. By contrast, when the two-body repulsive interaction is incorporated, it impedes the coalescence of the wavepackets and  preserves the two-body correlation. Under a strongly attractive external field at origin, the confined wavepackets still exhibit strong correlation. 

\begin{figure}[!h]
    \centering
    \subfigure[$t=0.5$a.u.]{
    \includegraphics[width=1.4in,height=1.0in]{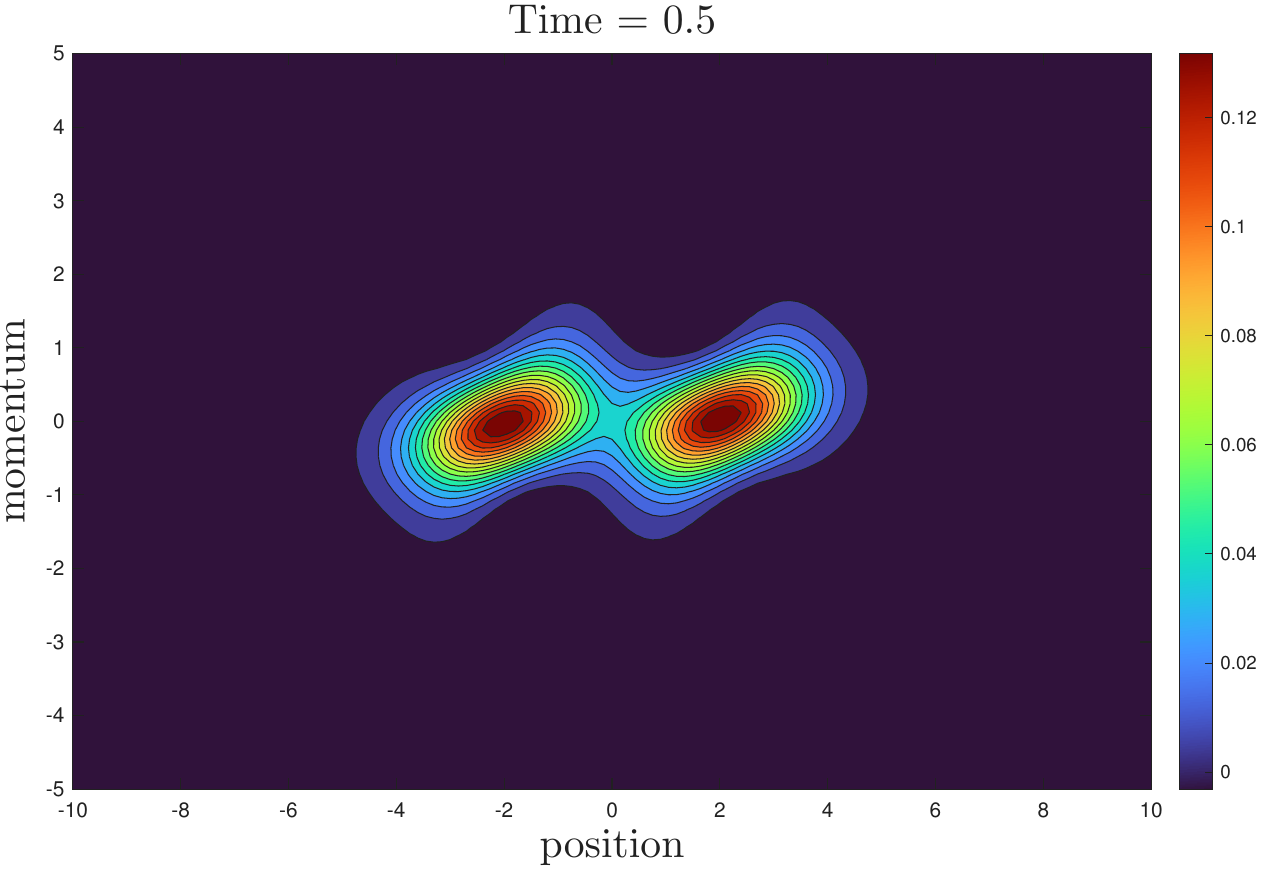}
    \includegraphics[width=1.4in,height=1.0in]{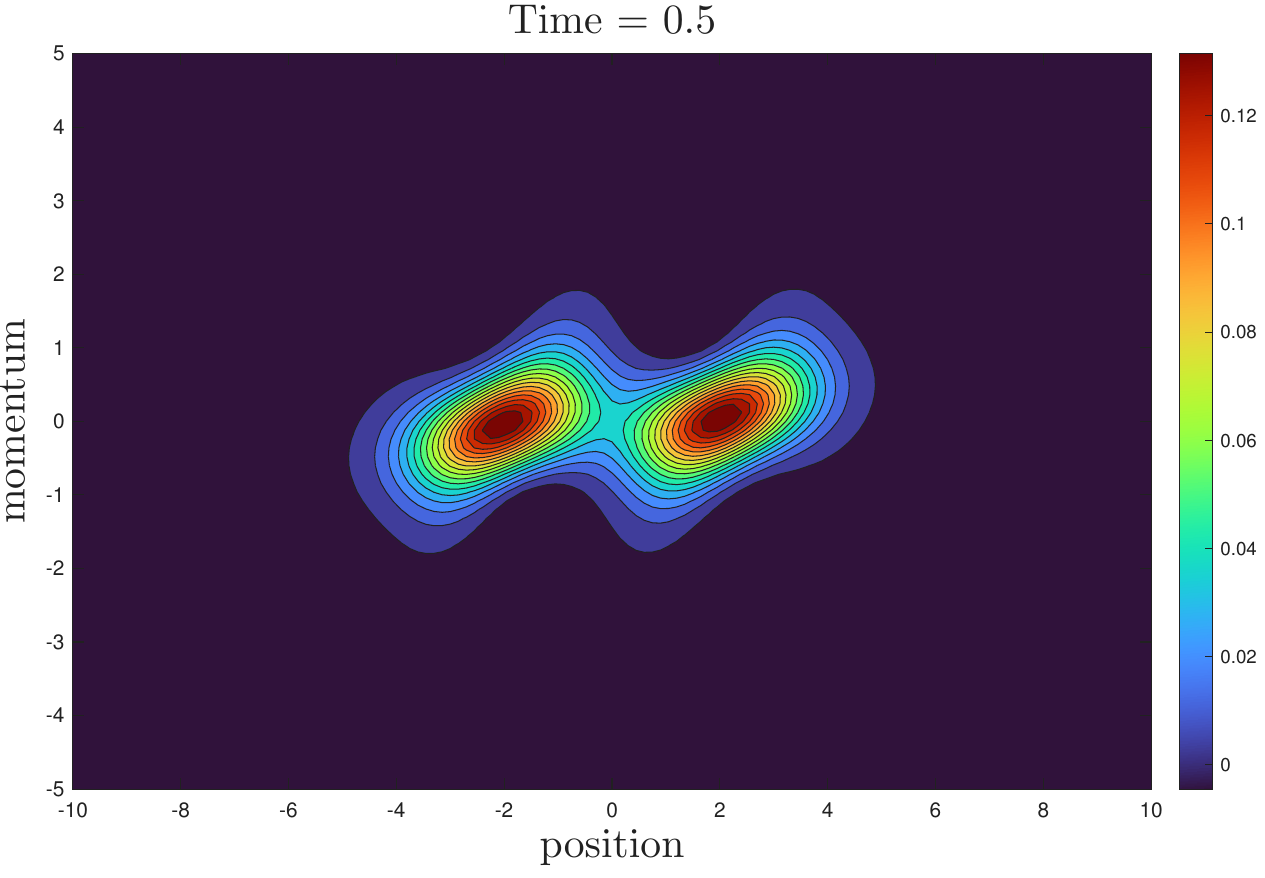}
    \includegraphics[width=1.4in,height=1.0in]{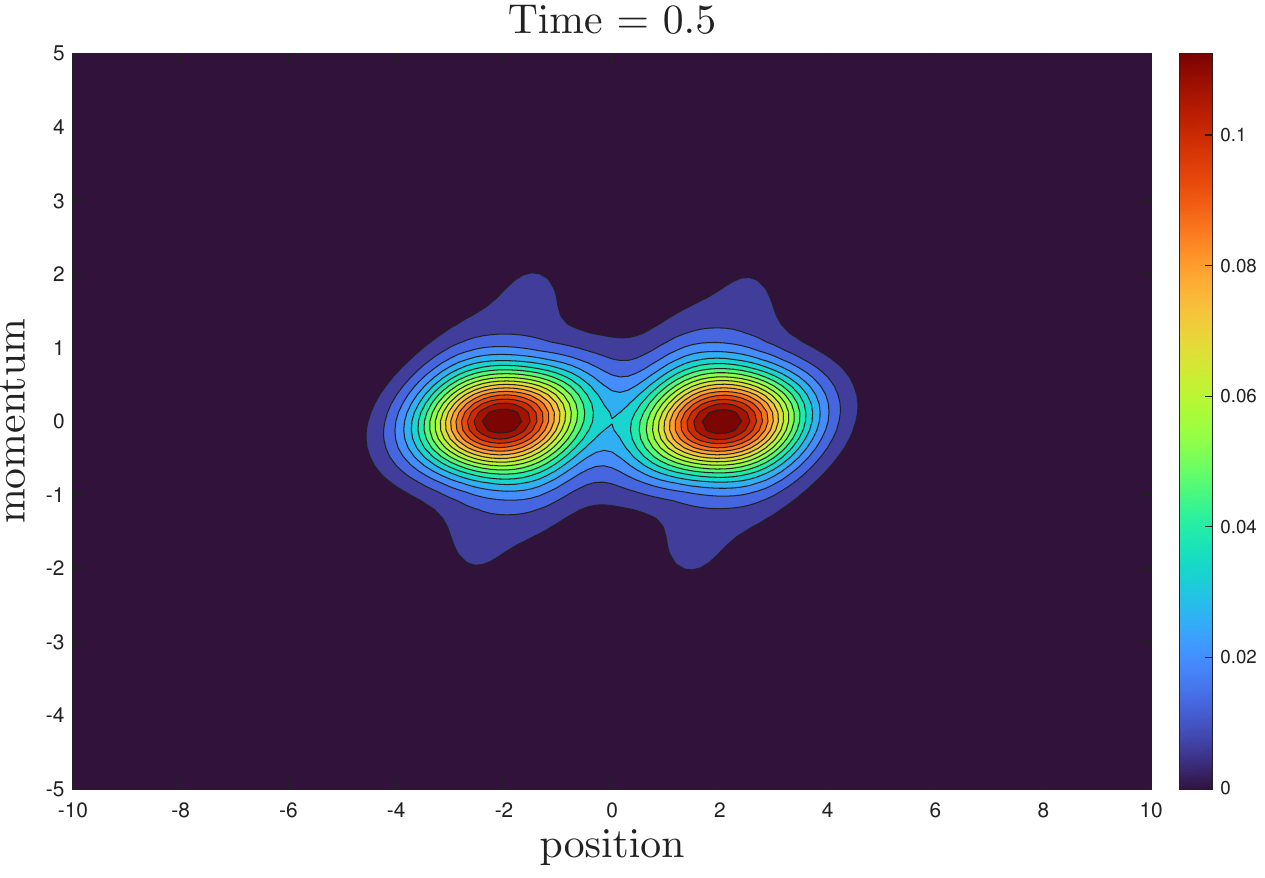}
     \includegraphics[width=1.4in,height=1.0in]{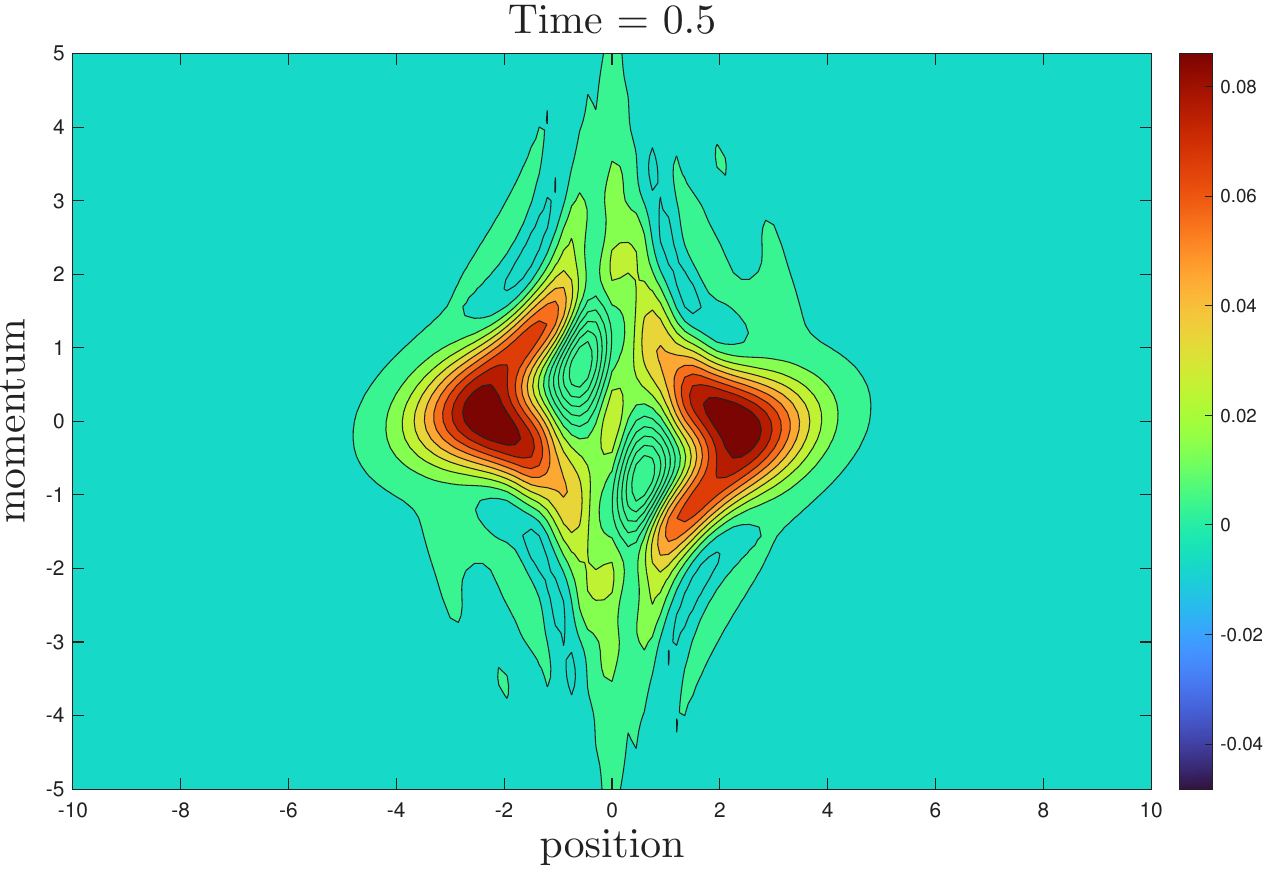}}
    \\
    \centering
    \subfigure[$t=1$a.u.]{
    \includegraphics[width=1.4in,height=1.0in]{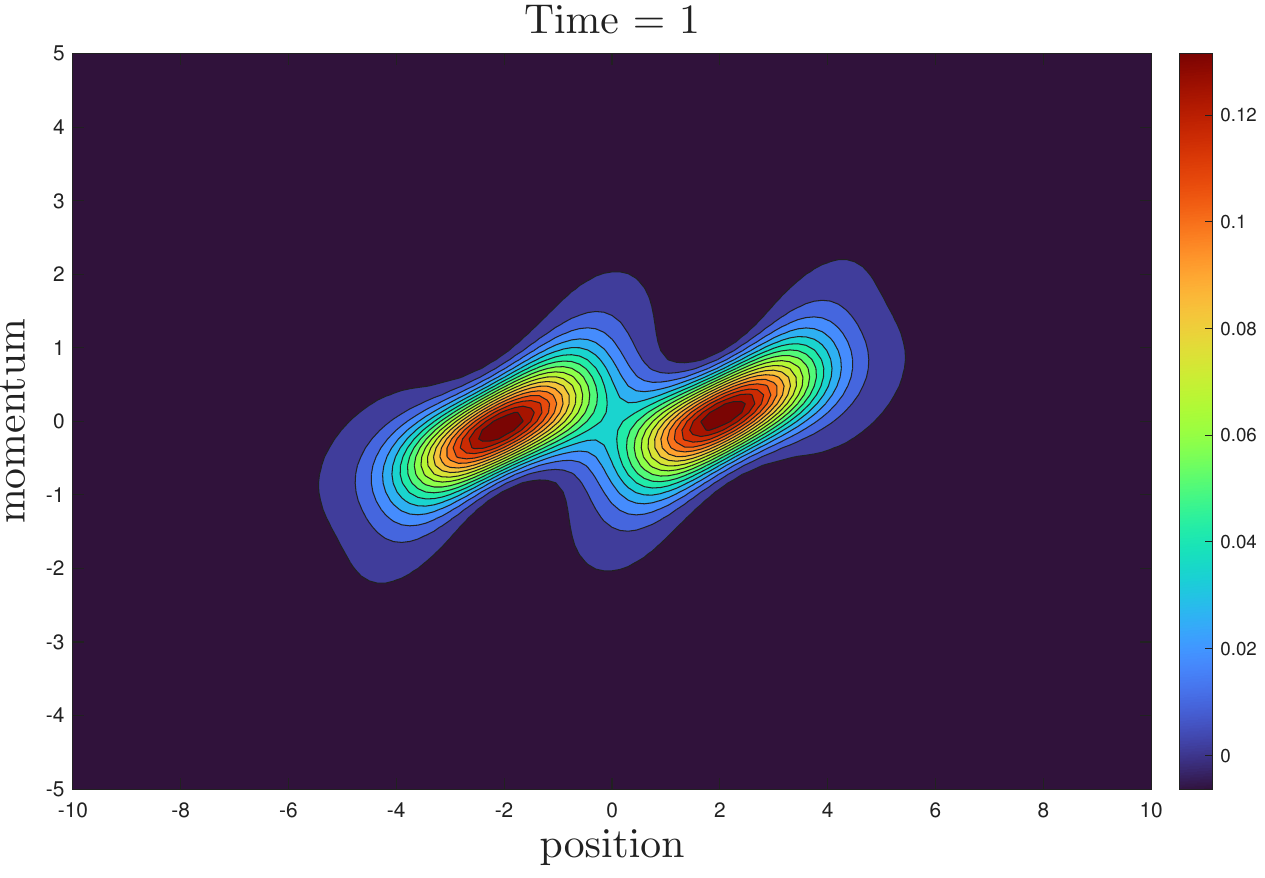}
    \includegraphics[width=1.4in,height=1.0in]{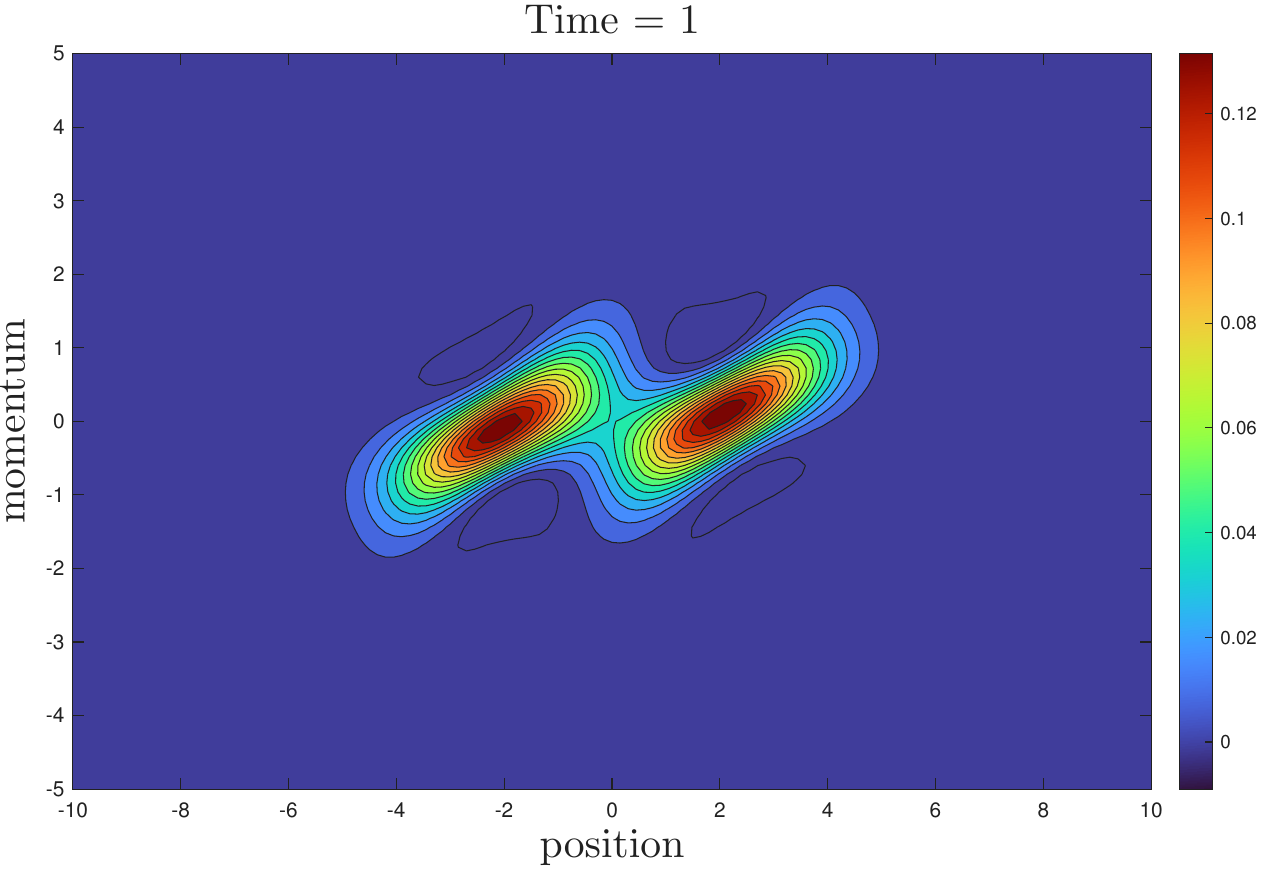}
    \includegraphics[width=1.4in,height=1.0in]{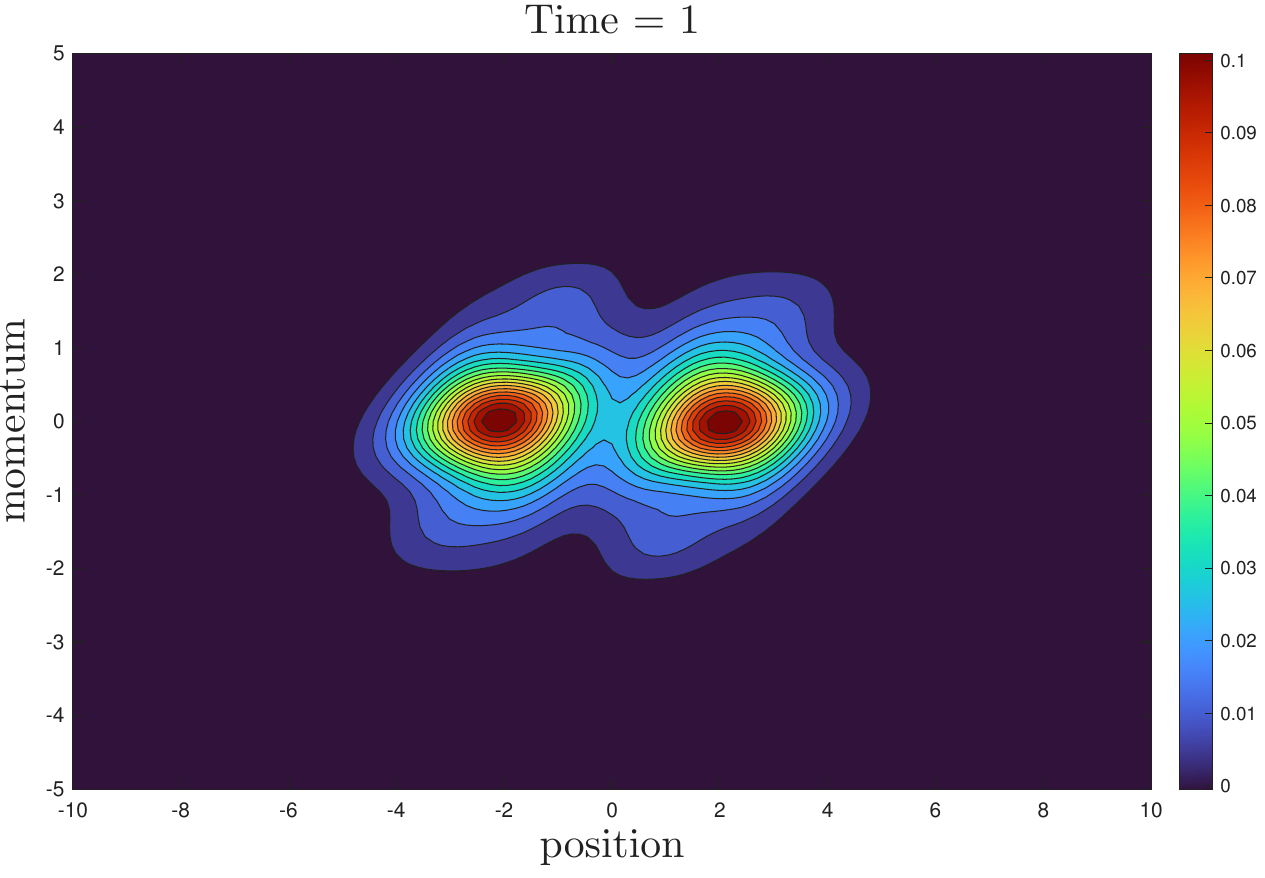}    
     \includegraphics[width=1.4in,height=1.0in]{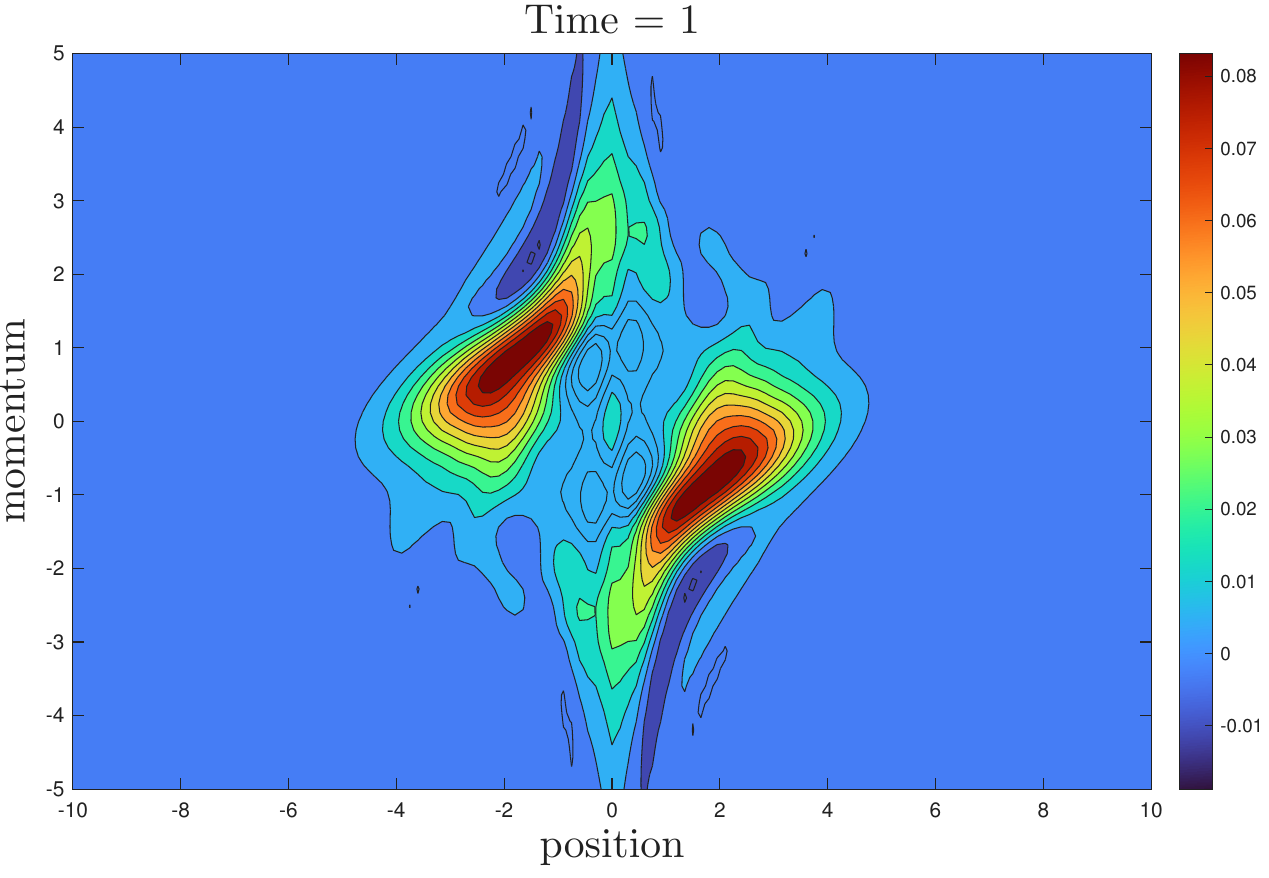}}
%     \\
%    \centering
%    \subfigure[$t=1.5$.]{
%    \includegraphics[width=1.4in,height=1.0in]{./SC_Free_redist_T15}
%    \includegraphics[width=1.4in,height=1.0in]{./SC_Hartree_redist_T15}
%    \includegraphics[width=1.4in,height=1.0in]{./SC_HXC_redist_T15}
%    \includegraphics[width=1.4in,height=1.0in]{./SC_HXC2B_redist_T15}}      
     \\
    \centering
    \subfigure[$t=2$a.u.]{
    \includegraphics[width=1.4in,height=1.0in]{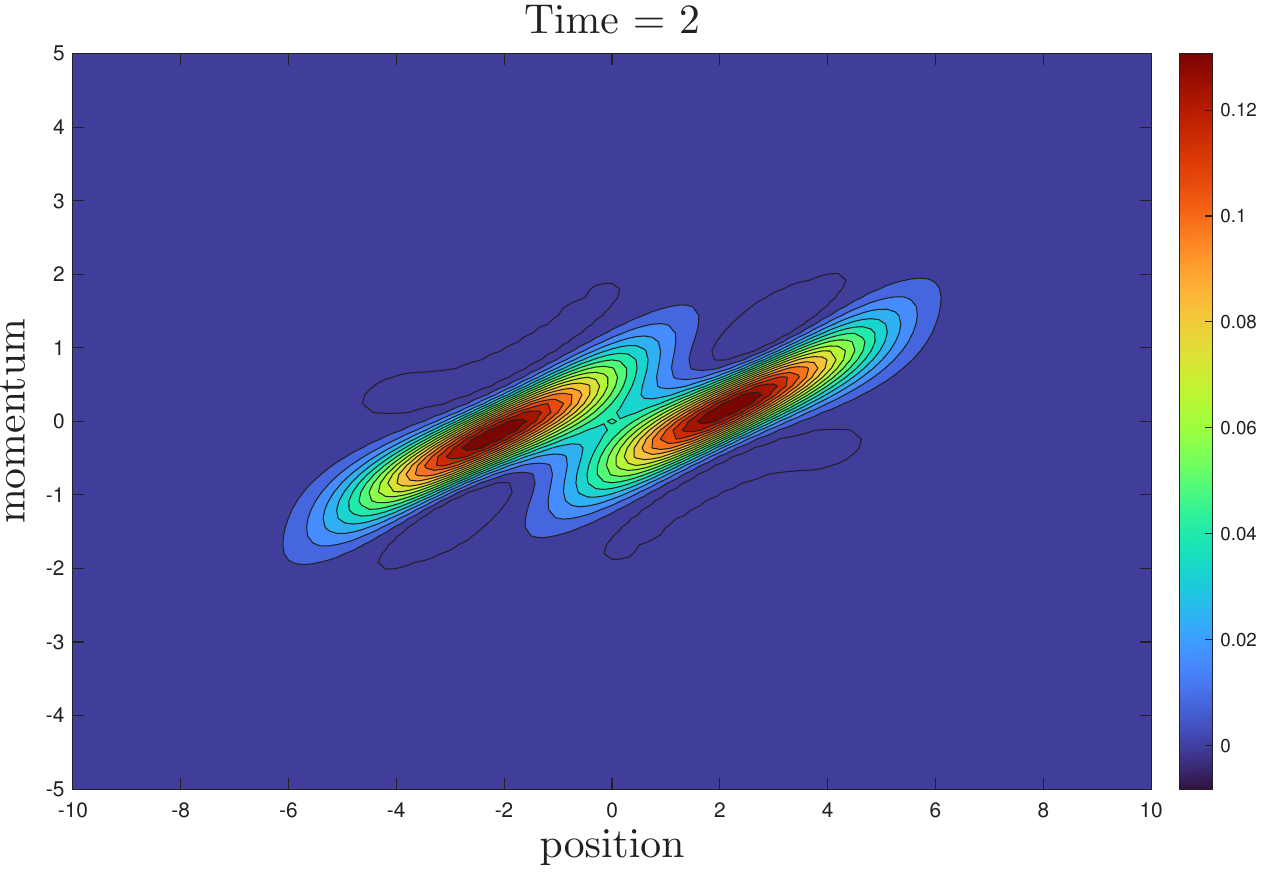}
    \includegraphics[width=1.4in,height=1.0in]{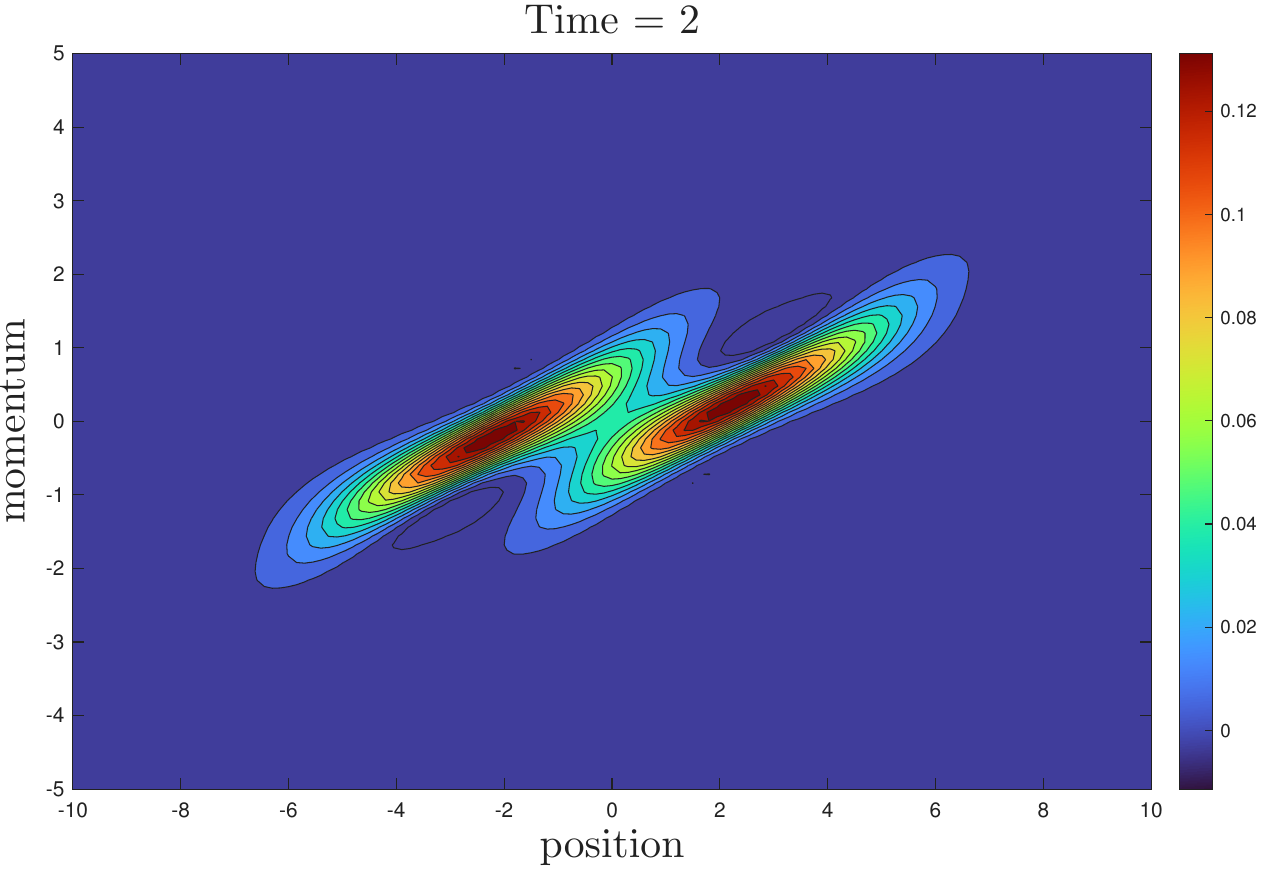}
    \includegraphics[width=1.4in,height=1.0in]{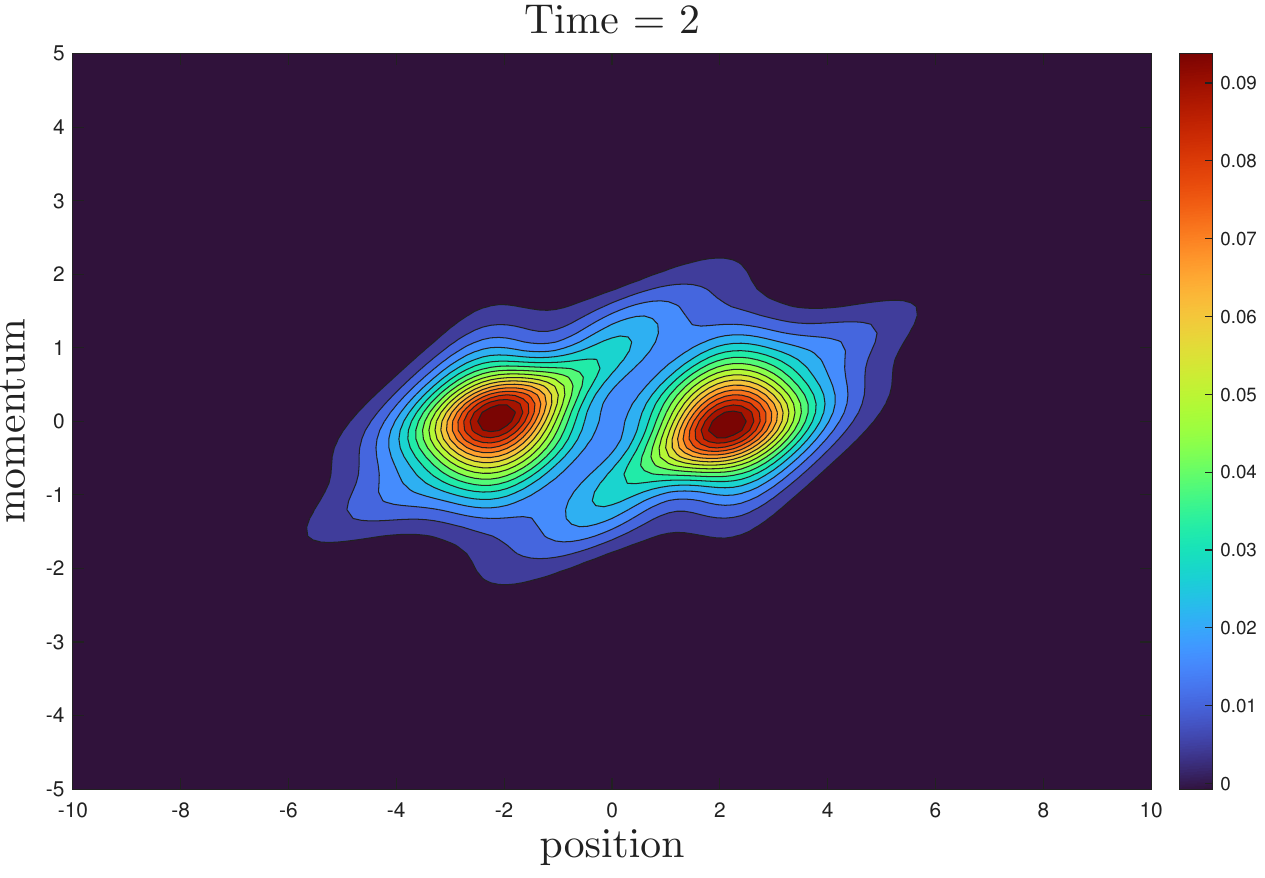}          
      \includegraphics[width=1.4in,height=1.0in]{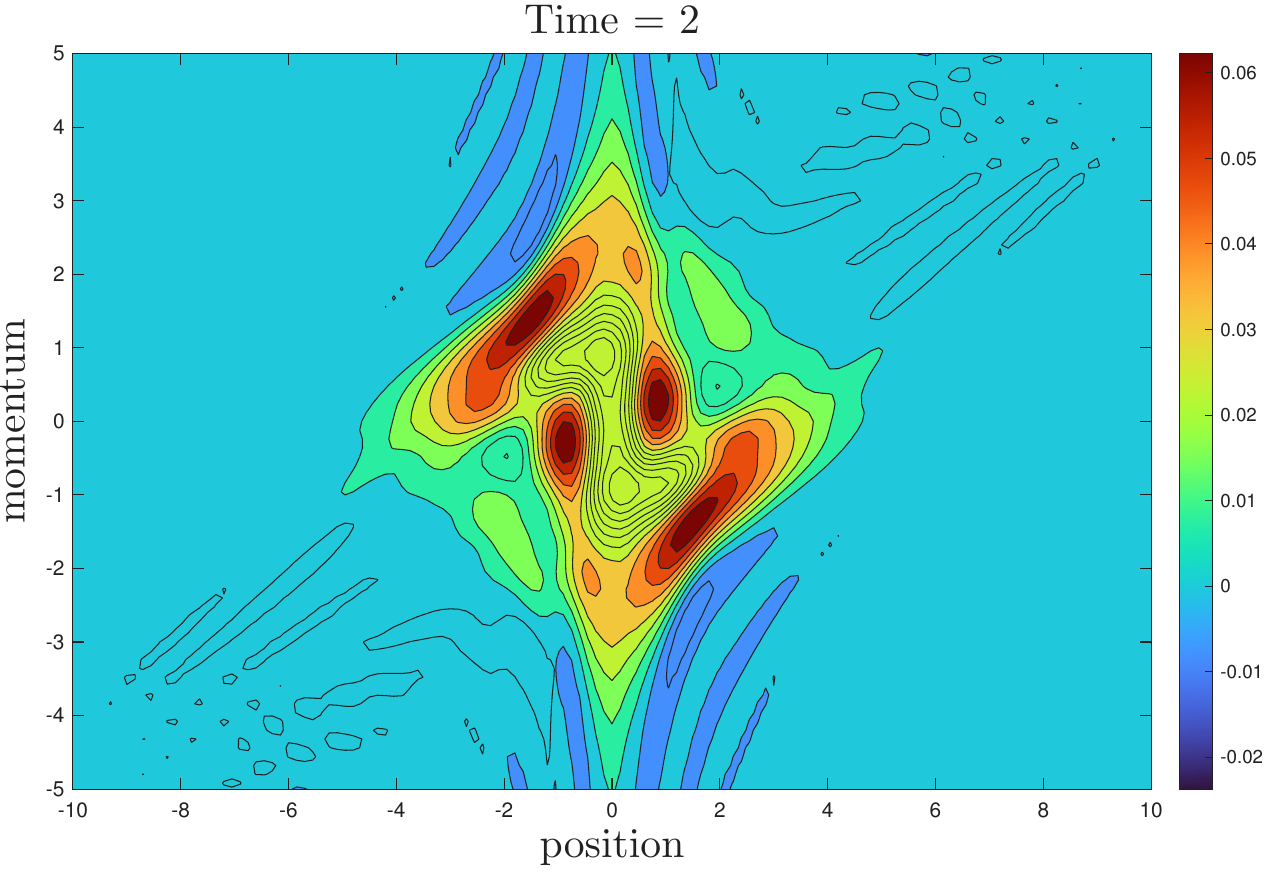}}
     \\
    \centering
    \subfigure[$t=3$a.u.]{
    \includegraphics[width=1.4in,height=1.0in]{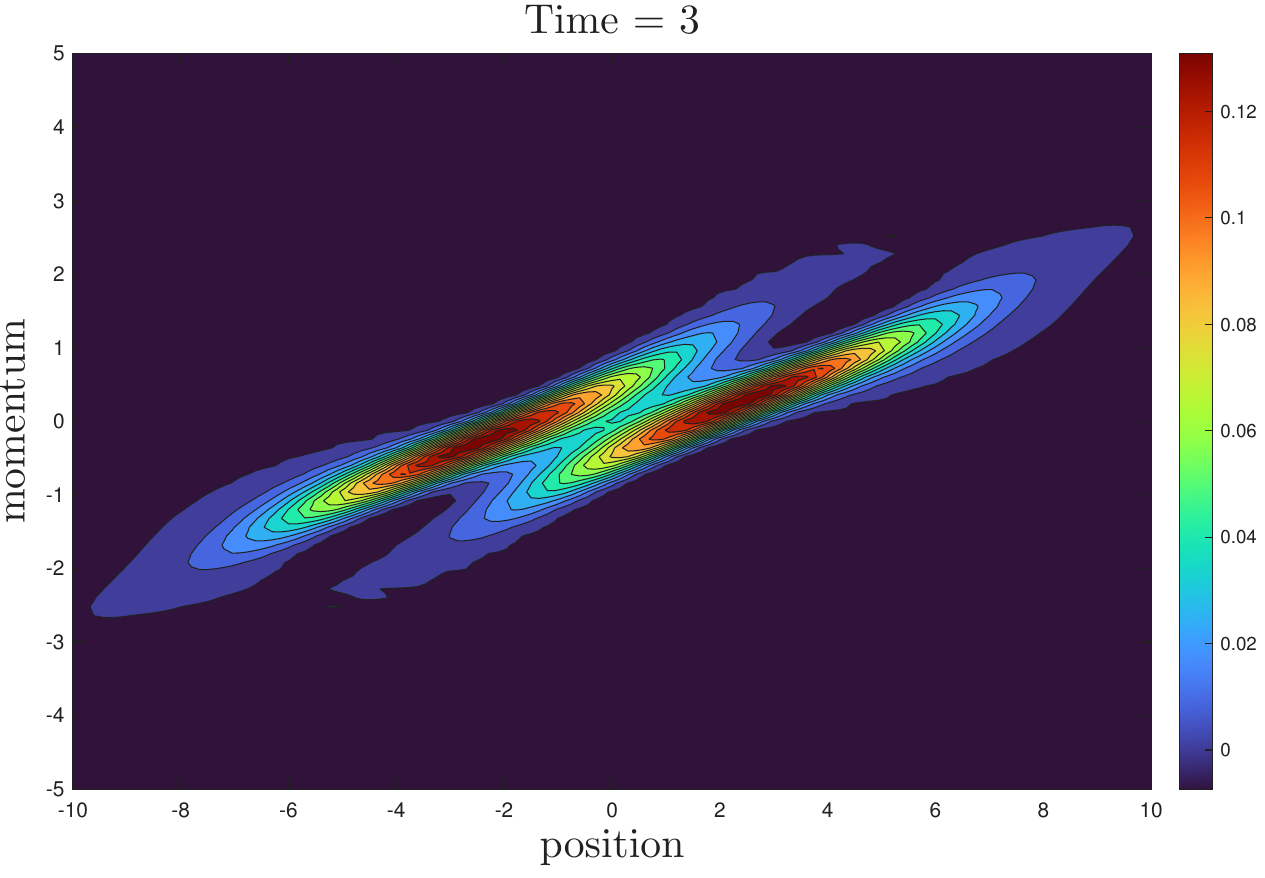}
    \includegraphics[width=1.4in,height=1.0in]{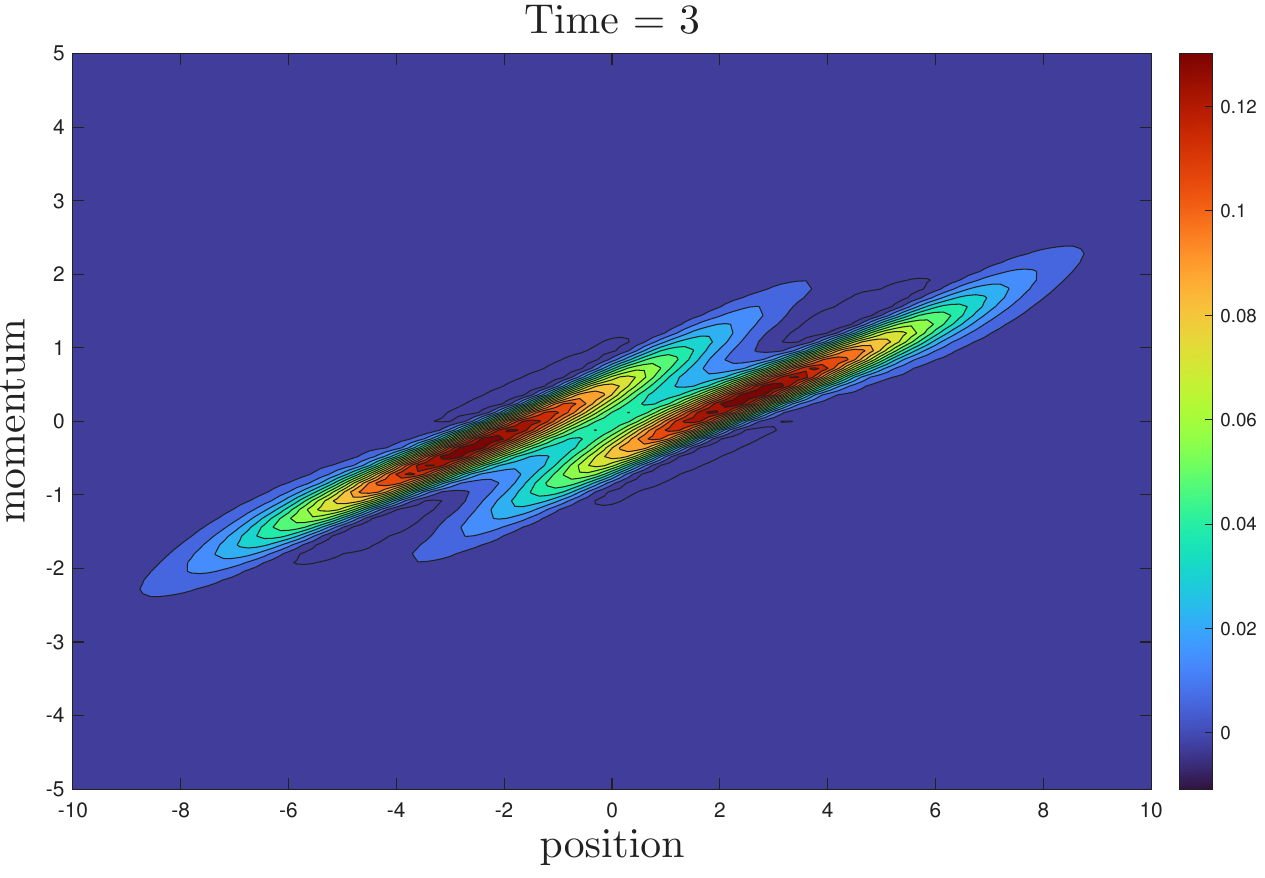}
    \includegraphics[width=1.4in,height=1.0in]{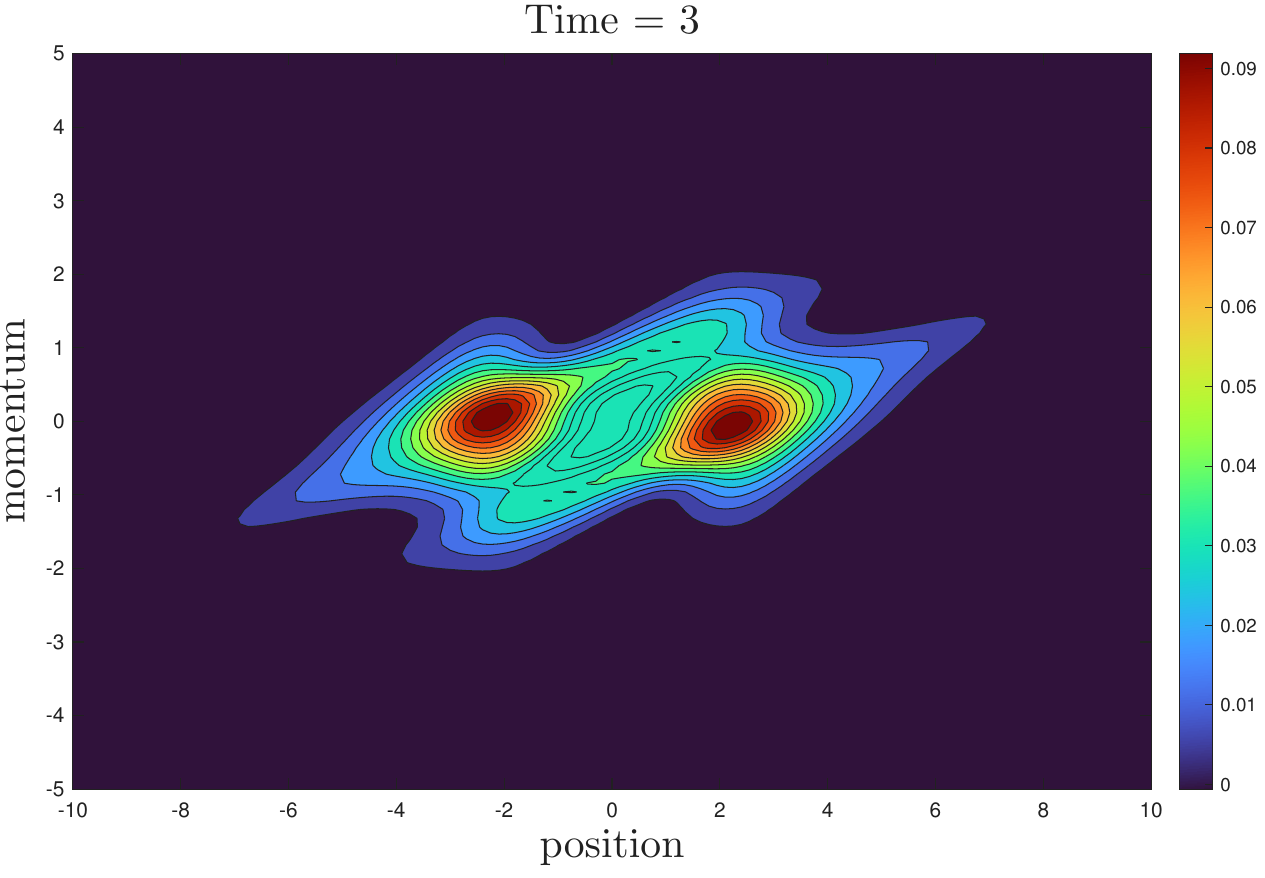}       
     \includegraphics[width=1.4in,height=1.0in]{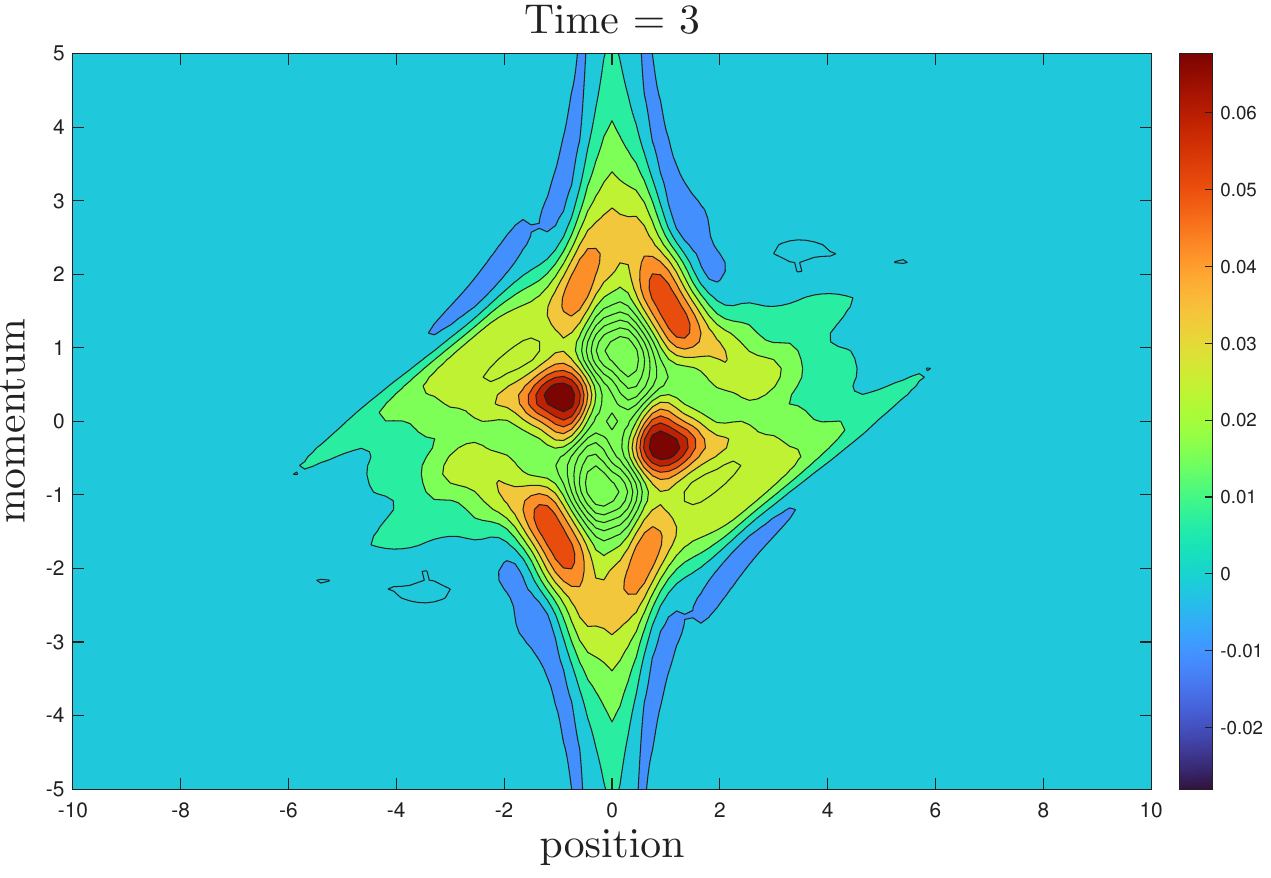}}           
     \caption{Non-periodic, strongly correlated system: The evolution of the reduced Wigner function. The cases from left side to right side correspond to the Hamiltonians (I) to (IV) in Eq.~\eqref{Hamiltonian}. Here the smoothing parameter is $\epsilon = 0.01$. \label{Wigner_correlated}}  
\end{figure}

\begin{figure}[!h]
\centering
\subfigure[The time evolution of correlation entropy (left: $\epsilon=1$, right: $\epsilon=0.01$)]{
\includegraphics[width=0.48\textwidth,height=0.27\textwidth]{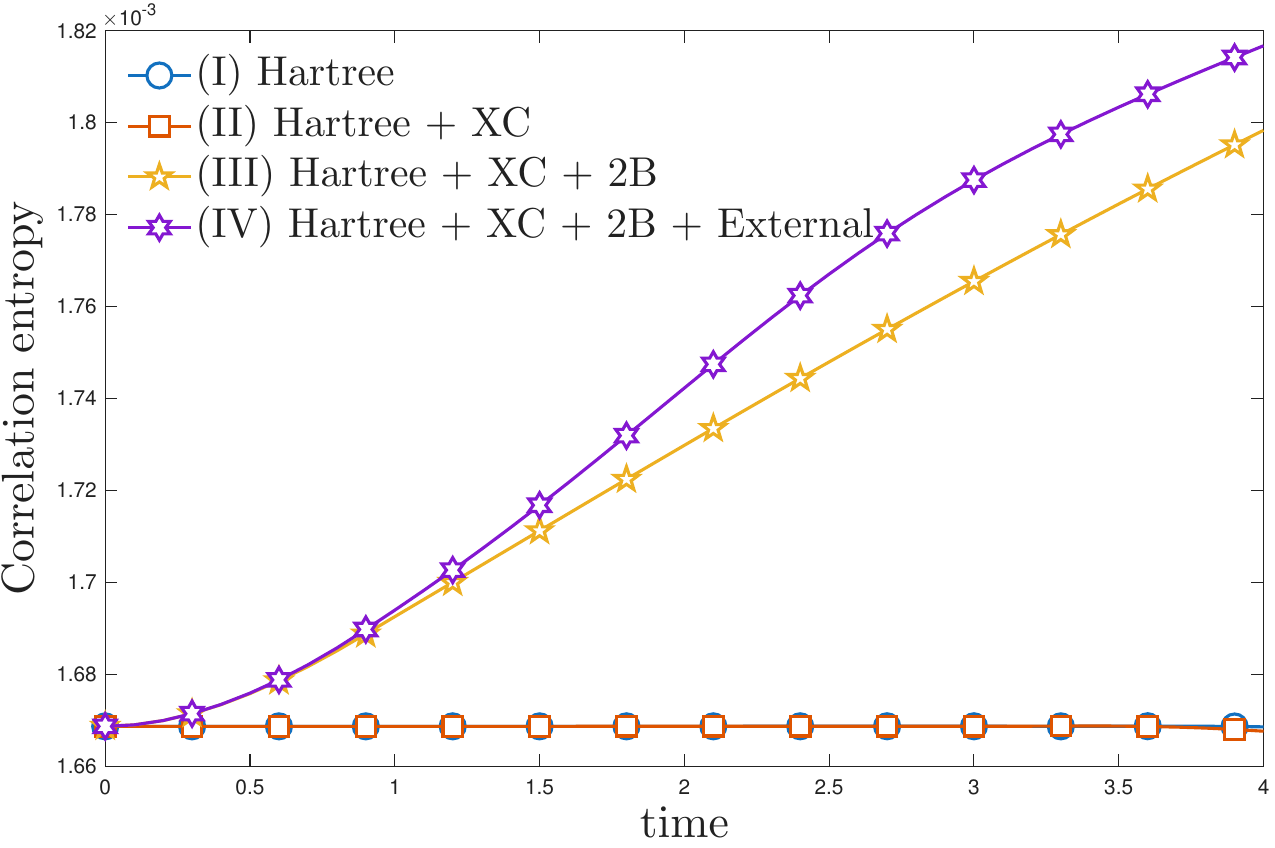}
\includegraphics[width=0.48\textwidth,height=0.27\textwidth]{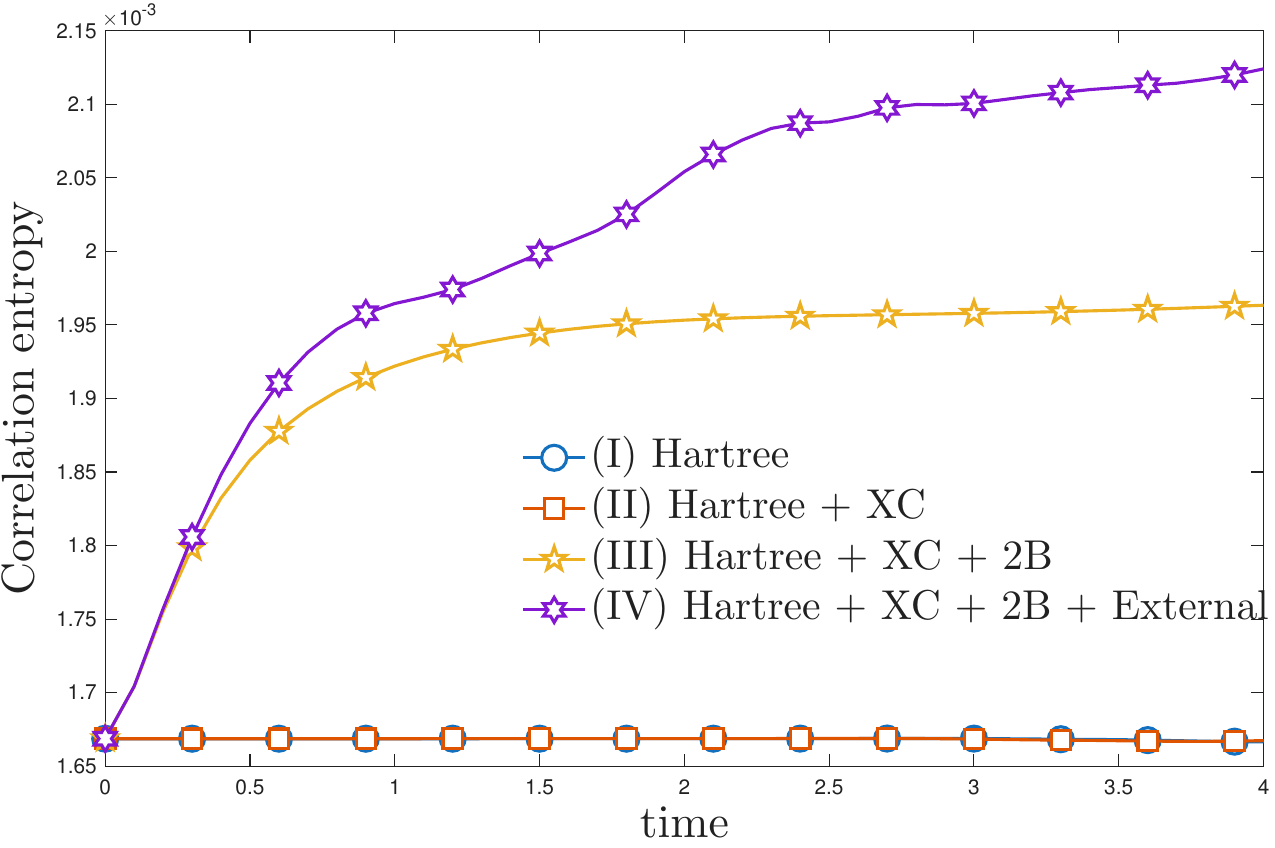}}
\centering
\subfigure[The occupation numbers at $t = 1$a.u. (left: $\epsilon=1$, right: $\epsilon=0.01$)]{
\includegraphics[width=0.48\textwidth,height=0.27\textwidth]{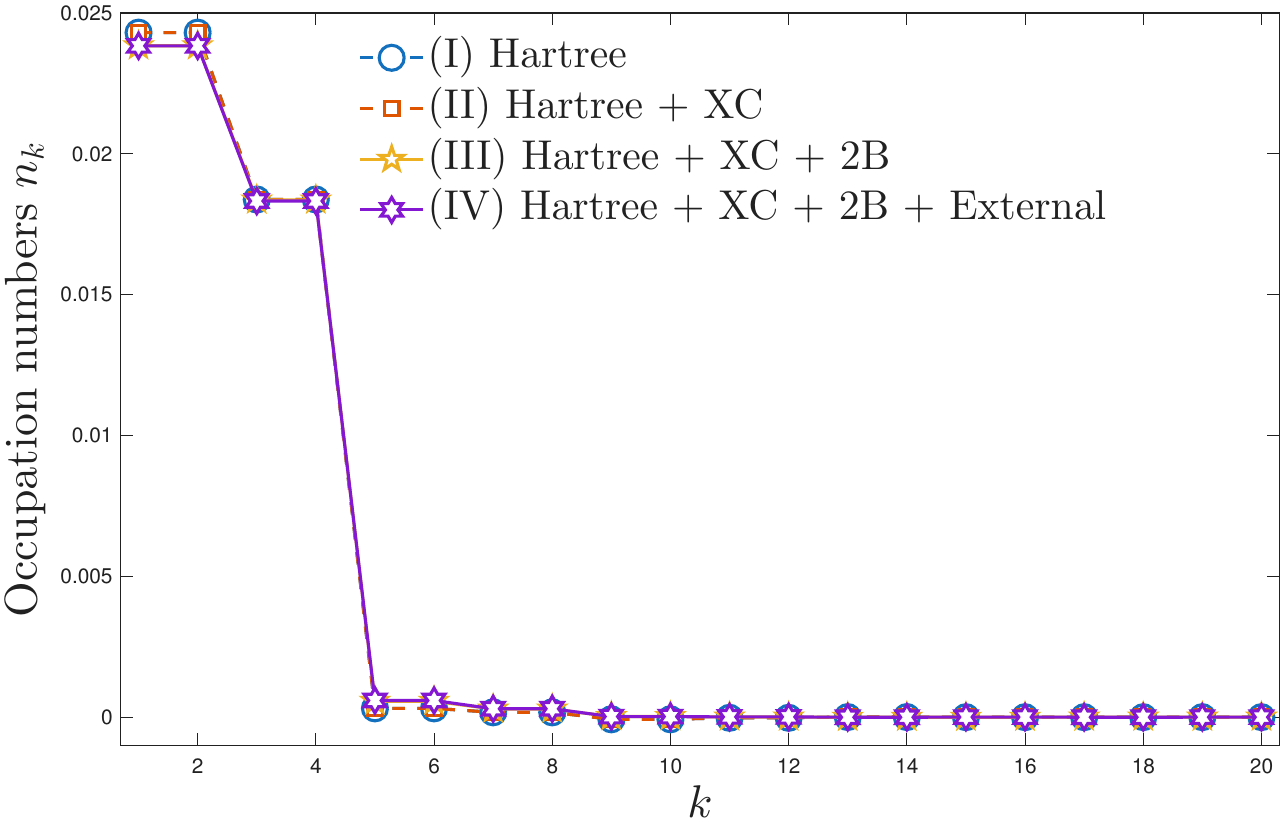}
\includegraphics[width=0.48\textwidth,height=0.27\textwidth]{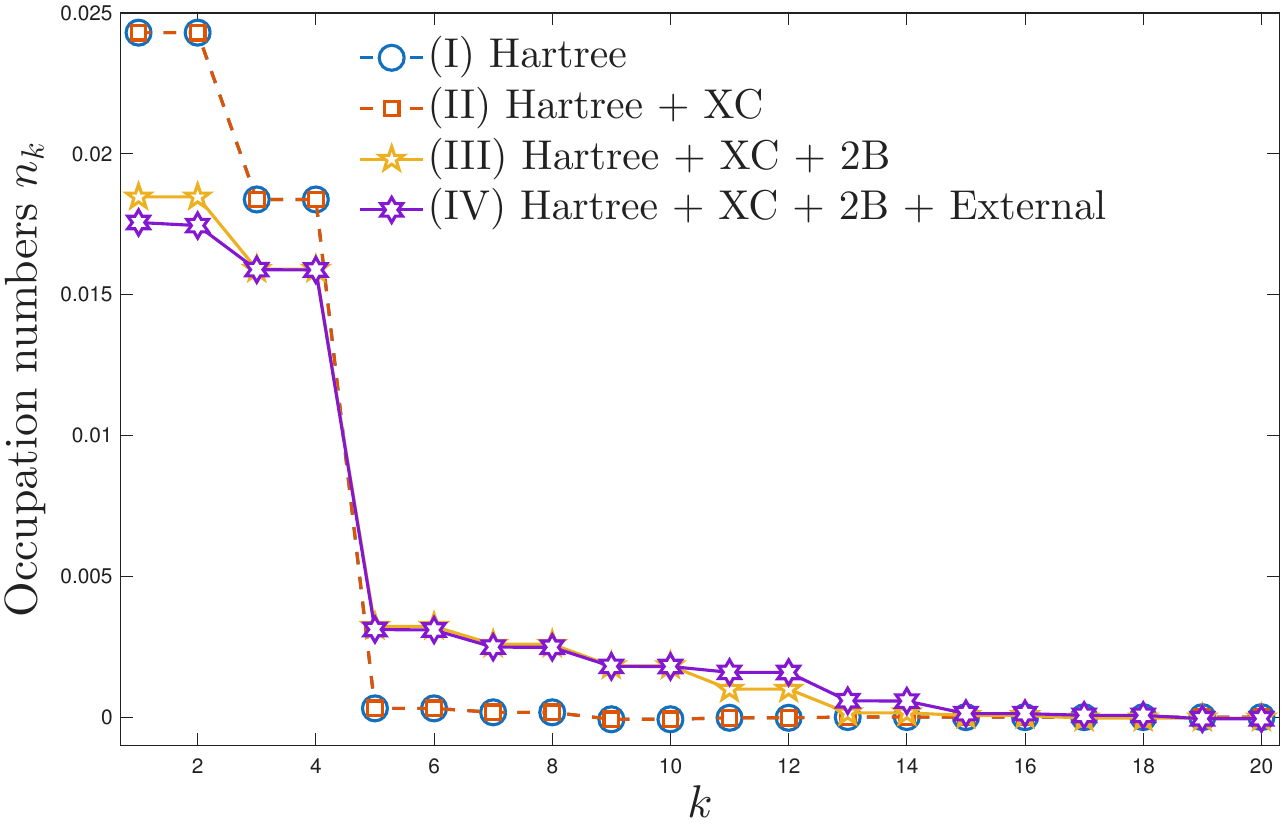}}
\centering
\subfigure[The occupation numbers at $t = 2$a.u. (left: $\epsilon=1$, right: $\epsilon=0.01$)]{
\includegraphics[width=0.48\textwidth,height=0.27\textwidth]{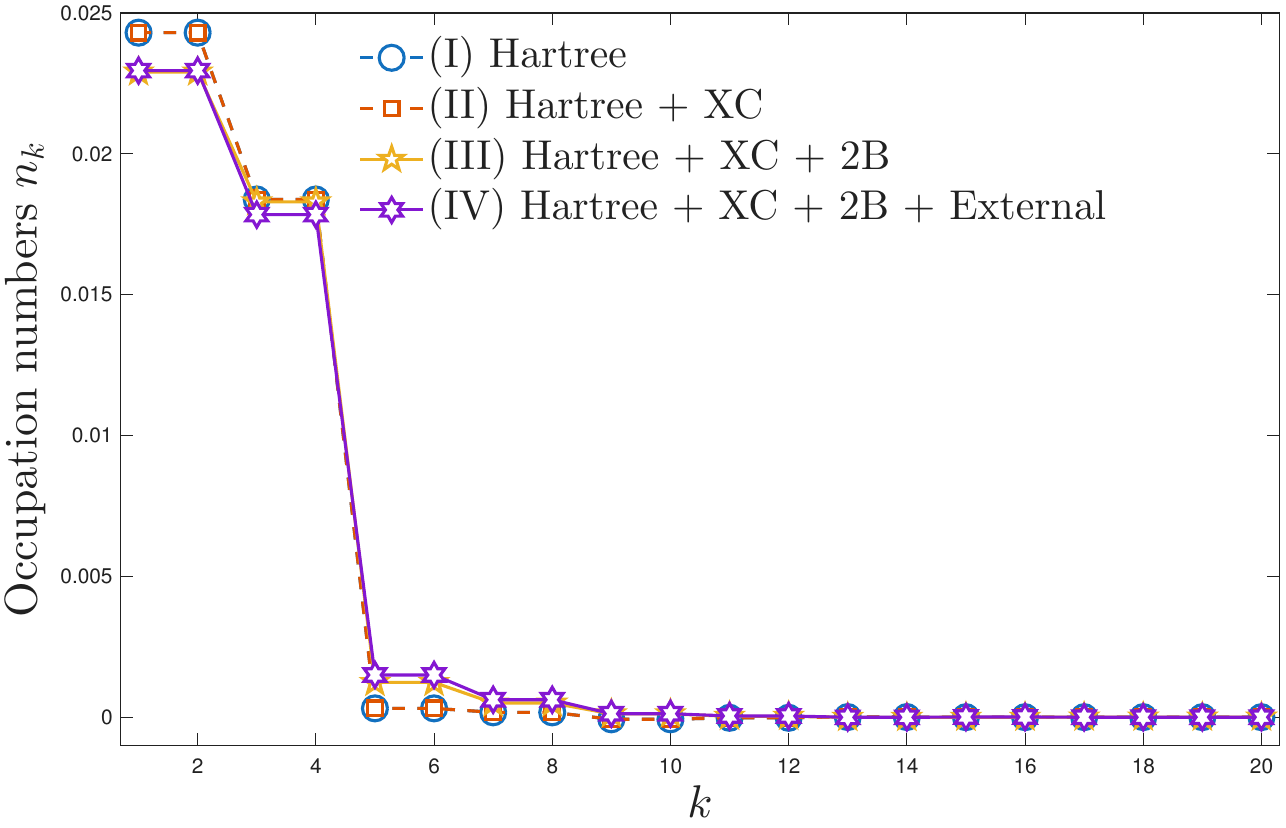}
\includegraphics[width=0.48\textwidth,height=0.27\textwidth]{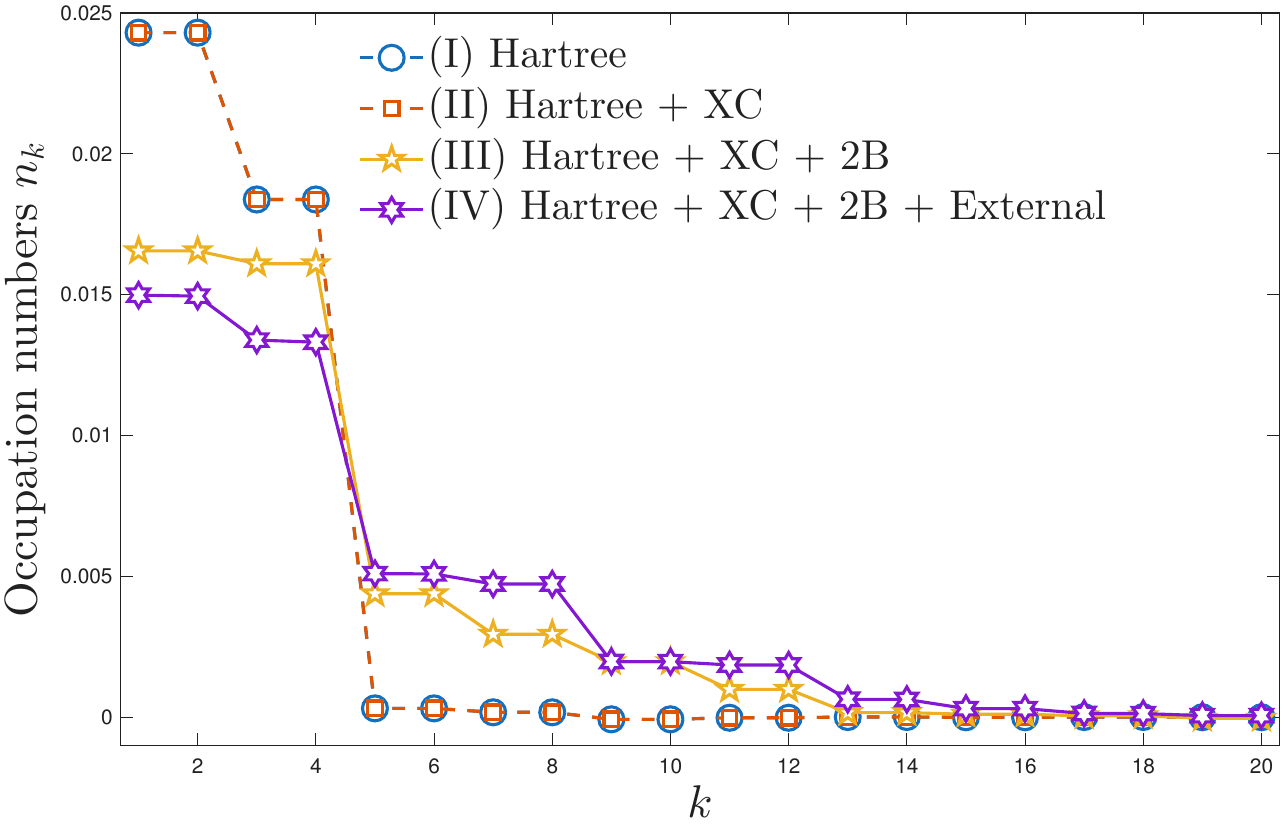}}
\centering
\subfigure[The occupation numbers at $t = 3$a.u. (left: $\epsilon=1$, right: $\epsilon=0.01$)]{
\includegraphics[width=0.48\textwidth,height=0.27\textwidth]{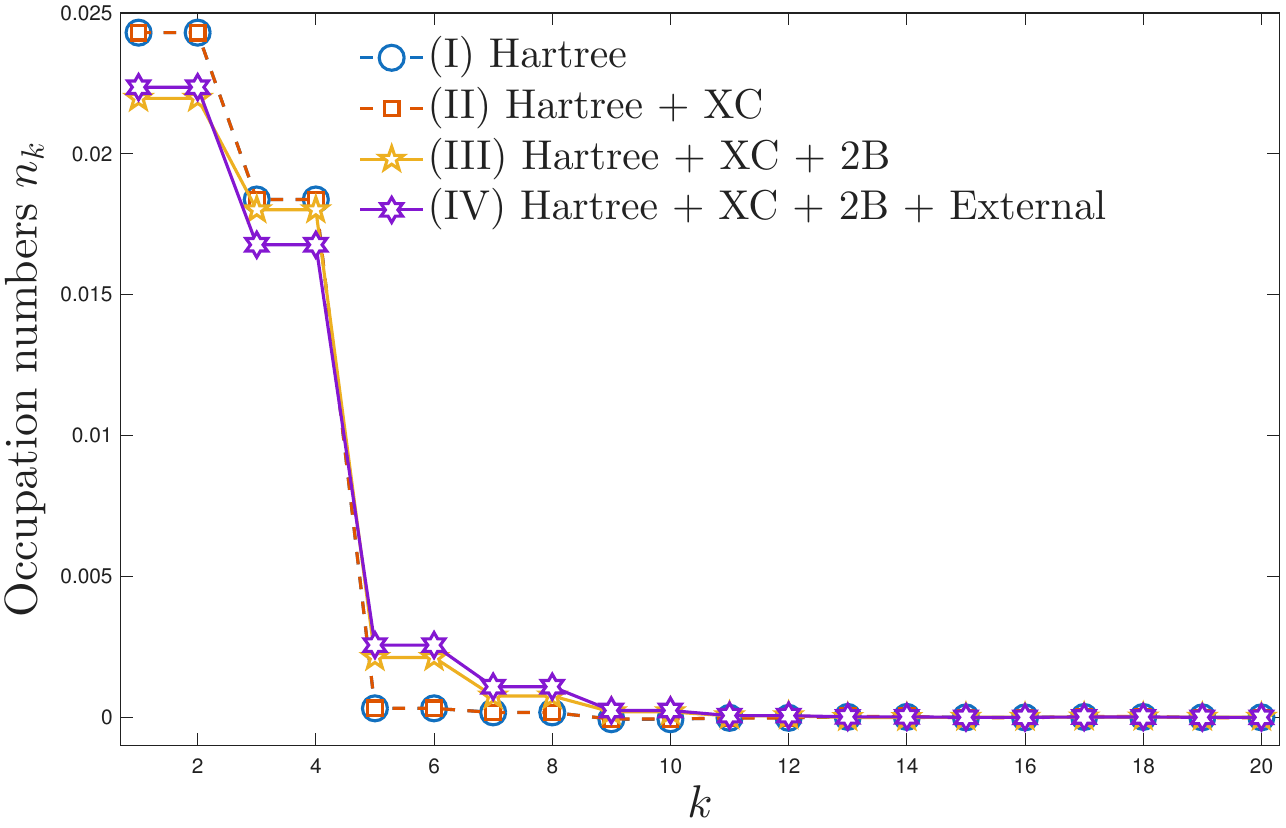}
\includegraphics[width=0.48\textwidth,height=0.27\textwidth]{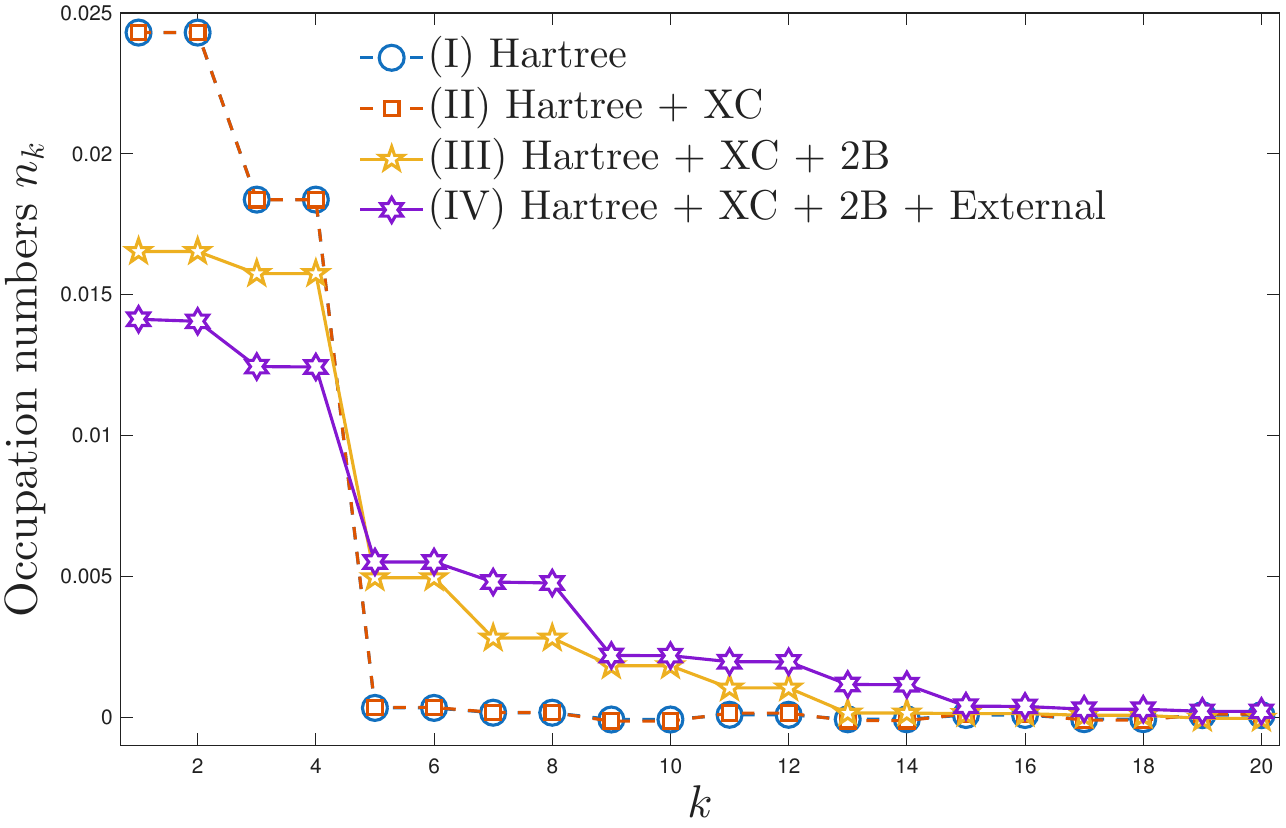}}
     \caption{Non-periodic, strongly correlated system: The evolution of the correlation entropy and the occupation numbers. It is seen the  correlation entropy is (nearly) invariant when two-body interaction is absent, while it increases when the two-body interaction is added and more orbitals are occupied. Moreover, the correlation is evidently stronger as the smoothing parameter $\epsilon$ increases. \label{correlated_entropy}} 
\end{figure}

The two-body correlations and uncertainty can be clearly visualized by the reduced Wigner function (see Figure \ref{Wigner_correlated}) and the correlation entropy (see Figure \ref{correlated_entropy}). Our main findings are summarized as follows. 
\begin{enumerate}

\item[(1)] When two-body repulsive interaction is absent, the dispersive behavior is dominated and  the Wigner function cannot approach the equilibrium like the BGK mode. 

\item[(2)]  The H-L potential only provides a correction on the uncertainty (see the negative part). 

\item[(3)]  The two-body scattering  provides a dissipative mechanism to suppress both dispersion and uncertainty. 

\item[(4)]  The external field contributes to the high oscillations and uncertainty in the phase space.

\item[(5)]  The two-body Coulomb interaction leads to an increase in the system's entropy and drives the evolution of the occupation numbers, gradually guiding the system toward its true equilibrium state, where the entropy ceases to change. In contrast, in the absence of the two-body Coulomb interaction, both the entropy and the occupation numbers remain constant.
\end{enumerate}

We believe the above results depict typical features of the complex dynamics in a strongly correlated quantum system, and have been enough to emphasize the importance of  the Coulomb repulsion.

\begin{remark}
In the above cases, we simply set $\gamma = 1$ in the potential $ \frac{\gamma}{\sqrt{|r_1 - r_2|^2 + \epsilon}}$, and the exchange-correlation effects seem negligible. In real problem, such XC term is over-simplified and the parameter $\gamma$ should be carefully tuned for a more physical interpretation \cite{VuSarma2020}, which needs to be investigated in our following work.  Nonetheless, the increase in the entropy, which is absent in the mean-field single-body approximation, has manifested the role of the two-body correction.
\end{remark}

\section{Conclusion and discussion}
\label{sec.conclusion}

This paper discusses the phase-space formulation of the extended time-dependent density functional theory (TDDFT) on multibody densities and proposes efficient numerical approximations. The main contribution is to address numerical challenges for the nonlocal pseudo-differential operators in the Wigner formalism. For spatial periodic case, a pseudo-difference operator approach is derived for both the Hartree-exchange-correlation term and two-body collision operator, while the discretization via the Chebyshev spectral element method is adopted for non-periodic case. Numerical experiments manifest the dissipation mechansim induced by the two-body interaction, which might have a non-negligible correction to the kinetic effects, the BGK equilibrium mode, dispersive wave and uncertainty in the phase space. It also provides numerical evidence that the Coulomb repulsion is much more dominant than the exchange energy. This work may pave the way for an accurate description and a practical application of many-body TDDFT.  

The grid-based scheme also provides a highly accurate reference solution for the Wigner Monte Carlo method \cite{KosinaNedjalkovSelberherr2003,MuscatoWagner2016,ShaoXiong2019}. The latter has the potential to break the curse of dimensionality and resolves the dynamics of 2-reduced density matrix in 12-dimensional phase space  for real quantum system. Combining  developments of efficient algorithms and modern computers, we believe that the direct usage of time-dependent many-body density matrix approach will be feasible in a near future, which lays the foundation for understanding the N-representability problem.

%We believe that the combination of modern computers and developments of efficient algorithms may finally 

%In the past, the numerical cost for solving 

\section*{Acknowledgement}
This research was supported by the National Key R\&D Program of China (No. 2024YFA1016100) and the National Natural Science Foundation of China (Nos.~12571413 and 12305271). The authors are thankful  to Prof. D. Wu (Dong Wu) and Dr. Tian-Xing Hu for the helpful discussions.

\bibliographystyle{elsarticle-num}
\bibliography{journalname,Bib_2024} 

\appendix

\section{Mean-field approximation to the triplet Wigner function}
\label{sec.appendix}

Here we give a form derivation of the Hartree part of 2-reduced Wigner dynamics. It starts from the $N$-body Hamiltonian with binary interaction and external potential
\begin{equation}
\hat{H}_{1,\dots N} = \sum_{i=1}^N \frac{\hat{\bp}_i^2}{2m}  + \frac{1}{2} \sum_{i=1}^N \sum_{j =1}^N  {V}_{i j}(\hat{\br}_i, \hat{\br}_j) +  W_{1,\dots N} (\hat{\br}_1, \dots, \hat{\br}_N),\quad V_{ij}(\br_i, \br_j) = V_{ee}(|\br_i - \br_j|).
\end{equation}
Under the indistinguishable assumption \eqref{def.indistinguishable}, we can reformulate Eq.~\eqref{N_body_Wigner} as 
\begin{equation}
\frac{\partial }{\partial t}f_{1, \dots, N}+  \frac{\bP}{m} \cdot \nabla_{\bR} f_{1, \dots, N}=\Theta_{V_{12}}[f_{1, \dots, N}] + (N-2) (\Theta_{V_{13}}+ \Theta_{V_{23}})[f_{1, \dots, N}] +  \Theta_{W_{1, \dots, N}}[f_{1, \dots, N}].
\end{equation}

For brevity, we use the simplest form of the external potential  composed of  one-body potentials,
\begin{equation}
\quad {W}_{1, \dots, N}({\br}_1, \dots, {\br}_N) = \sum_{i=1}^N W_i({\br}_i).
\end{equation}
By integrating the degrees of freedom of $3$ to $N$, it arrives at the dependence of the  2-reduced Wigner function $f_{12}$ on the triplet Wigner function $f_{123}$,
\begin{equation}
\begin{split}
\frac{\partial }{\partial t}f_{12}(\bR, \bP, t) +  \frac{\bP}{m} \cdot \nabla_{\bR} f_{12}&(\bR, \bP, t)= \int_{\mathbb{R}^{d}}  \D \br_3\int_{\mathbb{R}^{d}}  \D \bp_3(\Theta_{V_{13}} + \Theta_{V_{23}})[f_{123}](\bR, \br_3, \bP, \bp_3, t)  \\
&+\Theta_{V_{12}}[f_{12}](\bR, \bP, t) + \Theta_{W_1}[f_{12}](\bR, \bP, t) + \Theta_{W_2}[f_{12}](\bR, \bP, t),
\end{split}
\end{equation}
where  $(\bR, \bP) = (\br_1, \br_2, \bp_1, \bp_2)$, $\Theta_{V_{12}}[f_{12}](\bR, \bP, t)$ is given in Eq.~\eqref{two_body_scattering}, and
%\begin{equation}
%\begin{split}
% \Theta_{V_{13}}&[f_{123}](\br_1, \br_2, \br_3, \bp_1, \bp_2, \bp_3, t) = \frac{1}{\mi\hbar} \left(\frac{1}{\pi \hbar}\right)^d \int_{\mathbb{R}^d} \D \bp^{\prime}  \me^{-\frac{2\mi}{\hbar} \bp^{\prime} \cdot (\br_1 - \br_3)} \mathcal{F}_{\br \to \frac{2}{\hbar}\bp^{\prime}}{V_{ee}}(|\br|)\\
% &\times \left\{ f_{123}(\br_1, \br_2, \br_3, \bp_1 - \bp^{\prime}, \bp_2, \bp_3 + \bp^{\prime}, t) -  f_{123}(\br_1, \br_2, \br_3, \bp_1 + \bp^{\prime}, \bp_2, \bp_3-  \bp^{\prime}, t)\right\},
%\end{split}
%\end{equation}
\begin{equation*}
\begin{split}
 \Theta_{V_{13}}[f_{123}](\bR, &\br_3, \bP, \bp_3, t) = \frac{1}{\mi\hbar} \left(\frac{1}{2\pi\hbar}\right)^d \int_{\mathbb{R}^d} \D \bp_1^{\prime}  \int_{\mathbb{R}^d} \D \bp_3^{\prime} \int_{\mathbb{R}^d} \D \bri_1  \int_{\mathbb{R}^d} \D \bri_3 ~\me^{-\frac{\mi}{\hbar} \left[(\bp_1 - \bp_1^{\prime}) \cdot \bri_1 + (\bp_3 - \bp_3^{\prime}) \cdot \bri_3\right]}  \\
 & \times \left[V_{13}\left(\br_1 - \frac{\bri_1}{2}, \br_3 - \frac{\bri_3}{2}\right) - V_{13}\left(\br_1 + \frac{\bri_1}{2}, \br_3 + \frac{\bri_3}{2}\right)\right] f_{123}(\br_1, \br_2, \br_3, \bp_1^{\prime}, \bp_2, \bp_3^{\prime}, t).
\end{split}
\end{equation*}
%\begin{equation}
%\begin{split}
%f_{123}(\br_1, \br_2, \br_3, \bp_1, &\bp_2, \bp_3, t) \approx \frac{1}{3} f_{12}(\br_1, \br_2, \bp_1, \bp_2, t) f_1(\br_3, \bp_3, t) \\
%&+ \frac{1}{3} f_{12}(\br_1, \br_3, \bp_1, \bp_3, t) f_1(\br_2, \bp_2, t) + \frac{1}{3} f_{12}(\br_2, \br_3, \bp_2, \bp_3, t) f_1(\br_1, \bp_1, t). 
%\end{split}
%\end{equation}
%\begin{equation}
%f(123) \approx \frac{1}{3}f(12)f(3) + \frac{1}{3}f(13) f(2)+ \frac{1}{3} f(23)f(1)
%\end{equation}

In order to truncate the BBGKY hierarchy, we can make the following ansatz for the triplet Wigner function \cite{OlmstedCurtiss1975}
\begin{equation*}
\begin{split}
&f_{123}(\br_1, \br_2, \br_3, \bp_1, \bp_2, \bp_3, t) = f_{12}(\br_1, \br_2, \bp_1, \bp_2, t) f_1(\br_3, \bp_3, t) + f_{12}(\br_1, \br_3, \bp_1, \bp_3, t) f_1(\br_2, \bp_2, t) \\
& +  f_{12}(\br_2, \br_3, \bp_2, \bp_3, t) f_1(\br_1, \bp_1, t) - 2 f_1(\br_1, \bp_1) f_1(\br_2, \bp_2) f_1(\br_3, \bp_3) + U_{123}(\br_1, \br_2, \br_3, \bp_1, \bp_2, \bp_3, t).
\end{split}
\end{equation*}
or in the short-hand notation,
\begin{equation}
f(123) = f(12)f(3) + f(13) f(2)+f(23)f(1) - 2f(1)f(2)f(3) + U(123),
\end{equation}
so that
\begin{equation}
\begin{split}
\Theta_{V_{13}}[f(123)]  =~ &\Theta_{V_{13}}[f(12)f(3)] + \Theta_{V_{13}}[f(23)f(1) - f(1)f(2)f(3)]  \\
&+ \Theta_{V_{13}}[f(13)f(2) - f(1)f(2)f(3)] + \Theta_{V_{13}}[U(123)].
\end{split}
\end{equation}

We will show that  the first term $\Theta_{V_{13}}[f(12)f(3)]$ recovers the Hartree part of Eq.~\eqref{two_body_truncated_Wigner}. Since
\begin{equation*} 
\begin{split}
& \int_{\mathbb{R}^d} \D \bp_3 \int_{\mathbb{R}^d} \D \bp_3^{\prime} \int_{\mathbb{R}^d} \D \bri_3 ~ \me^{-\frac{\mi}{\hbar}(\bp_3 - \bp_3^{\prime}) \cdot \bri_3}  \left[V_{13}(\br_1 - \frac{\bri_1}{2}, \br_3 - \frac{\bri_3}{2}) - V_{13}(\br_1 + \frac{\bri_1}{2}, \br_3 + \frac{\bri_3}{2})\right] f_1(\br_3, \bp_3^{\prime}, t) \\
&= \left[V_{13}\left(\br_1 - \frac{\bri_1}{2}, \br_3\right) - V_{13}\left(\br_1 + \frac{\bri_1}{2}, \br_3 \right)\right] \times \int_{\mathbb{R}^d} \D \bp_3^{\prime} ~f_1(\br_3, \bp_3^{\prime}, t). 
\end{split}
\end{equation*}
 Thus it further yields that
 \begin{equation}
 \begin{split}
 \Theta_{V_{13}}[f_{123}](\bR, \br_3, \bP,& \bp_3, t) \approx \frac{1}{\mi\hbar} \left(\frac{1}{2\pi\hbar}\right)^d  \int_{\mathbb{R}^d} \D \bp_1^{\prime}   \int_{\mathbb{R}^d} \D \bri_1 ~  \me^{-\frac{\mi}{\hbar}(\bp_1 - \bp_1^{\prime}) \cdot \bri_1} f_{12}(\br_1, \br_2, \bp_1^{\prime}, \bp_2, t) \\
 & \times  \left\{\int_{\mathbb{R}^d}\D \br_3  \left[V_{13}\left(\br_1 - \frac{\bri_1}{2}, \br_3\right) - V_{13}\left(\br_1 + \frac{\bri_1}{2}, \br_3 \right)\right] n_1(\br_3, t)\right\}.
 \end{split}
 \end{equation} 
Finally, the contribution of two-body correlations $f(23) - f(2) f(3)$ and $f(13)-f(1)f(3)$ and the three-body interaction $U(123)$ should be put into the dynamical XC functional.

\section{Calculation of the correlation entropy via the Wigner function}
\label{sec.correlated}

Using the equivalence of 1-RDM and the corresponding Wigner function
\begin{equation}\label{DM_Wigner}
f_{1}(r, p, t) =  \frac{1}{2\pi}\int_{\mathbb{R}}  \rho_1(r - \frac{y}{2}, r + \frac{y}{2}, t)  \me^{\frac{\mi}{\hbar} p y } \D y,
\end{equation}
the 1-RDM can be recovered by 
\begin{equation}\label{DM_Wigner}
 \rho_1(r + \frac{\hbar y}{2}, r - \frac{\hbar y}{2}, t)   = \int_{\mathbb{R}} f_{1}(r, p, t)\me^{-\mi p \cdot y } \D y.
\end{equation}
Here $f_1(r, p, t)$ is assumed to be compactly supported in $[-L_x/2, L_x/2] \times [-L_p/2, L_p/2]$. One may choose $\Delta p = {\hbar \pi}/{L_x}$, $\Delta y = 2\Delta r/\hbar$ so that $\Delta p \Delta y = \frac{2\pi}{L_x}$. Then for $r = \mu\Delta r$, $y = \nu \Delta y$,
\begin{equation}
 \rho_1((\mu-\nu) \Delta r, (\mu+\nu) \Delta r, t) \approx \Delta y\sum_{n = -{N_x}/{2}}^{{N_x}/{2}} f_1(\mu\Delta r, n \Delta p, t)\me^{-\frac{2\pi \mi}{N_x} n \nu}.
\end{equation}
The occupation numbers $n_k(t)$ can be calculated by the eigenvalues of $\rho_1$ at grid points \cite{AppelGross2010}.

\end{document}